\documentclass[aps, prx, reprint, showpacs, superscriptaddress]{revtex4-2}
\usepackage{graphicx}
\usepackage{amsmath,amsfonts,amssymb,bm,amsthm,bbm}
\usepackage[hypertexnames=false]{hyperref}%
\usepackage{tikz}
\usepackage{placeins}
\usepackage{color}

\newcommand{\rd}{\textrm{d}}
\newcommand{\re}{\textrm{e}}
\newcommand{\inp}[2]{\langle#1|#2\rangle}
\newcommand{\ket}[1]{|#1\rangle}
\newcommand{\bra}[1]{\langle#1|}
\newcommand{\ave}[1]{\langle #1 \rangle}
\makeatletter
\newcommand*{\rom}[1]{\expandafter\@slowromancap\romannumeral #1@}
\makeatother

\begin{document}
\title{Electromagnetic responses of bilayer excitonic insulators: from exciton London equations to dipole and inverse dipole Hall effects}

\author{Yuelin Shao}
\affiliation{Department of Physics, The Hongkong University of Science and Technology,
Clear Water Bay, Kowloon 999077, Hong Kong, China}
\affiliation{Donostia International Physics Center (DIPC), Paseo Manuel de Lardizábal. 20018, San Sebastián, Spain}
\author{Hao Shi}
\affiliation{Department of Physics, The Hongkong University of Science and Technology,
Clear Water Bay, Kowloon 999077, Hong Kong, China}
\author{Xi Dai}
\email{daix@ust.hk}
\affiliation{Department of Physics, The Hongkong University of Science and Technology,
Clear Water Bay, Kowloon 999077, Hong Kong, China}


\begin{abstract}

We develop a microscopic theory of the linear electromagnetic response of bilayer excitonic insulators relevant to electron-hole double-layer systems.
Using a self-consistent Hartree-Fock description of the excitonic ground state and time-dependent Hartree-Fock for its dynamics, we compute the collective mode spectrum and the full first-order response to layer-symmetric (charge) and layer-antisymmetric (exciton) gauge fields. 
At zero magnetic field, we find that two gapped plasmon modes dominate the long-wavelength charge response, while the exciton channel is governed by a linearly dispersing phase (Goldstone) mode. 
From the Goldstone-dominated kernel, we derive London-like equations for the exciton condensate. The nondissipative acceleration under a layer-antisymmetric electric field provides a direct signature of exciton superfluidity; in contrast, a normal exciton fluid shows a Drude-like dissipative response.
In a perpendicular magnetic field, the Goldstone mode develops a magnetic-roton minimum that signals an instability toward a finite-momentum stripe-ordered excitonic insulator.
In addition, the field couples charge and exciton motion, giving rise to dipole and inverse dipole Hall effects in which a charge (exciton) bias induces a transverse exciton (charge) current.
In the condensate, these mixed Hall conductances approach finite values in the $\omega\to0$ linear-response limit.
Our findings provide concrete targets for microwave and transport probes of bilayer exciton superfluidity.

\end{abstract}
\maketitle

\section{Introduction}

The excitonic insulator (EI) is a correlated state in which bound electron-hole pairs (excitons) condense at low temperatures and spontaneously break the electron-hole $\mathrm{U}_{eh}(1)$ symmetry\cite{mottTransitionMetallicState1961,Jerome1967,kohnExcitonicPhases1967}.
As a Bose-Einstein condensate of charge-neutral excitons, the EI phase is expected to host a gapless Goldstone mode and exciton superfluidity\cite{littlewoodModelsCoherentExciton2004,balatskyDipolarSuperfluidityElectronHole2004,suHowMakeBilayer2008}.
Two-dimensional electron-hole bilayers provide an ideal platform for realizing EI physics, because spatially separated electrons and holes can form long-lived interlayer excitons while remaining individually addressable by layer-resolved probes\cite{zhuExcitonCondensateSemiconductor1995, foglerHightemperatureSuperfluidityIndirect2014a, wuTheoryTwodimensionalSpatially2015, zengElectricallyControlledTwodimensional2020a}.
In recent years, compelling experimental signatures of EI behavior have been reported in various bilayer systems, including semiconductor quantum wells \cite{duEvidenceTopologicalExcitonic2017, wuElectricallyTuningManybody2019, wuResistiveSignatureExcitonic2019} and transition metal dichalcogenide (TMD) double-layer structures \cite{wangEvidenceHightemperatureExciton2019a, maStronglyCorrelatedExcitonic2021, qiThermodynamicBehaviorCorrelated2023a, qiPerfectCoulombDrag2025, nguyenPerfectCoulombDrag2025, cutshallImagingInterlayerExciton2025}.

While the ground-state properties of bilayer EIs have been extensively studied at the mean-field level, the full structure of their collective modes and electromagnetic (EM) response functions is still not fully understood. 
Most previous works focused either on simplified phase-only actions\cite{kumarUnconventionalSuperconductivityMediated2024,panigrahiNonFermiLiquidsSubsystem2025}, or on amplitude/phase modes in lattice EI models\cite{golezNonlinearSpectroscopyCollective2020,murakamiCollectiveModesExcitonic2020,kanekoNewEraExcitonic2025}. 
Some also studied the Higgs-like\cite{xueHiggslikeModesTwodimensional2020} and Bardasis-Schrieffer-like modes in bilayer EIs \cite{sunBardasisSchriefferPolaritonsExcitonic2020,shaoElectricalBreakdownExcitonic2024}, but did not fully analyze their contributions to EM responses.
In particular, a bilayer EI supports two distinct combinations of gauge fields: a layer-symmetric gauge field that couples to the total charge, and a layer-antisymmetric gauge field that couples to the exciton degree of freedom. 
A systematic, microscopic treatment of how collective modes respond to various external driving potentials in these two channels-both at zero and finite perpendicular magnetic field-remains lacking.

In this work, we address these problems by studying the full first-order EM response of a two-dimensional electron-hole bilayer EI using the time-dependent Hartree-Fock (TDHF) formalism. 
We treat the EI ground state within self-consistent Hartree-Fock and then linearize the dynamics of the single-particle density matrix around this state. 
The TDHF approach naturally resums an infinite series of ladder-bubble diagrams, preserves charge conservation via Ward identities\cite{dassarmaCommentsTimedependentHartreeFock1983}, and yields a spectrum of collective modes together with their coupling vertices to external fields. 
This allows us to compute, on equal footing, the response to both layer-symmetric and layer-antisymmetric gauge fields, and to identify which modes dominate various experimentally relevant response functions.

Our main findings are summarized as follows:

1. \textit{Collective mode structure at zero magnetic field.}
We identify three types of collective modes that couple strongly to EM fields:
(i) a pair of gapped plasmon modes (dipole mode) that couple to the layer-symmetric gauge field and describe in-phase charge-density/current oscillations in the two layers, which is similar to the Bardasis-Schrieffer mode in superconductors\cite{bardasisExcitonsPlasmonsSuperconductors1961};
(ii) a gapless Goldstone mode and a gapped amplitude mode (both are monopole modes) that couple to the layer-antisymmetric gauge field and describe, respectively, phase and amplitude fluctuations of the EI order parameter;
(iii) a quadrupole mode that contributes to transverse exciton responses.
In the long-wavelength limit, the EM response is dominated by the dipole plasmon modes and the Goldstone mode.

2. \textit{Exciton London-like equations.}
By analyzing the response to a layer-antisymmetric gauge field, we re-derive a set of London-like equations describing the low-energy dynamics of the exciton superfluid
\begin{subequations}
  \begin{align}
  \partial_{t}\bm{j}^{-}_{L}(t)=&\frac{e^2n_X}{m_X}[-\nabla \phi^{-}(t)-\partial_t \bm{A}^{-}_{L}(t)],\\
  \bm{j}^{-}_{T}(t)=&-\frac{e^2 n_X}{m_X}\bm{A}^{-}_{T}(t).
\end{align}
\end{subequations}
where $\bm{j}^{-}$ is the exciton current, $n_X$ is the exciton density, the symbols in the subscripts L and T represent the longitudinal (curl-less) and transverse (divergence-less) parts of the fields, and $m_X$ is the exciton mass.
The first equation expresses a non-dissipative acceleration of excitons by an ``exciton electric field'' similar to the zero resistance in a superconductor. As will be explained in detail below, such a non-dissipative behavior described in the first equation is tied to an undamped Goldstone pole in the exciton density-density correlations; in contrast, a normal (non-condensed) exciton fluid exhibits a damped sound mode and a Drude-like, dissipative response.
The second equation describes the exciton Meissner effect under an ``exciton gauge field'', which, as first pointed out by P. Littlewood and his co-workers \cite{balatskyDipolarSuperfluidityElectronHole2004} is equivalent to the in-plane magnetic field. 
We also discuss how this effect can be measured experimentally in a bilayer EI.

3. \textit{Magnetic roton and Hall-like responses.}
In the presence of a perpendicular magnetic field $B$, the Goldstone mode develops a magnetic roton minimum at finite momentum due to the interplay between Landau quantization and interlayer Coulomb interactions, leading to an instability towards a finite-momentum stripe EI at strong fields. 
Furthermore, the magnetic field couples charge and exciton motion and generates mixed charge-exciton Hall responses.
We identify two novel effects: the dipole Hall effect, where a charge voltage induces a transverse exciton current, and the inverse dipole Hall effect, where an exciton voltage induces a transverse charge current.
The dipole Hall effect has also been discussed in twisted MoTe$_2$\cite{fanIntrinsicDipoleHall2024}.
Here we show that the inverse effect can also occur in bilayer EIs and that, in the condensate, both conductances approach finite values in the $\omega\to0$ linear-response limit.

4. \textit{Experimental consequences.}
We propose several experimental probes to detect the exciton London equations that directly distinguish the EI phase from a normal exciton fluid:
(i) The waveguide transmission experiment, where the undamped Goldstone mode can support a transverse magnetic (TM) waveguide mode that propagates without attenuation along the bilayer EI at very low frequency, whereas it is strongly damped in a normal exciton fluid with a finite decay length;
(ii) Microwave impedance microscopy (MIM), where the local admittance acquires a real part dominated by the Goldstone mode, producing a characteristic $\omega^3$ frequency dependence that is qualitatively different from the normal-fluid case;
(iii) Low-frequency inverse dipole Hall measurement. In the EI phase, the inverse dipole Hall conductance approaches a finite value in the $\omega\to0$ linear-response limit and can be viewed as a manifestation of exciton superfluidity in a magnetic field.
We propose probing this low-frequency coefficient in a Corbino-disk geometry by applying a weak radial exciton voltage and detecting the magnetic-flux change generated by the induced azimuthal charge current.

The remainder of this paper is organized as follows. 
In Sec. \ref{sec:model} we introduce the continuum bilayer model and the TDHF formalism for computing collective modes and EM response functions. 
In Sec. \ref{sec:zero-B} we analyze the collective-mode spectrum and EM responses at zero magnetic field, derive the London-like equations for the exciton superfluid, and discuss experimental signatures. 
In Sec. \ref{sec:finite-B} we extend the analysis to finite perpendicular magnetic field, uncovering a magnetic roton minimum in the Goldstone mode and characterizing the dipole and inverse dipole Hall effects. 
Section \ref{sec:conclusion} summarizes our main results and outlines several directions for future work. 
Technical details of the TDHF formalism, electromagnetics of the bilayer system, and response-function calculations are collected in the Supplemental Material (SM)\cite{SM}.

\section{Model and method}\label{sec:model}

\begin{figure}
  \centering
  \includegraphics[width=\linewidth]{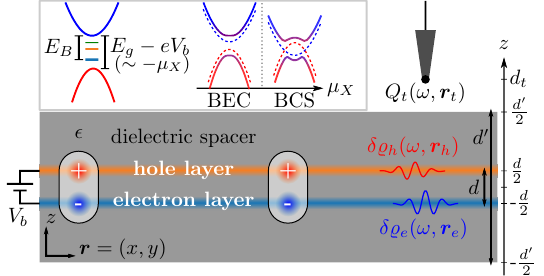}
  \caption{Setup of the bilayer EI. 
  The electron and hole layers are encapsulated in a dielectric environment with dielectric constant $\epsilon$. 
  A dielectric spacer is inserted between the two layers to suppress direct interlayer tunneling. 
  The interlayer band gap can be tuned by the bias voltage $V_b$.
  }
  \label{fig:bilayer_EI}
\end{figure}

The experimental setup of the electron-hole bilayer is illustrated in Fig.~\ref{fig:bilayer_EI}.
Within the effective-mass approximation, the conduction band in the electron layer and the valence band in the hole layer are both described by parabolic dispersions with effective masses $m_{e/h}$, separated by a band gap $E_g$.
Due to the weak screening in two dimensions, electron-hole pairs form a series of interlayer exciton levels below the band gap.
By inserting a dielectric barrier between the electron and hole layers, direct interlayer tunneling is exponentially suppressed, which substantially increases the exciton lifetime.
We denote the binding energy of the lowest interlayer exciton by $E_B$.
Typically, $E_B < E_g$, so that excitons are excited states in the absence of an external bias.
To inject excitons into the bilayer system, an interlayer bias potential $V_b$ is usually applied.
When interlayer tunneling is appreciable, the tunneling current drives the system into a nonequilibrium steady state~\cite{sunSecondOrderJosephsonEffect2021,sunDynamicalExcitonCondensates2024,zengKeldyshFieldTheory2024}.
However, when tunneling is negligible, the bias potential simply reduces the effective band gap to $E_g - eV_b$, and plays the role of an exciton chemical potential,
\begin{equation}
  \mu_X \equiv eV_b - E_g.
\end{equation}

Under this setup, the non-interacting Hamiltonian can be written in the layer basis $s=e,h$ as
\begin{equation}
  h^0_{\bm{k}}
  = \left(\frac{\hbar^2 k^2}{4m} - \frac{\mu_X}{2}\right)\sigma_z
    + \frac{\hbar^2 k^2\delta_m}{4m}\,\sigma_0,
\end{equation}
where $m \equiv m_e m_h/(m_e+m_h)$ is the reduced mass and 
$\delta_m \equiv (m_h-m_e)/(m_h+m_e)$ parametrizes the electron--hole mass asymmetry. 
The interaction Hamiltonian is
\begin{equation}
  \hat{H}_I
  = \frac{1}{2\mathcal{V}}
    \sum_{s,s'=e,h}
    \sum_{\bm{k}\bm{k}'\bm{q}}
    V_{ss'}(\bm{q})\,
    c^{\dagger}_{s\bm{k}}c^{\dagger}_{s'\bm{k}'}
    c_{s'\bm{k}'+\bm{q}}c_{s\bm{k}-\bm{q}},
\end{equation}
where $\mathcal{V}$ is the sample area and 
$[c^{\dagger}_{e\bm{k}},c^{\dagger}_{h\bm{k}}]$ are the electron creation operators in the two layers.
We take the interactions to be Coulomb,
\begin{equation}
  V_{s=s'}(\bm{q}) = V(\bm{q})=\frac{2\pi e^2}{\epsilon q},\;
  V_{s\neq s'}(\bm{q})=U(\bm{q}) = V(\bm{q})\re^{ - q d },
\end{equation}
where $\epsilon$ is the effective dielectric constant of the environment and $d$ is the interlayer distance.

In the presence of translational symmetry, the coupling Hamiltonian between an external field $f(t)$ and an operator $\hat{O}$ can be expressed in momentum space as
\begin{equation}
  \hat{H}_c = \frac{1}{\mathcal{V}}\sum_{\bm{q}} f(t,\bm{q})\,\hat{O}(-\bm{q}).
\end{equation}
Treating $\hat{H}_c$ as a perturbation, linear-response theory states that the expectation value of another operator $\hat{O}'(\bm{q})$ is
\begin{subequations}
  \begin{align}
    \langle\hat{O}'\rangle(t,\bm{q})
    &= \int\!\rd t'\;C_{\hat{O}'\hat{O}}(t,t',\bm{q})\,f(t',\bm{q}),
      \label{eq:linear-momentum-space}\\
    C_{\hat{O}'\hat{O}}(t,t',\bm{q})
    &\equiv -\frac{i}{\hbar\mathcal{V}}\Theta(t-t')\,
       \big\langle[\hat{O}'_{I}(t,\bm{q}),\hat{O}_I(t',-\bm{q})]\big\rangle,
  \end{align}
\end{subequations}
where $C_{\hat{O}'\hat{O}}$ is the retarded correlation function and 
$\hat{O}_{I}(t,\bm{q})\equiv \re^{i\hat{H}t/\hbar}\hat{O}(\bm{q})\re^{-i\hat{H}t/\hbar}$ is the operator in the interaction picture with respect to the unperturbed Hamiltonian $\hat{H}=\hat{H}_0+\hat{H}_I$.

To compute correlation functions and investigate the EM responses of the bilayer EI, we employ the time-dependent Hartree-Fock (TDHF) method and study the dynamics of the single-particle density matrix under the perturbation $\hat{H}_c$:
\begin{equation}
  i\hbar\,\partial_t\rho_{ij\bm{k}}(t,\bm{q})
  = \big\langle\big[c^{\dagger}_{j\bm{k}-\bm{q}/2}
                     c_{i\bm{k}+\bm{q}/2},
                     \hat{H}+\hat{H}_c\big]\big\rangle.
  \label{eq:tdhf-full}
\end{equation}
In the absence of $f(t)$, the stationary condition of the above equation gives the following HF mean field solution,
\begin{equation}
  [\rho^{X}_{\bm{k}},h^{\mathrm{MF}}_{\bm{k}}]=0,
\end{equation}
where $\rho^{X}_{ss'\bm{k}}\equiv \langle c^{\dagger}_{s'\bm{k}}c_{s\bm{k}}\rangle$ is the density matrix of the EI ground state. Defining $\zeta_{\bm{k}}\equiv\hbar^2k^2\delta_m/(4m)$, the mean-field Hamiltonian is
\begin{equation}
  h^{\mathrm{MF}}_{\bm{k}}
  = \zeta_{\bm{k}}\sigma_0 + \varepsilon_{\bm{k}}\sigma_z
    + \Delta_{\bm{k}}\sigma_{+}
    + \Delta^*_{\bm{k}}\sigma_{-},
\end{equation}
with $\sigma_{\pm}=(\sigma_x\pm i\sigma_y)/2$.
The renormalized band energy $\varepsilon_{\bm{k}}$ and EI order parameter $\Delta_{\bm{k}}$ are determined self-consistently from $\rho^{X}_{\bm{k}}$ as
\begin{subequations}
  \begin{align}
    \varepsilon_{\bm{k}}
    &\equiv \frac{\hbar^2k^2}{4 m}
       - \frac{\mu_X}{2}
       + \frac{2\pi e^2 d\, n_X}{\epsilon}
       - \frac{1}{\mathcal{V}}\sum_{\bm{k}'}
         V(\bm{k}-\bm{k}')\,\rho^{X}_{ee\bm{k}'},\\
    \Delta_{\bm{k}}
    &\equiv -\frac{1}{\mathcal{V}}\sum_{\bm{k}'}
      U(\bm{k}-\bm{k}')\,\rho^{X}_{eh\bm{k}'},
      \label{eq:order-parameter}
  \end{align}
  \label{eq:hartree-fock-parameters}%
\end{subequations}
where the third term in $\varepsilon_{\bm{k}}$ is the Hartree energy from the uniform exciton density
\begin{equation}
  \frac{1}{\mathcal{V}}\sum_{\bm{k}'}\lim_{q\to 0}\bigl[V(\bm{q})-U(\bm{q})\bigr]\,\rho^{X}_{ee\bm{k}'}=\frac{2\pi e^2 d\, n_X}{\epsilon}
\end{equation}

and $n_{X}\equiv\mathcal{V}^{-1}\sum_{\bm{k}}\rho^{X}_{ee\bm{k}}$ is the exciton density (charge density per layer).
In general, $\Delta_{\bm{k}}$ is complex-valued, and its phase labels the degenerate $\mathrm{U}_{eh}(1)$-symmetry-broken states.
In this paper we choose $\Delta_{\bm{k}}$ to be real and negative without loss of generality.
The mean-field Hamiltonian is diagonalized into quasiparticle conduction ($c$) and valence ($v$) bands with excitation energies $E_{c,\bm{k}}=\zeta_{\bm{k}}+\xi_{\bm{k}}$ and $E_{v,\bm{k}}=\zeta_{\bm{k}}-\xi_{\bm{k}}$, where $\xi_{\bm{k}}=\sqrt{\varepsilon^2_{\bm{k}}+\Delta^2_{\bm{k}}}$.
Because $\zeta_{\bm{k}}\sigma_0$ is proportional to the identity matrix, it shifts both quasiparticle energies equally without changing the eigenvectors.
The corresponding eigenvectors can be written as
\begin{equation}
  [\ket{v\bm{k}},\ket{c\bm{k}}]
  = \begin{bmatrix}
      \alpha_{\bm{k}} & \beta_{\bm{k}}\\
      \beta_{\bm{k}} & -\alpha_{\bm{k}}
    \end{bmatrix},
\end{equation}
where $\alpha_{\bm{k}}=\sqrt{\bigl(1-\varepsilon_{\bm{k}}/\xi_{\bm{k}}\bigr)/2}$ and 
$\beta_{\bm{k}}=\sqrt{\bigl(1+\varepsilon_{\bm{k}}/\xi_{\bm{k}}\bigr)/2}$.
In this paper we focus on zero temperature, so the quasiparticle density matrix is simply
\begin{equation}
  \rho^{X}_{\bm{k}}=\ket{v\bm{k}}\bra{v\bm{k}},
  \label{eq:new-dst}
\end{equation}
and the EI ground state is obtained by solving Eqs.~\eqref{eq:hartree-fock-parameters} and \eqref{eq:new-dst} self-consistently.

Near the EI ground state, the density matrix $\rho_{ij\bm{k}}(t,\bm{q})$ can be expanded in powers of the external field $f(t)$:
\begin{equation}
  \rho_{ij\bm{k}}(t,\bm{q}) = \sum_{n}\rho^{(n)}_{ij\bm{k}}(t,\bm{q}),
\end{equation}
where $\rho^{(n)}$ denotes the $n$-th order contribution.
It is convenient to work in the quasiparticle band basis.
In this basis, the zeroth-order density matrix is
$\rho^{(0)}_{ij\bm{k}}(t,\bm{q})=\rho^{X}_{ij\bm{k}}\delta_{\bm{q},\bm{0}}
=\delta_{ij}\delta_{iv}\delta_{\bm{q},\bm{0}}$.
The first-order correction has only off-diagonal components, 
$\rho^{(1)}_{cv\bm{k}}(t,\bm{q})$ and 
$\rho^{(1)}_{vc\bm{k}}(t,\bm{q})
= [\rho^{(1)}_{cv\bm{k}}(t,-\bm{q})]^*$.
Assuming that the operator $\hat{O}(\bm{q})$ can be expressed in terms of quasiparticles as
\begin{equation}
  \hat{O}(\bm{q})
  = \sum_{i,j=c,v}\sum_{\bm{k}}
    o_{ij\bm{k}}(\bm{q})\,
    c^{\dagger}_{i\bm{k}-\bm{q}/2}c_{j\bm{k}+\bm{q}/2},
\end{equation}
the TDHF equation \eqref{eq:tdhf-full} yields, to linear order in $f(t)$,
\begin{align}
  i\hbar\,\tau_z\,
  \partial_t
  \begin{bmatrix}
    \rho^{(1)}_{cv\bm{k}}(t,\bm{q})\\[2pt]
    \rho^{(1)}_{vc-\bm{k}}(t,\bm{q})
  \end{bmatrix}
  &= \sum_{\bm{k}'}
     \mathcal{H}_{\bm{k},\bm{k}'}(\bm{q})
     \begin{bmatrix}
       \rho^{(1)}_{cv\bm{k}'}(t,\bm{q})\\[2pt]
       \rho^{(1)}_{vc-\bm{k}'}(t,\bm{q})
     \end{bmatrix}\nonumber\\
  &\quad
     +\frac{1}{\mathcal{V}}
     \begin{bmatrix}
       o_{cv\bm{k}}(-\bm{q})\\[2pt]
       o_{vc-\bm{k}}(-\bm{q})
     \end{bmatrix}f(t,\bm{q}),
    \label{eq:dynamic_equation}
\end{align}
where $\tau_z$ is a Pauli matrix in the $(cv,vc)$ space and $\mathcal{H}_{\bm{k},\bm{k}'}(\bm{q})$ is the dynamical matrix. 
Detailed derivations and explicit expressions for $\mathcal{H}_{\bm{k},\bm{k}'}(\bm{q})$ are provided in SM Appendix C.1\cite{SM}.

To solve the dynamical equation~\eqref{eq:dynamic_equation}, we first determine the collective-mode eigenfunctions $\Phi_{n\bm{k}}(\bm{q})$ and corresponding excitation energies $\omega_n(\bm{q})$ by solving the generalized eigenvalue problem
\begin{equation}
  \sum_{\bm{k}'}\mathcal{H}_{\bm{k},\bm{k}'}(\bm{q})\,
    \Phi_{n\bm{k}'}(\bm{q})
  = \hbar\omega_n(\bm{q})\,\tau_z\,\Phi_{n\bm{k}}(\bm{q}).
  \label{eq:generalized_eigenvalue_equation}
\end{equation}
At each $\bm{k}$, the eigenfunction is a two-component vector, which we denote as
$\Phi_{n\bm{k}}(\bm{q})
= [\Phi_{n\bm{k}}^{cv}(\bm{q}),\Phi_{n\bm{k}}^{vc}(\bm{q})]^T$.

Performing a Fourier transform to frequency space, the dynamical equation~\eqref{eq:dynamic_equation} can be solved explicitly (see SM Appendix C.2\cite{SM}).
The solution is
\begin{subequations}
  \begin{align}
    \begin{bmatrix}
      \rho^{(1)}_{cv\bm{k}}(\omega,\bm{q})\\[2pt]
      \rho^{(1)}_{vc-\bm{k}}(\omega,\bm{q})
    \end{bmatrix}
    &= \frac{1}{\mathcal{V}}\sum_{\bm{k}'}
      \Pi_{\bm{k},\bm{k}'}(\omega,\bm{q})
      \begin{bmatrix}
        o_{cv\bm{k}'}(-\bm{q})\\[2pt]
        o_{vc-\bm{k}'}(-\bm{q})
      \end{bmatrix}
      f(\omega,\bm{q}),\\
    \Pi_{\bm{k},\bm{k}'}(\omega,\bm{q})
    &\equiv \sum_{n}
      \frac{\omega_{n}(\bm{q})\,
            \Phi_{n\bm{k}}(\bm{q})
            \Phi^{\dagger}_{n\bm{k}'}(\bm{q})}
           {\omega+i\eta-\omega_n(\bm{q})}.
  \end{align}
\end{subequations}
Thus, to linear order in the external field $f(\omega,\bm{q})$, the expectation value of $\hat{O}'(\bm{q})$ is
\begin{subequations}
  \begin{align}
    \langle\hat{O}'\rangle(\omega,\bm{q})
    &= \sum_{ij\bm{k}}o'_{ij\bm{k}}(\bm{q})\,
       \rho^{(1)}_{ji\bm{k}}(\omega,\bm{q})
       \equiv C_{\hat{O}'\hat{O}}(\omega,\bm{q})\,f(\omega,\bm{q}),
      \label{eq:linear-momentum-frequency-space}\\
    C_{\hat{O}'\hat{O}}(\omega,\bm{q})
    &= \frac{1}{\mathcal{V}}\sum_{n}
       \frac{\omega_n(\bm{q})\,
             [\mathrm{O}'_n(\bm{q})]^{*}\,
             \mathrm{O}_n(\bm{q})}
            {\omega+i\eta-\omega_n(\bm{q})},
      \label{eq:response_function}
  \end{align}
\end{subequations}
where $C_{\hat{O}'\hat{O}}(\omega,\bm{q})$ is the retarded correlation function in frequency-momentum space, and $\mathrm{O}_n(\bm{q})$ is the overlap between the vertex of $\hat{O}$ and the collective mode eigenfunction $\Phi_{n\bm{k}}(\bm{q})$:
\begin{equation}
  \mathrm{O}_n(\bm{q})
  \equiv \sum_{\bm{k}}
  \Phi^{\dagger}_{n\bm{k}}(\bm{q})
  \begin{bmatrix}
    o_{cv\bm{k}}(-\bm{q})\\[2pt]
    o_{vc-\bm{k}}(-\bm{q})
  \end{bmatrix}.\label{eq:overlap_vert}
\end{equation}

\section{Results at zero perpendicular magnetic field}\label{sec:zero-B}
It is convenient to work in the excitonic units, where the length and energy scales are defined as $a_B^*\equiv \epsilon \hbar^2/(m e^2)$ and $\mathrm{Ry}^*\equiv e^2/(2\epsilon a_B^*)$ respectively.
Then the only parameters in the many-body Hamiltonian are the exciton chemical potential $\mu_X/\mathrm{Ry}^*$, interlayer distance $d/a_{B}^*$, and electron--hole mass-asymmetry parameter $\delta_m\equiv (m_h-m_e)/(m_h+m_e)$.
In typical TMD bilayers such as the MoSe$_2$/WSe$_2$ heterostructure, the parameters are $m_e\approx 0.58 m_0$, $m_h=0.36 m_0$\cite{kormanyosKpTheoryTwodimensional2015} and $\epsilon\approx5$\cite{caiInfraredReflectanceSpectrum2007}.
{Within the minimal model, $\epsilon$ is the static dielectric constant of a homogeneous background medium, while electronic self-screening within the layers is not treated explicitly. The same $\epsilon$ enters the intralayer and interlayer interactions because both electric fields propagate through this medium; their geometric difference is described by the factor $\re^{-qd}$ in the interlayer interaction (see SM Appendix A\cite{SM}). For the numerical estimates, we use an effective value $\epsilon\approx 10$ to reproduce the measured interlayer-exciton binding-energy scale\cite{maStronglyCorrelatedExcitonic2021}. More general electronic screening or dielectric inhomogeneity would produce momentum- and frequency-dependent effective interactions and is beyond the present minimal model.}
Thus we have $m\approx 0.22 m_0$, $\mathrm{Ry}^*\approx 30\mathrm{meV}$ and $a_B^*\approx 2.4 \mathrm{nm}$.
With $5\sim 6$ layers hBN between the electron and hole layers, the interlayer distance is about $d/a_B^*=1$.
The corresponding mass-asymmetry magnitude is $|\delta_m|\approx 0.23$.
Unless stated otherwise, we set $\delta_m=0$ in the main text. An analytical treatment of electron--hole symmetry and asymmetry is provided in SM Appendix G, while numerical results for $\delta_m=0.2$ are presented in SM Appendices H.3 and H.4\cite{SM}.

\subsection{The collective mode spectrum}\label{sec:collective-mode-spectrum}
\begin{figure}
  \centering
  \includegraphics[width=\linewidth]{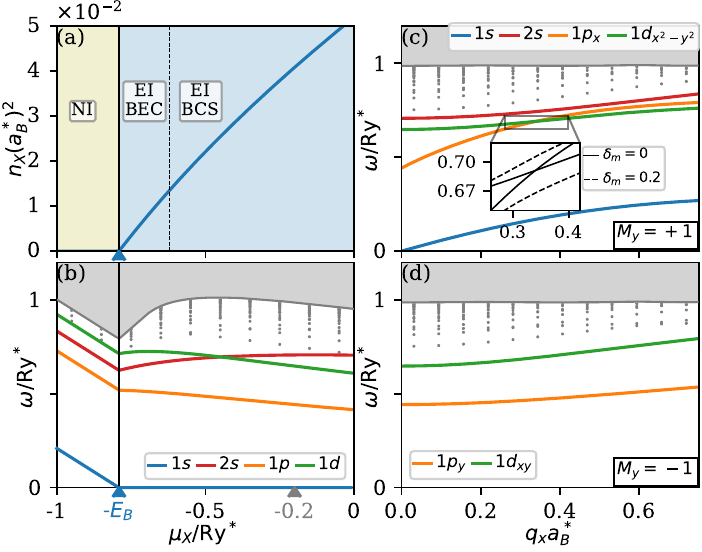}
  \caption{(a) Mean-field phase diagram at zero temperature as a function of the exciton chemical potential $\mu_X$. As $\mu_{X}$ increases, the ground state turns from the normal insulator (NI) to the EI.
  (b) Collective mode spectrum at zero momentum as a function of $\mu_X$.
  The shaded region represents the electron--hole continuum.
  A few lowest collective excitations are specially indicated by the color lines and labeled by their angular momentum in $z$ direction.
  For example, the modes labeled by $s,p,d$ have angular momentum $l_z=0,\pm1,\pm2$ respectively.
  (c)(d) Collective mode spectrum in momentum space along the $q_x$ axis ($q_x$ is the momentum in $x$ direction).
  Along this line, these modes can be distinguished by the mirror eigenvalue $M_y$.
  For clarity, the collective modes with $M_y=1$ are plotted in (c) and the modes with $M_y=-1$ are plotted in (d).
  In the inset of (c), we compare the spectrum near the level crossing for different electron--hole mass asymmetries $\delta_m$.
  }
  \label{fig:clc_modes}
\end{figure}

In Fig. \ref{fig:clc_modes}(a), we plot the ground-state exciton density $n_X$ (blue line) as a function of the exciton chemical potential $\mu_X$.
When the chemical potential satisfies $\mu_X=eV_b-E_g<-E_B$, the exciton levels are inside the band gap.
There are no exciton excitations at zero temperature and the ground state is a normal insulator (NI).
Here, the collective excitations correspond directly to interlayer exciton states, whose excitation energies depend linearly on the exciton chemical potential.
In Fig. \ref{fig:clc_modes}(b), we plot the lowest few exciton levels at zero momentum in the NI region.
Due to the rotational symmetry of the many-body Hamiltonian, these exciton states (at $\bm{q}=\bm{0}$) can be labeled by their angular momentum $l_z$ along the $z$-axis.
Specifically, the blue and red lines labeled ``1$s$'' and ``2$s$'' represent the exciton levels with $l_z=0$ (monopole mode); the orange line labeled ``1$p$'' denotes the doubly degenerate exciton levels with $l_z=\pm1$ (dipole mode); and the green line labeled ``1$d$'' corresponds to doubly degenerate exciton levels with $l_z=\pm2$ (quadrupole mode).
When the chemical potential increases beyond $\mu_X>-E_B$, the $1s$ exciton no longer remains an excited state and instead condenses at zero temperature.
As shown in Fig. \ref{fig:clc_modes}(a), the exciton density $n_{X}$ becomes nonzero in the ground state, signaling a transition into the excitonic insulator (EI) phase. 
Additionally, based on the sign of the renormalized band offset $\varepsilon_{\bm{k}=0}$, we mark the BEC-BCS crossover with a black dotted line in Fig. \ref{fig:clc_modes}(a).
Since the NI-EI phase transition does not break the rotational symmetry, collective modes in the EI phase can still be labeled by the angular momentum $l_z$.
Inspecting the collective mode spectrum in Fig. \ref{fig:clc_modes}(b), we observe that the 1$s$ exciton mode in the NI phase evolves continuously into a zero-energy mode (Goldstone mode) in the EI phase.

\begin{figure}
  \centering
  \includegraphics[width=\linewidth]{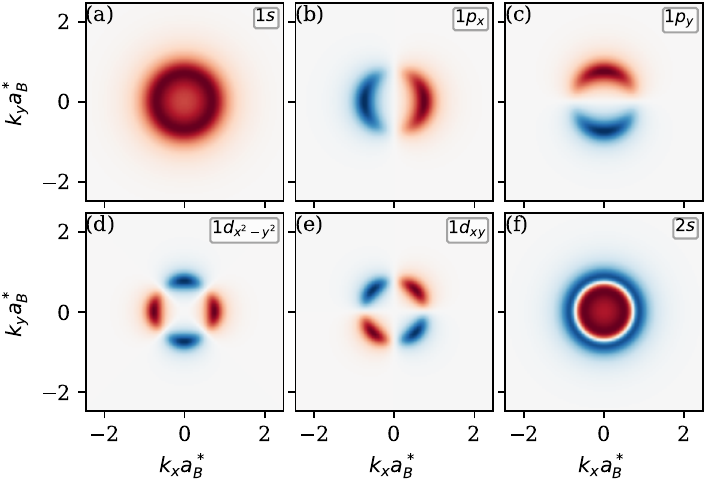}
  \caption{Typical wavefunctions of the collective modes at $\bm{q}=\bm{0}$ and $\mu_{X}=-0.2$. Here we only plot $\Phi_{n\bm{k}}^{cv}(\bm{q})$ in the two-component wavefunction $\Phi_{n\bm{k}}(\bm{q})=[\Phi_{n\bm{k}}^{cv}(\bm{q}),\Phi_{n\bm{k}}^{vc}(\bm{q})]$.}
  \label{fig:modes}
\end{figure}

In Fig. \ref{fig:clc_modes}(c)(d), we plot the dispersion relations of these collective modes along the $q_x$ axis at a representative chemical potential $\mu_X=-0.2$.
Along this line, the angular momentum is no longer a good quantum number, while the system still preserves reflection symmetry about the $x$-$z$ plane, denoted as $M_y$.
Consequently, the previously degenerate exciton levels split according to their $M_y$ eigenvalues.
Specifically, the $1p$ modes split into the $1p_x$ and $1p_y$ modes and the $1d$ modes split into the $1d_{x^2-y^2}$ and $1d_{xy}$ modes.
For clarity, the collective modes with $M_y=\pm1$ are plotted separately in Fig. \ref{fig:clc_modes}(c) and (d).
At $\bm{q}=\bm{0}$, typical wavefunctions of the collective modes are shown in Fig. \ref{fig:modes}.

In Fig. \ref{fig:clc_modes}(c), the linear dispersion of the $1s$ Goldstone mode is clearly illustrated by the blue line.
Additionally, a notable level crossing between the $1p_x$ mode (orange line) and $1d_{x^2-y^2}$ mode (green line) is observed.
In the inset of Fig. \ref{fig:clc_modes}(c), we further explore the effect of electron--hole asymmetry by plotting the collective mode spectrum near this crossing point with a finite electron--hole-asymmetry parameter $\delta_m=0.2$ (black dotted line).
As shown, introducing electron--hole asymmetry opens a gap at the crossing point, indicating that the degeneracy observed at $\delta_m=0$ is protected by electron--hole symmetry (see SM Appendix G.2\cite{SM}).
Away from this avoided crossing, weak mass asymmetry leaves the qualitative long-wavelength spectrum intact but produces the perturbative corrections derived in SM Appendix G.3\cite{SM}.
Thus, we use the electron--hole-symmetric limit to describe the qualitative collective-mode structure and the leading long-wavelength diagonal responses. Corrections due to finite mass asymmetry are analyzed in SM Appendices G and H\cite{SM}.

\begin{figure}
  \centering
  \includegraphics[width=\linewidth]{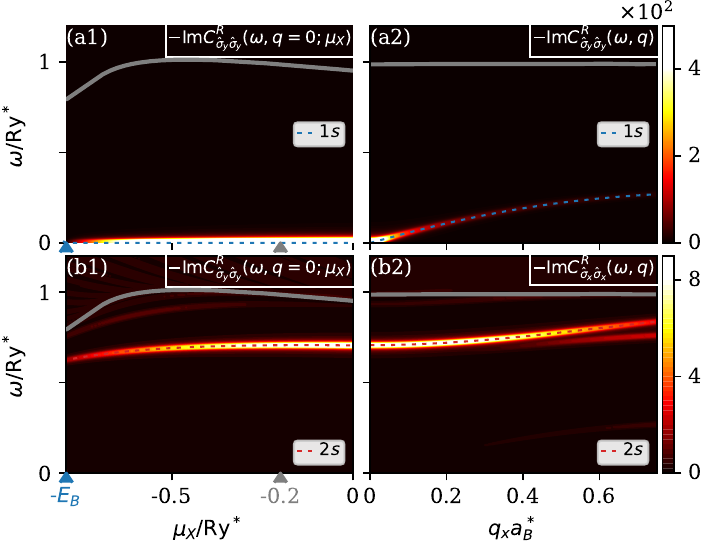}
  \caption{(a1)(a2) Imaginary part of the correlation function $C_{\hat{\sigma}_y\hat{\sigma}_y}(\omega+i\eta)$, accounting for the phase fluctuations of the EI order parameter.
  The dominant pole is the $1s$ Goldstone mode represented by the dashed blue line.
  (b1)(b2) Imaginary part of the correlation function $C_{\hat{\sigma}_x\hat{\sigma}_x}(\omega+i\eta)$, accounting for the amplitude fluctuations of the EI order parameter.
  The dominant pole is the $2s$ mode represented by the dashed red line.
  In (a1)(b1), the momentum is taken as $\bm{q}=\bm{0}$ and the horizontal axis is the exciton chemical potential $\mu_X$.
  In (a2)(b2), we take $\mu_X=-0.2$, $q_y=0$ and the horizontal axis is the momentum in $x$ direction $q_x$.
  }
  \label{fig:phase_amplitude}
\end{figure}

To identify the phase and amplitude modes in the EI phase, we can calculate the correlation functions associated with the phase and amplitude fluctuations.
In the ground state calculation, the EI order parameter $\Delta_{\bm{k}}$ defined by Eq.\eqref{eq:order-parameter} is chosen to be real.
Thus the phase- and amplitude-fluctuation operators can be constructed by the Pauli matrices $\sigma_y$ and $\sigma_x$ respectively:
\begin{equation}
  \hat{\sigma}_{i=x,y}=\sum_{\bm{k}}C^{\dagger}_{\bm{k}-\bm{q}/2}\sigma_{x,y} C_{\bm{k}+\bm{q}/2}
\end{equation}
where $C^{\dagger}_{\bm{k}}=[c^{\dagger}_{e\bm{k}},c^{\dagger}_{h\bm{k}}]$ is the layer basis electron creation operator.
Using the method introduced in Sec. \ref{sec:model}, we first calculate the correlation function $C_{\hat{\sigma}_y\hat{\sigma}_y}(\omega+i\eta)$, which describes the phase fluctuations, and plot its imaginary part in Fig. \ref{fig:phase_amplitude}(a1)(a2).
To generate these plots, a small imaginary broadening $\eta=0.01\mathrm{Ry}^*$  has been introduced to the frequency $\omega$.
In Fig. \ref{fig:phase_amplitude}(a1), we set the excitation momentum at $\bm{q}=\bm{0}$, and present the imaginary part of the correlation function as a function of exciton chemical potential $\mu_X$ (horizontal axis) and frequency $\omega$ (vertical axis).
In Fig. \ref{fig:phase_amplitude}(a2), we instead fix the exciton chemical potential at $\mu_X/\mathrm{Ry}^*=-0.2$ and change the horizontal axis to the momentum $q_x$ in the $x$ direction.
From these plots, we clearly identify the dominant pole corresponding to the $1s$ Goldstone mode (indicated by the dashed blue line).
Similarly, we calculate the correlation function $C_{\hat{\sigma}_x\hat{\sigma}_x}(\omega+i\eta)$ which describes the amplitude fluctuations.
Its imaginary part is plotted in Fig. \ref{fig:phase_amplitude}(b1)(b2).
This correlation function predominantly couples to the $2s$ mode (indicated by the dashed red line).
These results mean that the $1s$ and $2s$ modes can be viewed as the phase and amplitude modes respectively.

\subsection{The EM response functions}\label{sec:EM-response-function}

The EM field is directly coupled to the system through the four-vector gauge field
$A_{\mu}=(\phi,A_x,A_y,A_z)$, where $\phi$ is the scalar potential and $(A_x,A_y,A_z)$ is the vector potential.
In general, the gauge field has to be defined in the three-dimensional space.
However, due to the two-dimensional nature of the bilayer system, there is no current flow in the out-of-plane ($z$) direction.
Thus we can always make a gauge transformation to set $A_z=0$, and we will denote $\bm{A}=(A_x,A_y)$ in the following text.
In addition, as the bilayer system is located at $\pm d/2$, only the gauge field values at these two planes $A_{\mu}(\bm{r},z=\pm d/2)$ are relevant, which we denote as $A_{e\mu}\equiv A_{\mu}(z=-d/2)$ and $A_{h\mu}\equiv A_{\mu}(z=d/2)$ respectively.
These six components of the gauge field $A_{s\mu}(\bm{r})$ defined in the two dimensional plane, with $s=e,h$ (layer index) and $\mu=0,1,2$, expand the complete degrees of freedom of the EM field that couples to the bilayer system.

For convenience, we will rearrange the layer degree of freedom into layer-symmetric and layer-antisymmetric combinations:
\begin{equation}
  A^{+}_{\mu}=\frac{A_{e\mu}+A_{h\mu}}{2},\quad A^{-}_{\mu}={A_{e\mu}-A_{h\mu}}.
\end{equation}
As derived in SM Appendix C.3\cite{SM}, electron--hole symmetry makes these two combinations couple independently to the charge and exciton degrees of freedom at linear order.
And the paramagnetic linear coupling term is 
\begin{equation}
  \hat{H}_{c,p}=-\sum_{\bm{q}}A^{\sigma}_{\mu}(t,\bm{q})\hat{j}^{\sigma}_{p\mu}(-\bm{q}),\label{eq:gauge-field-coupling}
\end{equation}
where the repeated indices $\sigma=+,-$ and $\mu=0,1,2$ are summed over.
In the coupling Hamiltonian, $A^{\sigma}_{\mu}(t,\bm{q})$ is the Fourier transform of the gauge field $A^{\sigma}_{\mu}(t,\bm{r})$, and $\hat{j}^{\sigma}_{p\mu}(\bm{q})\equiv(-\hat{\varrho}^{\sigma}(\bm{q}),\hat{\bm{j}}^{\sigma}_p(\bm{q}))$ is the four-vector (here we only take $\mu=0,1,2$ since there is no current in $z$ direction) paramagnetic current operator defined as
\begin{equation}
  \hat{j}^{\sigma}_{p\mu}(\bm{q})=\sum_{\bm{k}}C^{\dagger}_{\bm{k}-\bm{q}/2}\gamma^{\sigma}_{\mu}(\bm{k},\bm{q})C_{\bm{k}+\bm{q}/2}.\label{eq:current-operator}
\end{equation}
where the bare vertex functions $\gamma^{\sigma}_{\mu}(\bm{k},\bm{q})$ are given by
\begin{subequations}
  \begin{align}
    \gamma^{+}_{0}(\bm{k},\bm{q})=&e\sigma_0,\quad \gamma^{+}_{a}(\bm{k},\bm{q})=-\frac{e\hbar}{2m}k_a\sigma_z,\\
    \gamma^{-}_{0}(\bm{k},\bm{q})=&\frac{e}{2}\sigma_z,\quad \gamma^{-}_{a}(\bm{k},\bm{q})=-\frac{e\hbar}{4m}k_a\sigma_0
  \end{align}\label{eq:bare_vertex}
\end{subequations}
where $a=1,2$ denotes the spatial components $x,y$ respectively.
Thus to first order in the gauge field, the expectation value of the paramagnetic current is written as
\begin{equation}
  j^{\sigma}_{p\mu}(\omega,\bm{q})=-C_{\hat{j}^{\sigma}_{p\mu}\hat{j}^{\sigma'}_{p\nu}}(\omega,\bm{q})A^{\sigma'}_{\nu}(\omega,\bm{q}),\label{eq:current-response-paramagnetic}
\end{equation}
where the retarded correlation function $C_{\hat{j}^{\sigma}_{p\mu}\hat{j}^{\sigma'}_{p\nu}}(\omega,\bm{q})$ can be calculated using the TDHF method introduced in Sec. \ref{sec:model}.
In addition to the paramagnetic current, there will also be a diamagnetic current term in each layer at finite vector potential $\bm{A}_s(t,\bm{r})$:
\begin{equation}
  \hat{\bm{j}}_{sd}(t,\bm{r})=-\frac{e^2}{2m}\Psi^{\dagger}_s(\bm{r})\Psi_s(\bm{r})\bm{A}_s(t,\bm{r}),
\end{equation}
where $\Psi^{\dagger}_{s}(\bm{r})=\mathcal{V}^{-1/2}\sum_{\bm{k}}\re^{-i\bm{k}\cdot\bm{r}}c^{\dagger}_{s\bm{k}}$ is the field operator in each layer.
In general, the diamagnetic current operator couldn't be simply decomposed into the charge and exciton channels.
However, at the charge neutral point, we find that its first order expectation value
\begin{equation}
  {\bm{j}}_{sd}(t,\bm{r})=\ave{\hat{\bm{j}}_{sd}(t,\bm{r})}=-\frac{e^2n_X}{2m}\bm{A}_{s}(t,\bm{r})
\end{equation}
could still be decomposed independently into the layer-symmetric charge channel and layer-antisymmetric exciton channel:
\begin{subequations}
  \begin{align}
  \bm{j}^{+}_{d}=&\bm{j}_{ed}+\bm{j}_{hd}=-\frac{e^2n_X}{m}\bm{A}^{+}\\
  \bm{j}^{-}_{d}=&\frac{1}{2}(\bm{j}_{ed}-\bm{j}_{hd})=-\frac{e^2n_X}{4m}\bm{A}^{-}
  \end{align}\label{eq:current-response-diamagnetic}
\end{subequations}
Summing Eqs. \eqref{eq:current-response-paramagnetic} and \eqref{eq:current-response-diamagnetic}, we obtain the total current response up to first order in the gauge field:
\begin{align}
  j^{\sigma}_{\mu}(\omega,\bm{q})\equiv& (1-\delta_{\mu0})j^{\sigma}_{d\mu}(\omega,\bm{q})+j^{\sigma}_{p\mu}(\omega,\bm{q})\nonumber\\
  =&K^{\sigma\sigma'}_{\mu\nu}(\omega,\bm{q})A^{\sigma'}_{\nu}(\omega,\bm{q}),\label{eq:current-response-total}
\end{align}
where the kernel $K^{\sigma\sigma'}_{\mu\nu}(\omega,\bm{q})$ describes the full EM response of the bilayer system.

As shown in SM Appendix B.2\cite{SM}, the TDHF approximation sums an infinite series of ladder-bubble diagrams for the two-particle correlation function.
Thus the response kernel $K^{\sigma\sigma'}_{\mu\nu}(\omega,\bm{q})$ will satisfy the Ward identity implied by the charge conservation law\cite{dassarmaCommentsTimedependentHartreeFock1983}: 
\begin{equation}
  q_{\mu}K^{\sigma\sigma'}_{\mu\nu}(\omega,\bm{q})=0,\quad K^{\sigma\sigma'}_{\mu\nu}(\omega,\bm{q})q_{\nu}=0,\label{eq:Ward-identify}
\end{equation}
where $q_{\mu}=(\omega,\bm{q})$ is the four-momentum.
Besides, due to the electron--hole and time-reversal symmetries, the response functions are diagonal in the charge and exciton channels, i.e., $K^{\sigma\sigma'}_{\mu\nu}(\omega,\bm{q})=K^{\sigma}_{\mu\nu}(\omega,\bm{q})\delta_{\sigma\sigma'}$.
Additionally, in the presence of rotational symmetry, the spatial components of the response function can be further decomposed into longitudinal and transverse parts as 
\begin{equation}
  K^{\sigma}_{ab}(\omega,\bm{q})=K^{\sigma}_L(\omega,q)\frac{q_a q_b}{q^2}+K^{\sigma}_T(\omega,q)\left(\delta_{ab}-\frac{q_a q_b}{q^2}\right).\label{eq:symmetric-decomposition}
\end{equation}
In summary, the full EM response of the bilayer system is characterized by four scalar functions: $K^{+}_L(\omega,q)$, $K^{+}_T(\omega,q)$, $K^{-}_L(\omega,q)$ and $K^{-}_T(\omega,q)$.

\subsubsection{Response to the layer-symmetric gauge field}

\begin{figure}
  \centering
  \includegraphics[width=\linewidth]{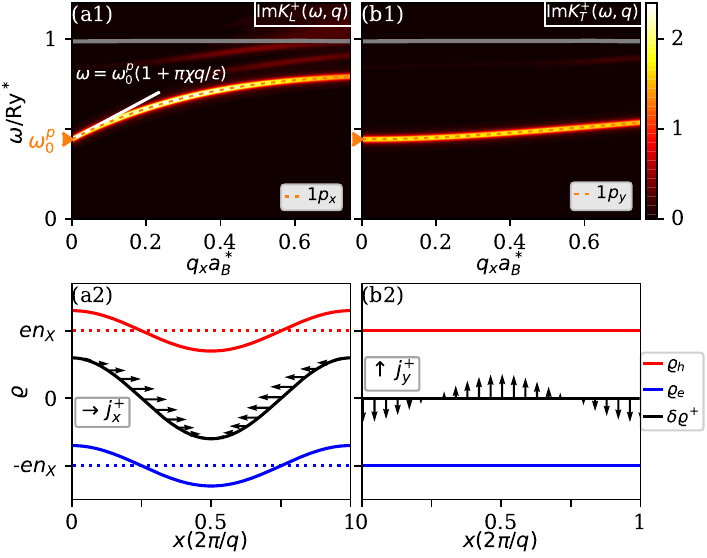}
  \caption{(a1) Imaginary part of the longitudinal response kernel to the layer-symmetric gauge field $K^{+}_{L}(\omega+i\eta,\bm{q})$ along the $q_x$ axis.
  The dominant pole is the $1p_x$ mode, which represents a plasmon mode with both charge and longitudinal charge current density fluctuations as illustrated by (a2).
  (b1) Imaginary part of the transverse response kernel to the layer-symmetric gauge field $K^{+}_{T}(\omega+i\eta,\bm{q})$ along the $q_x$ axis.
  The dominant pole is the $1p_y$ mode, which represents a plasmon mode with only transverse charge current density fluctuation as illustrated by (b2).
  }
  \label{fig:charge_responses}
\end{figure}

In Fig. \ref{fig:charge_responses}(a1)(b1), we plot the imaginary parts of the longitudinal and transverse electromagnetic response functions in the charge channel, along the positive $q_x$ axis.
For clarity, the dominant poles corresponding to the $1p_x$ and $1p_y$ modes, are also indicated by dashed orange lines in Fig. \ref{fig:charge_responses}(a1)(b1).
Since these two poles predominantly couple to the total charge current fluctuations within the bilayer EI, they can naturally be identified as plasmon modes.
When the longitudinal $1p_x$ mode is excited, the charge densities in electron and hole layers oscillate in-phase along the $x$ direction.
This collective oscillation generates a net charge density modulation and a corresponding longitudinal current fluctuation, as illustrated schematically in Fig. \ref{fig:charge_responses}(a2).
In contrast, the transverse $1p_y$ mode only couples to the transverse current fluctuation, which does not alter the charge distribution, as illustrated in Fig. \ref{fig:charge_responses}(b2).

In SM Appendix H.4\cite{SM}, we also provide the numerical results of the longitudinal and transverse response functions in the charge channel for a finite electron--hole mass asymmetry $\delta_m=0.2$. In the long-wavelength limit, the collective-mode spectrum is identical to that of the electron--hole-symmetric case. The mass-asymmetry-induced coupling amplitudes between the charge-current operators and additional collective modes are of order $\delta_m q$ [see SM Appendix G.3(b)], so the corresponding charge-response spectral weights are of order $\delta_m^2q^2$ and vanish as $q\to0$.

Although the two plasmon modes are degenerate at zero momentum, the long-range Coulomb interaction lifts this degeneracy due to the direct coupling between the longitudinal mode and charge density fluctuations. 
Consequently, the longitudinal plasmon mode exhibits a linear dispersion relation in the long-wavelength limit, arising from the same mechanism responsible for the splitting of longitudinal and transverse optical phonons in two-dimensional systems \cite{sohierBreakdownOpticalPhonons2017,riveraPhononPolaritonicsTwoDimensional2019,shiTwodimensionalMoirePhonon2025,liObservationNonanalyticBehavior2024}.
To see this, we need to find roots of the effective dielectric function
\begin{align}
  \epsilon_{\mathrm{eff}}(\omega,\bm{q})
  =&[1+\tilde{V}(q)C_{\hat{j}^{+}_{p0}\hat{j}^{+}_{p0}}(\omega,\bm{q})]^{-1},
  \label{eq:dielectric_function}
\end{align}
where $\tilde{V}(q)=e^{-2}V(q)=2\pi/\epsilon q$ is the Coulomb interaction in the long-wavelength limit ($e^2$ is divided out because the elementary charge $e$ has already been included in the definition of the current operator in Eq. \eqref{eq:current-operator}).
Equation \eqref{eq:dielectric_function} is derived in SM Appendix C.5\cite{SM}.
As detailed in SM Appendix C.4\cite{SM}, the density-density correlation function $C_{\hat{j}^{+}_{p0}\hat{j}^{+}_{p0}}(\omega,\bm{q})$ can be approximated as
\begin{equation}
  C_{\hat{j}^{+}_{p0}\hat{j}^{+}_{p0}}(\omega,\bm{q})\approx\frac{1}{\mathcal{V}}\sum_{\substack{\omega_n>0\\n\in\mathrm{dipole}}}\frac{2|\mathrm{J}^{+}_{0,n}|^2\omega^2_n(\bm{q})}{(\omega+i\eta)^2-\omega_n^2(\bm{q})},
\end{equation}
where $\mathrm{J}^{+}_{0,n}\sim \bm{q}\cdot \bm{p}_n$, as defined by Eq. \eqref{eq:overlap_vert}, is the overlap between the vertex function of the charge density operator (defined in Eq. \eqref{eq:bare_vertex}) and the collective mode wavefunction.
Here, $\bm{p}_n$ is a constant vector determined by the corresponding collective mode.
In the long-wavelength limit, the dominant poles are the degenerated $1p_x$ and $1p_y$ modes.
By retaining only the $1p$ modes and assuming rotational symmetry, the charge density-density response function can be expressed as:
\begin{align}
  C_{\hat{j}^{+}_{p0}\hat{j}^{+}_{p0}}(\omega,\bm{q})\approx&\frac{1}{2}\sum_{n=1p_x,1p_y}\frac{\chi q^2\omega_n^2(\bm{q})}{(\omega+i\eta)^2-\omega_n^2(\bm{q})}\nonumber\\
  =&\frac{\chi q^2(\omega_0^p)^2}{(\omega+i\eta)^2-(\omega_0^p)^2}+\mathcal{O}(q^3),\label{eq:charge-density-response}
\end{align}
where the coefficient
\begin{equation}
  \chi\equiv -\lim_{q\to 0}\lim_{\omega\to 0}\frac{1}{q^2}C_{\hat{j}^{+}_{p0}\hat{j}^{+}_{p0}}(\omega,\bm{q})\label{eq:susceptibility}
\end{equation}
has the physical meaning of 2D electrical polarizability.
In the long-wavelength limit, we have $\tilde{V}(q)C_{\hat{j}^{+}_{p0}\hat{j}^{+}_{p0}}(\omega,\bm{q})\sim q$.
Expanding Eq. \eqref{eq:dielectric_function} and keep up to the lowest order of $q$, the effective dielectric function is approximated as 
\begin{align}
  \epsilon_{\mathrm{eff}}(\omega,\bm{q})\approx&1-\tilde{V}(q)C_{\hat{j}^{+}_{p0}\hat{j}^{+}_{p0}}(\omega,\bm{q})\nonumber\\
  \approx &1-\frac{2\pi}{\epsilon q}\frac{\chi q^2(\omega_0^p)^2}{\omega^2-(\omega_0^p)^2}
\end{align}
where $\omega_0^p=\omega_{1p_x}(\bm{0})=\omega_{1p_y}(\bm{0})$ is the plasmon energy at zero momentum.
Solving the equation $\epsilon_{\mathrm{eff}}(\omega,\bm{q})=0$ yields the longitudinal plasmon dispersion relation:
\begin{equation}
  \omega=\omega_0^{p}\sqrt{1+2\pi \chi q/\epsilon}\approx\omega_0^{p}\left(1+\frac{\pi \chi}{\epsilon}q\right).\label{eq:longitudinal_plasmon_linear}
\end{equation}
In Fig. \ref{fig:charge_responses}(a1), the linear dispersion described by Eq. \eqref{eq:longitudinal_plasmon_linear} is depicted by the white line, which shows good agreement with the numerical results for the $1p_x$ mode in the long-wavelength regime.


\subsubsection{Response to the layer-antisymmetric gauge field}
\label{sec:layer-antisymmetric-gauge-field}

\begin{figure}
  \centering
  \includegraphics[width=\linewidth]{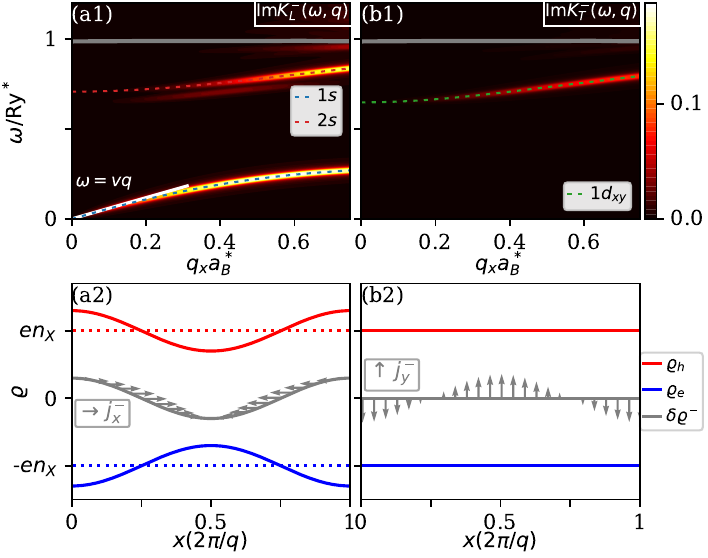}
  \caption{(a1) Imaginary part of the longitudinal response kernel to the layer-antisymmetric gauge field $K^{-}_{L}(\omega+i\eta,\bm{q})$ along the $q_x$ axis.
  The dominant poles are the $1s$ and $2s$ modes.
  These modes represent the phase and amplitude fluctuations of the EI order parameter and associate with both exciton and longitudinal exciton current density fluctuations as illustrated by (a2).
  (b1) Imaginary part of the transverse response kernel to the layer-antisymmetric gauge field $K^{-}_{T}(\omega+i\eta,\bm{q})$ along the $q_x$ axis.
  The dominant pole is the $1d_{xy}$ mode, which associates with transverse exciton current density fluctuations as illustrated by (b2).}
  \label{fig:exciton_responses}
\end{figure}

In Fig. \ref{fig:exciton_responses}(a1), the imaginary part of longitudinal response function $K^{-}_{L}(\omega+i\eta,\bm{q})$ is plotted along the $q_x$ axis ($\eta=0.01\mathrm{Ry}^*$).
The dominant poles correspond to the $1s$ and $2s$ monopole modes, represented by the dashed blue and red lines, respectively.
In real space, these modes correspond to charge fluctuations in the two layers that are out of phase. 
As a result, there is no net charge fluctuation, but rather an exciton density fluctuation accompanied by a longitudinal exciton current, as illustrated in Fig. \ref{fig:exciton_responses}(a2).
Similarly, in Fig. \ref{fig:exciton_responses}(b1), the imaginary part of the transverse response function $K^{-}_{T}(\omega+i\eta,\bm{q})$ is plotted along the $q_x$ axis.
The dominant pole in this case is the $1d_{xy}$ quadrupole mode, represented by the dashed blue line. 
The corresponding real-space configuration, characterized by transverse exciton current fluctuations without charge fluctuations, is illustrated in Fig. \ref{fig:exciton_responses}(b2).

In SM Appendix H.4\cite{SM}, we also provide the numerical results of the longitudinal and transverse response functions in the exciton channel for a finite electron--hole mass asymmetry $\delta_m=0.2$. Unlike the charge channel, the mass-asymmetry-induced coupling amplitudes between the longitudinal and transverse exciton-current operators and the $p_x$ and $p_y$ modes, respectively, are of order $\delta_m q^0$ [see SM Appendix G.3(b)], so the corresponding exciton-response spectral weights are of order $\delta_m^2q^0$ and remain finite as $q\to0$, producing visible plasmon poles.

According to the derivation in SM Appendix~C.4\cite{SM}, the response function $K^{-}_{00}(\omega,\bm{q})=-C_{\hat{j}^{-}_{p0}\hat{j}^{-}_{p0}}(\omega,\bm{q})$ takes the following form in the long-wavelength limit:
\begin{align}
  K^{-}_{00}(\omega,\bm{q})\approx-\frac{1}{\mathcal{V}}\sum_{\substack{\omega_n>0\\n\in\mathrm{monopole}}}\frac{2|\mathrm{J}^{-}_{0,n}|^2\omega^2_n(\bm{q})}{(\omega+i\eta)^2-\omega_n^2(\bm{q})},
\end{align}
where $\mathrm{J}^{-}_{0,n}$ is the overlap between the vertex function of the exciton density operator and the collective mode wavefunction.
As shown by Fig. \ref{fig:exciton_responses}(a1), the only dominant pole in the long-wavelength limit is the Goldstone mode with dispersion $\omega_{gs}(\bm{q})=vq$ (where $v$ is the Goldstone mode velocity).
Thus, we can take the single pole approximation and $K^{-}_{00}(\omega,\bm{q})$ can be written in the form as
\begin{equation}
  K^{-}_{00}(\omega,q)\approx \frac{-\kappa v^2 q^2}{(\omega+i\eta)^2-v^2q^2},\label{eq:K-00}
\end{equation}
where $\kappa$ is a constant.
In the static limit, $K^{-}_{00}$ has the physical meaning of isothermal exciton compressibility (or interlayer capacitance), given by
\begin{equation}
  \kappa=\lim_{q\to 0}\lim_{\omega\to 0}K^{-}_{00}(\omega,q)=e^2\left(\frac{\partial n_X}{\partial \mu_X}\right)_{T}.
\end{equation}
This indicates that the coefficient $\kappa$ in Eq. \eqref{eq:K-00} represents the exciton compressibility.
Assuming the momentum is along the $x$-direction, and using the Ward identity, the longitudinal response function $K^{-}_{11}[\omega,(q,0)]$ can be expressed as:
\begin{equation}
  K^{-}_{11}[\omega,(q,0)]\approx\frac{\omega^2}{q^2}K^{-}_{00}(\omega,q)\approx \frac{-\kappa v^2\omega^2}{(\omega+i\eta)^2-v^2q^2}.\label{eq:exciton-current-correlation1}
\end{equation}
On the other hand, as derived in SM Appendix~C.4\cite{SM}, in the finite-frequency $\omega\ne 0$ and long-wavelength limit, we have 
\begin{align}
  K^{-}_{aa}[\omega,(q,0)]=&-\frac{e^2n_X}{4m}-C_{\hat{j}^{-}_{pa}\hat{j}^{-}_{pa}}[\omega,(q,0)]\nonumber\\
  =&-\frac{e^2 n_X}{4m}+\mathcal{O}(q^2).\label{eq:exciton-current-correlation2}
\end{align}
By comparing Eq. \eqref{eq:exciton-current-correlation1} and Eq. \eqref{eq:exciton-current-correlation2}, we obtain:
\begin{equation}
  \kappa v^2=\frac{e^2 n_X}{4m}\implies v=\sqrt{\frac{e^2n_X}{4m\kappa}}.\label{eq:sound-velocity}
\end{equation}
In Fig. \ref{fig:exciton_responses}(a1), the dispersion relation $\omega=vq$ (with $v$ determined by Eq. \eqref{eq:sound-velocity}) is represented by the white line. 
This result shows excellent agreement with the Goldstone mode dispersion in the long-wavelength regime.

In fact, this velocity is nothing but the first sound velocity of the exciton condensate.
The first sound velocity is usually defined as $v=\sqrt{K_S/\rho_m}$, where $K_S\equiv -\mathcal{V}(\partial P/\partial \mathcal{V})_S$ is the isentropic bulk modulus and $\rho_m$ is the mass density [for the bilayer system, $\rho_m=m_Xn_X$, where $m_X=m_e+m_h=4m/(1-\delta_m^2)$ is the exciton mass].
Since we are considering the system at zero temperature, there is no difference between the isothermal and isentropic process, i.e. $K_S=-\mathcal{V}(\partial P/\partial \mathcal{V})_T$.
On the other hand, the thermodynamic variables satisfy the relation\cite{landauStatisticalPhysicsThird1980}
\begin{equation}
  \left(\frac{\partial \mathcal{V}}{\partial P}\right)_{T,N_X}=-\frac{\mathcal{V}}{n_X^2}\left(\frac{\partial n_X}{\partial \mu_X}\right)_{T,\mathcal{V}}=-\frac{\mathcal{V}}{n_X^2}\frac{\kappa}{e^2}.
\end{equation}
Thus the sound velocity can be expressed as
\begin{align}
  v=&\sqrt{\frac{K_S}{\rho_m}}=\sqrt{\frac{e^2n_X^2/\kappa}{m_Xn_X}}=\sqrt{\frac{e^2n_X}{m_X\kappa}},\label{eq:sound-velocity-thermodynamics}
\end{align}
which is exactly the same as Eq. \eqref{eq:sound-velocity} in the electron--hole-symmetric case $\delta_m=0$.

{
At zero temperature, the mean-field self-consistency equation shows that the static exciton compressibility depends only on the reduced mass $m$, and is therefore independent of the electron--hole mass asymmetry $\delta_m$ at fixed $m$. Since $m_X=4m/(1-\delta_m^2)$, Eq. \eqref{eq:sound-velocity-thermodynamics} gives
\[
  \tilde v=v\sqrt{1-\delta_m^2}.
\]
The Goldstone density response retains the pole form of Eq. \eqref{eq:K-00}, as shown in SM Appendix G.3(c)\cite{SM}, with the velocity $v$ replaced by $\tilde v$ according to the analysis above.
}

\subsection{The exciton London equations}

The interlayer exciton condensate is analogous to a superconductor in that its response obeys two London-like equations. However, its physical electromagnetic response differs from that of a charged superconductor. The broken symmetry is the relative $\mathrm U_{eh}(1)$ phase of the neutral bilayer condensate. Coupling the condensate to the layer-antisymmetric electromagnetic field therefore does not gap the Goldstone mode through the Anderson--Higgs mechanism. The Goldstone mode remains accessible to low-frequency probes in this channel. Because exciton resistance is difficult to measure directly, we propose waveguide-transmission and microwave-impedance probes of the nondissipative acceleration relation, together with a static magnetic probe of the exciton Meissner relation.

Consider a layer-antisymmetric gauge field $A^{-}_{\mu}(\omega,\bm{q})=(\phi^-(\omega,\bm{q}),\bm{A}^-(\omega,\bm{q}))$ applied to the bilayer EI system.
For simplicity, we choose $\bm{q}=q\hat{x}$, so the longitudinal and transverse directions are $x$ and $y$, respectively.
The induced exciton current is
\begin{subequations}
  \begin{align}
  i\omega j^{-}_{x}(\omega,\bm{q})=&i\omega[K^{-}_{10}(\omega,\bm{q})\phi^{-}(\omega,\bm{q})+K^{-}_{11}(\omega,\bm{q})A^{-}_{x}(\omega,\bm{q})]\nonumber\\
  =&K^{-}_{11}(\omega,q)\left[-iq\phi^{-}(\omega,\bm{q})+i\omega A^{-}_{x}(\omega,\bm{q}) \right]\\
  i\omega j^{-}_{y}(\omega,\bm{q})=&i\omega K^{-}_{22}(\omega,\bm{q})A^{-}_{y}(\omega,\bm{q})
\end{align}\label{eq:current-electrical-field}
\end{subequations}

For the exciton condensation phase,
in the long-wavelength limit, substituting Eq. \eqref{eq:exciton-current-correlation2} into Eq. \eqref{eq:current-electrical-field}, we obtain
\begin{subequations}
  \begin{align}
  \partial_{t}\bm{j}^{-}_{L}(t)=&\frac{e^2n_X}{m_X}[-\nabla \phi^{-}(t)-\partial_t \bm{A}^{-}_{L}(t)]\label{eq:London1}\\
  \bm{j}^{-}_{T}(t)=&-\frac{e^2 n_X}{m_X}\bm{A}^{-}_{T}(t)\label{eq:London2}
\end{align}\label{eq:London-like-equation}
\end{subequations}
These equations are analogous to the London equations in superconductors.
The first equation describes the nondissipative acceleration of the exciton condensate under a layer-antisymmetric electric field and can be interpreted as vanishing exciton resistance.
The second equation describes the exciton Meissner effect under an ``exciton gauge field''.

This non-dissipative feature is unique to the condensate phase of the EI and is absent in the normal exciton fluid.
For comparison, we consider a dilute exciton fluid without condensation and assume the excitons have a finite momentum relaxation time $\tau$ due to impurity scattering.
The response function in the long-wavelength limit can be derived as (see SM Appendix F for details\cite{SM}):
\begin{subequations}
  \begin{align}
  K^{-}_{11}(\omega,q)=&\frac{-\kappa v^2 \omega^2}{\omega(\omega+i\tau^{-1})-v^2 q^2},\\
  K^{-}_{22}(\omega,q)=&-\frac{e^2 n_X}{m_X}\frac{\omega}{\omega+i\tau^{-1}}\\
  K^-_{00}(\omega,q)=&\frac{q^2}{\omega^2}K^{-}_{11}(\omega,q)=-\frac{\kappa v^2 q^2}{\omega(\omega+i\tau^{-1})-v^2 q^2}.\label{eq:K-00-damped}
\end{align}\label{eq:response-scattering}
\end{subequations}
Note that in Eq. \eqref{eq:response-scattering}, $\tau$ has the real physical meaning of the exciton momentum relaxation time.
While in the condensation phase, the parameter $\eta$ introduced in Eq. \eqref{eq:K-00} is just an infinitesimal positive number to ensure causality.
In the long-wavelength limit, substituting Eq. \eqref{eq:response-scattering} into Eq. \eqref{eq:current-electrical-field} gives the induced exciton current in a normal exciton fluid:
\begin{equation}
  \bm{j}^{-}(\omega)=-\frac{e^2 n_X}{m_X}\frac{1}{i\omega-\tau^{-1}}[-\nabla \phi^{-}(t)+i\omega \bm{A}^{-}(t)](\omega),\label{eq:normal-exciton-response}
\end{equation}
which is a dissipative current similar to the Drude model in normal metals.

The fundamental difference between Eq. \eqref{eq:London-like-equation} and Eq. \eqref{eq:normal-exciton-response} lies in the different form of the response functions Eq. \eqref{eq:K-00} and Eq. \eqref{eq:K-00-damped}.
In the condensate phase, the response function exhibits a pole structure associated with the Goldstone mode, leading to a non-dissipative exciton current.
In contrast, in the normal exciton fluid, the corresponding phonon mode is damped due to scattering processes, resulting in a dissipative response.
Based on this distinction, we propose two experimental methods to detect the exciton superfluid effect in the bilayer EI systems.

\subsubsection{Detecting the exciton superfluid effect by waveguide transmission}\label{sec:waveguide}

\begin{figure}[h]
  \centering
  \includegraphics[width=\linewidth]{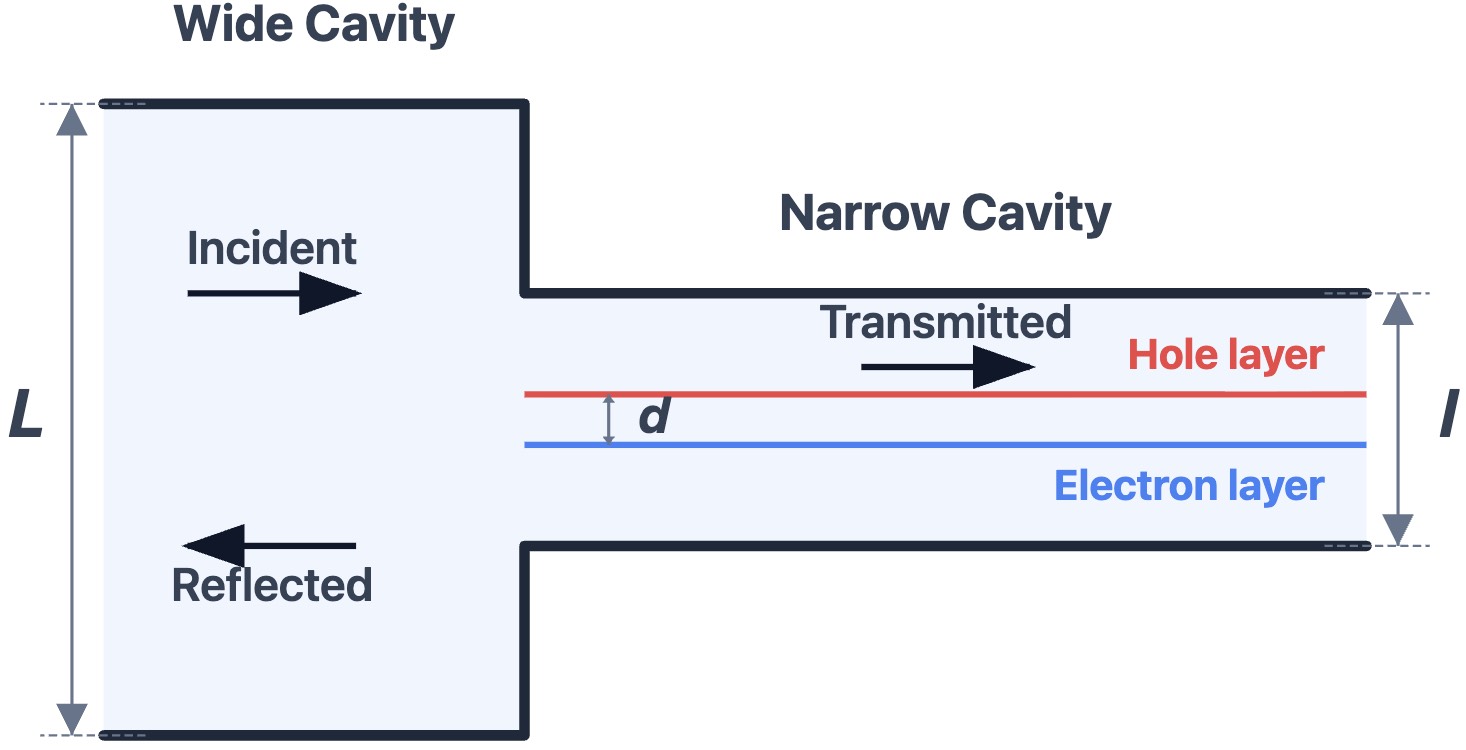}
  \caption{Schematic illustration of a bilayer EI sample embedded in a waveguide.
  In the EI phase, self-consistent coupling between the Goldstone response and the electromagnetic field produces a transverse-magnetic (TM) Goldstone-polariton branch below the cutoff frequency $c\pi/l$ of the empty narrow waveguide.
  In contrast, the corresponding TM mode in the normal exciton fluid is dissipative and has a complex wave vector.
  }
  \label{fig:waveguide_EI}
\end{figure}
The first method involves using a bilayer EI sample embedded in a waveguide, as illustrated in Fig. \ref{fig:waveguide_EI}.
The device is composed of a wide waveguide with width $L$ and a narrow waveguide with width $l$.
We assume that $L\gg l$, so the cutoff frequency of the wide waveguide, $\omega_{c}^{(L)}=c_L\pi/L$, is much smaller than that of the narrow waveguide, $\omega_c^{(R)}=c\pi/l$ ($c_L$ and $c$ are the light speeds in the wide and narrow cavities, respectively, and need not be equal).
For empty waveguides without bilayer EI sample, when the frequency of the incident mode $\omega$ satisfies $\omega_c^{(L)}<\omega<\omega_c^{(R)}$, no transmission occurs from the wide waveguide to the narrow waveguide.
However, when the bilayer EI sample (with interlayer distance $d\ll l$) is placed in the narrow waveguide, coupling the TDHF matter response self-consistently to Maxwell's equations produces a transverse-magnetic Goldstone polariton (see SM Appendix D.3\cite{SM}).
Its matter-dominated branch exists below the cutoff frequency of the empty narrow waveguide $\omega_c^{(R)}$.
In contrast, in the normal exciton fluid this TM mode is dissipative, with a complex wave vector and a finite attenuation length at nonzero frequency.
The decay length can be estimated from
\begin{equation}
  \omega(\omega+i\tau^{-1})=v^2 q^2\implies q= \frac{\omega}{v}\sqrt{1+\frac{i}{\omega\tau}},
\end{equation}
and is given by 
\begin{equation}
  \lambda_{\mathrm{decay}}=[\mathrm{Im}(q)]^{-1}=\frac{\sqrt{2}v}{\sqrt{\omega(\sqrt{\omega^2+\tau^{-2}}-\omega)}}.\label{eq:decay-length}
\end{equation}
For low frequency $\omega\ll \tau^{-1}$, the decay length reduces to $\lambda_{\mathrm{decay}}=\sqrt{2v^2\tau/\omega}$, which increases with decreasing frequency.
For high frequency $\omega\gg \tau^{-1}$, the decay length approaches a constant value $\lambda_{\mathrm{decay}}=2v\tau$.

{
Here we focus on the exciton condensation phase, where the Goldstone mode response is dissipationless.
Assume the wide cavity extends from $z=-L/2$ to $z=L/2$.
Because of the boundary conditions along the $z$ direction, the incident TM mode in the wide cavity will be labeled by the mode index $n$ with $B_y$ field distribution proportional to 
\begin{equation}
  \psi_n^{(L)}(z)=\sqrt{\frac{2}{L}}\cos\left[nk_c^{(L)}(z+L/2)\right],\;n=1,2,\cdots
\end{equation}
where $k_c^{(L)}=\pi/L$ is the cutoff wave vector of the wide waveguide.
In the $n$th branch, the cutoff frequency is $\omega_{n,c}^{(L)}=nc_L k_c^{(L)}$.
In the narrow cavity, the TM eigenmodes are no longer the bare cosine functions because the electromagnetic field hybridizes with the Goldstone response of the bilayer EI.
In SM Appendix D.4\cite{SM}, we solve the coupled Maxwell--matter problem and obtain the dressed TM eigenmodes in the narrow waveguide.
For an incident mode $n$ in the wide waveguide, we expand the reflected and transmitted fields in the eigenmodes of the wide and narrow waveguides and apply the interface boundary conditions. This procedure gives the transmission rate $\mathcal{T}_n(\omega)$ derived in SM Appendix D.5\cite{SM}.

For simplicity, we consider the most symmetric configuration where the narrow waveguide is located at $z\in[-l/2,l/2]$ and the bilayer EI sample is located at $z=\pm d/2$.
As the $B_y$ profile of the Goldstone mode is symmetric with respect to $z=0$, the incident TM mode with odd $n$ will not couple to the Goldstone mode, and thus the transmission rate $\mathcal{T}_n(\omega)$ for odd $n$ is zero.
Thus, in the most symmetric configuration, the lowest incident mode that can couple to the Goldstone-polariton branch is the $n=2$ TM mode, which has a cutoff frequency $\omega_{2,c}^{(L)}=2c_L\pi/L$.

For the typical parameters, we take $v\approx 0.6a_B^*\mathrm{Ry}^*/\hbar$ as indicated in Fig.~\ref{fig:exciton_responses}(a1).
The light speed in the narrow waveguide with $\epsilon=10$ is $c\approx 866.7a_B^*\mathrm{Ry}^*/\hbar$.
According to Fig.~\ref{fig:exciton_responses}(a1), when the in-plane momentum of the Goldstone mode is $q_xa_B^*\in[0,0.8]$, its frequency range is $\omega/\mathrm{Ry}^*\in[0,0.48]$.
To ensure that the cutoff frequency of the incident mode falls within the Goldstone-mode frequency range, we take $L=3000a_B^*$ and $c_L=c/\sqrt{40}$. Equivalently, the dielectric constant of the wide waveguide is $\epsilon_L=40\epsilon=400$. Relative permittivities of this order have been measured in SrTiO$_3$ at sub-THz frequencies\cite{dupasHighPermittivityProcessed2018}.
The cutoff frequency of the $n=2$ incident mode is then $\omega_{2,c}^{(L)}\approx 0.287\mathrm{Ry}^*\approx 2.08\mathrm{THz}$, within the Goldstone-mode frequency range.
By taking $l=12a_B^*\approx 29\mathrm{nm}$, we plot the transmission rate $\mathcal{T}_2(\omega)$ for the $n=2$ incident mode in Fig.~\ref{fig:waveguide-transmission}.

Near the cutoff, the transmission rate takes the form $\mathcal{T}_2\propto\kappa_2^{(L)}/|\kappa_2^{(L)}+\alpha_0|^2$, as derived in SM Appendix D.5(b)\cite{SM}. Here $\kappa_2^{(L)}\simeq c_L^{-1}\sqrt{2\omega_{2,c}^{(L)}\delta\omega}\propto\sqrt{\delta\omega}$ is the in-plane wave vector of the incident mode, and $\delta\omega=\omega-\omega_{2,c}^{(L)}$ is the detuning frequency. The transmission rate therefore vanishes as $\sqrt{\delta\omega}$ at the cutoff, reaches a maximum, and then decreases as $1/\sqrt{\delta\omega}$ with increasing detuning. Because this maximum is narrow, its numerical value on a finite frequency grid depends on the frequency resolution. At $\delta\omega=10\,\mathrm{kHz}$, the transmission rate is of order $10^{-6}$. The transmission is small and strongly frequency dependent, so this calculation should be viewed as a proof of principle; assessing experimental feasibility requires optimized mode coupling and an explicit treatment of propagation and material losses.
\begin{figure}
  \centering
  \includegraphics[width=0.8\linewidth]{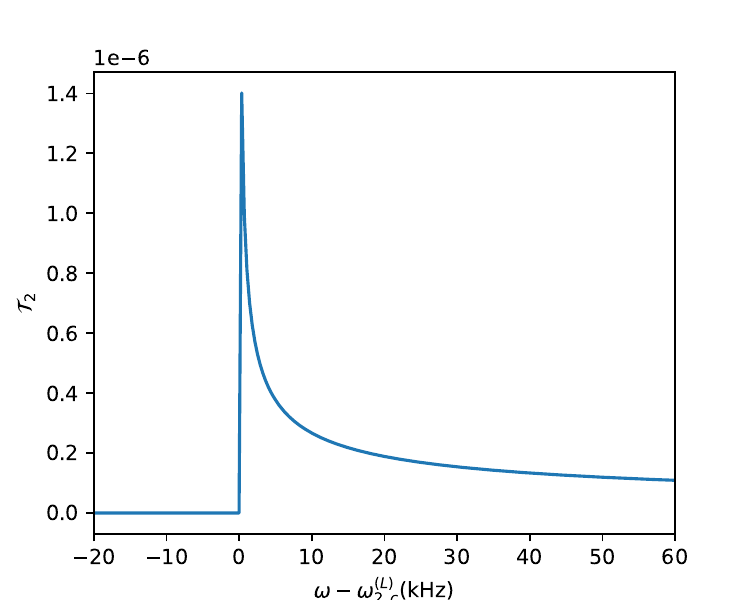}
  \caption{Transmission rate $\mathcal{T}_2(\omega)$ for the $n=2$ incident TM mode branch in the wide waveguide.
  The parameters are taken as $l=12a_B^*$, $\epsilon=10$, $L=3000a_B^*$, and $\epsilon_L=400$.
  The transmission rate is nonzero only for $\omega>\omega_{2,c}^{(L)}\approx 0.287\mathrm{Ry}^*=2.08\mathrm{THz}$, which is within the frequency range of the Goldstone mode.
  }
  \label{fig:waveguide-transmission}
\end{figure}
}

\subsubsection{Detecting the exciton superfluid effect by the microwave impedance microscopy}
\label{sec:mim}
Another method to detect the exciton superfluid effect is to use microwave impedance microscopy (MIM) to probe the local charge response of the bilayer EI system.
The experimental setup is illustrated in Fig. \ref{fig:bilayer_EI}.
Assume that the tip is located at $(\bm{r}_t,d_t)$ and the tip charge $Q_t(\omega)$ oscillates at the frequency $\omega$.
In momentum space, the Coulomb interaction between the tip and the two layers is given by $V_{ts}(\bm{q})=4\pi e^2 \re^{-q(d_t-d_s)}/[(1+\epsilon)q]$, where $d_e=-d/2$ and $d_h=d/2$ represent the positions of the electron and hole layers along the $z$-direction (see details in SM Appendix A\cite{SM}).
The real-space tip-layer interaction is then expressed as $V_{ts}(\bm{r})=\mathcal{V}^{-1}\sum_{\bm{q}}V_{ts}(\bm{q})\re^{i\bm{q}\cdot\bm{r}}$.
Define $\tilde{V}^{+}(\bm{r})=e^{-2}[V_{te}(\bm{r})+V_{th}(\bm{r})]/2$ and 
$\tilde{V}^{-}(\bm{r})=e^{-2}[V_{te}(\bm{r})-V_{th}(\bm{r})]$, then $\phi^{\sigma=\pm}(\omega,\bm{r})=\tilde{V}^{\sigma=\pm}(\bm{r}-\bm{r}_t)Q_{t}(\omega)$ are just the layer-symmetric ($\sigma=+$) and layer-antisymmetric ($\sigma=-$) scalar potential at point $\bm{r}$ induced by the tip charge.
The scalar potentials lead to total charge density fluctuation $\varrho^{+}(\omega,\bm{r})$ and exciton density fluctuation $\varrho^{-}(\omega,\bm{r})$ as
\begin{align}
  \delta\varrho^{\sigma}(\omega,\bm{r})
  =&-\int \rd\bm{r}'\;K^{\sigma}_{00}(\omega,\bm{r}-\bm{r}')\phi^{\sigma}(\omega,\bm{r}')\nonumber\\
  =&-\mathcal{V}^{-1}\sum_{\bm{q}}K_{00}^{\sigma}(\omega,\bm{q})\tilde{V}^{\sigma}(\bm{q})Q_{t}(\omega)\re^{i\bm{q}\cdot(\bm{r}-\bm{r}_t)}.
\end{align}
The charge density fluctuations in each layer can then be expressed in terms of $\delta\varrho^{\pm}(\omega,\bm{r})$ as $\delta\varrho_e(\omega,\bm{r})=\delta\varrho^{+}(\omega,\bm{r})/2+\delta\varrho^{-}(\omega,\bm{r})$ and $\delta\varrho_h(\omega,\bm{r})=\delta\varrho^{+}(\omega,\bm{r})/2-\delta\varrho^{-}(\omega,\bm{r})$,
which generate a potential feedback to the tip 
\begin{align}
  \delta U(\omega)
  =&\sum_{s=eh}\int\rd\bm{r}\; e^{-2}V_{ts}(\bm{r}_t-\bm{r})\delta\varrho_{s}(\omega,\bm{r})\nonumber\\
  =&\sum_{\sigma=\pm}\int\rd\bm{r}\; \tilde{V}^{\sigma}(\bm{r}_t-\bm{r})\delta\varrho^{\sigma}(\omega,\bm{r})\nonumber\\
  =&-\mathcal{V}^{-1}\sum_{\sigma=\pm,\bm{q}}[\tilde{V}^{\sigma}(\bm{q})]^{2}K^{\sigma}_{00}(\omega,\bm{q})Q_t(\omega).
\end{align}
Assuming the geometric capacitance between the tip and bilayer system is $C_t$, the total electrical potential at the tip is given by $U_{t}(\omega)=Q_t(\omega)/C_t+\delta U(\omega)$.
The admittance measured by MIM is $Y_t(\omega)\equiv I_{t}(\omega)/U_{t}(\omega)=-i\omega Q_t(\omega)/U_{t}(\omega)$.
Substituting for $U_{t}(\omega)$, we have
\begin{align}
  Y_t(\omega)=&-i\omega\left[C_{t}^{-1}-\mathcal{V}^{-1}\sum_{\sigma=\pm,\bm{q}}[\tilde{V}^{\sigma}(\bm{q})]^{2}K^{\sigma}_{00}(\omega,\bm{q})\right]^{-1}\nonumber\\
  \approx&-i\omega C_t\left[1+C_{t}\mathcal{V}^{-1}\sum_{\sigma=\pm,\bm{q}}[\tilde{V}^{\sigma}(\bm{q})]^{2}K^{\sigma}_{00}(\omega,\bm{q})\right].
\end{align}

Conventionally, when the distance between the tip and the bilayer sample is sufficiently large to prevent direct current tunneling, the tip and the sample form a capacitive load, resulting in a purely imaginary admittance. 
However, the electromagnetic responses of the sample introduce a quantum correction to the admittance, which includes both imaginary and real components:
\begin{subequations}
  \begin{align}
    \mathrm{Im}[\delta Y_t(\omega)]=&-\omega C_t^2\mathcal{V}^{-1}\sum_{\sigma=\pm,\bm{q}}[\tilde{V}^{\sigma}(\bm{q})]^{2}\mathrm{Re}[K^{\sigma}_{00}(\omega,\bm{q})],\\
    \mathrm{Re}[\delta Y_t(\omega)]=&\omega C_t^2\mathcal{V}^{-1}\sum_{\sigma=\pm,\bm{q}}[\tilde{V}^{\sigma}(\bm{q})]^{2}\mathrm{Im}[K^{\sigma}_{00}(\omega,\bm{q})].
  \end{align}
\end{subequations}
While the imaginary part of the correction can exist at any frequency, the real part becomes nonzero only when the collective modes are excited.
In MIM experiments, the operating frequency lies in the sub-THz range (below $1\mathrm{meV}$), which is lower than both the single-particle gap and the plasmon energy $\omega_0^{p}$ (on the order of $10\mathrm{meV}$ in the bilayer EI).
Consequently, only the gapless Goldstone mode can be excited at these frequencies, and it is the sole contributor to the real part of the admittance.
According to Eq. \eqref{eq:K-00}, in the long-wavelength and low frequency limit, the exciton response function takes the form:
\begin{equation}
  K^{-}_{00}(\omega,\bm{q})\approx
  \frac{\kappa v^2q^2}{v^2q^2-\omega^2}+\frac{i\pi\kappa vq}{2}[\delta(\omega-vq)-\delta(\omega+vq)].
\end{equation}
Additionally, the interaction between the tip charge and the exciton density fluctuation, $\tilde{V}^{-}(\bm{q})$, can also be approximated as 
\begin{equation}
  \tilde{V}^{-}(\bm{q})=\frac{4\pi \re^{-q d_t}(\re^{-qd/2}-\re^{qd/2})}{(1+\epsilon)q}\approx -\frac{4\pi d \re^{-q d_t}}{(1+\epsilon)}.
\end{equation}
Thus, the real admittance contributed by the Goldstone mode can be calculated as:
\begin{align}
  \mathrm{Re}[\delta Y_t(\omega)]=&\frac{\omega C_t^2(4\pi d)^2}{(1+\epsilon)^2\mathcal{V}}\sum_{\bm{q}}\re^{-2qd_t}\frac{\pi\kappa vq}{2}\delta(\omega-vq)\nonumber\\
  =&\frac{\omega C_t^2(4\pi d)^2\pi \kappa v}{2(1+\epsilon)^2}\int_0^{\infty}\frac{q\rd q}{2\pi}\;\re^{-2q d_t}q\delta(\omega-vq)\nonumber\\
  =&\frac{4\pi^2d^2C_t^2\kappa}{(1+\epsilon)^2v^2}\omega^3\re^{-2d_t\omega/v}.
\end{align}
From this expression, we can see that the linear dispersion Goldstone mode will yield a real admittance with cubic frequency dependence in MIM measurements. 
Furthermore, by tuning the tip-layer distance $d_t$, the velocity $v$ of the Goldstone mode can be determined from the exponential decay factor $\re^{-2d_t \omega/v}$.

For comparison, in a normal exciton fluid with finite momentum relaxation time $\tau$, the exciton response function is given by Eq. \eqref{eq:K-00-damped} and 
\begin{equation}
  \mathrm{Im}K^{-}_{00}(\omega,q)=\frac{\kappa v^2q^2 \omega\tau^3 }{\omega^2\tau^2+(v^2q^2\tau^2-\omega^2\tau^2)^2}.
\end{equation}
Then the real admittance contributed by the damped sound mode in a normal exciton fluid is calculated as
\begin{align}
  \mathrm{Re}[\delta Y_t(\omega)]\propto& \omega^2\int_0^{\infty}\rd q\;\frac{\re^{-2q d_t}q^3}{\omega^2\tau^2+(v^2q^2\tau^2-\omega^2\tau^2)^2}\nonumber\\
  \sim& \omega^2\left[2\gamma+\log(4d_t^2\omega/v^2\tau)\right]+\mathcal{O}(\omega^{5/2}),
\end{align}
where $\gamma$ is the Euler-Mascheroni constant.
We can clearly see that, in the normal exciton fluid phase, the real admittance exhibits a different frequency dependence compared to that in the condensate phase.

\subsubsection{Possible ways to detect the second London equation for exciton superfluid (DC Meissner effect)}

\begin{figure}[h]
  \centering
  \includegraphics[width=\linewidth]{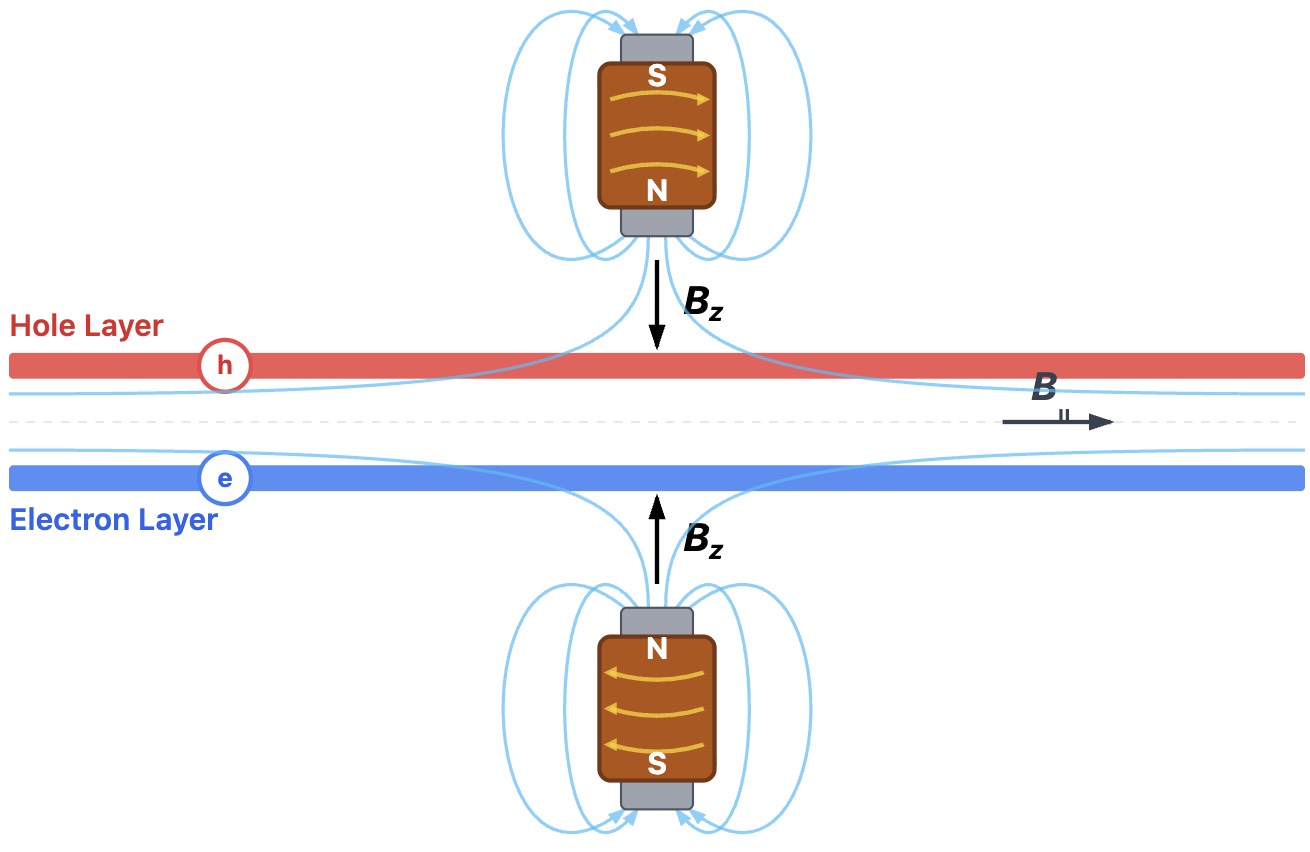}
  \caption{Schematic illustration of the exciton Meissner effect.
  The in-plane magnetic field $B_{\parallel}$ is applied by the coils surrounding the bilayer EI sample.
  The in-plane magnetic field is associated with a layer-antisymmetric out-of-plane magnetic field $B_{z}^{-}$, which is expelled from the bulk of the bilayer EI due to the exciton Meissner effect.
  }\label{fig:exciton-Meissner}
\end{figure}

The second London equation describes the static magnetic response associated with the exciton Meissner effect. 
As discussed above, the in-plane magnetic field $B_{\parallel}$ is directly related to the layer-antisymmetric gauge field $\bm{A}^{-}$, which directly couples to the exciton condensate and causes the exciton Meissner effect. 
Then it is easy to derive that the corresponding flux is just the spatial integration of the layer-antisymmetric $B_z^{-}$ as illustrated in Fig.\ref{fig:exciton-Meissner}. 
The effective layer-antisymmetric diamagnetic susceptibility contributed by the exciton condensate can then be detected by careful measurement of the change of layer-antisymmetric magnetic flux before and after the superfluid transition temperature.  
Analogous to the 2D superconductors, the effective 2D penetration depth (Pearl length) is\cite{pearlCURRENTDISTRIBUTIONSUPERCONDUCTING1964}
\begin{equation}
  \Lambda_P=\frac{m_X}{e^2 n_X}
\end{equation}
In the typical TMD bilayer system, the exciton density is about $n_X\sim 10^{11}\mathrm{cm}^{-2}$\cite{maStronglyCorrelatedExcitonic2021,nguyenPerfectCoulombDrag2025}.
Then the Pearl length is about $\Lambda_P\sim 0.1\mathrm{m}$, which is much larger than the typical sample size (about $10\mu m$).
Although the signal must be small, the state-of-the-art scanning SQUID technique can still have a chance to measure it directly.

\section{Results at finite perpendicular magnetic field}\label{sec:finite-B}

When a perpendicular magnetic field $\bm{B}=B\hat{z}$ is applied to the electron--hole bilayer, we work in the Landau-level (LL) basis.
The Hartree--Fock self-consistency equation was derived in Ref.~\onlinecite{shaoQuantumOscillationsExcitonic2024} and is detailed in SM Appendix E.2\cite{SM}.
The TDHF equation and the dynamical matrix in the LL basis are provided in SM Appendix E.3\cite{SM}.
The current operators and electromagnetic response functions in the LL basis are given in SM Appendices E.4 and E.5, respectively\cite{SM}.
In the following, the unit of magnetic field strength is chosen as $B^*=\hbar/[2(a_B^*)^2e]$, which is about $56.9\mathrm{T}$ for the parameters taken in last section.
Throughout this section, we present only the electron--hole-symmetric results ($\delta_m=0$) in the main text; numerical results for $\delta_m=0.2$ are provided in SM Appendix I\cite{SM}.

\subsection{The collective modes spectrum: magnetic roton}
\begin{figure}[h]
  \centering
  \includegraphics[width=\linewidth]{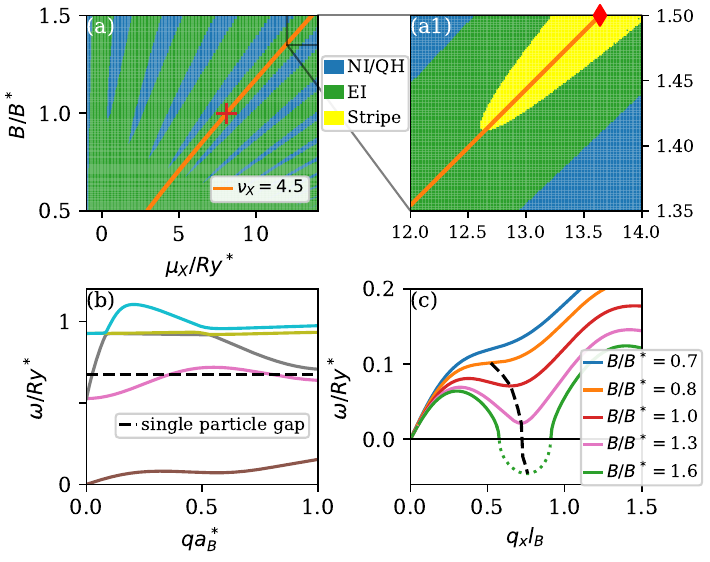}
  \caption{(a) Phase diagram of the electron--hole bilayer in a perpendicular magnetic field. The axes are the exciton chemical potential $\mu_X$ and magnetic field $B$, respectively. 
  The orange line represents the states where the LL filling in each layer is $\nu_X=4.5$.
  (b) Collective modes spectrum at $B/B^*=1$ and $\nu_X=4.5$ (the red cross in (a)). The Goldstone mode develops a magnetic roton minimum at finite momentum.
  (c) Goldstone mode dispersion at different magnetic field strengths along the line of $\nu_X=4.5$ (the orange line in (a)).
  The roton minimum is indicated by the black dashed line.
  (a1) Zoomed-in view of the phase diagram in the high-magnetic-field regime.
  A non-uniform stripe-ordered state emerges in the EI phase due to the softening of the roton minimum.}
  \label{fig:magnetic_roton}
\end{figure}

We first calculate the phase diagram of the electron--hole bilayer in a perpendicular magnetic field, as shown in Fig.~\ref{fig:magnetic_roton}(a).
The green area represents the excitonic insulator (EI) phase, while the blue area represents the normal phase without interlayer coherence.
For $\mu_X$ negative, the normal phase corresponds to the normal insulator (NI) phase, with no charge carriers in either layer.
For $\mu_X$ positive, the normal phase corresponds to the quantum Hall (QH) phase, where both layers host finite and opposite charge carriers occupying the hybridized electron and hole LLs.
Similar phase diagrams have also been reported in previous studies\citep{shaoQuantumOscillationsExcitonic2024,zouElectricalControlTwodimensional2024,nguyenQuantumOscillationsDipolar2026,qiCompetitionExcitonicInsulators2026}.

Define the LL filling factor in each layer as $\nu_X=2\pi l_B^2 n_X$, where $l_B=\sqrt{\hbar/eB}$ is the magnetic length and $n_X$ is the exciton density (charge number density per layer).
Then at $B/B^*=1$ and $\nu_X=4.5$ (marked by the red cross symbol in Fig. \ref{fig:magnetic_roton}(a)), the collective mode spectrum is calculated and shown in Fig. \ref{fig:magnetic_roton}(b).
Compared with the case without magnetic field shown in Fig. \ref{fig:clc_modes}(c)(d), there are two main differences.
First, the perpendicular magnetic field breaks reflection symmetry at finite momentum $\bm{q}$.
For example, in the absence of magnetic field, when $\bm{q}$ is along $x$ direction, the model has a reflection symmetry $M_y$ under $y\to-y$, and the collective modes can be classified by their mirror eigenvalues, as shown in Fig. \ref{fig:clc_modes}(c)(d).
However, when the magnetic field is applied, reflection symmetry is broken and the $M_y=\pm$ modes hybridize and form many gapped crossing points in the spectrum shown in Fig. \ref{fig:magnetic_roton}(b).
Second, the Goldstone mode develops a magnetic roton minimum at finite momentum.
To see this more clearly, we calculate the Goldstone mode dispersion at different magnetic field strengths along the line of $\nu_X=4.5$ (orange line in Fig. \ref{fig:magnetic_roton}(a)) and plot the results in Fig. \ref{fig:magnetic_roton}(c).
The roton minima are indicated by the black dashed line.
As the magnetic field strength increases, the roton minimum gradually softens.
This softening indicates an instability of the uniform EI toward a nonuniform stripe phase with finite-momentum electron--hole pairing at strong magnetic field.

Using the self-consistency equation for the stripe phase derived in SM Appendix E.2\cite{SM}, we compared the ground-state energies of the uniform and stripe EI phases. The region in which the stripe phase is energetically favorable is shown in the enlarged high-field portion of Fig.~\ref{fig:magnetic_roton}(a1).

\begin{figure}[h]
  \centering
  \includegraphics[width=\linewidth]{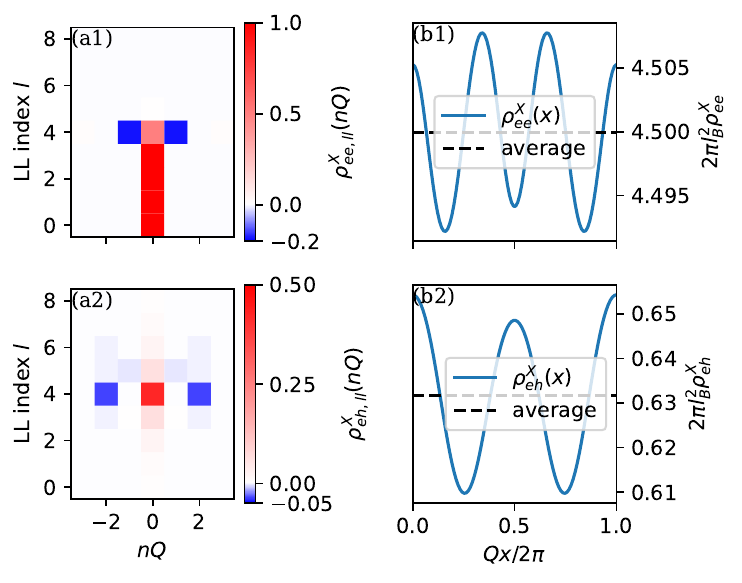}
  \caption{Density matrix of the stripe phase at $B/B^*=1.5$ and $\nu_X=4.5$ (the red diamond in Fig. \ref{fig:magnetic_roton}(a1)).
  (a1,a2) Density-matrix components $\rho^X_{ee,ll}(nQ)$ and $\rho^X_{eh,ll}(nQ)$ in the LL and momentum representations, respectively, where $n$ is an integer, $Q$ is the stripe-ordering wave vector, and $l$ is the LL index.
  (b1,b2) Corresponding real-space distributions of the exciton density $\rho^{X}_{ee}(\bm{r})$ and interlayer coherence $\rho^{X}_{eh}(\bm{r})$, respectively.
  }
  \label{fig:stripe}
\end{figure}

For the stripe phase, we retain only interlayer coherence between Landau levels with the same index and take the ordering wave vector $Q$ along the $x$ direction, corresponding to the dominant instability indicated by the roton minimum in Fig.~\ref{fig:magnetic_roton}(c).
The corresponding nonzero density matrix elements in the LL basis are
\begin{equation}
  \rho^X_{ss',ii'}(nQ)=\ave{l^{\dagger}_{s'i'k_x-nQ/2}l_{si k_x+nQ/2}}
\end{equation}
where $s,s'=e/h$ are the layer indices, $i,i'$ are the LL indices and $n$ is an integer.
Here, $l_{sik_x}$ denotes the annihilation operator for the state in layer $s$ with LL index $i$ and momentum $k_x$; the LL basis and density matrix are defined in SM Appendices E.1 and E.2\cite{SM}.
Taking $B/B^*=1.5$ and $\nu_X=4.5$ (the red diamond in Fig. \ref{fig:magnetic_roton}(a1)), we plot the density matrix of the stripe phase in Fig. \ref{fig:stripe}.
By comparing the ground-state energy for different values of $Q$, we obtain the optimal stripe-ordering wave vector $Ql_B\approx 0.75$, close to the roton-minimum momentum $q_x l_B\approx 0.7$ in Fig.~\ref{fig:magnetic_roton}(c).
In the stripe phase, the exciton density characterized by $\rho^X_{ee,ll}(nQ)$ exhibits a spatial modulation with wave vector $Q$, as shown in Fig.~\ref{fig:stripe}(b1).
This modulation is analogous to the stripe phase of a single-layer electron gas generated by the LL form factor\citep{foglerGroundStateTwodimensional1996}.
In addition to the density modulation, the solution retains a nonzero interlayer-coherence component, including a finite spatial average, as shown in Fig.~\ref{fig:stripe}(a2,b2).
Thus, within the present mean-field treatment, the stripe solution simultaneously breaks translation symmetry and the relative-layer $\mathrm U_{eh}(1)$ symmetry and can be identified as an exciton-supersolid candidate.

{A roton minimum can also occur at zero magnetic field. For $d/a_B^*=2$ and $n_X(a_B^*)^2=0.032$, we find a maxon--roton structure whose roton wave vector agrees with the shortest reciprocal-lattice vector $G_{\triangle}\propto\sqrt{n_X}$ of a triangular crystal at the same density. The corresponding numerical results are presented in SM Appendix H.5\cite{SM}.}

\subsection{The EM response functions}
In the presence of a perpendicular magnetic field, the paramagnetic current operators are modified to include the background vector potential and can be written in real space as
\begin{subequations}
  \begin{align}
    \hat{\bm{j}}^{+}_{p}(\bm{r})=&-\frac{e}{4m}\Psi^{\dagger}(\bm{r})\left(-i\hbar\nabla_{\bm{r}}+e\bm{A}^0\right)\sigma_z\Psi(\bm{r})+h.c.,\\
    \hat{\bm{j}}^{-}_{p}(\bm{r})=&-\frac{e}{8m}\Psi^{\dagger}(\bm{r})\left(-i\hbar\nabla_{\bm{r}}+e\bm{A}^0\right)\Psi(\bm{r})+h.c.,
  \end{align}
\end{subequations}
where $\bm{A}^0$ is the vector potential corresponding to the uniform perpendicular magnetic field $\bm{B}=\nabla\times \bm{A}^0$; the superscripts $+$ and $-$ of $\bm{j}^{\pm}$ denote the charge and exciton channels, respectively.

\begin{figure}
  \centering
  \includegraphics[width=\linewidth]{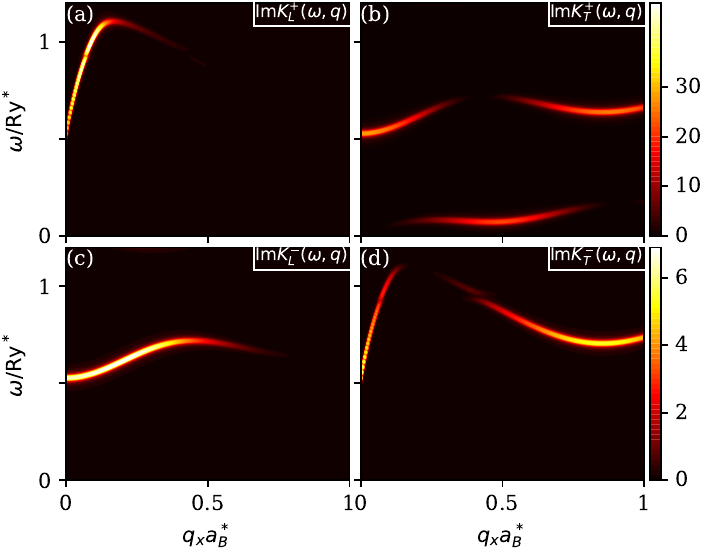}
  \caption{Diagonal response functions. (a)(b) are the longitudinal and transverse charge-current responses, and (c)(d) are the longitudinal and transverse exciton-current responses. The parameters are set to $B/B^*=1$ and $\nu_X=4.5$ (the red cross in Fig. \ref{fig:magnetic_roton}(a)).}
  \label{fig:symmetric_response_magnetic}
\end{figure}

\begin{figure}
  \centering
  \includegraphics[width=\linewidth]{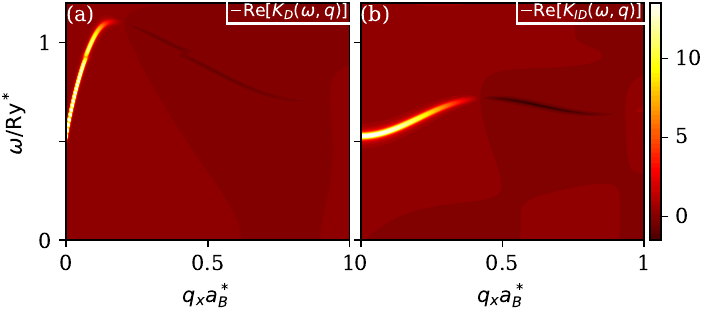}
  \caption{Mixed Hall response functions. (a) is the dipole Hall response and (b) is the inverse dipole Hall response. The parameters are set to $B/B^*=1$ and $\nu_X=4.5$ (the red cross in Fig. \ref{fig:magnetic_roton}(a)).}
  \label{fig:antisymmetric_response_magnetic}
\end{figure}

After transforming to the LL basis, we calculate the full response kernel $K^{\sigma\sigma'}_{\mu\nu}(\omega,\bm{q})$ as described in Sec.~\ref{sec:EM-response-function}; details are given in SM Appendices E.4 and E.5\cite{SM}. At $\delta_m=0$, the diagonal charge and exciton kernels retain the longitudinal--transverse decomposition in Eq.~\eqref{eq:symmetric-decomposition}. Unlike the zero-field case, however, the perpendicular magnetic field breaks time-reversal symmetry and generates a mixed response between the charge and exciton channels. As shown in SM Appendix E.5\cite{SM}, the mixed kernels satisfy $K^{+-}_{ab}=-K^{-+}_{ba}$. We define their exchange-antisymmetric combination as $K^{+-,A}_{ab}=[K^{+-}_{ab}-K^{-+}_{ba}]/2$. Rotational symmetry then gives the decomposition:
\begin{equation}
  \begin{aligned}
  K^{+-,A}_{ab}(\omega,\bm{q})={}&K_{D}(\omega,\bm{q})\frac{q_a\epsilon_{bc}q_c}{q^2}\\
  &-K_{ID}(\omega,\bm{q})\left(\epsilon_{ab}+\frac{q_a\epsilon_{bc}q_c}{q^2}\right)
  \end{aligned}\label{eq:Hall_response_function}
\end{equation}
where $\epsilon_{ab}$ is the Levi-Civita symbol in two dimensions. We refer to $K_D$ and $K_{ID}$ as the dipole and inverse dipole Hall response functions, respectively; their physical interpretation is discussed in Sec.~\ref{sec:hall}.
In summary, in the electron--hole-symmetric case, there will be six independent response functions in total: $K^{+}_{L}(\omega,\bm{q})$, $K^{+}_{T}(\omega,\bm{q})$, $K^{-}_{L}(\omega,\bm{q})$, $K^{-}_{T}(\omega,\bm{q})$, $K_{D}(\omega,\bm{q})$ and $K_{ID}(\omega,\bm{q})$.
In Figs.~\ref{fig:symmetric_response_magnetic} and \ref{fig:antisymmetric_response_magnetic}, we plot the diagonal charge/exciton responses and the mixed Hall responses at $B/B^*=1$ and $\nu_X=4.5$ (the red cross in Fig.~\ref{fig:magnetic_roton}(a)).

{
In the $q\to0$ limit, all six response functions have finite spectral weight at the $l_z=\pm1$ plasmon poles. At finite momentum, their dominant plasmon poles fall into two groups. The longitudinal charge response $K_L^+$, transverse exciton response $K_T^-$, and dipole Hall response $K_D$ follow the branch evolving from the longitudinal plasmon, whereas the transverse charge response $K_T^+$, longitudinal exciton response $K_L^-$, and inverse dipole Hall response $K_{ID}$ follow the branch evolving from the transverse plasmon. This grouping results from the coupling induced by the perpendicular magnetic field: longitudinal charge motion is coupled to transverse exciton motion, while transverse charge motion is coupled to longitudinal exciton motion. The mixed Hall response functions therefore probe the same collective modes as the diagonal responses rather than introducing additional modes.
}

{
At fixed reduced mass, electron--hole mass asymmetry splits opposite-angular-momentum collective modes at $q=0$ according to $\omega_{+|l|}-\omega_{-|l|}=|l|\delta_m\,\mathrm{Ry}^*B/B^*$. The numerical mode spectra are shown in SM Appendix I.2\cite{SM}. Mass asymmetry also activates four additional response functions: the charge and exciton Hall responses $K_H^{++}$ and $K_H^{--}$, where $K_H^{\sigma\sigma}\equiv[K_{xy}^{\sigma\sigma}-K_{yx}^{\sigma\sigma}]/2$, and the longitudinal and transverse components $K_L^{+-}$ and $K_T^{+-}$ of the exchange-symmetric mixed response $K^{+-,S}_{ab}\equiv[K^{+-}_{ab}+K^{-+}_{ba}]/2$. These four response functions are odd functions of $\delta_m$ and vanish in the electron--hole-symmetric case. Their numerical results are shown in SM Appendix I.3\cite{SM}.
}

\subsection{Dipole and inverse dipole Hall effect}\label{sec:hall}

\begin{figure}[h]
  \centering
  \includegraphics[width=\linewidth]{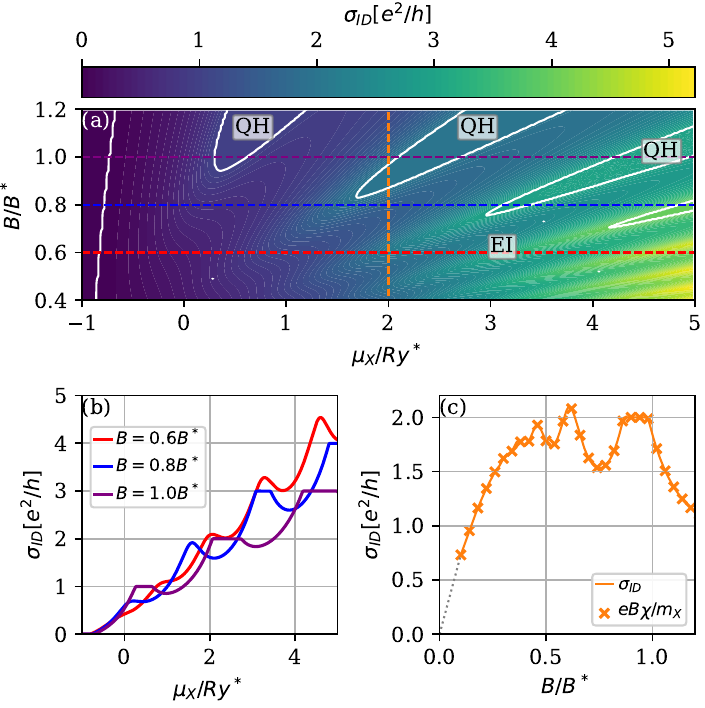}
  \caption{(a) The low-frequency linear-response (inverse) dipole Hall conductance as a function of exciton chemical potential $\mu_X$ and magnetic field strength $B$.
  The white line marks the boundary between the normal phase and the EI phase.
  (b) The same conductance as a function of exciton chemical potential $\mu_X$ at several fixed magnetic fields $B$ (the red, blue, and purple lines in (a)).
  Quantized plateaus appear in the QH phases.
  (c) Comparison of the low-frequency conductance derived from the Hall response function (solid lines) and from the semiclassical picture (crosses) at $\mu_X/\mathrm{Ry}^*=2$ (the orange dashed line in (a)).}
  \label{fig:dipole_Hall}
  
\end{figure}

We first discuss the physical meaning of the response functions $K_{D}(\omega,\bm{q})$ and $K_{ID}(\omega,\bm{q})$.
Assuming that the momentum is along the $x$ direction, $\bm{q}=(q,0)$, Eq. \eqref{eq:Hall_response_function} reduces to
\begin{equation}
  K^{-+}_{\mu\nu}=-K^{+-}_{\nu\mu}=\begin{bmatrix}
    0 & 0 & \frac{q}{\omega}K_{ID}\\
    0 & 0 & -K_{ID}\\
    -\frac{q}{\omega}K_{D} & K_{D} & 0
  \end{bmatrix}
\end{equation}
When a layer-symmetric gauge field $A^{+}_{\mu}(\omega,\bm{q})$ is applied, the corresponding electric field is $\bm{E}^{+}(\omega,\bm{q})=i[\omega \bm{A}^{+}(\omega,\bm{q})-\bm{q}\phi^{+}(\omega,\bm{q})]$, and the induced exciton current is
\begin{subequations}
  \begin{align}
  j^{-}_{y}(\omega,\bm{q})=&K_{D}(\omega,\bm{q})\left[-\frac{q}{\omega}\phi^{+}(\omega,\bm{q})+A^{+}_{x}(\omega,\bm{q})\right]\nonumber\\
  =&-\frac{i}{\omega}K_{D}(\omega,\bm{q})E^{+}_{x}(\omega,\bm{q}),\\
  j^{-}_{x}(\omega,\bm{q})=&-K_{ID}(\omega,\bm{q})A^{+}_{y}(\omega,\bm{q})\nonumber\\
  =&\frac{i}{\omega}K_{ID}(\omega,\bm{q})E^{+}_{y}(\omega,\bm{q}).
\end{align}\label{eq:dipole_Hall}
\end{subequations}
Similarly, when a layer-antisymmetric gauge field $A^{-}_{\mu}(\omega,\bm{q})$ is applied, the induced charge current is related to the layer-antisymmetric electric field by
\begin{subequations}
  \begin{align}
    j^{+}_y(\omega,\bm{q})=&-\frac{i}{\omega}K_{ID}(\omega,\bm{q})E^{-}_{x}(\omega,\bm{q}),\label{eq:inverse_dipole_Hall_1}\\
    j^{+}_x(\omega,\bm{q})=&\frac{i}{\omega}K_{D}(\omega,\bm{q})E^{-}_{y}(\omega,\bm{q}).
  \end{align}\label{eq:inverse_dipole_Hall}
\end{subequations}
For a longitudinal electric field along the $x$ direction, $K_D$ determines the transverse exciton current $j_y^-$ induced by $E_x^+$, whereas $K_{ID}$ determines the transverse charge current $j_y^+$ induced by $E_x^-$. We therefore refer to them as the dipole Hall and inverse dipole Hall response functions, respectively.
In the long-wavelength limit $\bm{q}\to 0$, we define the low-frequency linear-response dipole and inverse dipole Hall conductances by
\begin{subequations}
  \begin{align}
    \sigma_{D}=&\lim_{\omega\to 0}\lim_{\bm{q}\to 0}\frac{1}{i\omega}K_{D}(\omega,\bm{q}),\\
    \sigma_{ID}=&\lim_{\omega\to 0}\lim_{\bm{q}\to 0}\frac{1}{i\omega}K_{ID}(\omega,\bm{q}).
  \end{align}
\end{subequations}
Due to rotational symmetry, we must have $\sigma_{D}=\sigma_{ID}$.
In Fig.~\ref{fig:dipole_Hall}(a), we plot the low-frequency linear-response inverse dipole Hall conductance $\sigma_{ID}$ as a function of exciton chemical potential $\mu_X$ and magnetic field $B$.
The white line marks the boundary between the normal phase and the EI phase.
We also plot several line cuts of $\sigma_{ID}$ at fixed magnetic fields in Fig.~\ref{fig:dipole_Hall}(b).
In the quantum Hall (QH) phases, the nonzero low-frequency inverse dipole Hall conductance reflects the quantized Hall conductance of each layer.
The resulting plateaus, shown in Fig.~\ref{fig:dipole_Hall}(b), arise from the chiral edge states in each layer.
More notably, the low-frequency inverse dipole Hall conductance remains finite in the EI phase, although the bilayer is a bulk insulator without chiral edge states.
This response has the following semiclassical interpretation in terms of dissipationless exciton flow.

Consider an exciton with center-of-mass velocity $\bm{v}_{X}$.
The layer-antisymmetric electric field exerts the force $\bm{F}_{X}=-e\bm{E}^{-}$, giving the damped equation of motion
\begin{equation}
  \partial_t \bm{v}_{X}(t)=-\frac{e\bm{E}^{-}(t)}{m_X}-\frac{\bm{v}_{X}(t)}{\tau}
\end{equation}
where $m_X$ is the exciton mass and $\tau$ is the exciton momentum-relaxation time.
In frequency domain, the exciton velocity is given by
\begin{equation}
  \bm{v}_{X}(\omega)=\frac{-e\bm{E}^{-}(\omega)}{(-i\omega+\tau^{-1})m_X}.
\end{equation}
Because the electron and hole in an exciton carry opposite charges, they experience opposite Lorentz forces in a magnetic field along the $z$ axis:
$\bm{F}_{e}=-\bm{F}_h=-e\bm{v}_{X}\times\bm{B}$.
This Lorentz force acts on the relative motion of the electron--hole pair and produces the electric polarization
\begin{equation}
  \bm{p}_{X}(\omega)=\chi_{X}\bm{v}_{X}(\omega)\times\bm{B}=\frac{-e\chi_{X}\bm{E}^{-}(\omega)\times\bm{B}}{(-i\omega+\tau^{-1})m_X}.
\end{equation}
where $\chi_{X}$ is the electric susceptibility of a single exciton and is related to the exciton-fluid susceptibility $\chi$, defined by Eq.~\eqref{eq:susceptibility}, through $\chi_{X}=\chi/n_X$.
Finally, the total current contributed by all excitons is
\begin{equation}
  \bm{j}^{+}(\omega)=-i\omega n_X \bm{p}_{X}(\omega)=\frac{i\omega e \chi B}{(i\omega-\tau^{-1})m_X}\hat{z}\times\bm{E}^{-}(\omega).\label{eq:inverse_dipole_Hall_semi_classical}
\end{equation}

For a normal exciton fluid with finite momentum-relaxation time $\tau$, the inverse dipole Hall current vanishes as $\omega\to0$.
However, in the superfluid phase with infinite relaxation time $\tau\to \infty$, Eq. \eqref{eq:inverse_dipole_Hall_semi_classical} reduces to
\begin{equation}
  \bm{j}^{+}(\omega)=\frac{e\chi B}{m_X}\hat{z}\times\bm{E}^{-}(\omega)
\end{equation}
which gives a finite inverse dipole Hall conductance $\sigma_{ID}=eB\chi/m_X$. 
This finite value is the $\omega\to0$ limit of the linear-response coefficient, not a steady response to a finite static voltage, which would continuously accelerate the condensate. It should therefore be probed using a weak low-frequency AC field or a short pulse, with the induced counterflow kept below the critical current.
In Fig.~\ref{fig:dipole_Hall}(c), we compare the above semiclassical result with the value obtained from the full quantum-mechanical Kubo formula and find excellent agreement.


{
At fixed reduced mass, we also examine the effect of electron--hole mass asymmetry on the inverse dipole Hall response. In the weak-field limit, the susceptibility obeys $\chi(B,\delta_m)/\chi(B,0)=1+\mathcal{O}[\delta_m^2(B/B^*)^2]$. Using the semiclassical relation $\sigma_{ID}=eB\chi/m_X$ and $m_X=4m/(1-\delta_m^2)$, we then obtain $\sigma_{ID}(B,\delta_m)/\sigma_{ID}(B,0)=1-\delta_m^2+\mathcal{O}[\delta_m^2(B/B^*)^2]$. Our numerical results agree with this weak-field form and are shown in SM Appendix I.4\cite{SM}. For the representative magnitude $|\delta_m|=0.2$, the conductance is reduced by approximately $4\%$ from its equal-mass value in the weak-field regime. Electron--hole mass asymmetry therefore changes the response quantitatively, while the qualitative inverse dipole Hall signature remains robust.
}

\begin{figure}[h]
  \centering
  \includegraphics[width=\linewidth]{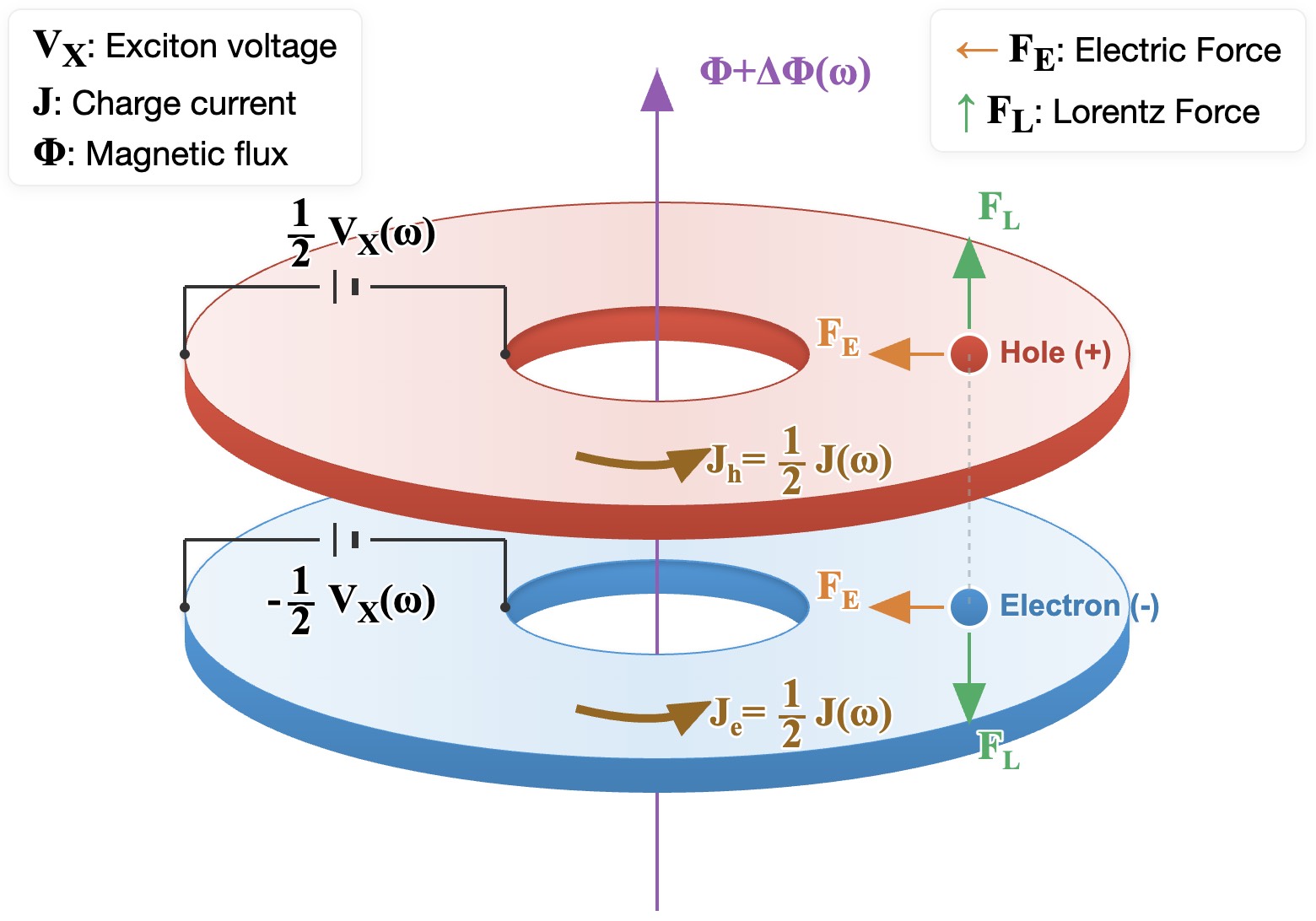}
  \caption{Schematic illustration of the Corbino geometry to measure the inverse dipole Hall effect in the bilayer EI system.
  The radial exciton voltage $V_X(\omega)$ induces a tangential charge current in each layer due to the inverse dipole Hall effect, resulting in a measurable flux change $\Delta\Phi(\omega)$ through the center of the Corbino disk.
  {In the condensate, the inverse dipole Hall conductance remains finite as $\omega\to0$ and can be probed using a weak low-frequency AC or pulsed voltage.}
  }
  \label{fig:corbino_inverse_exciton_Hall}
\end{figure}

Experimentally, the non-dissipative current in the inverse dipole Hall effect can generate magnetic flux, which can be detected by the Corbino geometry illustrated in Fig.~\ref{fig:corbino_inverse_exciton_Hall}.
{
In the Corbino setup, the two layers are driven with opposite ac voltages, $+V_X(\omega)/2$ and $-V_X(\omega)/2$, between their inner and outer contacts. The resulting radial layer-antisymmetric field induces an azimuthal charge current. The associated ac magnetic flux through the central aperture can be detected inductively using a pickup coil. The frequency dependence distinguishes the two phases: for a normal exciton fluid with finite momentum relaxation the inverse Hall response vanishes as $\omega\to0$, whereas in the condensate it approaches a nonzero constant.
}

\section{Conclusion}\label{sec:conclusion}

In this work, we have developed a unified microscopic theory of the linear electromagnetic response of bilayer excitonic insulators, treating charge and exciton degrees of freedom on equal footing within a time-dependent Hartree-Fock framework. By explicitly resolving the coupling to layer-symmetric and layer-antisymmetric gauge fields, we clarified how collective modes govern experimentally measurable response functions in both zero and finite magnetic fields.

At zero magnetic field, we showed that the long-wavelength charge response is dominated by gapped plasmon modes, while the exciton channel is controlled by a gapless Goldstone mode associated with spontaneous $\mathrm{U}_{eh}(1)$ symmetry breaking. From the Goldstone-mode-dominated response kernel, we derived London-like equations for the exciton condensate, demonstrating non-dissipative acceleration under a layer-antisymmetric electric field and a transverse exciton Meissner effect under an effective exciton gauge field. These results provide a sharp and operational definition of exciton superfluidity at the level of linear response, and clearly distinguish the condensate phase from a normal exciton fluid with dissipative, Drude-like dynamics.

In a perpendicular magnetic field, we uncovered two qualitatively new phenomena. First, the Goldstone mode develops a magnetic roton minimum at finite momentum, signaling an instability toward a stripe-ordered excitonic insulator at strong fields. Second, the magnetic field intrinsically couples charge and exciton motion, giving rise to mixed Hall responses: the dipole Hall effect and inverse dipole Hall effect, in which charge and exciton currents transmute into one another. {In the condensate, both mixed Hall conductances approach finite values in the $\omega\to0$ linear-response limit, reflecting the exciton superfluid stiffness.}

{Beyond establishing the fundamental response theory, we propose waveguide transmission and microwave impedance microscopy to probe the zero-field exciton superfluid response, and weak low-frequency or pulsed inverse dipole Hall measurements in Corbino geometries to probe the field-induced mixed Hall response. These measurements provide complementary signatures of exciton superfluidity in electron--hole bilayers and moir\'e heterostructures.}

Our results open several directions for future work. Extensions to finite temperature will allow one to address the fate of exciton superfluidity across the Berezinskii-Kosterlitz-Thouless transition and the role of vortex dynamics in electromagnetic response. Incorporating disorder, interlayer tunneling, and strong mass asymmetry will further refine the connection to realistic materials. More broadly, the framework developed here can be generalized to multicomponent condensates and topological excitonic phases, where the interplay between symmetry, topology, and electromagnetic response is expected to yield even richer collective phenomena.

Taken together, this work establishes a comprehensive and experimentally grounded theory of electromagnetic responses in bilayer excitonic insulators, providing clear benchmarks for identifying exciton superfluidity.

\begin{acknowledgments}
  This work was supported by a fellowship and a CRF award from the Research Grants Council of the Hong Kong Special Administrative Region, China (Projects No. HKUST SRFS2324-6S01 and No. C703722GF). We also acknowledge the support from the New Cornerstone Science Foundation.
\end{acknowledgments}


%
\clearpage
\onecolumngrid
\begin{center}
\textbf{\large Supplemental Material: Electromagnetic responses of bilayer excitonic insulators: from exciton London equations to dipole and inverse dipole Hall effects}
\end{center}

\setcounter{equation}{0}
\setcounter{figure}{0}
\setcounter{table}{0}
\setcounter{section}{0}
\setcounter{page}{1}
\makeatletter
\renewcommand{\theequation}{S\arabic{equation}}
\renewcommand{\thefigure}{S\arabic{figure}}
\appendix
\counterwithin{figure}{section}
\numberwithin{equation}{section}
\tableofcontents

\section{The Coulomb potentials between layers and tip}\label{app:Coulomb}

For a point charge at $(\bm{0},z_0)$, the Poisson equation of the electrical potential $\varphi(\bm{r},z;z_0)$ reads
\begin{equation}
  \epsilon(z)\nabla_{\bm{r}}^2\varphi(\bm{r},z;z_0)+\partial_{z}[\epsilon(z)\partial_{z}\varphi(\bm{r},z;z_0)]=-4\pi e\delta(\bm{r})\delta(z-z_0)
\end{equation}
where the dielectric profile is
\begin{equation}
  \epsilon(z)=
  \begin{cases}
    1, & |z|>d'/2,\\
    \epsilon, & |z|<d'/2.
  \end{cases}
\end{equation}
We define the two-dimensional Fourier transform of $\varphi(\bm{r},z;z_0)$ by
\begin{equation}
  \tilde{\varphi}(\bm{q},z;z_0)=\int d\bm{r}\;\varphi(\bm{r},z;z_0)e^{-i\bm{q}\cdot\bm{r}}
\end{equation}
The transformed potential satisfies
\begin{equation}
  \partial_z[\epsilon(z)\partial_z\tilde{\varphi}(\bm{q},z;z_0)]-q^2\epsilon(z)\tilde{\varphi}(\bm{q},z;z_0)=-4\pi e\delta(z-z_0)
\end{equation}
For $z_0\ne \pm d'/2$, the general solution is written as 
\begin{equation}
  \tilde{\varphi}(\bm{q},z;z_0)=\frac{2\pi e}{q}[c_1 e^{-q|z-d'/2|}+c_2 e^{-q|z+d'/2|}+\epsilon^{-1}(z_0)\re^{-q|z-z_0|}]
\end{equation}
The corresponding displacement field is
\begin{align}
  D_z(\bm{q},z;z_0)\equiv-\epsilon(z)\partial_z\tilde{\varphi}(\bm{q},z;z_0)=&2\pi e\epsilon(z)\left\{c_1[2\Theta(z-d'/2)-1] e^{-q|z-d'/2|}+c_2[2\Theta(z+d'/2)-1] e^{-q|z+d'/2|}\right.\nonumber\\
  &\left.+\epsilon^{-1}(z_0)[2\Theta(z-z_0)-1]e^{-q|z-z_0|}\right\}
\end{align}
Continuity of the displacement field at $z=\pm d'/2$ gives
\begin{align}
    &c_1+c_2 e^{-qd'}+\epsilon^{-1}(z_0)[2\Theta(d'/2-z_0)-1]e^{-q|d'/2-z_0|}\nonumber\\
    =&-\epsilon c_1+\epsilon c_2 e^{-qd'}+\epsilon \epsilon^{-1}(z_0)[2\Theta(d'/2-z_0)-1]e^{-q|d'/2-z_0|}\\
    &-\epsilon c_1 e^{-qd'}+\epsilon c_2+\epsilon\epsilon^{-1}(z_0)[2\Theta(-d'/2-z_0)-1]e^{-q|d'/2+z_0|}\nonumber\\
    =&-c_1e^{-q(d+2d')}-c_2+\epsilon^{-1}(z_0)[2\Theta(-d'/2-z_0)-1]e^{-q|d'/2+z_0|}
\end{align}
These boundary conditions reduce to
\begin{align}
  (\epsilon+1)c_1-(\epsilon-1)c_2e^{-qd'}=f(z_0)\\
  (\epsilon-1)c_1e^{-qd'}-(\epsilon+1)c_2=-f(-z_0)
\end{align}
where
\begin{equation}
  f(z_0)=(\epsilon-1) \epsilon^{-1}(z_0)[2\Theta(d'/2-z_0)-1]e^{-q|d'/2-z_0|}
\end{equation}
Solving for $c_1$ and $c_2$ gives
\begin{gather}
  c_1=\frac{(\epsilon+1)f(z_0)e^{qd'}+(\epsilon-1)f(-z_0)}{(\epsilon+1)^2e^{qd'}-(\epsilon-1)^2 e^{-qd'}}\\
  c_2=\frac{(\epsilon+1)f(-z_0)e^{qd'}+(\epsilon-1)f(z_0)}{(\epsilon+1)^2e^{qd'}-(\epsilon-1)^2 e^{-qd'}}
\end{gather}
\begin{itemize}
  \item \textit{The intralayer potential}: Set $z=z_0=d/2$. Then
  \begin{gather}
    f(z_0)=f(d/2)=(\epsilon-1)\epsilon^{-1}e^{-q(d'-d)/2}\\
    f(-z_0)=f(-d/2)=(\epsilon-1)\epsilon^{-1}e^{-q(d'+d)/2}
  \end{gather}
  and 
  \begin{gather}
    c_1=\frac{\epsilon-1}{\epsilon}\frac{(\epsilon+1)e^{q(d'+d)/2}+(\epsilon-1)e^{-q(d'+d)/2}}{(\epsilon+1)^2e^{qd'}-(\epsilon-1)^2 e^{-qd'}}\\
    c_2=\frac{\epsilon-1}{\epsilon}\frac{(\epsilon+1)e^{q(d'-d)/2}+(\epsilon-1)e^{-q(d'-d)/2}}{(\epsilon+1)^2e^{qd'}-(\epsilon-1)^2 e^{-qd'}}
  \end{gather}
  Thus the intralayer interaction is 
  \begin{align}
    V_{ee}(\bm{q})\equiv e\tilde{\varphi}(\bm{q},d/2;d/2)=&\frac{2\pi e^2}{q}[c_1e^{-q(d'-d)/2}+c_2e^{-q(d'+d)/2}+\epsilon^{-1}]\nonumber\\
    =&\frac{2\pi e^2}{\epsilon q}\left[1+2(\epsilon-1)\frac{(\epsilon+1)\cosh(qd)+(\epsilon-1)e^{-qd'}}{(\epsilon+1)^2e^{qd'}-(\epsilon-1)^2 e^{-qd'}} \right]
  \end{align}
  \item \textit{The interlayer potential}: Set $z_0=d/2$ and $z=-d/2$:
  \begin{align}
    V_{he}(\bm{q})\equiv e\tilde{\varphi}(\bm{q},-d/2;d/2)=&\frac{2\pi e^2}{q}[c_1e^{-q(d'+d)/2}+c_2e^{-q(d'-d)/2}+\epsilon^{-1}e^{-qd}]\nonumber\\
    =&\frac{2\pi e^2}{\epsilon q}\left[e^{-qd}+2(\epsilon-1)\frac{(\epsilon+1)+(\epsilon-1)e^{-qd'}\cosh(qd)}{(\epsilon+1)^2e^{qd'}-(\epsilon-1)^2 e^{-qd'}} \right]
  \end{align}
  \item \textit{Tip potential induced by layer charge}: The potentials at the tip due to point charges in the electron and hole layers are, respectively,
  \begin{align}
    V_{th}(\bm{q})\equiv e\tilde{\varphi}(\bm{q},d_t;d/2)=&\frac{2\pi e^2}{q}[c_1e^{-q(d_t-d'/2)}+c_2e^{-q(d_t+d'/2)}+\epsilon^{-1}e^{-q(d_t-d/2)}]\nonumber\\
    =&\frac{2\pi e^2 e^{-q(d_t-d/2)}}{\epsilon q}\left[1+(\epsilon-1)\frac{(\epsilon+1) e^{qd'}+(\epsilon-1)e^{-qd'}+2\epsilon e^{-qd}}{(\epsilon+1)^2e^{qd'}-(\epsilon-1)^2 e^{-qd'}} \right]\\
    V_{te}(\bm{q})\equiv e\tilde{\varphi}(\bm{q},d_t;-d/2)&=e\tilde{\varphi}(\bm{q},-d_t;d/2)=\frac{2\pi e^2}{q}[c_1e^{-q(d_t+d'/2)}+c_2e^{-q(d_t-d'/2)}+\epsilon^{-1}e^{-q(d_t+d/2)}]\nonumber\\
    =&\frac{2\pi e^2 e^{-q(d_t+d/2)}}{\epsilon q}\left[1+(\epsilon-1)\frac{(\epsilon+1) e^{qd'}+(\epsilon-1)e^{-qd'}+2\epsilon e^{-qd}}{(\epsilon+1)^2e^{qd'}-(\epsilon-1)^2 e^{-qd'}} \right]\\
  \end{align}

\end{itemize}

For a dielectric slab much thicker than the layer separation, $d'\gg d$, the interactions reduce to
\begin{align}
  V_{ee}(\bm{q})=V_{hh}(\bm{q})\approx&\frac{2\pi e^2}{\epsilon q}\\
  V_{eh}(\bm{q})=V_{he}(\bm{q})\approx&\frac{2\pi e^2\re^{-qd}}{\epsilon q}\\
  V_{th}(\bm{q})=V_{ht}(\bm{q})\approx&\frac{2\pi e^2\re^{-q(d_t-d/2)}}{\epsilon q}\left(1+\frac{\epsilon-1}{\epsilon+1}\right)=\frac{4\pi e^2\re^{-q(d_t-d/2)}}{(\epsilon+1) q}\\
  V_{te}(\bm{q})=V_{et}(\bm{q})\approx&\frac{2\pi e^2\re^{-q(d_t+d/2)}}{\epsilon q}\left(1+\frac{\epsilon-1}{\epsilon+1}\right)=\frac{4\pi e^2\re^{-q(d_t+d/2)}}{(\epsilon+1) q}
\end{align}

\section{The time-dependent Hartree-Fock method: general formulation}
\subsection{Dynamics equation of the density matrix}\label{app:tdhf}
Consider the many-body Hamiltonian
\begin{equation}
  \hat{H}=\sum_{ij}[h^0_{ij}+o_{ij}f(t)]c^{\dagger}_ic_j+\frac{1}{2}\sum_{ijkl}V_{ijkl}c^{\dagger}_{i}c^{\dagger}_jc_{l}c_{k}
\end{equation}
where matrix elements are defined as
\begin{equation}
  h^0_{ij}\equiv \bra{i}h^0\ket{j}, \;o_{ij}\equiv \bra{i}o\ket{j},\; V_{ijkl}\equiv\bra{i,j}V\ket{k,l}
\end{equation}
The Hermiticity of the Hamiltonian requires that the matrix elements satisfy the following relations:
\begin{equation}
  h^0_{ij}=(h_{ji}^0)^*,\;o_{ij}=o_{ji}^*,\;V_{ijkl}=V_{klij}^*
\end{equation}
The fermionic anticommutation relations also imply
\begin{equation}
  V_{ijkl}=V_{jilk}
\end{equation}

The equation of motion of the density matrix $\rho_{mn}=\ave{c^{\dagger}_nc_m}$ is written as 
\begin{align}
  i\hbar\partial_t\rho_{mn}=&\ave{[c^{\dagger}_nc_m,H]}\nonumber\\
  =&\ave{c^{\dagger}_n[c_m,H]}-\ave{[H,c^{\dagger}_n]c_m}\nonumber\\
  =&[h^0_{mj}+o_{mj}f(t)]\ave{c^{\dagger}_nc_j}+\frac{1}{2}V_{mjkl}\ave{c^{\dagger}_nc^{\dagger}_jc_lc_k}-\frac{1}{2}V_{imkl}\ave{c^{\dagger}_nc^{\dagger}_ic_lc_k}\nonumber\\
  &-[h^0_{in}+o_{in}f(t)]\ave{c^{\dagger}_ic_m}-\frac{1}{2}V_{ijnl}\ave{c^{\dagger}_ic^{\dagger}_jc_lc_m}+\frac{1}{2}V_{ijkn}\ave{c^{\dagger}_ic^{\dagger}_jc_kc_m}\nonumber\\
  =&[h^0_{mj}+o_{mj}f(t)]\ave{c^{\dagger}_nc_j}+V_{mikl}\ave{c^{\dagger}_nc^{\dagger}_ic_lc_k}-[h^0_{in}+o_{in}f(t)]\ave{c^{\dagger}_ic_m}-V_{ijnl}\ave{c^{\dagger}_ic^{\dagger}_jc_lc_m}
\end{align}
We use Einstein summation for repeated indices throughout this appendix. Within the TDHF approximation, $\ave{c^{\dagger}_ic^{\dagger}_jc_lc_k}\approx \rho_{ki}\rho_{lj}-\rho_{kj}\rho_{li}$, so
\begin{align}
  i\hbar\partial_t\rho_{mn}=&[h^0_{mj}+o_{mj}f(t)]\rho_{jn}+V_{mikl}\rho_{li}\rho_{kn}-V_{mikl}\rho_{ki}\rho_{ln}\nonumber\\
  &-\rho_{mi}[h^0_{in}+o_{in}f(t)]-\rho_{mi}V_{ijnl}\rho_{lj}+\rho_{mj}V_{ijnl}\rho_{li}
\end{align}
Define the Hartree and Fock Hamiltonians by
\begin{equation}
  h^{H}_{ik}=V_{ijkl}\rho_{lj},\;h^{F}_{il}=-V_{ijkl}\rho_{kj}
\end{equation}
The equation of motion becomes
\begin{equation}
  i\hbar\partial_t\rho_{mn}=[h^0+h^H+h^F+of(t),\rho]_{mn}
\end{equation}

We expand the density matrix in powers of $f(t)$:
\begin{equation}
  \rho=\sum_{n}\rho^{(n)}
\end{equation}
where $\rho^{(n)}$ is of order $n$ in $f(t)$. The zeroth-order equation is
\begin{equation}
  i\hbar\partial_t\rho^{(0)}_{mn}=[h^0+h^H+h^F,\rho^{(0)}]_{mn}.
\end{equation}
The static Hartree--Fock state satisfies
\begin{subequations}
  \begin{gather}
    \rho^{(0)}=\sum_{v}\ket{v}\bra{v}\\
    (h^0+h^H+h^F)\ket{v}=\xi_{v}\ket{v},\;(h^0+h^H+h^F)\ket{c}=\xi_{c}\ket{c},\xi_{c}>\mu>\xi_v\label{eq:static-hf}
  \end{gather}
\end{subequations}
where $\mu$ is the chemical potential.
Because the density matrix of a pure state satisfies $\rho^2=\rho$, its first-order expansion gives
\begin{equation}
  [\rho^{(0)}]^2+\rho^{(0)}\rho^{(1)}+\rho^{(1)}\rho^{(0)}=\rho^{(0)}+\rho^{(1)}\implies \rho^{(0)}\rho^{(1)}+\rho^{(1)}\rho^{(0)}-\rho^{(1)}=0
\end{equation}
Taking matrix elements between the unoccupied states $\ket{c}$, $\ket{c'}$ and the occupied states $\ket{v}$, $\ket{v'}$ gives
\begin{subequations}
  \begin{gather}
    \bra{c}[\rho^{(0)}\rho^{(1)}+\rho^{(1)}\rho^{(0)}-\rho^{(1)}]\ket{c'}=-\rho^{(1)}_{cc'}=0,\\
    \bra{v}[\rho^{(0)}\rho^{(1)}+\rho^{(1)}\rho^{(0)}-\rho^{(1)}]\ket{v'}=\rho^{(1)}_{vv'}=0.\label{eq:rho-cc'-vv'}
  \end{gather}
\end{subequations}
Thus, only matrix elements connecting states with different occupations contribute at first order in $f(t)$. The first-order TDHF equation for $\rho^{(1)}_{cv}$ is
\begin{align}
  i\hbar\partial_t\rho_{cv}^{(1)}=&[h^0+h^{H}(\rho^{(0)})+h^{F}(\rho^{(0)}),\rho^{(1)}]_{cv}+f(t)[o,\rho^{(0)}]_{cv}+[h^{H}(\rho^{(1)})+h^{F}(\rho^{(1)}),\rho^{(0)}]_{cv}\nonumber\\
  =&\bra{c}[h^0+h^{H}(\rho^{(0)})+h^{F}(\rho^{(0)}),\rho^{(1)}]+f(t)[o,\rho^{(0)}]+[h^{H}(\rho^{(1)})+h^{F}(\rho^{(1)}),\rho^{0}]\ket{v}\nonumber\\
  =&(\xi_c-\xi_v)\rho^{(1)}_{cv}+o_{cv}f(t)+\bra{c}h^{H}(\rho^{(1)})+h^{F}(\rho^{(1)})\ket{v}\nonumber\\
  =&(\xi_c-\xi_v)\rho^{(1)}_{cv}+V_{cv'vc'}\rho^{(1)}_{c'v'}+V_{cc'vv'}\rho^{(1)}_{v'c'}-V_{cv'c'v}\rho^{(1)}_{c'v'}-V_{cc'v'v}\rho^{(1)}_{v'c'}+o_{cv}f(t)\nonumber\\
  =&\mathcal{E}_{cv,c'v'}\rho^{(1)}_{c'v'}+\Gamma_{cv,c'v'}\rho^{(1)}_{v'c'}+o_{cv}f(t)\label{eq:dynamic_cv}
\end{align}
where the dynamical-matrix elements are defined as
\begin{subequations}
  \begin{gather}
    \mathcal{E}_{cv,c'v'}=\delta_{(cv),(c'v')}(\xi_c-\xi_v)+V_{cv'vc'}-V_{cv'c'v}\\
    \Gamma_{cv,c'v'}=V_{cc'vv'}-V_{cc'v'v}
  \end{gather}
\end{subequations}
where $\delta_{(cv),(c'v')}\equiv\delta_{cc'}\delta_{vv'}$.
Hermiticity gives
\begin{subequations}
  \begin{gather}
    \mathcal{E}_{c'v',cv}=\delta_{(cv),(c'v')}(\xi_c-\xi_v)+V_{c'vv'c}-V_{c'vcv'}=\mathcal{E}^*_{cv,c'v'}\\
    \Gamma_{c'v',cv}=V_{c'cv'v}-V_{c'cvv'}=V_{cc'vv'}-V_{cc'v'v}=\Gamma_{cv,c'v'}
  \end{gather}
\end{subequations}
Taking the complex conjugate of Eq. \eqref{eq:dynamic_cv} gives
\begin{align}
  -i\hbar\partial_t\rho^{(1)}_{vc}=&(\xi_c-\xi_v)\rho^{(1)}_{vc}+V_{vc'cv'}\rho^{(1)}_{v'c'}+V_{vv'cc'}\rho^{(1)}_{c'v'}-V_{c'vcv'}\rho^{(1)}_{v'c'}-V_{v'vcc'}\rho^{(1)}_{c'v'}+o_{vc}f(t)\nonumber\\
  =&\mathcal{E}^*_{cv,c'v'}\rho^{(1)}_{v'c'}+\Gamma^*_{cv,c'v'}\rho^{(1)}_{c'v'}+o_{vc}f(t)\label{eq:dynamic_vc}
\end{align}
After Fourier transformation, Eqs. \eqref{eq:dynamic_cv} and \eqref{eq:dynamic_vc} take the compact form
\begin{equation}
  \hbar\omega^+\tau_z\begin{bmatrix}
    \rho^{(1)}_{cv}(\omega)\\\rho^{(1)}_{vc}(\omega)
  \end{bmatrix}=\mathcal{H}_{cv,c'v'}\begin{bmatrix}
    \rho^{(1)}_{c'v'}(\omega) \\ \rho^{(1)}_{v'c'}(\omega) 
  \end{bmatrix}+\begin{bmatrix}
    o_{cv}\\o_{vc}
  \end{bmatrix}f(\omega)\label{eq:dynamic-equation}
\end{equation}
where $\tau_z$ is the Pauli matrix, $\omega^{+}\equiv \omega+i\eta$ and $\eta$ is a small positive number to account for the retarded effect.
Here $\mathcal{H}$ denotes the dynamical matrix
\begin{equation}
  \mathcal{H}_{cv,c'v'}=\begin{bmatrix}
    \mathcal{E}_{cv,c'v'} & \Gamma_{cv,c'v'}\\
    \Gamma^*_{cv,c'v'} & \mathcal{E}^*_{cv,c'v'}
  \end{bmatrix}
\end{equation}
Then Eq. \eqref{eq:dynamic-equation} can be formally solved by
\begin{equation}
  \begin{bmatrix}
    \rho^{(1)}_{cv}\\\rho^{(1)}_{vc}
  \end{bmatrix}=\left(\hbar\omega^+\tau_z-\mathcal{H}_{cv,c'v'}\right)^{-1}\begin{bmatrix}
    o_{c'v'}\\o_{v'c'}
  \end{bmatrix}f(\omega)\label{eq:dynamic-solution}
\end{equation}

\subsection{Feynman diagrammatic representation}\label{app:feynman-diagram}
\begin{figure}[h]
  \centering
  \includegraphics[width=0.8\textwidth]{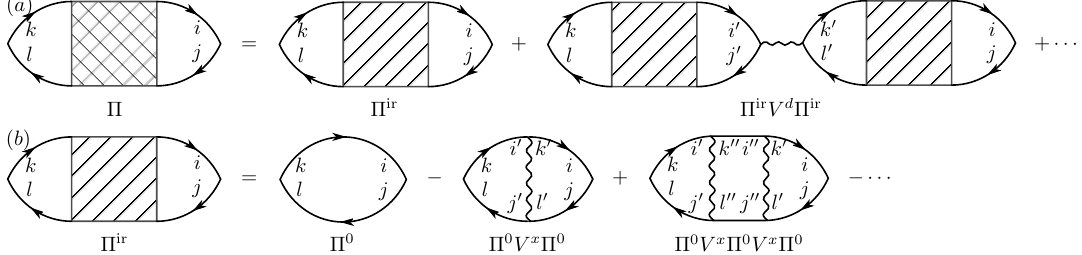}
  \caption{Feynman diagrammatic representation of the TDHF approximation.}
  \label{fig:feynman-polarization}
\end{figure}

The TDHF approximation can also be represented diagrammatically, as shown in Fig. \ref{fig:feynman-polarization}.
Consider the time-ordered two-particle correlation function defined as 
\begin{equation}
  \Pi_{ij,kl}(\tau)\equiv-\ave{T c^{\dagger}_j(\tau)c_{i}(\tau)c^{\dagger}_{k}(0)c_{l}(0)}
\end{equation}
where $T$ is the time-ordering operator and $\tau$ is the imaginary time.
Its Fourier transform is
\begin{equation}
  \Pi_{ij,kl}(i\nu_n)\equiv\int_0^{\beta}\rd\tau\;\re^{i\nu_n\tau}\Pi_{ij,kl}(\tau)
\end{equation}
where $\nu_n=2\pi n/\beta$ is the Matsubara frequency and $\beta=1/k_BT$ is the inverse temperature.
As shown in Fig. \ref{fig:feynman-polarization}(a), the two-particle correlation function $\Pi$ has the series expansion
\begin{equation}
  \Pi=\Pi^{\mathrm{ir}}+\Pi^{\mathrm{ir}}V^{d}\Pi^{\mathrm{ir}}+\Pi^{\mathrm{ir}}V^{d}\Pi^{\mathrm{ir}}V^{d}\Pi^{\mathrm{ir}}+\cdots=\Pi^{\mathrm{ir}}(1+V^{d}\Pi)=(1+\Pi V^{d})\Pi^{\mathrm{ir}}\label{eq:polarization-irreducible}
\end{equation}
where $V^d$ is the direct interaction such that
\begin{equation}
  V^{d}_{k'l',i'j'}=\bra{k',j'}V\ket{l',i'}=V_{k'j'l'i'}
\end{equation}
Within TDHF, the irreducible two-particle correlation function $\Pi^{\mathrm{ir}}$ is obtained by summing the ladder diagrams in Fig. \ref{fig:feynman-polarization}(b):
\begin{equation}
  \Pi^{\mathrm{ir}}=\Pi^0-\Pi^0V^{x}\Pi^0+\Pi^0V^{x}\Pi^0V^{x}\Pi^0-\cdots=\Pi^0(1-V^x\Pi^{\mathrm{ir}})=(1-\Pi^{\mathrm{ir}}V^{x})\Pi^0\label{eq:polarization-irreducible-ladder}
\end{equation}
where $V^x$ is the exchange interaction such that
\begin{equation}
  V^{x}_{k'l',i'j'}=\bra{k',j'}V\ket{i',l'}=V_{k'j'i'l'}
\end{equation}
and $\Pi^0$ is the two-particle bubble.

Under the Hartree-Fock approximation, the single-particle Green function $\mathcal{G}$ is
\begin{equation}
  \mathcal{G}_{ij}(i\omega_m)=\frac{\delta_{ij}}{i\omega_m-\xi_i}
\end{equation}
where $\xi_i$ is the single-particle energy given by Eq. \ref{eq:static-hf}, and $\omega_m=(2m+1)\pi/\beta$ is the fermionic Matsubara frequency.
The bubble diagram $\Pi^0$ is
\begin{align}
  \Pi^0_{ij,kl}(i\nu_n)=&\frac{1}{\beta}\sum_{i\omega_m}\mathcal{G}_{ik}(i\omega_m+i\nu_n)\mathcal{G}_{lj}(i\omega_m)\nonumber\\
  =&\frac{1}{\beta}\sum_{i\omega_m}\frac{\delta_{(ij)(kl)}}{(i\omega_m+i\nu_n-\xi_i)(i\omega_m-\xi_j)}\nonumber\\
  =&\delta_{(ij)(kl)}\frac{n_F(\xi_j)-n_F(\xi_i-i\nu_n)}{i\nu_n-\xi_i+\xi_j}\nonumber\\
  =&\frac{\delta_{(ij)(kl)}f_{ji}}{i\nu_n-(\xi_i-\xi_j)}.
\end{align}
In the above equation, $\delta_{(ij)(kl)}\equiv \delta_{ik}\delta_{jl}$, $f_{ij}\equiv f_i-f_j$ and  $f_i\equiv n_F(\xi)=1/[1+\exp(\beta\xi)]$ is the occupation number of the state $\ket{i}$.
At zero temperature, we have $f_{cv}=-f_{vc}=-1$ and $f_{cc'}=f_{vv'}=0$.
Thus, $\Pi^0_{ij,kl}\ne 0$ only when the index pairs $(ij)$ and $(kl)$ are taken from $(cv)$ or $(vc)$.
According to Eq. \eqref{eq:polarization-irreducible} and \eqref{eq:polarization-irreducible-ladder}, this property also holds for $\Pi^{\mathrm{ir}}$ and $\Pi$.

Equations \eqref{eq:polarization-irreducible} and \eqref{eq:polarization-irreducible-ladder} imply
\begin{subequations}
  \begin{align}
    &(1-\Pi^{\mathrm{ir}}V^d)\Pi=\Pi^{\mathrm{ir}}\implies \Pi^{-1}=[\Pi^{\mathrm{ir}}]^{-1}-V^d\\
    &(1+\Pi^0V^x)\Pi^{\mathrm{ir}}=\Pi^0\implies [\Pi^{\mathrm{ir}}]^{-1}=[\Pi^0]^{-1}+V^x
  \end{align}
\end{subequations}
Restricting all matrices to this nonzero $(cv)/(vc)$ subspace, Eqs. \eqref{eq:polarization-irreducible} and \eqref{eq:polarization-irreducible-ladder} imply
\begin{align}
  [\Pi(\omega)]^{-1}_{ij,kl}=&[\Pi^0(i\nu_n\to \hbar\omega^{+})]^{-1}_{ij,kl}-V^d_{ij,kl}+V^x_{ij,kl}\nonumber\\
  =&\delta_{(ij)(kl)}\frac{\hbar\omega^{+}-(\xi_i-\xi_j)}{f_{ji}}-V_{iljk}+V_{ilkj}\nonumber\\
  =&\delta_{(ij)(kl)}f_{ji}[\hbar\omega^{+}-(\xi_i-\xi_j)]-V_{iljk}+V_{ilkj}.
\end{align}
where the last equality uses $f_{ji}^{-1}=f_{ji}$ in this subspace.
Explicitly,
\begin{subequations}
  \begin{align}
    [\Pi(\omega)]^{-1}_{cv,c'v'}=&\delta_{(cv)(c'v')}\hbar\omega^{+}-[\delta_{(cv)(c'v')}(\xi_c-\xi_v)+V_{cv'vc'}-V_{cv'c'v}]=\delta_{(cv)(c'v')}\hbar\omega^{+}-\mathcal{E}_{cv,c'v'}\\
    [\Pi(\omega)]^{-1}_{cv,v'c'}=&-(V_{cc'v'v}-V_{cc'vv'})=-\Gamma_{cv,v'c'}\\
    [\Pi(\omega)]^{-1}_{vc,c'v'}=&-(V_{vv'c'c}-V_{vv'cc'})=-\Gamma^*_{cv,v'c'}\\
    [\Pi(\omega)]^{-1}_{vc,v'c'}=&-\delta_{(cv)(c'v')}\hbar\omega^{+}-[\delta_{(cv)(c'v')}(\xi_c-\xi_v)+V_{vc'cv'}-V_{vc'v'c}]\nonumber\\
    =&-\delta_{(cv)(c'v')}\hbar\omega^{+}-[\delta_{(cv)(c'v')}(\xi_c-\xi_v)+V_{c'vv'c}-V_{c'vcv'}]=-\delta_{(cv)(c'v')}\hbar\omega^{+}-\mathcal{E}^*_{cv,c'v'}
  \end{align}
\end{subequations}
In other words,
\begin{equation}
  [\Pi(\omega)]^{-1}=\hbar\omega^{+}\tau_z-\mathcal{H}
\end{equation}
which is consistent with Eq. \eqref{eq:dynamic-solution}.

\section{TDHF: Application to the bilayer system}\label{app:tdhf-bilayer}
\subsection{The TDHF equation of the bilayer EI}\label{app:tdhf-bilayer-equation}
The first-quantization form of the mean-field Hamiltonian in the layer space is formally written as
\begin{equation}
  h^{MF}_{\bm{k}}=\begin{bmatrix}
    \varepsilon_{\bm{k}} & \Delta_{\bm{k}}\\
    \Delta_{\bm{k}} & -\varepsilon_{\bm{k}}
  \end{bmatrix}.
\end{equation}
where $\Delta_{\bm{k}}$ is chosen to be real and negative.
The eigenstates of the mean-field Hamiltonian satisfy
\begin{equation}
  h^{MF}_{\bm{k}}\ket{c\bm{k}}=\xi_{c\bm{k}}\ket{c\bm{k}},\;h^{MF}_{\bm{k}}\ket{v\bm{k}}=\xi_{v\bm{k}}\ket{v\bm{k}}
\end{equation}
where
\begin{subequations}
  \begin{align}
    &\xi_{c\bm{k}}=-\xi_{v\bm{k}}=\xi_{\bm{k}}=\sqrt{\varepsilon_{\bm{k}}^2+\Delta^2_{\bm{k}}}\\
    &\ket{c\bm{k}}=\begin{bmatrix}
    \beta_{\bm{k}}\\-\alpha_{\bm{k}}
  \end{bmatrix},\;\ket{v\bm{k}}=\begin{bmatrix}
    \alpha_{\bm{k}}\\\beta_{\bm{k}}
  \end{bmatrix}\label{eq:quasi-particle-wavefunction}\\
  &\alpha_{\bm{k}}=\sqrt{(1-\varepsilon_{\bm{k}}/\xi_{\bm{k}})/2},\;\beta_{\bm{k}}=\sqrt{(1+\varepsilon_{\bm{k}}/\xi_{\bm{k}})/2}.
  \end{align}\label{eq:quasi-particle-wavefunction-full}
\end{subequations}

In the bilayer system, the EI ground has translation symmetry, thus the density matrix can be labeled by the total momentum $\bm{q}$,
\begin{equation}
  \rho_{ij\bm{k}}(\bm{q})\equiv \ave{c^{\dagger}_{j\bm{k}-\bm{q}/2}c_{i\bm{k}+\bm{q}/2}}\label{eq:rho-momentum-space}
\end{equation}
Thus the TDHF equation Eq. \eqref{eq:dynamic_cv} of $\rho^{(1)}_{cv}$ is rewritten as
\begin{align}
  i\hbar\partial_t\rho^{(1)}_{cv\bm{k}}(\bm{q})=&i\hbar\partial_t\rho^{(1)}_{c=(c\bm{k}+\bm{q}/2)v=(v\bm{k}-\bm{q}/2)}\nonumber\\
  =&\sum_{\substack{c'=(c\bm{k}'+\bm{q}/2)\\v'=(v\bm{k}'-\bm{q}/2)}}\mathcal{E}_{cv,c'v'}\rho^{(1)}_{c'v'}+\sum_{\substack{v'=(v\bm{k}'+\bm{q}/2)\\c'=(c\bm{k}'-\bm{q}/2)}}\Gamma_{cv,c'v'}\rho^{(1)}_{v'c'}+o_{cv}f(t)\nonumber\\
  =&\sum_{\bm{k}'}\mathcal{E}_{\bm{k},\bm{k}'}(\bm{q})\rho^{(1)}_{cv\bm{k}'}(\bm{q})+\sum_{\bm{k}'}\Gamma_{\bm{k},\bm{k}'}(\bm{q})\rho^{(1)}_{vc\bm{k}'}(\bm{q})+\frac{1}{\mathcal{V}}o_{cv\bm{k}}(-\bm{q})f(t,\bm{q})\label{eq:tdhf-bilayer-cv}
\end{align}
where $o_{ij\bm{k}}(-\bm{q})\equiv \bra{i\bm{k}+\bm{q}/2}o_{\bm{k}}\ket{j\bm{k}-\bm{q}/2}$ and $o_{\bm{k}}$ is the bare vertex in layer space---for example, $\gamma^{\sigma}_{\mu,\bm{k}}$ defined in Eq. \eqref{eq:vertex-function}---while $f(t,\bm{q})$ is the corresponding gauge field coupled to $o$.
The matrix elements $\mathcal{E}_{\bm{k},\bm{k}'}(\bm{q})$ and $\Gamma_{\bm{k},\bm{k}'}(\bm{q})$ are defined by
\begin{subequations}
  \begin{align}
    \mathcal{E}_{\bm{k},\bm{k}'}(\bm{q})\equiv &\mathcal{E}_{cv,c'v'}\big|_{c=(c\bm{k}+\bm{q}/2),v=(v\bm{k}-\bm{q}/2),c'=(c\bm{k}'+\bm{q}/2),v'=(v\bm{k}'-\bm{q}/2)}\nonumber\\
    =&[\delta_{(cv),(c'v')}(\xi_c-\xi_v)+V_{cv'vc'}-V_{cv'c'v}]\big|_{c=(c\bm{k}+\bm{q}/2),v=(v\bm{k}-\bm{q}/2),c'=(c\bm{k}'+\bm{q}/2),v'=(v\bm{k}'-\bm{q}/2)}\nonumber\\
    =&\delta_{\bm{k},\bm{k}'}(\xi_{c\bm{k}+\bm{q}/2}-\xi_{v\bm{k}'-\bm{q}/2})+\bra{c\bm{k}+\bm{q}/2,v\bm{k}'-\bm{q}/2}V\ket{v\bm{k}-\bm{q}/2,c\bm{k}'+\bm{q}/2}\nonumber\\
    &-\bra{c\bm{k}+\bm{q}/2,v\bm{k}'-\bm{q}/2}V\ket{c\bm{k}'+\bm{q}/2,v\bm{k}-\bm{q}/2}\\
    \Gamma_{\bm{k},\bm{k}'}(\bm{q})\equiv &\Gamma_{cv,c'v'}\big|_{c=(c\bm{k}+\bm{q}/2),v=(v\bm{k}-\bm{q}/2),v'=(v\bm{k}'+\bm{q}/2),c'=(c\bm{k}'-\bm{q}/2)}\nonumber\\
    =&(V_{cc'vv'}-V_{cc'v'v})\big|_{c=(c\bm{k}+\bm{q}/2),v=(v\bm{k}-\bm{q}/2),v'=(v\bm{k}'+\bm{q}/2),c'=(c\bm{k}'-\bm{q}/2)}\nonumber\\
    =&\bra{c\bm{k}+\bm{q}/2,c\bm{k}'-\bm{q}/2}V\ket{v\bm{k}-\bm{q}/2,v\bm{k}'+\bm{q}/2}-\bra{c\bm{k}+\bm{q}/2,c\bm{k}'-\bm{q}/2}V\ket{v\bm{k}'+\bm{q}/2,v\bm{k}-\bm{q}/2}
  \end{align}
\end{subequations}
Equation \eqref{eq:tdhf-bilayer-cv} shows that the dynamics of $\rho_{cv\bm{k}}(\bm{q})$ is coupled to $\rho_{vc\bm{k}'}(\bm{q})$. To derive the equation of motion for $\rho_{vc\bm{k}}(\bm{q})$, we use $\rho_{vc\bm{k}}(\bm{q})=[\rho_{cv\bm{k}}(-\bm{q})]^*$, which follows from Eq. \eqref{eq:rho-momentum-space}. The resulting equation is
\begin{align}
  -i\hbar\partial_t\rho^{(1)}_{vc\bm{k}}(\bm{q})=&[i\hbar\partial_t\rho^{(1)}_{cv\bm{k}}(-\bm{q})]^*\nonumber\\
  =&\left[\sum_{\bm{k}'}\mathcal{E}_{\bm{k},\bm{k}'}(-\bm{q})\rho^{(1)}_{cv\bm{k}'}(-\bm{q})+\sum_{\bm{k}'}\Gamma_{\bm{k},\bm{k}'}(-\bm{q})\rho^{(1)}_{vc\bm{k}'}(-\bm{q})+\frac{1}{\mathcal{V}}o_{cv\bm{k}}(\bm{q})f(t,-\bm{q})\right]^*\nonumber\\
  =&\sum_{\bm{k}'}\mathcal{E}^*_{\bm{k},\bm{k}'}(-\bm{q})\rho^{(1)}_{vc\bm{k}'}(\bm{q})+\sum_{\bm{k}'}\Gamma^*_{\bm{k},\bm{k}'}(-\bm{q})\rho^{(1)}_{cv\bm{k}'}(\bm{q})+\frac{1}{\mathcal{V}}o_{vc\bm{k}}(-\bm{q})f(t,\bm{q})\label{eq:tdhf-bilayer-vc}
\end{align}
Here $f(t,\bm{r})$ is real, so $f(t,-\bm{q})=f(t,\bm{q})^*$. The matrix elements obey
\begin{subequations}
  \begin{align}
    \mathcal{E}_{\bm{k}',\bm{k}}(\bm{q})=&\delta_{\bm{k}',\bm{k}}(\xi_{c\bm{k}+\bm{q}/2}-\xi_{v\bm{k}-\bm{q}/2})\nonumber\\
    &+\bra{c\bm{k}'+\bm{q}/2,v\bm{k}-\bm{q}/2}V\ket{v\bm{k}'-\bm{q}/2,c\bm{k}+\bm{q}/2}-\bra{c\bm{k}'+\bm{q}/2,v\bm{k}-\bm{q}/2}V\ket{c\bm{k}+\bm{q}/2,v\bm{k}'-\bm{q}/2}\nonumber\\
    =&\delta_{\bm{k}',\bm{k}}(\xi_{c\bm{k}+\bm{q}/2}-\xi_{v\bm{k}-\bm{q}/2})\nonumber\\
    &+[\bra{v\bm{k}'-\bm{q}/2,c\bm{k}+\bm{q}/2}V\ket{c\bm{k}'+\bm{q}/2,v\bm{k}-\bm{q}/2}-\bra{c\bm{k}+\bm{q}/2,v\bm{k}'-\bm{q}/2}V\ket{c\bm{k}'+\bm{q}/2,v\bm{k}-\bm{q}/2}]^*\nonumber\\
    =&\delta_{\bm{k}',\bm{k}}(\xi_{c\bm{k}+\bm{q}/2}-\xi_{v\bm{k}-\bm{q}/2})\nonumber\\
    &+[\bra{c\bm{k}+\bm{q}/2,v\bm{k}'-\bm{q}/2}V\ket{v\bm{k}-\bm{q}/2,c\bm{k}'+\bm{q}/2}-\bra{v\bm{k}'-\bm{q}/2,c\bm{k}+\bm{q}/2}V\ket{v\bm{k}-\bm{q}/2,c\bm{k}'+\bm{q}/2}]^*\nonumber\\
    =&[\mathcal{E}_{\bm{k},\bm{k}'}(\bm{q})]^*\\
    \Gamma_{\bm{k}',\bm{k}}(\bm{q})=&\bra{c\bm{k}'+\bm{q}/2,c\bm{k}-\bm{q}/2}V\ket{v\bm{k}'-\bm{q}/2,v\bm{k}+\bm{q}/2}-\bra{c\bm{k}'+\bm{q}/2,c\bm{k}-\bm{q}/2}V\ket{v\bm{k}+\bm{q}/2,v\bm{k}'-\bm{q}/2}\nonumber\\
    =&\bra{c\bm{k}-\bm{q}/2,c\bm{k}'+\bm{q}/2}V\ket{v\bm{k}+\bm{q}/2,v\bm{k}'-\bm{q}/2}-\bra{c\bm{k}-\bm{q}/2,c\bm{k}'+\bm{q}/2}V\ket{v\bm{k}'-\bm{q}/2,v\bm{k}+\bm{q}/2}\nonumber\\
    =&\Gamma_{\bm{k},\bm{k}'}(-\bm{q})
  \end{align}\label{eq:hermiticity}
\end{subequations}
With inversion symmetry, $\xi_{i,-\bm{k}}=\xi_{i\bm{k}}$, and we choose a gauge in which $\ket{i,-\bm{k}}=\ket{i\bm{k}}$.
Thus 
\begin{subequations}
  \begin{align}
    \mathcal{E}_{-\bm{k},-\bm{k}'}(-\bm{q})=&\delta_{\bm{k},\bm{k}'}(\xi_{c-\bm{k}-\bm{q}/2}-\xi_{v-\bm{k}+\bm{q}/2})+\bra{c-\bm{k}-\bm{q}/2,v-\bm{k}'+\bm{q}/2}V\ket{v-\bm{k}+\bm{q}/2,c-\bm{k}'-\bm{q}/2}\nonumber\\
    &-\bra{c-\bm{k}-\bm{q}/2,v-\bm{k}'+\bm{q}/2}V\ket{c-\bm{k}'-\bm{q}/2,v-\bm{k}+\bm{q}/2}\nonumber\\
    =&\delta_{\bm{k},\bm{k}'}(\xi_{c\bm{k}+\bm{q}/2}-\xi_{v\bm{k}-\bm{q}/2})+\bra{c\bm{k}+\bm{q}/2,v\bm{k}'-\bm{q}/2}V\ket{v\bm{k}-\bm{q}/2,c\bm{k}'+\bm{q}/2}\nonumber\\
    &-\bra{c\bm{k}+\bm{q}/2,v\bm{k}'-\bm{q}/2}V\ket{c\bm{k}'+\bm{q}/2,v\bm{k}-\bm{q}/2}=\mathcal{E}_{\bm{k},\bm{k}'}(\bm{q})\\
    \Gamma_{-\bm{k},-\bm{k}'}(-\bm{q})=&\bra{c-\bm{k}-\bm{q}/2,c-\bm{k}'+\bm{q}/2}V\ket{v-\bm{k}+\bm{q}/2,v-\bm{k}'-\bm{q}/2}\nonumber\\
    &-\bra{c-\bm{k}-\bm{q}/2,c-\bm{k}'+\bm{q}/2}V\ket{v-\bm{k}'-\bm{q}/2,v-\bm{k}+\bm{q}/2}\nonumber\\
    =&\bra{c\bm{k}+\bm{q}/2,c\bm{k}'-\bm{q}/2}V\ket{v\bm{k}-\bm{q}/2,v\bm{k}'+\bm{q}/2}\nonumber\\
    &-\bra{c\bm{k}+\bm{q}/2,c\bm{k}'-\bm{q}/2}V\ket{v\bm{k}'+\bm{q}/2,v\bm{k}-\bm{q}/2}=\Gamma_{\bm{k},\bm{k}'}(\bm{q})
  \end{align}\label{eq:inversion-symmetry}
\end{subequations}
By rearranging the $\bm{k}$ summation in Eq. \eqref{eq:tdhf-bilayer-cv} and Eq. \eqref{eq:tdhf-bilayer-vc}, the dynamic equations can be written in a more compact form:
\begin{align}
  i\hbar\tau_z\partial_t\begin{bmatrix}
    \rho^{(1)}_{cv\bm{k}}(\bm{q})\\\rho^{(1)}_{vc-\bm{k}}(\bm{q})
  \end{bmatrix}=&\sum_{\bm{k}'}\mathcal{H}_{\bm{k},\bm{k}'}(\bm{q})\begin{bmatrix}
    \rho^{(1)}_{cv\bm{k}'}(\bm{q})\\\rho^{(1)}_{vc-\bm{k}'}(\bm{q})
  \end{bmatrix}+\frac{1}{\mathcal{V}}\begin{bmatrix}
    o_{cv\bm{k}}(-\bm{q})\\o_{vc-\bm{k}}(-\bm{q})
  \end{bmatrix}f(t,\bm{q})\label{eq:tdhf-bilayer-compact}
\end{align}
where $\tau_z$ is the Pauli matrix and $\mathcal{H}_{\bm{k},\bm{k}'}(\bm{q})$ is the dynamical matrix defined as
\begin{equation}
    \mathcal{H}_{\bm{k},\bm{k}'}(\bm{q})=\begin{bmatrix}
      \mathcal{E}_{\bm{k},\bm{k}'}(\bm{q}) & \Gamma_{\bm{k},-\bm{k}'}(\bm{q})\\
      \Gamma^*_{-\bm{k},\bm{k}'}(-\bm{q}) & \mathcal{E}^*_{-\bm{k},-\bm{k}'}(-\bm{q})
    \end{bmatrix}=\begin{bmatrix}
      \mathcal{E}_{\bm{k},\bm{k}'}(\bm{q}) & \Gamma_{\bm{k},-\bm{k}'}(\bm{q})\\
      \Gamma^*_{\bm{k},-\bm{k}'}(\bm{q}) & \mathcal{E}^*_{\bm{k}',\bm{k}}(\bm{q})
    \end{bmatrix}
\end{equation}
In the above equation, we have used Eq. \eqref{eq:inversion-symmetry} to get the second equality.
Equation \eqref{eq:hermiticity} shows that $\mathcal{H}_{\bm{k},\bm{k}'}(\bm{q})$ is Hermitian. When the EI order parameter $\Delta_{\bm{k}}$ is chosen to be real and negative, the dynamical matrix is also real. Using the quasiparticle wavefunctions in Eq. \eqref{eq:quasi-particle-wavefunction}, the explicit expressions for $\mathcal{E}_{\bm{k},\bm{k}'}$ and $\Gamma_{\bm{k},\bm{k}'}$ are
\begin{align}
  \mathcal{E}_{\bm{k},\bm{k}'}(\bm{q})=&\delta_{\bm{k},\bm{k}'}(\xi_{c\bm{k}+\bm{q}/2}-\xi_{v\bm{k}-\bm{q}/2})+\frac{1}{\mathcal{V}}V(\bm{q})[\beta_{\bm{k}+\bm{q}/2}\alpha_{\bm{k}-\bm{q}/2}\alpha_{\bm{k}'-\bm{q}/2}\beta_{\bm{k}'+\bm{q}/2}+(-\alpha_{\bm{k}+\bm{q}/2})\beta_{\bm{k}-\bm{q}/2}\beta_{\bm{k}'-\bm{q}/2}(-\alpha_{\bm{k}'+\bm{q}/2})]\nonumber\\
  &+\frac{1}{\mathcal{V}}U(\bm{q})[\beta_{\bm{k}+\bm{q}/2}\alpha_{\bm{k}-\bm{q}/2}\beta_{\bm{k}'-\bm{q}/2}(-\alpha_{\bm{k}'+\bm{q}/2})+(-\alpha_{\bm{k}+\bm{q}/2})\beta_{\bm{k}-\bm{q}/2}\alpha_{\bm{k}'-\bm{q}/2}\beta_{\bm{k}'+\bm{q}/2}]\nonumber\\
  &-\frac{1}{\mathcal{V}}V(\bm{k}-\bm{k}')[\beta_{\bm{k}+\bm{q}/2}\beta_{\bm{k}'+\bm{q}/2}\alpha_{\bm{k}'-\bm{q}/2}\alpha_{\bm{k}-\bm{q}/2}+(-\alpha_{\bm{k}+\bm{q}/2})(-\alpha_{\bm{k}'-\bm{q}/2})\beta_{\bm{k}'-\bm{q}/2}\beta_{\bm{k}-\bm{q}/2}]\nonumber\\
  &-\frac{1}{\mathcal{V}}U(\bm{k}-\bm{k}')[\beta_{\bm{k}+\bm{q}/2}\beta_{\bm{k}'+\bm{q}/2}\beta_{\bm{k}'-\bm{q}/2}\beta_{\bm{k}-\bm{q}/2}+(-\alpha_{\bm{k}+\bm{q}/2})(-\alpha_{\bm{k}'-\bm{q}/2})\alpha_{\bm{k}'-\bm{q}/2}\alpha_{\bm{k}-\bm{q}/2}]\nonumber\\
  =&\delta_{\bm{k},\bm{k}'}(\xi_{c\bm{k}+\bm{q}/2}-\xi_{v\bm{k}-\bm{q}/2})\nonumber\\
  &+\frac{1}{\mathcal{V}}[V(\bm{q})-V(\bm{k}-\bm{k}')][(\beta_{\bm{k}+\bm{q}/2}\alpha_{\bm{k}-\bm{q}/2})(\beta_{\bm{k}'+\bm{q}/2}\alpha_{\bm{k}'-\bm{q}/2})+(\alpha_{\bm{k}+\bm{q}/2}\beta_{\bm{k}-\bm{q}/2})(\alpha_{\bm{k}'+\bm{q}/2}\beta_{\bm{k}'-\bm{q}/2})]\nonumber\\
  &-\frac{1}{\mathcal{V}}U(\bm{q})[(\beta_{\bm{k}+\bm{q}/2}\alpha_{\bm{k}-\bm{q}/2})(\alpha_{\bm{k}'+\bm{q}/2}\beta_{\bm{k}'-\bm{q}/2})+(\alpha_{\bm{k}+\bm{q}/2}\beta_{\bm{k}-\bm{q}/2})(\beta_{\bm{k}'+\bm{q}/2}\alpha_{\bm{k}'-\bm{q}/2})]\nonumber\\
  &-\frac{1}{\mathcal{V}}U(\bm{k}-\bm{k}')[(\beta_{\bm{k}+\bm{q}/2}\beta_{\bm{k}-\bm{q}/2})(\beta_{\bm{k}'+\bm{q}/2}\beta_{\bm{k}'-\bm{q}/2})+(\alpha_{\bm{k}+\bm{q}/2}\alpha_{\bm{k}-\bm{q}/2})(\alpha_{\bm{k}'+\bm{q}/2}\alpha_{\bm{k}'-\bm{q}/2})]\label{eq:E-explicit}\\
  \Gamma_{\bm{k},\bm{k}'}(\bm{q})=&\frac{1}{\mathcal{V}}V(\bm{q})[\beta_{\bm{k}+\bm{q}/2}\alpha_{\bm{k}-\bm{q}/2}\beta_{\bm{k}'-\bm{q}/2}\alpha_{\bm{k}'+\bm{q}/2}+(-\alpha_{\bm{k}+\bm{q}/2})\beta_{\bm{k}-\bm{q}/2}(-\alpha_{\bm{k}'-\bm{q}/2})\beta_{\bm{k}'+\bm{q}/2}]\nonumber\\
  &+\frac{1}{\mathcal{V}}U(\bm{q})[\beta_{\bm{k}+\bm{q}/2}\alpha_{\bm{k}-\bm{q}/2}(-\alpha_{\bm{k}'-\bm{q}/2})\beta_{\bm{k}'+\bm{q}/2}+(-\alpha_{\bm{k}+\bm{q}/2})\beta_{\bm{k}-\bm{q}/2}\beta_{\bm{k}'-\bm{q}/2}\alpha_{\bm{k}'+\bm{q}/2}]\nonumber\\
  &-\frac{1}{\mathcal{V}}V(\bm{k}-\bm{k}')[\beta_{\bm{k}+\bm{q}/2}\alpha_{\bm{k}'+\bm{q}/2}\beta_{\bm{k}'-\bm{q}/2}\alpha_{\bm{k}-\bm{q}/2}+(-\alpha_{\bm{k}+\bm{q}/2})\beta_{\bm{k}'+\bm{q}/2}(-\alpha_{\bm{k}'-\bm{q}/2})\beta_{\bm{k}-\bm{q}/2}]\nonumber\\
  &-\frac{1}{\mathcal{V}}U(\bm{k}-\bm{k}')[\beta_{\bm{k}+\bm{q}/2}\alpha_{\bm{k}'+\bm{q}/2}(-\alpha_{\bm{k}'-\bm{q}/2})\beta_{\bm{k}-\bm{q}/2}+(-\alpha_{\bm{k}+\bm{q}/2})\beta_{\bm{k}'+\bm{q}/2}\beta_{\bm{k}'-\bm{q}/2}\alpha_{\bm{k}-\bm{q}/2}]\nonumber\\
  =&\frac{1}{\mathcal{V}}[V(\bm{q})-V(\bm{k}-\bm{k}')][(\beta_{\bm{k}+\bm{q}/2}\alpha_{\bm{k}-\bm{q}/2})(\alpha_{\bm{k}'+\bm{q}/2}\beta_{\bm{k}'-\bm{q}/2})+(\alpha_{\bm{k}+\bm{q}/2}\beta_{\bm{k}-\bm{q}/2})(\beta_{\bm{k}'+\bm{q}/2}\alpha_{\bm{k}'-\bm{q}/2})]\nonumber\\
  &-\frac{1}{\mathcal{V}}U(\bm{q})[(\beta_{\bm{k}+\bm{q}/2}\alpha_{\bm{k}-\bm{q}/2})(\beta_{\bm{k}'+\bm{q}/2}\alpha_{\bm{k}'-\bm{q}/2})+(\alpha_{\bm{k}+\bm{q}/2}\beta_{\bm{k}-\bm{q}/2})(\alpha_{\bm{k}'+\bm{q}/2}\beta_{\bm{k}'-\bm{q}/2})]\nonumber\\
  &+\frac{1}{\mathcal{V}}U(\bm{k}-\bm{k}')[(\beta_{\bm{k}+\bm{q}/2}\beta_{\bm{k}-\bm{q}/2})(\alpha_{\bm{k}'+\bm{q}/2}\alpha_{\bm{k}'-\bm{q}/2})+(\alpha_{\bm{k}+\bm{q}/2}\alpha_{\bm{k}-\bm{q}/2})(\beta_{\bm{k}'+\bm{q}/2}\beta_{\bm{k}'-\bm{q}/2})]\label{eq:gamma-explicit}
\end{align}
where $V(q)$ and $U(q)$ are the intralayer and interlayer Coulomb interactions, respectively.
The explicit expression also shows that $\Gamma_{\bm{k}',\bm{k}}(\bm{q})=\Gamma_{\bm{k},\bm{k}'}(\bm{q})$, so $\Gamma(\bm{q})$ is symmetric.

\subsection{Solving the TDHF equation}\label{app:tdhf-bilayer-solution}
Equation \eqref{eq:dynamic-solution} gives the formal frequency-space solution of Eq. \eqref{eq:tdhf-bilayer-compact}:
\begin{equation}
  \begin{bmatrix}
    \rho^{(1)}_{cv\bm{k}}(\omega,\bm{q})\\\rho^{(1)}_{vc-\bm{k}}(\omega,\bm{q})
  \end{bmatrix}=\frac{1}{\mathcal{V}}\sum_{\bm{k}'}[\hbar\omega^{+}\tau_z-\mathcal{H}(\bm{q})]^{-1}_{\bm{k},\bm{k}'}\begin{bmatrix}
    o_{cv\bm{k}'}(-\bm{q})\\o_{vc-\bm{k}'}(-\bm{q})
  \end{bmatrix}f(\omega,\bm{q})\label{eq:tdhf-bilayer-solution}
\end{equation}
where the matrix inversion could be evaluated in the generalized-eigenstate basis of $\mathcal{H}(\bm{q})$.

The generalized eigenvalue equation of $\mathcal{H}(\bm{q})$ is given by
\begin{equation}
  \sum_{\bm{k}'}\mathcal{H}_{\bm{k},\bm{k}'}(\bm{q})\Phi_{n\bm{k}'}(\bm{q})=\hbar\omega_{n}(\bm{q})\tau_z\Phi_{n\bm{k}}(\bm{q})\label{eq:generalized-eigenvalue}
\end{equation}
To solve the generalized eigenvalue equation, we first define two auxiliary matrices, $\mathcal{K}^{(\pm)}_{\bm{k},\bm{k}'}(\bm{q})\equiv \mathcal{E}_{\bm{k},\bm{k}'}(\bm{q})\pm \Gamma_{\bm{k},-\bm{k}'}(\bm{q})$. Equations \eqref{eq:E-explicit} and \eqref{eq:gamma-explicit} show that both matrices are real and symmetric.
For convenience, we will omit the $\bm{k}$ subscripts in the following.
They are the Hessian matrices of the Hartree--Fock total-energy functional for the amplitude and phase fluctuations of the EI order parameter, respectively \cite{wuTheoryTwodimensionalSpatially2015,shaoElectricalBreakdownExcitonic2024}. Ground-state stability makes both matrices nonnegative, so $\sqrt{\mathcal{K}^{(+)}(\bm{q})}$ is well defined. We choose this square root to be real, symmetric, and nonnegative.
Define
\begin{equation}
  \mathcal{D}(\bm{q})=\sqrt{\mathcal{K}^{(+)}(\bm{q})}\mathcal{K}^{(-)}(\bm{q})\sqrt{\mathcal{K}^{(+)}(\bm{q})}
\end{equation}
Then $\mathcal{D}(\bm{q})$ is also real, symmetric, and nonnegative. We will now show that its eigenvalues are the squared generalized eigenfrequencies of $\mathcal{H}(\bm{q})$:
\begin{equation}
  \mathcal{D}(\bm{q})u_{n}(\bm{q})=\hbar^2\omega^2_{n}(\bm{q})u_{n}(\bm{q})
\end{equation}
At $\bm{q}\ne\bm{0}$ with no zero modes, $\mathcal{K}^{(\pm)}(\bm{q})$ are positive definite. Thus we can take the square root of $\mathcal{K}^{(+)}(\bm{q})$ and define the auxiliary vectors $x_n(\bm{q})$ and $y_n(\bm{q})$ as 
\begin{equation}
  x_{n}(\bm{q})=[\mathcal{K}^{(+)}(\bm{q})]^{-1/2}u_{n}(\bm{q}),\;y_{n}(\bm{q})=[i\hbar\omega_n(\bm{q})]^{-1}[\mathcal{K}^{(+)}(\bm{q})]^{1/2}u_{n}(\bm{q})\label{eq:x-y}
\end{equation}
We can then verify that
\begin{align}
  \mathcal{K}^{(+)}(\bm{q})x_{n}(\bm{q})=&[\mathcal{K}^{(+)}(\bm{q})]^{1/2}u_n(\bm{q})=i\hbar\omega_n(\bm{q})y_n(\bm{q})\\
  \mathcal{K}^{(-)}(\bm{q})y_n(\bm{q})=&[i\hbar\omega_n(\bm{q})]^{-1}[\mathcal{K}^{(+)}(\bm{q})]^{-1/2}\mathcal{D}(\bm{q})u_n(\bm{q})=-i\hbar\omega_n(\bm{q})[\mathcal{K}^{(+)}(\bm{q})]^{-1/2}u_n(\bm{q})=-i\hbar\omega_n(\bm{q})x_n(\bm{q})
\end{align}
The generalized eigenvectors of $\mathcal{H}(\bm{q})$ with positive and negative eigenvalues can therefore be constructed as
\begin{equation}
  \Phi_{\pm n}(\bm{q})=\frac{1}{2}\begin{bmatrix}
    x_{n}(\bm{q})\pm iy_{n}(\bm{q})\\
    x_{n}(\bm{q})\mp iy_{n}(\bm{q})
  \end{bmatrix}\label{eq:generalized-eigenvector}
\end{equation}
One can verify that 
\begin{align}
  \mathcal{H}(\bm{q})\Phi_{+n}(\bm{q})=&\frac{1}{2}\begin{bmatrix}
    \mathcal{E}(\bm{q}) & \Gamma(\bm{q})\\ \Gamma(\bm{q}) & \mathcal{E}(\bm{q})
  \end{bmatrix}\begin{bmatrix}
    x_{n}(\bm{q})+iy_{n}(\bm{q})\\
    x_{n}(\bm{q})-iy_{n}(\bm{q})
  \end{bmatrix}\nonumber\\
  =&\frac{1}{2}\begin{bmatrix}
    \mathcal{K}^{(+)}(\bm{q})x_n(\bm{q})+i\mathcal{K}^{(-)}(\bm{q})y_n(\bm{q})\\
    \mathcal{K}^{(+)}(\bm{q})x_n(\bm{q})-i\mathcal{K}^{(-)}(\bm{q})y_n(\bm{q})\\
  \end{bmatrix}\nonumber\\
  =&\frac{1}{2}\begin{bmatrix}
    i\hbar\omega_n(\bm{q})y_n(\bm{q})+\hbar\omega_n(\bm{q})x_n(\bm{q})\\
    i\hbar\omega_n(\bm{q})y_n(\bm{q})-\hbar\omega_n(\bm{q})x_n(\bm{q})
  \end{bmatrix}\nonumber\\
  =&\hbar\omega_n(\bm{q})\tau_z\Phi_{+n}(\bm{q})
\end{align}
Similarly, we also have $\mathcal{H}(\bm{q})\Phi_{-n}(\bm{q})=-\hbar\omega_n(\bm{q})\tau_z\Phi_{-n}(\bm{q})$.
Thus, Eqs. \eqref{eq:x-y} and \eqref{eq:generalized-eigenvector} construct the generalized eigenvectors of $\mathcal{H}(\bm{q})$ from the eigenvectors of $\mathcal{D}(\bm{q})$. The corresponding generalized eigenvalues are the signed square roots of the eigenvalues of $\mathcal{D}(\bm{q})$.

As a real-symmetric matrix, the eigenvectors of $\mathcal{D}(\bm{q})$ can be taken to be real and orthonormal, i.e.
\begin{equation}
  u^{\dagger}_{m}(\bm{q})u_{n}(\bm{q})=\delta_{mn},\; \mathrm{Im}u_n(\bm{q})=0.
\end{equation}
According to the construction in Eqs. \eqref{eq:x-y} and \eqref{eq:generalized-eigenvector}, $\Phi_n(\bm{q})$ can also be chosen real and satisfies a generalized orthogonality relation with respect to $\tau_z$:
\begin{equation}
  \Phi^{\dagger}_{m}(\bm{q})\mathcal{H}(\bm{q})\Phi_n(\bm{q})=\hbar\omega_n(\bm{q})\Phi^{\dagger}_{m}(\bm{q})\tau_z\Phi_n(\bm{q})=\hbar\omega_m(\bm{q})\Phi^{\dagger}_{m}(\bm{q})\tau_z\Phi_n(\bm{q})\implies \Phi^{\dagger}_{m}(\bm{q})\tau_z\Phi_n(\bm{q})\propto \delta_{mn}
\end{equation}
However, the generalized eigenvectors are not normalized with respect to $\tau_z$:
\begin{align}
  &\Phi^{\dagger}_{n}(\bm{q})\tau_z\Phi_{n}(\bm{q})\nonumber\\
  =&\frac{1}{4}u_n^{\dagger}(\bm{q})\left\{
    \{[\mathcal{K}^{(+)}(\bm{q})]^{-1/2}+[\hbar\omega_n(\bm{q})]^{-1}[\mathcal{K}^{(+)}(\bm{q})]^{1/2}\}^2-\{[\mathcal{K}^{(+)}(\bm{q})]^{-1/2}-[\hbar\omega_n(\bm{q})]^{-1}[\mathcal{K}^{(+)}(\bm{q})]^{1/2}\}^2
    \right\}u_{n}(\bm{q})\nonumber\\
  =&\frac{1}{4}u_{n}^{\dagger}(\bm{q})4[\hbar\omega_n(\bm{q})]^{-1}u_{n}(\bm{q})=\frac{1}{\hbar\omega_n(\bm{q})}
\end{align}
Instead, they are normalized with respect to the inner product defined by $\mathcal{H}(\bm{q})$:
\begin{equation}
  \Phi^{\dagger}_n(\bm{q})\mathcal{H}(\bm{q})\Phi_n(\bm{q})=\hbar\omega_n(\bm{q})\Phi^{\dagger}_n(\bm{q})\tau_z\Phi_n(\bm{q})=1.
\end{equation}
In summary, we have
\begin{equation}
  \Phi^{\dagger}_m(\bm{q})\mathcal{H}(\bm{q})\Phi_n(\bm{q})=\delta_{mn},\;\Phi^{\dagger}_m(\bm{q})\tau_z\Phi_n(\bm{q})=\frac{\delta_{mn}}{\hbar\omega_n(\bm{q})},\;\mathrm{Im}\Phi_n(\bm{q})=0\label{eq:orthonormal-real}
\end{equation}

Since $\Phi_n(\bm{q})$ are orthonormal with respect to the inner product defined by $\mathcal{H}(\bm{q})$, $\{[\mathcal{H}(\bm{q})]^{1/2}\Phi_n(\bm{q})\}$ will form a complete basis set, such that 
\begin{equation}
  \sum_{n}[\mathcal{H}(\bm{q})]^{1/2}\Phi_n(\bm{q})\Phi^{\dagger}_n(\bm{q})[\mathcal{H}(\bm{q})]^{1/2}=\mathbbm{1}\label{eq:identity}
\end{equation}
This implies two other identities:
\begin{subequations}
  \begin{align}
  \sum_{n}\Phi_n(\bm{q})\Phi^{\dagger}_n(\bm{q})\mathcal{H}(\bm{q})=[\mathcal{H}(\bm{q})]^{-1/2}\left\{\sum_{n}[\mathcal{H}(\bm{q})]^{1/2}\Phi_n(\bm{q})\Phi^{\dagger}_n(\bm{q})[\mathcal{H}(\bm{q})]^{1/2}\right\}[\mathcal{H}(\bm{q})]^{1/2}=\mathbbm{1}\\
  \sum_{n}\mathcal{H}(\bm{q})\Phi_n(\bm{q})\Phi^{\dagger}_n(\bm{q})=[\mathcal{H}(\bm{q})]^{1/2}\left\{\sum_{n}[\mathcal{H}(\bm{q})]^{1/2}\Phi_n(\bm{q})\Phi^{\dagger}_n(\bm{q})[\mathcal{H}(\bm{q})]^{1/2}\right\}[\mathcal{H}(\bm{q})]^{-1/2}=\mathbbm{1}
  \end{align}
\end{subequations}

Define 
\begin{equation}
  \Pi(\omega,\bm{q})=\sum_{n}\frac{\omega_n(\bm{q})\Phi_{n}(\bm{q})\Phi^{\dagger}_{n}(\bm{q})}{\omega^+-\omega_n(\bm{q})}\label{eq:polarization}
\end{equation}
where $n$ runs over all the positive and negative eigenvalues of $\mathcal{H}(\bm{q})$.
Then we can verify that 
\begin{align}
  [\hbar\omega^+\tau_z-\mathcal{H}(\bm{q})]\Pi(\omega,\bm{q})=&\sum_{n}\frac{\omega_n(\bm{q})[\hbar\omega^+\tau_z-\hbar\omega_n\tau_z]\Phi_n(\bm{q})\Phi^{\dagger}_n(\bm{q})}{\omega^+-\omega_n(\bm{q})}\nonumber\\
  =&\sum_{n}\hbar\omega_n(\bm{q})\tau_z\Phi_n(\bm{q})\Phi^{\dagger}_n(\bm{q})\nonumber\\
  =&\sum_{n}\mathcal{H}(\bm{q})\Phi_n(\bm{q})\Phi_n^{\dagger}(\bm{q})=\mathbbm{1}
\end{align}
This implies that $[\hbar\omega^{+}\tau_z-\mathcal{H}(\bm{q})]^{-1}=\Pi(\omega,\bm{q})$ and the formal solution Eq. \eqref{eq:tdhf-bilayer-solution} can be explicitly written as 
\begin{equation}
  \begin{bmatrix}
    \rho^{(1)}_{cv\bm{k}}(\omega,\bm{q})\\ \rho^{(1)}_{vc-\bm{k}}(\omega,\bm{q})
  \end{bmatrix}=\frac{1}{\mathcal{V}}\sum_{n}\frac{\omega_n(\bm{q})\Phi_{n\bm{k}}(\bm{q})\mathrm{O}_n(\bm{q})}{\omega^+-\omega_n(\bm{q})}f(\omega,\bm{q})\label{eq:density-matrix-first-order}
\end{equation}
where 
\begin{equation}
  \mathrm{O}_n(\bm{q})\equiv \sum_{\bm{k}}\Phi^{\dagger}_{n\bm{k}}(\bm{q})\begin{bmatrix}
    o_{cv\bm{k}}(-\bm{q})\\o_{vc-\bm{k}}(-\bm{q})
  \end{bmatrix}
\end{equation}
is the vertex overlap between the $n$-th collective-mode wavefunction and the bare vertex of the operator $\hat{O}$.

\subsection{Charge- and exciton-channel density and current operators}\label{app:gauge-field-coupling}


Under $k\cdot p$ approximation, the many-body Hamiltonian in momentum space is written as
\begin{subequations}
  \begin{align}
    \hat{H}_0=&\sum_{\bm{k}}[c^{\dagger}_{e\bm{k}},c^{\dagger}_{h\bm{k}}]\begin{bmatrix}
      \frac{\hbar^2 k^2}{2m_e}-\frac{\mu_X}{2} & 0 \\
      0& -\frac{\hbar^2 k^2}{2m_h}+\frac{\mu_X}{2}
    \end{bmatrix}\begin{bmatrix}
      c_{e\bm{k}}\\c_{h\bm{k}}
    \end{bmatrix}\\
    \hat{H}_I=&\frac{1}{2\mathcal{V}}\sum_{ss'=eh}\sum_{\bm{k}\bm{k}'\bm{q}}V_{ss'}(\bm{q})c^{\dagger}_{s\bm{k}}c^{\dagger}_{s'\bm{k}'}c_{s'\bm{k}'+\bm{q}}c_{s\bm{k}-\bm{q}}
  \end{align}
\end{subequations}
where $c^{\dagger}_{e\bm{k}}$ ($c_{e\bm{k}}$) and $c^{\dagger}_{h\bm{k}}$ ($c_{h\bm{k}}$) create (annihilate) electrons in the electron and hole layers, respectively.
In the absence of the interlayer bias, the ground state of the bilayer system is defined as $\ket{G_{\text{unbiased}}}=\prod_{\bm{k}}c^{\dagger}_{h\bm{k}}\ket{\text{vac.}}$, where $\ket{\text{vac.}}$ is the vacuum state.
To avoid double counting, the many-body Hamiltonian should be normal ordered with respect to $\ket{G_{\text{unbiased}}}$, i.e., $\hat{H}\equiv:\hat{H}:$. The normal-ordering rules for the creation and annihilation operators are
\begin{subequations}
  \begin{align}
    :c^{\dagger}_{e\bm{k}}c_{s\bm{k}'}:=-:c_{s\bm{k}'}c^{\dagger}_{e\bm{k}}:=c^{\dagger}_{e\bm{k}}c_{s\bm{k}'}\\
    :c_{s\bm{k}}c^{\dagger}_{h\bm{k}'}:=-:c^{\dagger}_{h\bm{k}'}c_{s\bm{k}}:=c_{s\bm{k}}c^{\dagger}_{h\bm{k}'}
  \end{align}\label{eq:normal-order-rules}
\end{subequations}
For simplicity, we do not write the Hamiltonian explicitly in normal-ordered form. We also omit the normal-ordering symbols $:\cdots:$ except where they are needed. The Hamiltonian and the layer charge-current operators defined below are nevertheless normal ordered with respect to $\ket{G_{\text{unbiased}}}$.

In real space, the many-body Hamiltonian of the bilayer system is written as
\begin{subequations}
  \begin{align}
    \hat{H}_0=&\int \rd\bm{r}\;\Psi^{\dagger}(\bm{r})\begin{bmatrix}
      \frac{p^2}{2m_e}-\frac{\mu_X}{2} & 0 \\
      0 & -\frac{p^2}{2m_h}+\frac{\mu_X}{2}
    \end{bmatrix}\Psi(\bm{r})\\
    \hat{H}_I=&\frac{1}{2}\sum_{ss'=eh}\int\rd\bm{r}\rd\bm{r}'\;\Psi_s^{\dagger}(\bm{r})\Psi_s'^{\dagger}(\bm{r}')V_{ss'}(\bm{r}-\bm{r}')\Psi_{s'}(\bm{r}')\Psi_{s}(\bm{r})
  \end{align}
\end{subequations}
where $\Psi^{\dagger}(\bm{r})\equiv[\Psi_e^{\dagger}(\bm{r}),\Psi^{\dagger}_h(\bm{r})]$ and $\Psi_s^{\dagger}(\bm{r})=\mathcal{V}^{-1/2}\sum_{\bm{k}}\re^{-i\bm{k}\cdot\bm{r}}c^{\dagger}_{s\bm{k}}$ is the field operator.
When a gauge field $A_{s\mu}(t,\bm{r})=(\phi_{s}(t,\bm{r}),\bm{A}_{s}(t,\bm{r}))$ is applied to each layer, the Hamiltonian is modified by the Peierls substitution $\bm{p}\to \bm{p}+e\bm{A}$ ($e=|e|$). The noninteracting Hamiltonian then becomes
\begin{equation}
  \hat{H}_0'=\int\rd\bm{r}\;\Psi^{\dagger}(\bm{r})\begin{bmatrix}
    \frac{|\bm{p}+e\bm{A}_e|^2}{2m_e}-\frac{\mu_X}{2}-e\phi_e & 0 \\
    0 & -\frac{|\bm{p}+e\bm{A}_h|^2}{2m_h}+\frac{\mu_X}{2}-e\phi_h
  \end{bmatrix}\Psi(\bm{r})\label{eq:hamiltonian-gauge-field}
\end{equation}
The charge-density ($\mu=0$) and charge-current ($\mu=1,2$) operators in each layer are defined as
\begin{equation}
  \hat{j}_{s\mu}(t,\bm{r})=-\frac{\delta\hat{H}_0'}{\delta A_{s\mu}(t,\bm{r})}
\end{equation}
Explicitly,
\begin{subequations}
  \begin{align}
    -\hat{\varrho}_e(t,\bm{r})\equiv&\hat{j}_{e\mu=0}(t,\bm{r})=e\Psi^{\dagger}_{e}(\bm{r})\Psi_{e}(\bm{r})\\
    \hat{\bm{j}}_e(t,\bm{r})\equiv&\hat{j}_{e\mu=12}(t,\bm{r})=-\frac{e}{2m_e}\Psi_e^{\dagger}(\bm{r})\left[-i\hbar\nabla_{\bm{r}}+e\bm{A}_{e}(t,\bm{r})\right]\Psi_e(\bm{r})+\mathrm{h.c.}\\
    -\hat{\varrho}_h(t,\bm{r})\equiv&\hat{j}_{h\mu=0}(t,\bm{r})=e\Psi^{\dagger}_{h}(\bm{r})\Psi_{h}(\bm{r})\\
    \hat{\bm{j}}_h(t,\bm{r})\equiv&\hat{j}_{h\mu=12}(t,\bm{r})=\frac{e}{2m_h}\Psi_h^{\dagger}(\bm{r})\left[-i\hbar\nabla_{\bm{r}}+e\bm{A}_{h}(t,\bm{r})\right]\Psi_h(\bm{r})+\mathrm{h.c.}
  \end{align}
\end{subequations}
In the expression of the layer charge-current operators $\hat{\bm{j}}_{s}$, there are two terms: the paramagnetic charge current, which is independent of the gauge field,
\begin{subequations}
  \begin{align}
    \hat{\bm{j}}_{e,p}(t,\bm{r})=&\frac{ie\hbar}{2m_e}\Psi^{\dagger}_{e}(\bm{r})\nabla_{\bm{r}}\Psi_{e}(\bm{r})+\mathrm{h.c.}\\
  \hat{\bm{j}}_{h,p}(t,\bm{r})=&-\frac{ie\hbar}{2m_h}\Psi^{\dagger}_{h}(\bm{r})\nabla_{\bm{r}}\Psi_{h}(\bm{r})+\mathrm{h.c.}
  \end{align}
\end{subequations}
and the diamagnetic charge current, which is proportional to the vector potential $\bm{A}_{s}(t,\bm{r})$,
\begin{subequations}
  \begin{align}
    \hat{\bm{j}}_{e,d}(t,\bm{r})=&-\frac{e^2}{m_e}\Psi^{\dagger}_{e}(\bm{r})\Psi_{e}(\bm{r})\bm{A}_{e}(t,\bm{r})\\
    \hat{\bm{j}}_{h,d}(t,\bm{r})=&\frac{e^2}{m_h}\Psi^{\dagger}_{h}(\bm{r})\Psi_{h}(\bm{r})\bm{A}_{h}(t,\bm{r})
  \end{align}
\end{subequations}
To linear order in the gauge field $A_{s\mu}(t,\bm{r})$, the perturbed Hamiltonian can be written as $\hat{H}'_0=\hat{H}_0+\hat{H}_c$, where the linear coupling term is
\begin{equation}
  \hat{H}_c=\sum_{s=eh}\int\rd\bm{r}\;[\hat{\varrho}_{s}(\bm{r})\phi_{s}(t,\bm{r})-\hat{\bm{j}}_{s,p}(\bm{r})\cdot \bm{A}_{s}(t,\bm{r})]
\end{equation}
Here we drop the $t$ variable in $\hat{\varrho}(t,\bm{r})$ and $\hat{\bm{j}}_{s,p}(t,\bm{r})$ since they are time-independent according to their definitions.

We can define the total charge-density and paramagnetic charge-current operators as
\begin{subequations}
  \begin{align}
    \hat{\varrho}^{+}(\bm{r})\equiv&\hat{\varrho}_e(\bm{r})+\hat{\varrho}_h(\bm{r})=-e\Psi^{\dagger}(\bm{r})\Psi(\bm{r})\\
    \hat{\bm{j}}_p^{+}(\bm{r})\equiv&\hat{\bm{j}}_{e,p}(\bm{r})+\hat{\bm{j}}_{h,p}(\bm{r})=\frac{ie\hbar}{4m}\Psi^{\dagger}(\bm{r})\sigma_z\nabla_{\bm{r}}\Psi(\bm{r}) +\frac{ie\hbar \delta_m}{4m}\Psi^{\dagger}(\bm{r})\nabla_{\bm{r}}\Psi(\bm{r}) 
    +h.c.
  \end{align}
\end{subequations}
where $\sigma_z$ is the Pauli matrix in the layer space, $m=m_em_h/(m_e+m_h)$ is the reduced mass and $\delta_m=(m_h-m_e)/(m_e+m_h)$ is the electron--hole asymmetry strength.
Similarly, we can define the exciton-density and paramagnetic exciton-current operators (multiplied by $-e$)
\begin{subequations}
  \begin{align}
    \hat{\varrho}^{-}(\bm{r})\equiv&\frac{1}{2}[\hat{\varrho}_e(\bm{r})-\hat{\varrho}_h(\bm{r})]=-\frac{e}{2}\Psi^{\dagger}(\bm{r})\sigma_z\Psi(\bm{r})\\
    \hat{\bm{j}}^{-}_{p}(\bm{r})\equiv&\frac{1}{2}[\hat{\bm{j}}_{e,p}(\bm{r})-\hat{\bm{j}}_{h,p}(\bm{r})]=\frac{ie\hbar}{8m}\Psi^{\dagger}(\bm{r})\nabla_{\bm{r}}\Psi(\bm{r})+\frac{ie\hbar \delta_m}{8m}\Psi^{\dagger}(\bm{r})\sigma_z\nabla_{\bm{r}}\Psi(\bm{r}) +h.c.
  \end{align}
\end{subequations}
The coupling term becomes
\begin{equation}
  \hat{H}_c=\sum_{\sigma=\pm}\int\rd\bm{r}\;[\hat{\varrho}^{\sigma}(\bm{r})\phi^{\sigma}(t,\bm{r})-\hat{\bm{j}}_{p}^{\sigma}(\bm{r})\cdot \bm{A}^{\sigma}(t,\bm{r})]\label{eq:coupling-term}
\end{equation}
where $A_{\mu}^{+}(t,\bm{r})\equiv[A_{e\mu}(t,\bm{r})+A_{h\mu}(t,\bm{r})]/2$ and $A_{\mu}^{-}(t,\bm{r})\equiv A_{e\mu}(t,\bm{r})-A_{h\mu}(t,\bm{r})$ are the layer-symmetric and layer-antisymmetric gauge fields, respectively.
The linear coupling Hamiltonian in Eq. \eqref{eq:coupling-term} shows that the layer-symmetric and layer-antisymmetric gauge fields couple to the total-charge and exciton degrees of freedom, respectively.
For the charge and exciton diamagnetic currents, we have

\begin{subequations}
  \begin{align}
    \ave{\hat{\bm{j}}_{e,d}(t,\bm{r})}=&-\frac{e^2(1+\delta_m)}{2m}\ave{\Psi^{\dagger}_{e}(\bm{r})\Psi_e(\bm{r})}\bm{A}_{e}(t,\bm{r})=-\frac{e^2n_X(1+\delta_m)}{2m}\bm{A}_{e}(t,\bm{r})\\
    \ave{\hat{\bm{j}}_{h,d}(t,\bm{r})}=&\frac{e^2(1-\delta_m)}{2m}\ave{:\Psi^{\dagger}_{h}(\bm{r})\Psi_h(\bm{r}):}\bm{A}_{h}(t,\bm{r})=-\frac{e^2(1-\delta_m)}{2m}\ave{\Psi_h(\bm{r})\Psi^{\dagger}_{h}(\bm{r})}\bm{A}_h(t,\bm{r})=-\frac{e^2n_X(1-\delta_m)}{2m}\bm{A}_h(t,\bm{r})\label{eq:current-diamagnetic-hole-layer}
  \end{align}
\end{subequations}
In Eq. \eqref{eq:current-diamagnetic-hole-layer}, we have used the normal-ordering rules in Eq. \eqref{eq:normal-order-rules}. We have also used charge neutrality in the EI phase.
The charge diamagnetic current is
\begin{align}
  \ave{\hat{\bm{j}}_{d}^{+}(t,\bm{r})}\equiv &\ave{\hat{\bm{j}}_{e,d}(t,\bm{r})}+\ave{\hat{\bm{j}}_{h,d}(t,\bm{r})}\nonumber\\
  =&-\frac{e^2n_X}{2m}[\bm{A}_e(t,\bm{r})+\bm{A}_h(t,\bm{r})]-\frac{e^2n_X\delta_m}{2m}[\bm{A}_e(t,\bm{r})-\bm{A}_h(t,\bm{r})]\nonumber\\
  =&-\frac{e^2n_X}{m}\bm{A}^{+}(t,\bm{r})-\frac{e^2n_X\delta_m}{2m}\bm{A}^{-}(t,\bm{r})
  \label{eq:current-diamagnetic-charge}
\end{align}
while the exciton diamagnetic current is
\begin{align}
  \ave{\hat{\bm{j}}_{d}^{-}(t,\bm{r})}\equiv&\frac{1}{2}[\ave{\hat{\bm{j}}_{e,d}(t,\bm{r})}-\ave{\hat{\bm{j}}_{h,d}(t,\bm{r})}] \nonumber\\
  =&-\frac{e^2n_X}{4m}[\bm{A}_e(t,\bm{r})-\bm{A}_h(t,\bm{r})]-\frac{e^2n_X\delta_m}{4m}[\bm{A}_e(t,\bm{r})+\bm{A}_h(t,\bm{r})]\nonumber\\
  =&-\frac{e^2n_X}{4m}\bm{A}^{-}(t,\bm{r})-\frac{e^2n_X\delta_m}{2m}\bm{A}^{+}(t,\bm{r})
  \label{eq:current-diamagnetic-exciton}
\end{align}

  If the system has exact electron--hole symmetry such that $\delta_m=0$, the charge (exciton) diamagnetic current is only proportional to the layer-symmetric (antisymmetric) gauge field.
  Then the layer-symmetric and layer-antisymmetric gauge fields couple only to the charge and exciton degrees of freedom, respectively, for both the corresponding charge- and exciton-channel paramagnetic and diamagnetic currents.

The linear coupling Hamiltonian $\hat{H}_c$ can also be written in momentum space. We define the Fourier transform of the gauge field by
\begin{equation}
  A^{\sigma}_{\mu}(t,\bm{r})=\frac{1}{\mathcal{V}}\sum_{\bm{q}}\re^{i\bm{q}\cdot\bm{r}}A^{\sigma}_{\mu}(t,\bm{q})\Longleftrightarrow A^{\sigma}_{\mu}(t,\bm{q})=\int \rd\bm{r}\;\re^{-i\bm{q}\cdot\bm{r}}A^{\sigma}_{\mu}(t,\bm{r})
\end{equation}
The coupling Hamiltonian becomes
\begin{align}
  \hat{H}_c=&\sum_{\sigma=\pm}\int\rd\bm{r}\;[\hat{\varrho}^{\sigma}(\bm{r})\phi^{\sigma}(t,\bm{r})-\hat{\bm{j}}_{p}^{\sigma}(\bm{r})\cdot \bm{A}^{\sigma}(t,\bm{r})]\nonumber\\
  =&\sum_{\sigma=\pm}\frac{1}{\mathcal{V}}\sum_{\bm{q}}\int\rd\bm{r}\;\re^{i\bm{q}\cdot\bm{r}}[\hat{\varrho}^{\sigma}(\bm{r})\phi^{\sigma}(t,\bm{q})-\hat{\bm{j}}_{p}^{\sigma}(\bm{r})\cdot \bm{A}^{\sigma}(t,\bm{q})]\nonumber\\
  =&\sum_{\sigma=\pm}\frac{1}{\mathcal{V}}\sum_{\bm{q}}[\hat{\varrho}^{\sigma}(-\bm{q})\phi^{\sigma}(t,\bm{q})-\hat{\bm{j}}_{p}^{\sigma}(-\bm{q})\cdot \bm{A}^{\sigma}(t,\bm{q})]
\end{align}
where 
\begin{subequations}
  \begin{gather}
    \hat{\varrho}^{\sigma}(\bm{q})=\int\rd\bm{r}\;\re^{-i\bm{q}\cdot\bm{r}}\hat{\varrho}^{\sigma}(\bm{r})\Longleftrightarrow \hat{\varrho}^{\sigma}(\bm{r})=\frac{1}{\mathcal{V}}\sum_{\bm{q}}\re^{i\bm{q}\cdot\bm{r}}\hat{\varrho}^{\sigma}(\bm{q})\\
    \hat{\bm{j}}_{p}^{\sigma}(\bm{q})=\int\rd\bm{r}\;\re^{-i\bm{q}\cdot\bm{r}}\hat{\bm{j}}_{p}^{\sigma}(\bm{r})\Longleftrightarrow \hat{\bm{j}}_{p}^{\sigma}(\bm{r})=\frac{1}{\mathcal{V}}\sum_{\bm{q}}\re^{i\bm{q}\cdot\bm{r}}\hat{\bm{j}}_{p}^{\sigma}(\bm{q})
\end{gather}
\end{subequations}
Explicitly,
\begin{subequations}
  \begin{align}
    \hat{\varrho}^{+}(\bm{q})=&-\frac{e}{\mathcal{V}}\int\rd\bm{r}\;\re^{-i\bm{q}\cdot\bm{r}}\sum_{\bm{k}\bm{k}'}\re^{-i(\bm{k}-\bm{k}')\cdot\bm{r}}C^{\dagger}_{\bm{k}}C_{\bm{k}'}=\sum_{\bm{k}}-eC^{\dagger}_{\bm{k}-\bm{q}/2}C_{\bm{k}+\bm{q}/2}\\
    \hat{\bm{j}}_{p}^{+}(\bm{q})=&\frac{1}{\mathcal{V}}\int\rd\bm{r}\;\re^{-i\bm{q}\cdot\bm{r}}\sum_{\bm{k}\bm{k}'}\re^{-i(\bm{k}-\bm{k}')\cdot\bm{r}}\left[\frac{ie\hbar}{4m}C^{\dagger}_{\bm{k}}\sigma_z(i\bm{k}'+i\bm{k})C_{\bm{k}'}+\frac{ie\hbar \delta_m}{4m}C^{\dagger}_{\bm{k}}(i\bm{k}'+i\bm{k})C_{\bm{k}'}
    \right]\nonumber\\
    =&\sum_{\bm{k}}-\frac{e\hbar\bm{k}}{2m}C^{\dagger}_{\bm{k}-\bm{q}/2}\sigma_z C_{\bm{k}+\bm{q}/2}
    +\sum_{\bm{k}}-\frac{e\hbar\bm{k}\delta_m}{2m}C^{\dagger}_{\bm{k}-\bm{q}/2} C_{\bm{k}+\bm{q}/2}
    \\
    \hat{\varrho}^{-}(\bm{q})=&-\frac{e}{2\mathcal{V}}\int\rd\bm{r}\;\re^{-i\bm{q}\cdot\bm{r}}\sum_{\bm{k}\bm{k}'}\re^{-i(\bm{k}-\bm{k}')\cdot\bm{r}}C^{\dagger}_{\bm{k}}\sigma_z C_{\bm{k}'}=\sum_{\bm{k}}-\frac{e}{2}C^{\dagger}_{\bm{k}-\bm{q}/2}\sigma_z C_{\bm{k}+\bm{q}/2}\\
    \hat{\bm{j}}_{p}^{-}(\bm{q})=&\frac{1}{\mathcal{V}}\int\rd\bm{r}\;\re^{-i\bm{q}\cdot\bm{r}}\sum_{\bm{k}\bm{k}'}\re^{-i(\bm{k}-\bm{k}')\cdot\bm{r}}\left[\frac{ie\hbar}{8m}C^{\dagger}_{\bm{k}}(i\bm{k}'+i\bm{k})C_{\bm{k}'}+\frac{ie\hbar\delta_m}{8m}C^{\dagger}_{\bm{k}}\sigma_z(i\bm{k}'+i\bm{k})C_{\bm{k}'}\right]\nonumber\\
    =&\sum_{\bm{k}}-\frac{e\hbar\bm{k}}{4m}C^{\dagger}_{\bm{k}-\bm{q}/2}C_{\bm{k}+\bm{q}/2}+\sum_{\bm{k}}-\frac{e\hbar\bm{k}\delta_m}{4m}C^{\dagger}_{\bm{k}-\bm{q}/2}\sigma_z C_{\bm{k}+\bm{q}/2}
  \end{align}
  \label{eq:current-density-momentum-space}
\end{subequations}
where $C^{\dagger}_{\bm{k}}=[c^{\dagger}_{e\bm{k}},c^{\dagger}_{h\bm{k}}]$ is the operator spinor.
It is convenient to define the bare vertices as
\begin{equation}
  \gamma^{+}_{\mu=0,\bm{k}}=e\sigma_0,\;\gamma^{+}_{\mu=12,\bm{k}}=-\frac{e\hbar\bm{k}}{2m}\sigma_z-\frac{e\hbar\bm{k}\delta_m}{2m}\sigma_0,\;\gamma^{-}_{\mu=0,\bm{k}}=\frac{e}{2}\sigma_z,\;\gamma^{-}_{\mu=12,\bm{k}}=-\frac{e\hbar\bm{k}}{4m}\sigma_0-\frac{e\hbar\bm{k}\delta_m}{4m}\sigma_z\label{eq:vertex-function}
\end{equation}
The charge- and exciton-channel paramagnetic four-current operator $\hat{j}^{\sigma}_{p\mu}(\bm{q})\equiv (-\hat{\varrho}^{\sigma}(\bm{q}),\hat{\bm{j}}^{\sigma}_{p}(\bm{q}))$ is then
\begin{equation}
  \hat{j}^{\sigma}_{p\mu}(\bm{q})=\sum_{\bm{k}}C^{\dagger}_{\bm{k}-\bm{q}/2}\gamma^{\sigma}_{\mu,\bm{k}}C_{\bm{k}+\bm{q}/2}
\end{equation}

\subsection{\texorpdfstring{The electromagnetic response kernel in the long-wavelength limit}{The electromagnetic response kernel in the long-wavelength limit}}\label{app:response-kernel}
{
\raggedbottom

In this subsection, we focus on the electron--hole-symmetric limit $\delta_m=0$ and derive the electromagnetic response kernel in the long-wavelength limit. The effect of unequal electron and hole masses is discussed in Appendix \ref{app:eh_sy}. For clarity, we call $\gamma^{\sigma}_{\mu,\bm{k}}$ in Eq. \eqref{eq:vertex-function} the \emph{bare vertex} and $\mathrm J^{\sigma}_{\mu,n}$ in Eq. \eqref{eq:collective-mode-vertex-overlap} the \emph{collective-mode vertex overlap}. The bare vertex describes the single-particle coupling to the external field, while the collective-mode vertex overlap determines how strongly the external field couples to a given collective mode.

\subsubsection{\texorpdfstring{Collective-mode spectral representation}{Collective-mode spectral representation}}

Let $\sigma=+$ and $\sigma=-$ denote the charge and exciton channels, respectively. For a collective mode $n$, we define its vertex overlap with the operator $\hat j^{\sigma}_{p\mu}$ as
{
\begin{equation}
  \mathrm{J}^{\sigma}_{\mu,n}(\bm{q})
  \equiv\sum_{\bm{k}}\Phi^{\dagger}_{n\bm{k}}(\bm{q})
  \begin{bmatrix}
    \gamma^{\sigma}_{\mu,cv\bm{k}}(-\bm{q})\\
    \gamma^{\sigma}_{\mu,vc-\bm{k}}(-\bm{q})
  \end{bmatrix}
  =\sum_{\bm{k}}\Phi_{n\bm{k}}^{\dagger}(\bm{q})
  \begin{bmatrix}
    \bra{c\bm{k}+\bm{q}/2}\gamma^{\sigma}_{\mu,\bm{k}}\ket{v\bm{k}-\bm{q}/2}\\
    \bra{v-\bm{k}+\bm{q}/2}\gamma^{\sigma}_{\mu,-\bm{k}}\ket{c-\bm{k}-\bm{q}/2}
  \end{bmatrix}.
  \label{eq:collective-mode-vertex-overlap}
\end{equation}
}
Equation \eqref{eq:collective-mode-vertex-overlap} contracts the two components of the collective-mode wavefunction with the corresponding interband matrix elements of the bare vertex. In the electron--hole-symmetric case, the quasiparticle wavefunctions, collective-mode wavefunctions, and bare vertices can all be chosen real, as shown in Eqs. \eqref{eq:orthonormal-real}, \eqref{eq:quasi-particle-wavefunction-full}, and \eqref{eq:vertex-function}. Thus $\mathrm J^{\sigma}_{\mu,n}$ is real, and the correlation function is unchanged when the two operators are interchanged:
{
\begin{equation}
  \begin{aligned}
  C_{\hat{j}^{\sigma}_{p\mu}\hat{j}^{\sigma'}_{p\nu}}(\omega,\bm{q})
  =&\frac{1}{\mathcal{V}}\sum_n
  \frac{\omega_n(\bm{q})[\mathrm{J}^{\sigma}_{\mu,n}(\bm{q})]^*
  \mathrm{J}^{\sigma'}_{\nu,n}(\bm{q})}
  {\omega^+-\omega_n(\bm{q})}\\
  =&\frac{1}{\mathcal{V}}\sum_n
  \frac{\omega_n(\bm{q})[\mathrm{J}^{\sigma'}_{\nu,n}(\bm{q})]^*
  \mathrm{J}^{\sigma}_{\mu,n}(\bm{q})}
  {\omega^+-\omega_n(\bm{q})}
  =C_{\hat{j}^{\sigma'}_{p\nu}\hat{j}^{\sigma}_{p\mu}}(\omega,\bm{q}).
  \end{aligned}
  \label{eq:correlation-spectral-representation}
\end{equation}
}

\subsubsection{\texorpdfstring{Regular gapped modes}{Regular gapped modes}}

We first consider the regular gapped modes, whose excitation energies remain finite at $q=0$. To determine their coupling to the charge and exciton operators in the long-wavelength limit, we expand the interband matrix elements of the bare vertices in powers of $q$. At $\delta_m=0$, the bare vertices in Eq. \eqref{eq:vertex-function} are proportional to either $\sigma_0$ or $\sigma_z$, and the corresponding matrix elements are
{
\begin{subequations}\label{eq:band-basis-bare-vertex-expansion}
  \begin{align}
    \bra{c\bm{k}+\bm{q}/2}\sigma_0\ket{v\bm{k}-\bm{q}/2}
    =&-\bm{q}\cdot\inp{c\bm{k}}{\nabla_{\bm{k}}v\bm{k}}+\mathcal{O}(q^2),\\
    \bra{v-\bm{k}+\bm{q}/2}\sigma_0\ket{c-\bm{k}-\bm{q}/2}
    =&-\bm{q}\cdot\inp{\nabla_{\bm{k}}v\bm{k}}{c\bm{k}}+\mathcal{O}(q^2),\\
    \bra{c\bm{k}+\bm{q}/2}\sigma_z\ket{v\bm{k}-\bm{q}/2}
    =&2\alpha_k\beta_k+\mathcal{O}(q),\\
    \bra{v-\bm{k}+\bm{q}/2}\sigma_z\ket{c-\bm{k}-\bm{q}/2}
    =&2\alpha_k\beta_k+\mathcal{O}(q).
  \end{align}
\end{subequations}
}
Since the band eigenstates are chosen to be real, we have $\inp{c\bm{k}}{\nabla_{\bm{k}}v\bm{k}}=\inp{\nabla_{\bm{k}}v\bm{k}}{c\bm{k}}$. Substituting Eq. \eqref{eq:band-basis-bare-vertex-expansion} into Eq. \eqref{eq:collective-mode-vertex-overlap}, we obtain
{
\begin{subequations}\label{eq:regular-mode-vertex-overlap-expansion}
  \begin{align}
    \mathrm{J}^{+}_{0,n}(\bm{q})
    =&e\bm{q}\cdot\sum_{\bm{k}}\inp{c\bm{k}}{\nabla_{\bm{k}}v\bm{k}}
    \Phi^{\dagger}_{n\bm{k}}(\bm{0})
    \begin{bmatrix}1\\1\end{bmatrix}+\mathcal{O}(q^2),
    \label{eq:regular-mode-charge-density-overlap}\\
    \mathrm{J}^{+}_{\mu,n}(\bm{q})
    =&-\frac{e\hbar}{m}\sum_{\bm{k}}k_{\mu}\alpha_k\beta_k
    \Phi^{\dagger}_{n\bm{k}}(\bm{0})
    \begin{bmatrix}1\\-1\end{bmatrix}+\mathcal{O}(q),
    \qquad \mu\in\{1,2\},
    \label{eq:regular-mode-charge-current-overlap}\\
    \mathrm{J}^{-}_{0,n}(\bm{q})
    =&-e\sum_{\bm{k}}\alpha_k\beta_k
    \Phi^{\dagger}_{n\bm{k}}(\bm{0})
    \begin{bmatrix}1\\1\end{bmatrix}+\mathcal{O}(q),
    \label{eq:regular-mode-exciton-density-overlap}\\
    \mathrm{J}^{-}_{\mu,n}(\bm{q})
    =&\frac{e\hbar}{4m}\bm{q}\cdot\sum_{\bm{k}}k_{\mu}
    \inp{c\bm{k}}{\nabla_{\bm{k}}v\bm{k}}
    \Phi^{\dagger}_{n\bm{k}}(\bm{0})
    \begin{bmatrix}1\\-1\end{bmatrix}+\mathcal{O}(q^2),
    \qquad \mu\in\{1,2\}.
    \label{eq:regular-mode-exciton-current-overlap}
  \end{align}
\end{subequations}
}
From Eq. \eqref{eq:regular-mode-vertex-overlap-expansion}, we can identify the angular-momentum sectors that couple to each operator. The two components of $\inp{c\bm{k}}{\nabla_{\bm{k}}v\bm{k}}$ and $k_{\mu}\alpha_k\beta_k$ transform as $p_x$ and $p_y$, while $\alpha_k\beta_k$ is rotationally symmetric and belongs to the $s$ sector. For $\mu\ne\nu$, the product $k_{\mu}\inp{c\bm{k}}{\partial_{k_{\nu}}v\bm{k}}$ has $d_{xy}$ symmetry. When $\mu=\nu$, it contains both $s$ and $d_{x^2-y^2}$ components. These selection rules apply to the regular gapped modes; the Goldstone mode will be discussed separately.

It is convenient to further decompose the current into the longitudinal and transverse components with respect to $\bm q$. For $\bm q\ne\bm0$, we define $q=|\bm q|$, $\hat{\bm q}=\bm q/q$, and $\hat t_a=-\epsilon_{ab}\hat q_b$, where $\epsilon_{12}=+1$. Here and below, repeated spatial indices $a,b\in\{1,2\}$ are summed. The corresponding current operators and vertex overlaps are
{
\begin{equation}
  \begin{gathered}
    \hat{j}^{\sigma}_{pL}=\hat{q}_a\hat{j}^{\sigma}_{pa},\qquad
    \hat{j}^{\sigma}_{pT}=\hat{t}_a\hat{j}^{\sigma}_{pa},\\
    \mathrm{J}^{\sigma}_{L,n}=\hat{q}_a\mathrm{J}^{\sigma}_{a,n},\qquad
    \mathrm{J}^{\sigma}_{T,n}=\hat{t}_a\mathrm{J}^{\sigma}_{a,n}.
  \end{gathered}
  \label{eq:longitudinal-transverse-projections}
\end{equation}
}
In the absence of interlayer tunneling, the particle number in each layer is separately conserved. Therefore, the vertex overlaps in both the charge and exciton channels satisfy the mode-resolved Ward identity
{
\begin{equation}
  \omega_n(q)\mathrm{J}^{\sigma}_{0,n}(q)+q\mathrm{J}^{\sigma}_{L,n}(q)=0.
  \label{eq:mode-resolved-ward-identity}
\end{equation}
}
The plus sign in Eq. \eqref{eq:mode-resolved-ward-identity} follows from our convention $\hat j^{\sigma}_{p0}=-\hat\varrho^{\sigma}$. In the following, we take $\bm q=q\hat x$, so the longitudinal and transverse directions correspond to $L=1$ and $T=2$, respectively. For the exciton-current vertex in Eq. \eqref{eq:regular-mode-exciton-current-overlap}, the longitudinal component contains the angular factor $\cos^2\theta_{\bm k}=(1+\cos2\theta_{\bm k})/2$ and therefore couples to the $s$ and $d_{x^2-y^2}$ modes. The transverse component contains $\sin\theta_{\bm k}\cos\theta_{\bm k}=\sin2\theta_{\bm k}/2$ and couples to the $d_{xy}$ modes. The leading power of $q$ for each allowed coupling is summarized in Table \ref{tab:vertex-overlap-power-counting}. In the table, $s\setminus\{G\}$ denotes the regular gapped $s$-wave modes, and $G$ denotes the positive-frequency Goldstone mode.

For a regular gapped mode, Eq. \eqref{eq:regular-mode-exciton-current-overlap} gives $\mathrm{J}^{-}_{L,n}=\mathcal{O}(q)$. Since $\omega_n(0)\ne0$, the Ward identity then requires $\mathrm{J}^{-}_{0,n}=\mathcal{O}(q^2)$. Thus, both the constant term shown explicitly in Eq. \eqref{eq:regular-mode-exciton-density-overlap} and its linear correction in $q$ must vanish. For modes outside the $s$ sector, the constant term vanishes by angular symmetry. For a regular $s$-wave mode, its cancellation follows from the generalized orthogonality to the Goldstone mode, as shown below.
The ground-state density matrix in the layer basis is
{
\[
  \rho^X_{\bm k}=\ket{v\bm k}\bra{v\bm k}
  =\begin{bmatrix}
    \alpha_k^2 & \alpha_k\beta_k\\
    \alpha_k\beta_k & \beta_k^2
  \end{bmatrix}.
\]
}
An infinitesimal relative phase rotation of the two layers, $c_e\rightarrow e^{i\theta/2}c_e$ and $c_h\rightarrow e^{-i\theta/2}c_h$, changes this density matrix by
{
\[
  \delta\rho^G_{\bm k}
  =\frac{i\theta}{2}[\sigma_z,\rho^X_{\bm k}]
  =i\theta\alpha_k\beta_k
  \begin{bmatrix}0&1\\-1&0\end{bmatrix}.
\]
}
Projecting $\delta\rho^G_{\bm k}$ onto the quasiparticle states in Eq. \eqref{eq:quasi-particle-wavefunction} gives
{
\[
  \delta\rho^G_{cv,\bm k}=i\theta\alpha_k\beta_k,
  \qquad
  \delta\rho^G_{vc,\bm k}=-i\theta\alpha_k\beta_k.
\]
}
Therefore, up to an overall phase and a nonzero normalization constant $\mathcal N_G$, the Goldstone eigenvector at $\bm q=\bm0$ is
{
\[
  \Phi_{G\bm k}(\bm0)=\mathcal N_G\alpha_k\beta_k
  \begin{bmatrix}1\\-1\end{bmatrix}.
\]
}
This is the phase fluctuation of the condensate: because $\Delta_k$ is real and negative in our convention, $\alpha_k\beta_k=-\Delta_k/(2\xi_k)$. The Goldstone mode satisfies $\mathcal H(\bm0)\Phi_G(\bm0)=0$, whereas a regular gapped mode satisfies $\mathcal H(\bm0)\Phi_n(\bm0)=\hbar\omega_n(0)\tau_z\Phi_n(\bm0)$ with $\omega_n(0)\ne0$. Since $\mathcal H(\bm0)$ is Hermitian,
{
\begin{align*}
  0
  &=\Phi_G^\dagger(\bm0)\mathcal H(\bm0)\Phi_n(\bm0)\\
  &=\hbar\omega_n(0)\Phi_G^\dagger(\bm0)\tau_z\Phi_n(\bm0).
\end{align*}
}
It follows that $\Phi_G^\dagger(\bm0)\tau_z\Phi_n(\bm0)=0$. Substituting the Goldstone eigenvector above and taking the complex conjugate gives
{
\begin{equation}
  \sum_{\bm{k}}\alpha_k\beta_k\Phi^{\dagger}_{n\bm{k}}(\bm{0})
  \begin{bmatrix}1\\1\end{bmatrix}=0,
  \qquad n\in s\setminus\{G\}.
  \label{eq:regular-s-mode-overlap-cancellation}
\end{equation}
}

\begin{table}[!ht]
  
  \centering
  \setlength{\tabcolsep}{4pt}
  \renewcommand{\arraystretch}{1.35}
  \begin{tabular}{c c c c p{0.34\textwidth}}
    \hline\hline
    Channel and mode sector
    & $\mathrm{J}^{\sigma}_{0,n}$
    & $\mathrm{J}^{\sigma}_{L,n}$
    & $\mathrm{J}^{\sigma}_{T,n}$
    & Formula or constraint \\
    \hline
    $+,\ n\in p_x$
    & $\mathcal{O}(q)$
    & $\mathcal{O}(q^0)$
    & $0$
    & Eqs. \eqref{eq:regular-mode-charge-density-overlap} and \eqref{eq:regular-mode-charge-current-overlap}; angular selection \\
    $+,\ n\in p_y$
    & $0$
    & $0$
    & $\mathcal{O}(q^0)$
    & Eq. \eqref{eq:regular-mode-charge-current-overlap}; angular and reflection selection \\
    $-,\ n\in\bigl(s\setminus\{G\}\bigr)\cup d_{x^2-y^2}$
    & $\mathcal{O}(q^2)$
    & $\mathcal{O}(q)$
    & $0$
    & Eq. \eqref{eq:regular-mode-exciton-density-overlap}; angular selection for $d_{x^2-y^2}$. Eq. \eqref{eq:regular-s-mode-overlap-cancellation}; Goldstone-mode orthogonality for $s\setminus\{G\}$. Eqs. \eqref{eq:regular-mode-exciton-current-overlap} and \eqref{eq:mode-resolved-ward-identity}; angular and reflection selection and the Ward identity \\
    $-,\ n\in d_{xy}$
    & $0$
    & $0$
    & $\mathcal{O}(q)$
    & Eq. \eqref{eq:regular-mode-exciton-current-overlap}; angular and reflection selection \\
    $-,\ n=G$
    & $\mathcal{O}(q^0)$
    & $\mathcal{O}(q^0)$
    & $0$
    & Eqs. \eqref{eq:mode-resolved-ward-identity} and \eqref{eq:Goldstone-vertex-overlaps}; static compressibility sum rule and the Ward identity \\
    \hline\hline
  \end{tabular}
  \caption{Leading powers of $q$ in the collective-mode vertex overlaps for $\bm{q}=q\hat{x}$. A zero denotes a coupling forbidden by angular-momentum or reflection symmetry. Terms not listed vanish or enter at higher order. The numerical results are given in Appendix \ref{app:numerical-overlap-power-counting}.}
  \label{tab:vertex-overlap-power-counting}
\end{table}

To calculate the regular-mode contribution to the correlation function, we pair each positive-frequency mode with its negative-frequency partner. Equation \eqref{eq:generalized-eigenvector} gives, for general $\bm q$,
{
\begin{equation}
  \begin{gathered}
    \Phi^{\dagger}_{n\bm{k}}(\bm{q})\begin{bmatrix}1\\1\end{bmatrix}
    =\Phi^{\dagger}_{-n\bm{k}}(\bm{q})\begin{bmatrix}1\\1\end{bmatrix}
    =x_{n\bm{k}}^*(\bm{q}),\\
    \Phi^{\dagger}_{n\bm{k}}(\bm{q})\begin{bmatrix}1\\-1\end{bmatrix}
    =-\Phi^{\dagger}_{-n\bm{k}}(\bm{q})\begin{bmatrix}1\\-1\end{bmatrix}=-iy_n^*(\bm{q}).
  \end{gathered}
  \label{eq:regular-gapped-eigenvector}
\end{equation}
}
The two interband matrix elements in Eq. \eqref{eq:collective-mode-vertex-overlap} have the same sign for a density vertex and opposite signs for a current vertex. Equation \eqref{eq:regular-gapped-eigenvector} therefore gives $\mathrm J^{\sigma}_{0,-n}(\bm q)=\mathrm J^{\sigma}_{0,n}(\bm q)$ and $\mathrm J^{\sigma}_{L,-n}(\bm q)=-\mathrm J^{\sigma}_{L,n}(\bm q)$. Together with $\omega_{-n}(\bm q)=-\omega_n(\bm q)$, these relations preserve the Ward identity at finite $q$:
{
\[
  \omega_{-n}\mathrm J^{\sigma}_{0,-n}+q\mathrm J^{\sigma}_{L,-n}
  =-\left(\omega_n\mathrm J^{\sigma}_{0,n}+q\mathrm J^{\sigma}_{L,n}\right)=0.
\]
}
Thus, the products entering the diagonal correlation functions are even under $n\to-n$, while $(\mathrm{J}^{\sigma}_{L,n})^*\mathrm{J}^{\sigma}_{0,n}$ is odd. After summing over the two branches, the diagonal correlation functions contain the factor $2\omega_n^2/[(\omega^+)^2-\omega_n^2]$, while the density--longitudinal-current correlation function contains $2\omega^+\omega_n/[(\omega^+)^2-\omega_n^2]$. We denote the contribution from the regular gapped modes by $C'$, where the Goldstone mode is excluded. Taking $p_L=p_x$ and $p_T=p_y$, we obtain
{
\begin{subequations}\label{eq:correlation-function-long-wavelength}
  \begin{align}
    C'_{\hat{j}^{+}_{p0}\hat{j}^{+}_{p0}}(\omega,q)
    =&\frac{1}{\mathcal{V}}\sum_{\substack{\omega_n>0\\ n\in p_x}}
    \frac{2|\mathrm{J}^{+}_{0,n}|^2\omega_n^2(q)}{(\omega^+)^2-\omega_n^2(q)}
    =\mathcal{O}(q^2),\\
    C'_{\hat{j}^{+}_{pL}\hat{j}^{+}_{p0}}(\omega,q)
    =&-\frac{2q\omega^+}{\mathcal{V}}\sum_{\substack{\omega_n>0\\ n\in p_x}}
    \frac{|\mathrm{J}^{+}_{L,n}|^2}{(\omega^+)^2-\omega_n^2(q)}
    =\mathcal{O}(q),\\
    C'_{\hat{j}^{+}_{p\lambda}\hat{j}^{+}_{p\lambda}}(\omega,q)
    =&\frac{1}{\mathcal{V}}\sum_{\substack{\omega_n>0\\ n\in p_\lambda}}
    \frac{2|\mathrm{J}^{+}_{\lambda,n}|^2\omega_n^2(q)}{(\omega^+)^2-\omega_n^2(q)}
    =\mathcal{O}(q^0),\qquad \lambda\in\{L,T\},\displaybreak[2]\\
    C'_{\hat{j}^{-}_{p0}\hat{j}^{-}_{p0}}(\omega,q)
    =&\frac{2q^2}{\mathcal{V}}\sum_{\substack{\omega_n>0\\ n\in (s\setminus\{G\})\cup d_{x^2-y^2}}}
    \frac{|\mathrm{J}^{-}_{L,n}|^2}{(\omega^+)^2-\omega_n^2(q)}
    =\mathcal{O}(q^4),\\
    C'_{\hat{j}^{-}_{pL}\hat{j}^{-}_{p0}}(\omega,q)
    =&-\frac{2q\omega^+}{\mathcal{V}}\sum_{\substack{\omega_n>0\\ n\in (s\setminus\{G\})\cup d_{x^2-y^2}}}
    \frac{|\mathrm{J}^{-}_{L,n}|^2}{(\omega^+)^2-\omega_n^2(q)}
    =\mathcal{O}(q^3),\\
    C'_{\hat{j}^{-}_{pL}\hat{j}^{-}_{pL}}(\omega,q)
    =&\frac{1}{\mathcal{V}}\sum_{\substack{\omega_n>0\\ n\in (s\setminus\{G\})\cup d_{x^2-y^2}}}
    \frac{2|\mathrm{J}^{-}_{L,n}|^2\omega_n^2(q)}{(\omega^+)^2-\omega_n^2(q)}
    =\mathcal{O}(q^2),\\
    C'_{\hat{j}^{-}_{pT}\hat{j}^{-}_{pT}}(\omega,q)
    =&\frac{1}{\mathcal{V}}\sum_{\substack{\omega_n>0\\ n\in d_{xy}}}
    \frac{2|\mathrm{J}^{-}_{T,n}|^2\omega_n^2(q)}{(\omega^+)^2-\omega_n^2(q)}
    =\mathcal{O}(q^2),\\
    C'_{\hat{j}^{\sigma}_{pT}\hat{j}^{\sigma}_{p0}}(\omega,q)
    =&C'_{\hat{j}^{\sigma}_{pT}\hat{j}^{\sigma}_{pL}}(\omega,q)=0,
    \qquad \sigma\in\{+,-\}.
  \end{align}
\end{subequations}
}
The last line of Eq. \eqref{eq:correlation-function-long-wavelength} follows from reflection symmetry about the propagation direction. Under this reflection, the transverse current is mirror-odd, while the density and longitudinal current are mirror-even, so their correlation functions vanish. In addition, for a regular gapped exciton mode, $\mathrm{J}^{-}_{L,n}=\mathcal{O}(q)$ and $\mathrm{J}^{-}_{0,n}=\mathcal{O}(q^2)$. Therefore, the exciton density--density and longitudinal-current--density correlation functions start from $\mathcal{O}(q^4)$ and $\mathcal{O}(q^3)$, respectively.

\subsubsection{\texorpdfstring{Goldstone-mode contribution}{Goldstone-mode contribution}}

The above power counting for the exciton-density vertex assumes a finite excitation energy at $q=0$ and therefore does not apply to the Goldstone mode. In the long-wavelength limit, the Goldstone mode has the linear dispersion
{
\begin{equation}
  \omega_G(q)=vq+\mathcal{O}(q^2).
  \label{eq:Goldstone-dispersion}
\end{equation}
}
Combining the positive- and negative-frequency Goldstone branches in Eq. \eqref{eq:correlation-spectral-representation}, we obtain their contribution to the diagonal correlation functions:
{
\begin{equation}
  C^{G}_{\hat{j}^{-}_{p\mu}\hat{j}^{-}_{p\mu}}(\omega,q)
  =\frac{2}{\mathcal{V}}
  \frac{|\mathrm{J}^{-}_{\mu,G}(q)|^2\omega_G^2(q)}
  {(\omega^+)^2-\omega_G^2(q)},
  \qquad \mu\in\{0,L\}.
  \label{eq:Goldstone-spectral-representation}
\end{equation}
}
There is no diamagnetic contact term in the exciton-density response. Thus, in the static limit, the response can be separated into the regular-mode and Goldstone-mode contributions as
{
\begin{equation}
  K^{-}_{00}(0,q)
  =-C'_{\hat{j}^{-}_{p0}\hat{j}^{-}_{p0}}(0,q)
  +\frac{2}{\mathcal{V}}\left|\mathrm{J}^{-}_{0,G}(q)\right|^2.
  \label{eq:Goldstone-static-sum-rule}
\end{equation}
}
As $q\to0$, the regular-mode contribution vanishes as $q^4$, while $K^{-}_{00}(0,q)$ approaches the finite exciton compressibility $\kappa$. Therefore, Eq. \eqref{eq:Goldstone-static-sum-rule} fixes the exciton-density vertex overlap of the Goldstone mode. The corresponding longitudinal-current vertex overlap is then determined by the Ward identity:
{
\begin{equation}
  \lim_{q\to0}\frac{2}{\mathcal{V}}\left|\mathrm{J}^{-}_{0,G}(q)\right|^2=\kappa,
  \qquad
  \lim_{q\to0}\frac{2}{\mathcal{V}}\left|\mathrm{J}^{-}_{L,G}(q)\right|^2=\kappa v^2,
  \qquad
  \mathrm{J}^{-}_{L,G}(q)=-\frac{\omega_G(q)}{q}\mathrm{J}^{-}_{0,G}(q).
  \label{eq:Goldstone-vertex-overlaps}
\end{equation}
}
We can see from Eq. \eqref{eq:Goldstone-vertex-overlaps} that both Goldstone-mode vertex overlaps remain finite as $q\to0$. This is different from the longitudinal-current vertex overlap of a regular gapped exciton mode, which vanishes at least linearly with $q$. At a fixed nonzero frequency, the Goldstone contribution in Eq. \eqref{eq:Goldstone-spectral-representation} is of order $q^2$ because $\omega_G^2\propto q^2$, and it gives the leading exciton-density response in the long-wavelength limit. After including the longitudinal diamagnetic weight $D_-=e^2n_X/(4m)=\kappa v^2$, we obtain
{
\begin{subequations}\label{eq:Goldstone-response}
  \begin{align}
    K^{-}_{00}(\omega,q)=&-\frac{\kappa v^2q^2}{(\omega^+)^2-v^2q^2},\\
    K^{-}_{L}(\omega,q)=&-\frac{\kappa v^2\omega^2}{(\omega^+)^2-v^2q^2}.
  \end{align}
\end{subequations}
}

\subsubsection{\texorpdfstring{Full response kernel}{Full response kernel}}

Finally, the full response kernel is obtained by adding the diamagnetic contact term to the paramagnetic correlation function. We define $D_+=e^2n_X/m$ and $D_-=e^2n_X/(4m)$. Since the charge and exciton channels are decoupled at $\delta_m=0$, the response kernel can be written as
{
\begin{equation}
  \begin{aligned}
    K^{\sigma\sigma'}_{\mu\nu}(\omega,\bm{q})
    =&\,\delta_{\sigma\sigma'}K^{\sigma}_{\mu\nu}(\omega,\bm{q}),\\
    K^{\sigma}_{\mu\nu}(\omega,\bm{q})
    =&-C_{\hat{j}^{\sigma}_{p\mu}\hat{j}^{\sigma}_{p\nu}}(\omega,\bm{q})
    -(1-\delta_{\mu0})(1-\delta_{\nu0})D_{\sigma}\delta_{\mu\nu}.
  \end{aligned}
  \label{eq:response-kernel-from-correlation}
\end{equation}
}
Here the correlation function without a prime includes all collective modes, including the Goldstone mode in the exciton channel. Due to rotational symmetry, the spatial response can be decomposed into the longitudinal and transverse components $K^{\sigma}_{L}=\hat{q}_aK^{\sigma}_{ab}\hat{q}_b$ and $K^{\sigma}_{T}=\hat{t}_aK^{\sigma}_{ab}\hat{t}_b$. When $\bm{q}=q\hat{x}$, the Ward identity further relates the density and density--current response functions to $K^{\sigma}_{L}$, giving
{
\begin{equation}
  K^{\sigma}_{\mu\nu}(\omega,q\hat{x})=
  \begin{bmatrix}
    \dfrac{q^2}{\omega^2}K^{\sigma}_{L} & -\dfrac{q}{\omega}K^{\sigma}_{L} & 0\\
    -\dfrac{q}{\omega}K^{\sigma}_{L} & K^{\sigma}_{L} & 0\\
    0 & 0 & K^{\sigma}_{T}
  \end{bmatrix}.
  \label{eq:response-kernel-long-wavelength-matrix}
\end{equation}
}
\flushbottom


}
\subsection{The effective dielectric function}\label{app:effective-dielectric}

\begin{figure}
  \centering
  \includegraphics[width=0.8\textwidth]{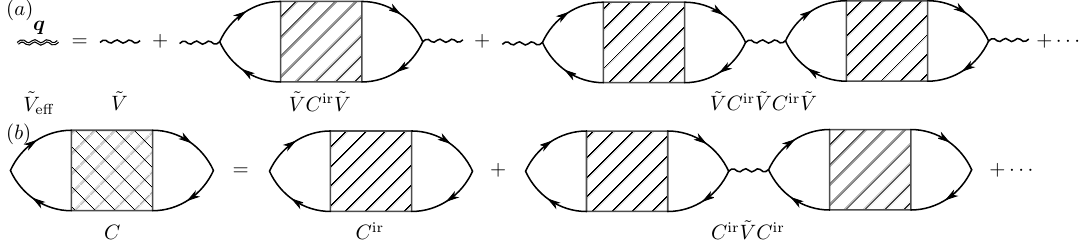}
  \caption{The Feynman diagrams for the effective interaction.}
  \label{fig:feynman-interaction}
\end{figure}

The effective dielectric function $\epsilon_{\mathrm{eff}}(\omega,\bm{q})$ is defined as 
\begin{equation}
  \epsilon_{\mathrm{eff}}(\omega,\bm{q})=\frac{\tilde{V}(\bm{q})}{\tilde{V}_{\mathrm{eff}}(\omega,\bm{q})}
\end{equation}
where $\tilde{V}(\bm{q})$ is the bare Coulomb interaction and $\tilde{V}_{\mathrm{eff}}(\omega,\bm{q})$ is the effective Coulomb interaction renormalized by charge-density fluctuations, as illustrated in Fig. \ref{fig:feynman-interaction}(a).
In Fig. \ref{fig:feynman-interaction}, $C\equiv C_{\hat{j}^{+}_{p0}\hat{j}^{+}_{p0}}(\omega,\bm{q})$ is the charge density--density correlation function and $C^{\mathrm{ir}}\equiv C^{\mathrm{ir}}_{\hat{j}^{+}_{p0}\hat{j}^{+}_{p0}}(\omega,\bm{q})$ is its irreducible counterpart. Within TDHF, $C^{\mathrm{ir}}$ is given by the ladder diagrams in Fig. \ref{fig:feynman-polarization}(b).
The bare Coulomb interaction is $\tilde{V}(\bm{q})=2\pi/\epsilon q$ for charge in the same layer and $\tilde{U}(\bm{q})=\tilde{V}(\bm{q})\re^{-qd}$ for charge in different layers.
In the long-wavelength limit $qd\to0$, $\tilde{U}(\bm{q})\approx\tilde{V}(\bm{q})$, so the intra- and interlayer Coulomb interactions need not be distinguished. The diagrams in Fig. \ref{fig:feynman-interaction}(a) then give
\begin{align}
  \tilde{V}_{\mathrm{eff}}(\omega,\bm{q})=&\tilde{V}(\bm{q})+\tilde{V}(\bm{q})C^{\mathrm{ir}}_{\hat{j}^{+}_{p0}\hat{j}^{+}_{p0}}(\omega,\bm{q})\tilde{V}(\bm{q})+\tilde{V}(\bm{q})C^{\mathrm{ir}}_{\hat{j}^{+}_{p0}\hat{j}^{+}_{p0}}(\omega,\bm{q})\tilde{V}(\bm{q})C^{\mathrm{ir}}_{\hat{j}^{+}_{p0}\hat{j}^{+}_{p0}}(\omega,\bm{q})\tilde{V}(\bm{q})+\cdots\nonumber\\
  =&\sum_{n=0}^{\infty}[\tilde{V}(\bm{q})C^{\mathrm{ir}}_{\hat{j}^{+}_{p0}\hat{j}^{+}_{p0}}(\omega,\bm{q})]^n\tilde{V}(\bm{q})\nonumber\\
  =&[1-\tilde{V}(\bm{q})C^{\mathrm{ir}}_{\hat{j}^{+}_{p0}\hat{j}^{+}_{p0}}(\omega,\bm{q})]^{-1}\tilde{V}(\bm{q})
\end{align}
On the other hand, according to Fig. \ref{fig:feynman-interaction}(b), the total charge density--density correlation function is given by
\begin{equation}
  C_{\hat{j}^{+}_{p0}\hat{j}^{+}_{p0}}(\omega,\bm{q})=C^{\mathrm{ir}}_{\hat{j}^{+}_{p0}\hat{j}^{+}_{p0}}(\omega,\bm{q})+\tilde{V}(\bm{q})C^{\mathrm{ir}}_{\hat{j}^{+}_{p0}\hat{j}^{+}_{p0}}(\omega,\bm{q})\tilde{V}(\bm{q})+\cdots
\end{equation}
which means the effective Coulomb interaction could also be written as
\begin{align}
  \tilde{V}_{\mathrm{eff}}(\omega,\bm{q})=&\tilde{V}(\bm{q})+\tilde{V}(\bm{q})[C^{\mathrm{ir}}_{\hat{j}^{+}_{p0}\hat{j}^{+}_{p0}}(\omega,\bm{q})+C^{\mathrm{ir}}_{\hat{j}^{+}_{p0}\hat{j}^{+}_{p0}}(\omega,\bm{q})\tilde{V}(\bm{q})C^{\mathrm{ir}}_{\hat{j}^{+}_{p0}\hat{j}^{+}_{p0}}(\omega,\bm{q})+\cdots]\tilde{V}(\bm{q})\nonumber\\
  =&\tilde{V}(\bm{q})+\tilde{V}(\bm{q})C_{\hat{j}^{+}_{p0}\hat{j}^{+}_{p0}}(\omega,\bm{q})\tilde{V}(\bm{q})\nonumber\\
  =&[1+\tilde{V}(\bm{q})C_{\hat{j}^{+}_{p0}\hat{j}^{+}_{p0}}(\omega,\bm{q})]\tilde{V}(\bm{q})
\end{align}
Thus the effective dielectric function is given by
\begin{equation}
  \epsilon_{\mathrm{eff}}(\omega,\bm{q})=\frac{\tilde{V}(\bm{q})}{\tilde{V}_{\mathrm{eff}}(\omega,\bm{q})}=1-\tilde{V}(\bm{q})C^{\mathrm{ir}}_{\hat{j}^{+}_{p0}\hat{j}^{+}_{p0}}(\omega,\bm{q})=[1+\tilde{V}(\bm{q})C_{\hat{j}^{+}_{p0}\hat{j}^{+}_{p0}}(\omega,\bm{q})]^{-1}\label{eq:effective-dielectric}
\end{equation}

Equating the last two expressions in Eq. \eqref{eq:effective-dielectric}, the irreducible charge-density correlation function is
\begin{equation}
  C^{\mathrm{ir}}_{\hat{j}^{+}_{p0}\hat{j}^{+}_{p0}}(\omega,\bm{q})=\frac{C_{\hat{j}^{+}_{p0}\hat{j}^{+}_{p0}}(\omega,\bm{q})}{1+\tilde{V}(\bm{q})C_{\hat{j}^{+}_{p0}\hat{j}^{+}_{p0}}(\omega,\bm{q})}\label{eq:irreducible-density-density}
\end{equation}
In the long-wavelength limit we have $C_{\hat{j}^{+}_{p0}\hat{j}^{+}_{p0}}\propto q^2$ and $\tilde{V}(q)\sim q^{-1}$, thus 
\begin{equation}
  C^{\mathrm{ir}}_{\hat{j}^{+}_{p0}\hat{j}^{+}_{p0}}(\omega,\bm{q})\approx C_{\hat{j}^{+}_{p0}\hat{j}^{+}_{p0}}(\omega,\bm{q})+\mathcal{O}(q^3)
\end{equation}

\section{The retarded effect and electromagnetic field associated with the collective modes}\label{app:retarded-effect}

In the main text, we have neglected the retarded effect of the electromagnetic field, i.e., we have taken the speed of light $c\to \infty$.
To include the retarded effect, we need to solve the full set of Maxwell equations together with the matter equations.

\subsection{Matter equations for the bilayer system}
For the bilayer system, the charge-density and charge-current-density sources are confined to the 2D planes at $z=\pm z_0=\pm d/2$,
\begin{equation}
  j_{\mu}(\omega,\bm{q},z)=j^{+}_{\mu}(\omega,\bm{q})[\delta(z-z_0)+\delta(z+z_0)]/2+j^{-}_{\mu}(\omega,\bm{q})[\delta(z-z_0)-\delta(z+z_0)].
\end{equation}
However, the gauge field is continuously defined in the full 3D space as $A_{\mu}(\omega,\bm{q},z)$, and the layer-symmetric and layer-antisymmetric components are defined as
\begin{equation}
  A^{+}_{\mu}(\omega,\bm{q})=[A_{\mu}(\omega,\bm{q},z_0)+A_{\mu}(\omega,\bm{q},-z_0)]/2,\;A^{-}_{\mu}(\omega,\bm{q})=A_{\mu}(\omega,\bm{q},z_0)-A_{\mu}(\omega,\bm{q},-z_0).
\end{equation}
The matter equations for the bilayer system are 
\begin{equation}
  j^{\sigma}_{\mu}(\omega,\bm{q})=\sum_{\nu}K^{\mathrm{ir},\sigma}_{\mu\nu}(\omega,\bm{q})A^{\sigma}_{\nu}(\omega,\bm{q})
\end{equation}
where $K^{\mathrm{ir},\sigma}_{\mu\nu}(\omega,\bm{q})$ is the irreducible response kernel (to include the retarded effect, the direct Coulomb interaction should be treated by the Maxwell equation).
For the charge response, the irreducible response kernel is given by
\begin{subequations}
  \begin{align}
  K^{\mathrm{ir},+}_{00}(\omega,\bm{q})=&-C^{\mathrm{ir}}_{\hat{j}^{+}_{p0}\hat{j}^{+}_{p0}}(\omega,\bm{q})\\
  K^{\mathrm{ir},+}_{L}(\omega,\bm{q})=&\frac{\omega^2}{q^2}K^{\mathrm{ir},+}_{00}(\omega,\bm{q})\\
  K^{\mathrm{ir},+}_{T}(\omega,\bm{q})=&K^{+}_{T}(\omega,\bm{q})\\
  K^{\mathrm{ir},+}_{a0}(\omega,q\hat{x})=&-\delta_{a1}\frac{\omega}{q}K^{\mathrm{ir},+}_{00}(\omega,\bm{q})
\end{align}
\end{subequations}
where $C^{\mathrm{ir}}_{\hat{j}^{+}_{p0}\hat{j}^{+}_{p0}}(\omega,\bm{q})$ is given by Eq. \eqref{eq:irreducible-density-density}.
For the exciton response, the results are similar,
\begin{subequations}
  \begin{align}
  K^{\mathrm{ir},-}_{00}(\omega,\bm{q})=&-C^{\mathrm{ir}}_{\hat{j}^{-}_{p0}\hat{j}^{-}_{p0}}(\omega,\bm{q})\\
  K^{\mathrm{ir},-}_{L}(\omega,\bm{q})=&\frac{\omega^2}{q^2}K^{\mathrm{ir},-}_{00}(\omega,\bm{q})\\
  K^{\mathrm{ir},-}_{T}(\omega,\bm{q})=&K^{-}_{T}(\omega,\bm{q})\\
  K^{\mathrm{ir},-}_{a0}(\omega,q\hat{x})=&-\delta_{a1}\frac{\omega}{q}K^{\mathrm{ir},-}_{00}(\omega,\bm{q})
  \end{align}
\end{subequations}
where 
\begin{equation}
  C^{\mathrm{ir}}_{\hat{j}^{-}_{p0}\hat{j}^{-}_{p0}}(\omega,\bm{q})=\frac{C_{\hat{j}^{-}_{p0}\hat{j}^{-}_{p0}}(\omega,\bm{q})}{1+\tilde{V}_X(\bm{q})C_{\hat{j}^{-}_{p0}\hat{j}^{-}_{p0}}(\omega,\bm{q})}
\end{equation}
and $\tilde{V}_X(\bm{q})=2[\tilde{V}(\bm{q})-\tilde{U}(\bm{q})]\approx 4\pi d/\epsilon=1/\kappa_g$ is the effective interaction between excitons in the long-wavelength limit and $\kappa_g=\epsilon/(4\pi d)$ is the geometric capacitance per unit area.

\subsection{The Lorenz gauge}
To solve the full set of Maxwell equations, we work in the Lorenz gauge defined by
\begin{equation}
  \nabla\cdot\bm{A}+\partial_z A_z+\frac{1}{c^2}\partial_t \phi=0
\end{equation}
where $\phi$ is the scalar potential, $\bm{A}=(A_x,A_y)$ is the in-plane vector potential, $A_z$ is the out-of-plane vector potential, and $c=1/\sqrt{\epsilon\epsilon_0\mu\mu_0}$ is the speed of light in the medium.
Since the system is translational invariant in the in-plane directions, we Fourier transform the fields in the in-plane directions and work in frequency-momentum space $(\omega,\bm{q},z)$.
In this gauge, the Maxwell equation is equivalent to the d'Alembert equation
\begin{equation}
  (-\omega^2/c^2+q^2-\partial_z^2)\begin{bmatrix}
    \phi(\omega,\bm{q},z) \\ \bm{A}(\omega,\bm{q},z) \\ A_z(\omega,\bm{q},z)
  \end{bmatrix}=\frac{4\pi}{\epsilon}\begin{bmatrix}
    \varrho(\omega,\bm{q},z) \\ \bm{j}(\omega,\bm{q},z)/c^2 \\ 0
  \end{bmatrix}.
\end{equation}
We first show that, subject to the Lorenz-gauge condition, a gauge transformation can always be made such that $A_z(\omega,\bm{q},z)=0$ for $\omega,q\ne0$.
Since the d'Alembert equation for $A_z(\omega,\bm{q},z)$ is homogeneous, the general solution is given by
\begin{equation}
  A_z(\omega,\bm{q},z)=A_{z,1}(\omega,\bm{q})\re^{i k_z z}+A_{z,2}(\omega,\bm{q})\re^{-i k_z z}
\end{equation}
where $k_z=\sqrt{\omega^2/c^2-q^2}$ ($k_z=i\sqrt{q^2-\omega^2/c^2}$ if $q^2>\omega^2/c^2$) is the out-of-plane momentum.
Then we can make a gauge transformation defined by the gauge function 
\begin{equation}
  \chi(\omega,\bm{q},z)=\frac{i}{k_z}[A_{z,1}(\omega,\bm{q})\re^{i k_z z}-A_{z,2}(\omega,\bm{q})\re^{-i k_z z}]
\end{equation}
such that
\begin{subequations}
  \begin{align}
  A'_z(\omega,\bm{q},z)=&A_z(\omega,\bm{q},z)+\partial_z \chi(\omega,\bm{q},z)=0\\
  \bm{A}'(\omega,\bm{q},z)=&\bm{A}(\omega,\bm{q},z)+i\bm{q}\chi(\omega,\bm{q},z)\\
  \phi'(\omega,\bm{q},z)=&\phi(\omega,\bm{q},z)+i\omega\chi(\omega,\bm{q},z)
\end{align}
\end{subequations}
The transformed fields continue to satisfy the Lorenz-gauge condition:
\begin{align}
  &i\bm{q}\cdot \bm{A}'(\omega,\bm{q},z)+\partial_z A'_z(\omega,\bm{q},z)-\frac{i\omega}{c^2}\phi'(\omega,\bm{q},z)\nonumber\\
  =&-q^2\chi(\omega,\bm{q},z)+\partial_z^2 \chi(\omega,\bm{q},z)+\frac{\omega^2}{c^2}\chi(\omega,\bm{q},z)\nonumber\\
  =&(-q^2-k_z^2+\frac{\omega^2}{c^2})\chi(\omega,\bm{q},z)=0
\end{align}
Thus we may work with $A_z(\omega,\bm{q},z)=0$ for $\omega,q\ne0$.

By writing the in-plane vector potential as $\bm{A}=A_L\bm{q}/q+A_T\hat{z}\times\bm{q}/q$, the Lorenz-gauge condition becomes
\begin{equation}
  iq A_L(\omega,\bm{q},z)-i\frac{\omega}{c^2}\phi(\omega,\bm{q},z)=0\implies \phi(\omega,\bm{q},z)=\frac{c^2 q}{\omega}A_L(\omega,\bm{q},z)\label{eq:scalar-longitudinal-vector}
\end{equation}
In the Lorenz gauge with $A_z=0$, the scalar potential is fixed by $A_L$, so only the in-plane components $A_{L/T}$ remain independent.
The in-plane electromagnetic fields are related to the gauge fields by
\begin{align}
  \bm{E}(\omega,\bm{q},z)=&i\omega\bm{A}(\omega,\bm{q},z)-i\bm{q}\phi(\omega,\bm{q},z)\nonumber\\
  =&i\left[\omega A_L(\omega,\bm{q},z)-\frac{c^2 q^2}{\omega}A_L(\omega,\bm{q},z)\right]\frac{\bm{q}}{q}+i\omega A_T(\omega,\bm{q},z)\frac{\hat{z}\times\bm{q}}{q}\nonumber\\
  =&i\frac{\omega^2-c^2 q^2}{\omega}A_L(\omega,\bm{q},z)\frac{\bm{q}}{q}+i\omega A_T(\omega,\bm{q},z)\frac{\hat{z}\times\bm{q}}{q}\label{eq:E-inplane}\\
  \bm{B}(\omega,\bm{q},z)=&\hat{z}\times\partial_z \bm{A}(\omega,\bm{q},z)\nonumber\\
  =&\hat{z}\times\left[\partial_z A_L(\omega,\bm{q},z)\frac{\bm{q}}{q}+\partial_z A_T(\omega,\bm{q},z)\frac{\hat{z}\times\bm{q}}{q}\right]\nonumber\\
  =&-\partial_z A_T(\omega,\bm{q},z)\frac{\bm{q}}{q}+\partial_z A_L(\omega,\bm{q},z)\frac{\hat{z}\times\bm{q}}{q}\label{eq:B-inplane}
\end{align}
The out-of-plane electromagnetic fields are given by
\begin{align}
  E_z(\omega,\bm{q},z)=&i\omega A_z(\omega,\bm{q},z)-\partial_z\phi(\omega,\bm{q},z)=- \partial_z\phi(\omega,\bm{q},z)=-\frac{c^2 q}{\omega}\partial_z A_L(\omega,\bm{q},z)\label{eq:Ez}\\
  B_z(\omega,\bm{q},z)=&i\bm{q}\times\bm{A}(\omega,\bm{q},z)\cdot \hat{z}=iq A_T(\omega,\bm{q},z)\label{eq:Bz}
\end{align}
When $A_T=0$, the magnetic field is perpendicular to $\bm q$, giving a transverse-magnetic (TM) mode. When $A_L=0$, the electric field is perpendicular to $\bm q$, giving a transverse-electric (TE) mode.

\subsection{Solution in free space}
Because the charge-density and charge-current-density sources are confined to the bilayers at $z=\pm z_0$, the electromagnetic fields vanish as $z\to \pm \infty$.
Then according to Eq. \eqref{eq:Bz}, we have $A_T(\omega,\bm{q},z\to \pm \infty)=0$.
Similarly, according to Eq. \eqref{eq:E-inplane}, we have $A_L(\omega,\bm{q},z\to \pm \infty)=0$.
Thus the general solution for the in-plane vector potential takes the form
\begin{equation}
  \bm{A}(\omega,\bm{q},z)=\begin{cases}
    \bm{A}(\omega,\bm{q},z_0)\re^{-\lambda(z-z_0)}, & z>z_0,\\
    \bm{A}(\omega,\bm{q},z_0)\dfrac{\sinh[\lambda(z+z_0)]}{\sinh(2\lambda z_0)}-\bm{A}(\omega,\bm{q},-z_0)\dfrac{\sinh[\lambda(z-z_0)]}{\sinh(2\lambda z_0)}, & |z|\le z_0,\\
    \bm{A}(\omega,\bm{q},-z_0)\re^{\lambda(z+z_0)}, & z<-z_0.
  \end{cases}
\end{equation}
Define $\bm{A}^{+}(\omega,\bm{q})=[\bm{A}(\omega,\bm{q},z_0)+\bm{A}(\omega,\bm{q},-z_0)]/2$ and $\bm{A}^{-}(\omega,\bm{q})=\bm{A}(\omega,\bm{q},z_0)-\bm{A}(\omega,\bm{q},-z_0)$, the solution can be written in a more compact form as
\begin{equation}
  \bm{A}(\omega,\bm{q},z)=\bm{A}^{+}(\omega,\bm{q})\frac{\re^{-\lambda|z-z_0|}+\re^{-\lambda|z+z_0|}}{1+\re^{-2\lambda z_0}}+\bm{A}^{-}(\omega,\bm{q})\frac{\re^{-\lambda|z-z_0|}-\re^{-\lambda|z+z_0|}}{2(1-\re^{-2\lambda z_0})}\label{eq:bounded-vector-potential}
\end{equation}
One can verify that 
\begin{align}
  \partial_z^2\bm{A}(\omega,\bm{q},z)=&\lambda^2\bm{A}(\omega,\bm{q},z)-\frac{4\lambda}{1+\re^{-2\lambda z_0}}\bm{A}^{+}(\omega,\bm{q})[\delta(z-z_0)+\delta(z+z_0)]/2\nonumber\\
  &-\frac{\lambda}{1-\re^{-2\lambda z_0}}\bm{A}^{-}(\omega,\bm{q})[\delta(z-z_0)-\delta(z+z_0)]\label{eq:vector_potential_derivatives}
\end{align}
Substituting into the d'Alembert equation, we get
\begin{subequations}
  \begin{align}
    -\omega^2/c^2+q^2-\lambda^2=&0\\
    \frac{4\lambda}{1+\re^{-2\lambda z_0}}\bm{A}^{+}(\omega,\bm{q})=&\frac{4\pi}{\epsilon}\frac{\bm{j}^{+}(\omega,\bm{q})}{c^2}\\
    \frac{\lambda}{1-\re^{-2\lambda z_0}}\bm{A}^{-}(\omega,\bm{q})=&\frac{4\pi}{\epsilon}\frac{\bm{j}^{-}(\omega,\bm{q})}{c^2}
  \end{align}\label{eq:2D-Maxwell}
\end{subequations}
Separating the longitudinal and transverse components, we have
\begin{align}
  j_{L}(\omega,q)=&K^{\mathrm{ir}}_{10}(\omega,q\hat{x})\phi(\omega,q)+K^{\mathrm{ir}}_{L}(\omega,q)A_{L}(\omega,q)\nonumber\\
  =&(1-c^2 q^2/\omega^2)K_{L}^{\mathrm{ir}}(\omega,q)A_{L}(\omega,q)\nonumber\\
  =&-\frac{c^2\lambda^2}{\omega^2}K_{L}^{\mathrm{ir}}(\omega,q)A_{L}(\omega,q)\\
  j_{T}(\omega,q)=&K^{\mathrm{ir}}_{T}(\omega,q)A_{T}(\omega,q)
\end{align}
Thus Eq. \eqref{eq:2D-Maxwell} becomes
\begin{subequations}
  \begin{align}
    \left[\frac{4\lambda}{1+\re^{-2\lambda z_0}}+\frac{4\pi}{\epsilon}\frac{\lambda^2}{\omega^2}K^{\mathrm{ir,+}}_{L}(\omega,q)\right]A^{+}_{L}(\omega,q)=&0 \label{eq:even-TM}\\
    \left[\frac{4\lambda}{1+\re^{-2\lambda z_0}}-\frac{4\pi}{\epsilon c^2}K^{\mathrm{ir},+}_{T}(\omega,q)\right]A^{+}_{T}(\omega,q)=&0\label{eq:even-TE}\\
    \left[\frac{\lambda}{1-\re^{-2\lambda z_0}}+\frac{4\pi}{\epsilon}\frac{\lambda^2}{\omega^2}K^{\mathrm{ir},-}_{L}(\omega,q)\right]A^{-}_{L}(\omega,q)=&0\label{eq:odd-TM}\\
    \left[\frac{\lambda}{1-\re^{-2\lambda z_0}}-\frac{4\pi}{\epsilon c^2}K^{\mathrm{ir},-}_{T}(\omega,q)\right]A^{-}_{T}(\omega,q)=&0\label{eq:odd-TE}
  \end{align}
\end{subequations}
The above four equations give four kinds of waveguide modes supported by the bilayer EI:
\subsubsection{The even TM mode}
  Eq. \eqref{eq:even-TM} describes the even TM mode, with dispersion relation determined by
  \begin{equation}
    \frac{4\lambda}{1+\re^{-2\lambda z_0}}+\frac{4\pi}{\epsilon}\frac{\lambda^2}{\omega^2}K^{\mathrm{ir},+}_{L}(\omega,q)=0\label{eq:even-TM-2}
  \end{equation}
  In the long-wavelength limit $q\to 0$, we have 
  \begin{align}
    \frac{q^2}{\omega^2}K_L^{\mathrm{ir},+}(\omega,q)=&K^{\mathrm{ir},+}_{00}(\omega,q)=-C^{\mathrm{ir}}_{\hat{j}^{+}_{p0}\hat{j}^{+}_{p0}}(\omega,q)\approx -C_{\hat{j}^{+}_{p0}\hat{j}^{+}_{p0}}\approx-\frac{\chi q^2(\omega_0^p)^2}{\omega^2-(\omega_0^p)^2}\nonumber\\
  \end{align}
  For $q<\omega/c$, we have $\mathrm{Im}\lambda\ne 0$ and 
  \begin{equation}
    \mathrm{Im}\left[\frac{4\lambda}{1+\re^{-2\lambda z_0}}+\frac{4\pi}{\epsilon}\frac{\lambda^2}{\omega^2}K^{\mathrm{ir},+}_{L}(\omega,q)\right]\ne 0
  \end{equation}
  thus there is no real solution for $\omega(q)$, indicating that the even TM mode is strongly damped by the light cone.
  For $q>\omega/c$, we have $\lambda=\sqrt{q^2-\omega^2/c^2}$. In the long-wavelength limit $q\to \omega/c$ and $\lambda\to 0$, Eq. \eqref{eq:even-TM-2} becomes
  \begin{equation}
    1-\frac{2\pi\lambda}{\epsilon q^2}\frac{\chi q^2(\omega_0^p)^2}{\omega^2-(\omega_0^p)^2}=0\implies \frac{\omega^2}{(\omega^p_0)^2}=1+\frac{2\pi\chi}{\epsilon} \sqrt{q^2-\frac{\omega^2}{c^2}},\;q>\omega/c\label{eq:even-TM-disp}
  \end{equation}
  In the non-retarded limit $c\to \infty$, we have $\omega(q)=\omega_0^p\sqrt{1+2\pi\chi q/\epsilon}$, which recovers the result in the main text.

  For the even TM mode propagating with in-plane momentum $\bm{q}=q\hat{x}$, the associated electromagnetic fields are
  \begin{align}
    \bm{A}(t,\bm{r},z)=&A^{+}_{L}\frac{\re^{-\lambda|z-z_0|}+\re^{-\lambda|z+z_0|}}{1+\re^{-2\lambda z_0}}\re^{i(qx-\omega t)}\hat{x}\\
    \bm{E}(t,\bm{r},z)=&-i\frac{c^2\lambda^2}{\omega}A^{+}_{L}\frac{\re^{-\lambda|z-z_0|}+\re^{-\lambda|z+z_0|}}{1+\re^{-2\lambda z_0}}\re^{i(qx-\omega t)}\hat{x}\nonumber\\
    &+\frac{c^2 q \lambda}{\omega}A^{+}_{L}\frac{\mathrm{sgn}(z-z_0)\re^{-\lambda|z-z_0|}+\mathrm{sgn}(z+z_0)\re^{-\lambda|z+z_0|}}{1+\re^{-2\lambda z_0}}\re^{i(qx-\omega t)}\hat{z}\\
    \bm{B}(t,\bm{r},z)=&-\frac{\lambda}{1+\re^{-2\lambda z_0}}A^{+}_{L}[\mathrm{sgn}(z-z_0)\re^{-\lambda|z-z_0|}+\mathrm{sgn}(z+z_0)\re^{-\lambda|z+z_0|}]\re^{i(qx-\omega t)}\hat{y}
  \end{align}

\subsubsection{The even TE mode}
Eq. \eqref{eq:even-TE} describes the even TE mode, with dispersion relation determined by
\begin{equation}
    \frac{4\lambda}{1+\re^{-2\lambda z_0}}-\frac{4\pi}{\epsilon c^2}K^{\mathrm{ir},+}_{T}(\omega,q)=0\label{eq:even-TE-2} 
\end{equation}
In the long-wavelength limit, we have 
\begin{equation}
  K^{\mathrm{ir},+}_{T}(\omega,q)\approx K^{+}_{T}(\omega,0)=-\frac{\chi\omega^2 (\omega_0^p)^2}{\omega^2-(\omega_0^p)^2}
\end{equation}
Similar to the even TM mode, for $q<\omega/c$ there is no real solution for $\omega(q)$, indicating that the even TE mode is also strongly damped by the light cone.
For $q>\omega/c$, when $\omega>\omega_0^p$, we have $K^{\mathrm{ir},+}_{T}(\omega,q)<0$ and 
\begin{equation}
  \frac{4\lambda}{1+\re^{-2\lambda z_0}}-\frac{4\pi}{\epsilon c^2}K^{\mathrm{ir},+}_{T}(\omega,q)>0
\end{equation}
always holds, thus there is no even TE mode above the dipole exciton gap.
When $\omega<\omega_0^p$, we have $K^{\mathrm{ir},+}_{T}(\omega,q)>0$ and the dispersion relation of the even TE mode is determined by
\begin{equation}
  \frac{\omega^2}{(\omega_0^p)}=\left[1+\frac{\pi\chi (\omega_0^p)^2}{\epsilon c^2}\frac{1+\re^{-\lambda d}}{\lambda}\right]^{-1},\;q>\omega/c\label{eq:even-TE-disp}
\end{equation}
Near the light cone $q\to \omega/c$ and $\lambda\to 0$, Eq. \eqref{eq:even-TE-disp} becomes
\begin{equation}
  \frac{\omega^2}{(\omega_0^p)^2}=\lambda\frac{\epsilon c^2}{2\pi\chi (\omega_0^p)^2}\implies \lambda^2=q^2-\frac{\omega^2}{c^2}=\left(\frac{\omega^2}{c^2}\frac{2\pi\chi}{\epsilon}\right)^2\implies q^2=\frac{\omega^2}{c^2}\left[1+\left(\frac{2\pi\chi \omega}{\epsilon c}\right)^2\right]
\end{equation}

For the even TE mode propagating with in-plane momentum $\bm{q}=q\hat{x}$, the associated electromagnetic fields are
\begin{align}
  \bm{A}(t,\bm{r},z)=&A^{+}_{T}\frac{\re^{-\lambda|z-z_0|}+\re^{-\lambda|z+z_0|}}{1+\re^{-2\lambda z_0}}\re^{i(qx-\omega t)}\hat{y}\\
  \bm{E}(t,\bm{r},z)=&i\omega A^{+}_{T}\frac{\re^{-\lambda|z-z_0|}+\re^{-\lambda|z+z_0|}}{1+\re^{-2\lambda z_0}}\re^{i(qx-\omega t)}\hat{y}\\
  \bm{B}(t,\bm{r},z)=&\frac{iq}{1+\re^{-2\lambda z_0}}A^{+}_{T}[\re^{-\lambda|z-z_0|}+\re^{-\lambda|z+z_0|}]\re^{i(qx-\omega t)}\hat{z}\nonumber\\
  &+\frac{\lambda}{1+\re^{-2\lambda z_0}}A^{+}_{T}[\mathrm{sgn}(z-z_0)\re^{-\lambda|z-z_0|}+\mathrm{sgn}(z+z_0)\re^{-\lambda|z+z_0|}]\re^{i(qx-\omega t)}\hat{x}
\end{align}

\subsubsection{The odd TM mode}\label{app:odd_TM}
Eq. \eqref{eq:odd-TM} describes the odd TM mode, with dispersion relation determined by
\begin{equation}
    \frac{\lambda}{1-\re^{-2\lambda z_0}}+\frac{4\pi}{\epsilon}\frac{\lambda^2}{\omega^2}K^{\mathrm{ir},-}_{L}(\omega,q)=0\label{eq:odd-TM-2}
\end{equation}
In the long-wavelength limit, we have
\begin{align}
  \frac{q^2}{\omega^2}K_L^{\mathrm{ir},-}(\omega,q)=&K^{\mathrm{ir},-}_{00}(\omega,q)=-C^{\mathrm{ir}}_{\hat{j}^{-}_{p0}\hat{j}^{-}_{p0}}(\omega,q)\nonumber\\
  =&-\{\tilde{V}_X(\bm{q})+[C_{\hat{j}^{-}_{p0}\hat{j}^{-}_{p0}}(\omega,q)^{-1}]\}^{-1}\nonumber\\
  =&-\left(\frac{1}{\kappa_g}+\frac{\omega^2-v^2q^2}{\kappa v^2q^2}\right)^{-1}\nonumber\\
  =&-\left(\frac{\omega^2}{\kappa v^2 q^2}-\frac{1}{\kappa}+\frac{1}{\kappa_g}\right)^{-1}\label{eq:KL_minus_ir}
\end{align}
For $\omega\to 0,\omega/c<q\to 0$, we have $\lambda\to 0$ and Eq. \eqref{eq:odd-TM-2} becomes
\begin{equation}
  \frac{1}{d}-\frac{4\pi}{\epsilon}\frac{\lambda^2}{q^2}\left(\frac{\omega^2}{\kappa v^2 q^2}-\kappa^{-1}+\kappa_g^{-1}\right)^{-1}=0\implies\frac{\omega^2}{\kappa v^2 q^2}=\kappa^{-1}+\kappa_g^{-1}\left(\frac{\lambda^2}{q^2}-1\right)=\kappa^{-1}-\frac{\omega^2}{\kappa_g c^2 q^2}
\end{equation}
and the dispersion relation of the odd TM mode is given by
\begin{equation}
  \omega(q)=vq\left(1+\frac{v^2}{c^2}\frac{\kappa}{\kappa_g}\right)^{-1/2}\label{eq:odd-TM-disp}
\end{equation}
In the non-retarded limit $c\to \infty$, we have $\omega(q)=vq$, which recovers the Goldstone mode dispersion in the main text.

For the odd TM mode propagating with in-plane momentum $\bm{q}=q\hat{x}$, the associated electromagnetic fields are
\begin{align}
  \bm{A}(t,\bm{r},z)=&A^{-}_{L}\frac{\re^{-\lambda|z-z_0|}-\re^{-\lambda|z+z_0|}}{2(1-\re^{-2\lambda z_0})}\re^{i(qx-\omega t)}\hat{x}\\
  \bm{E}(t,\bm{r},z)=&-i\frac{c^2\lambda^2}{\omega}A^{-}_{L}\frac{\re^{-\lambda|z-z_0|}-\re^{-\lambda|z+z_0|}}{2(1-\re^{-2\lambda z_0})}\re^{i(qx-\omega t)}\hat{x}\nonumber\\
  &+\frac{c^2 q \lambda}{\omega}A^{-}_{L}\frac{\mathrm{sgn}(z-z_0)\re^{-\lambda|z-z_0|}-\mathrm{sgn}(z+z_0)\re^{-\lambda|z+z_0|}}{2(1-\re^{-2\lambda z_0})}\re^{i(qx-\omega t)}\hat{z}\\
  \bm{B}(t,\bm{r},z)=&-\frac{\lambda}{2(1-\re^{-2\lambda z_0})}A^{-}_{L}[\mathrm{sgn}(z-z_0)\re^{-\lambda|z-z_0|}-\mathrm{sgn}(z+z_0)\re^{-\lambda|z+z_0|}]\re^{i(qx-\omega t)}\hat{y}\label{eq:odd_TM_By}
\end{align}

\subsubsection{The odd TE mode}
Eq. \eqref{eq:odd-TE} describes the odd TE mode, with dispersion relation determined by
\begin{equation}
    \frac{\lambda}{1-\re^{-2\lambda z_0}}-\frac{4\pi}{\epsilon c^2}K^{\mathrm{ir},-}_{T}(\omega,q)=0\label{eq:odd-TE-2}
\end{equation}
In the long-wavelength limit we have
\begin{equation}
  K^{\mathrm{ir},-}_T(\omega,q)=K^{-}_T(\omega,q)=-\frac{e^2n_X}{4m}-\frac{\alpha q^2(\omega_0^q)^2}{\omega^2-(\omega_0^q)^2}=-\kappa v^2 -\frac{\alpha q^2(\omega_0^q)^2}{\omega^2-(\omega_0^q)^2} +O(q^3)
\end{equation}
where $\omega_0^q$ is the quadrupole mode excitation energy at $\bm{q}=0$ and $\alpha$ is a positive constant.
For $\omega>\omega_0^q$, we always have $K^{\mathrm{ir},-}_T(\omega,q)<0$ and
\begin{equation}
  \frac{\lambda}{1-\re^{-2\lambda z_0}}-\frac{4\pi}{\epsilon c^2}K^{\mathrm{ir},-}_{T}(\omega,q)>0 
\end{equation}
which means that Eq. \eqref{eq:odd-TE-2} has no solution for $\omega>\omega_0^q$, and hence no odd TE mode exists above the quadrupole gap.
For $\omega<\omega_0^q$, near the light cone $q\to \omega/c+0^+$ and $\lambda\to 0$, Eq. \eqref{eq:odd-TE-2} becomes
\begin{equation}
  \kappa_gc^2=-\kappa v^2-\frac{\alpha q^2 (\omega_0^q)^2}{\omega^2-(\omega_0^q)^2}\implies \frac{\omega^2}{(\omega_0^q)^2}=1-\frac{\alpha q^2}{\kappa_gc^2+\kappa v^2},\;q>\omega/c\label{eq:odd-TE-disp}
\end{equation}

For the odd TE mode propagating with in-plane momentum $\bm{q}=q\hat{x}$, the associated electromagnetic fields are
\begin{align}
  \bm{A}(t,\bm{r},z)=&A^{-}_{T}\frac{\re^{-\lambda|z-z_0|}-\re^{-\lambda|z+z_0|}}{2(1-\re^{-2\lambda z_0})}\re^{i(qx-\omega t)}\hat{y}\\
  \bm{E}(t,\bm{r},z)=&i\omega A^{-}_{T}\frac{\re^{-\lambda|z-z_0|}-\re^{-\lambda|z+z_0|}}{2(1-\re^{-2\lambda z_0})}\re^{i(qx-\omega t)}\hat{y}\\
  \bm{B}(t,\bm{r},z)=&\frac{iq}{2(1-\re^{-2\lambda z_0})}A^{-}_{T}[\re^{-\lambda|z-z_0|}-\re^{-\lambda|z+z_0|}]\re^{i(qx-\omega t)}\hat{z}\nonumber\\
  &+\frac{\lambda}{2(1-\re^{-2\lambda z_0})}A^{-}_{T}[\mathrm{sgn}(z-z_0)\re^{-\lambda|z-z_0|}-\mathrm{sgn}(z+z_0)\re^{-\lambda|z+z_0|}]\re^{i(qx-\omega t)}\hat{x}
\end{align}

{
\subsection{\texorpdfstring{Solution in metal cavity}{Solution in metal cavity}}
Assume the bilayer EI is placed in a metal cavity with thickness $l$ in the $z$ direction, and the metal surfaces are located at $\pm z_1=\pm l/2$.
We ignore the in-plane boundary conditions in the $x$ and $y$ directions, and only consider the boundary conditions in the $z$ direction, i.e., the tangential electric field Eq. \eqref{eq:E-inplane} and normal magnetic field Eq. \eqref{eq:Bz} should vanish at the metal surfaces.
Thus, we have $A_T(\omega,\bm{q},\pm z_1)=0$, $A_L(\omega,\bm{q},\pm z_1)=0$ and $\bm{A}(\omega,\bm{q},\pm z_1)=0$.
The general solution for the in-plane vector potential is
\begin{equation}
  \bm{A}(\omega,\bm{q},z)=\begin{cases}
    \bm{A}(\omega,\bm{q},z_0)\dfrac{\sinh[\lambda(z_1-z)]}{\sinh[\lambda(z_1-z_0)]}, & z_0<z<z_1,\\
    \bm{A}(\omega,\bm{q},z_0)\dfrac{\sinh[\lambda(z+z_0)]}{\sinh(2\lambda z_0)}-\bm{A}(\omega,\bm{q},-z_0)\dfrac{\sinh[\lambda(z-z_0)]}{\sinh(2\lambda z_0)}, & |z|\le z_0,\\
    \bm{A}(\omega,\bm{q},-z_0)\dfrac{\sinh[\lambda(z+z_1)]}{\sinh[\lambda(z_1-z_0)]}, & -z_1<z<-z_0.
  \end{cases}
\end{equation}
Define $\bm{A}^{+}(\omega,\bm{q})=[\bm{A}(\omega,\bm{q},z_0)+\bm{A}(\omega,\bm{q},-z_0)]/2$ and $\bm{A}^{-}(\omega,\bm{q})=\bm{A}(\omega,\bm{q},z_0)-\bm{A}(\omega,\bm{q},-z_0)$, the solution can be written in a more compact form as
\begin{align}
  \bm{A}(\omega,\bm{q},z)=&\bm{A}^{+}(\omega,\bm{q})\frac{\re^{-\lambda|z-z_0|}+\re^{-\lambda|z+z_0|}-2\frac{\re^{-\lambda z_1}\cosh(\lambda z_0)\cosh(\lambda z)}{\cosh (\lambda z_1)}}{1+\re^{-2\lambda z_0}-2\frac{\re^{-\lambda z_1}\cosh^2(\lambda z_0)}{\cosh(\lambda z_1)}}\nonumber\\
  &+\bm{A}^{-}(\omega,\bm{q})\frac{\re^{-\lambda|z-z_0|}-\re^{-\lambda|z+z_0|}-2\frac{\re^{-\lambda z_1}\sinh(\lambda z_0)\sinh(\lambda z)}{\sinh(\lambda z_1)}}{2[1-\re^{-2\lambda z_0}-2\frac{\re^{-\lambda z_1}\sinh^2(\lambda z_0)}{\sinh(\lambda z_1)}]}\label{eq:bounded-vector-potential-cavity}
\end{align}
In the region $|z|<z_1$, one can verify that
\begin{align}
  \partial_z^2\bm{A}(\omega,\bm{q},z)=&\lambda^2\bm{A}(\omega,\bm{q},z)-\frac{4\lambda}{1+\re^{-2\lambda z_0}-2\frac{\re^{-\lambda z_1}\cosh^2(\lambda z_0)}{\cosh(\lambda z_1)}}\bm{A}^{+}(\omega,\bm{q})[\delta(z-z_0)+\delta(z+z_0)]/2\nonumber\\
  &-\frac{\lambda}{1-\re^{-2\lambda z_0}-2\frac{\re^{-\lambda z_1}\sinh^2(\lambda z_0)}{\sinh(\lambda z_1)}}\bm{A}^{-}(\omega,\bm{q})[\delta(z-z_0)-\delta(z+z_0)]\label{eq:vector_potential_derivatives_cavity}
\end{align}
Unlike in free space, $\lambda$ can take both real and imaginary values.

Substituting into the d'Alembert equation, we get
\begin{subequations}
  \begin{align}
    -\omega^2/c^2+q^2-\lambda^2=&0\\
    \frac{4\lambda}{1+\re^{-2\lambda z_0}-2\frac{\re^{-\lambda z_1}\cosh^2(\lambda z_0)}{\cosh(\lambda z_1)}}\bm{A}^{+}(\omega,\bm{q})=&\frac{4\pi}{\epsilon}\frac{\bm{j}^{+}(\omega,\bm{q})}{c^2}\\
    \frac{\lambda}{1-\re^{-2\lambda z_0}-2\frac{\re^{-\lambda z_1}\sinh^2(\lambda z_0)}{\sinh(\lambda z_1)}}\bm{A}^{-}(\omega,\bm{q})=&\frac{4\pi}{\epsilon}\frac{\bm{j}^{-}(\omega,\bm{q})}{c^2}
  \end{align}\label{eq:2D-Maxwell-cavity}
\end{subequations}
Separating the longitudinal and transverse components, we have
\begin{subequations}
  \begin{align}
    \left[\frac{4\lambda}{1+\re^{-2\lambda z_0}-2\frac{\re^{-\lambda z_1}\cosh^2(\lambda z_0)}{\cosh(\lambda z_1)}}+\frac{4\pi}{\epsilon}\frac{\lambda^2}{\omega^2}K^{\mathrm{ir},+}_L(\omega,q) \right]A^{+}_L(\omega,q)=&0\\
    \left[\frac{4\lambda}{1+\re^{-2\lambda z_0}-2\frac{\re^{-\lambda z_1}\cosh^2(\lambda z_0)}{\cosh(\lambda z_1)}}-\frac{4\pi}{\epsilon c^2}K^{\mathrm{ir},+}_T(\omega,q) \right]A^{+}_T(\omega,q)=&0\\
    \left[\frac{\lambda}{1-\re^{-2\lambda z_0}-2\frac{\re^{-\lambda z_1}\sinh^2(\lambda z_0)}{\sinh(\lambda z_1)}}+\frac{4\pi}{\epsilon}\frac{\lambda^2}{\omega^2}K^{\mathrm{ir},-}_L(\omega,q) \right]A^{-}_L(\omega,q)=&0\\
    \left[\frac{\lambda}{1-\re^{-2\lambda z_0}-2\frac{\re^{-\lambda z_1}\sinh^2(\lambda z_0)}{\sinh(\lambda z_1)}}-\frac{4\pi}{\epsilon c^2}K^{\mathrm{ir},-}_T(\omega,q) \right]A^{-}_T(\omega,q)=&0
  \end{align}
\end{subequations}
which determine the dispersion relations of the four kinds of waveguide modes supported by the bilayer EI in the metal cavity.

\subsubsection{\texorpdfstring{The odd TM mode}{The odd TM mode}}\label{app:odd_TM_cavity}
Here we focus on the odd TM mode characterized by $A_L^{-}$, which is associated with the Goldstone mode of the bilayer EI.
The dispersion relation of the odd TM mode is determined by
\begin{equation}
  \frac{\lambda}{1-\re^{-2\lambda z_0}-2\frac{\re^{-\lambda z_1}\sinh^2(\lambda z_0)}{\sinh(\lambda z_1)}}+\frac{4\pi}{\epsilon}\frac{\lambda^2}{\omega^2}K^{\mathrm{ir},-}_L(\omega,q)=0\label{eq:odd-TM-cavity}
\end{equation}
Using hyperbolic function identities, we have
\begin{align}
  & 1-\re^{-2\lambda z_0}-2\frac{\re^{-\lambda z_1}\sinh^2(\lambda z_0)}{\sinh(\lambda z_1)}\nonumber\\
=&2\re^{-\lambda z_0}\sinh(\lambda z_0)-2\frac{\re^{-\lambda z_1}\sinh^2(\lambda z_0)}{\sinh(\lambda z_1)}\nonumber\\
=&2\frac{\sinh(\lambda z_0)}{\sinh(\lambda z_1)}[\re^{-\lambda z_0}\sinh(\lambda z_1)-\re^{-\lambda z_1}\sinh(\lambda z_0)]\nonumber\\
=&2\frac{\sinh(\lambda z_0)}{\sinh(\lambda z_1)}\{\sinh(\lambda z_1)[\cosh(\lambda z_0)-\sinh(\lambda z_0)]-\sinh(\lambda z_0)[\cosh(\lambda z_1)-\sinh(\lambda z_1)] \}\nonumber\\
=&2\frac{\sinh(\lambda z_0)}{\sinh(\lambda z_1)}\sinh[\lambda(z_1-z_0)]
\end{align}
and 
\begin{align}
  &\frac{1}{1-\re^{-2\lambda z_0}-2\frac{\re^{-\lambda z_1}\sinh^2(\lambda z_0)}{\sinh(\lambda z_1)}}\nonumber\\
  =&\frac{1}{2}\frac{\sinh(\lambda z_1)}{\sinh(\lambda z_0)\sinh[\lambda(z_1-z_0)]}\nonumber\\
  =&\frac{1}{2}\frac{\sinh[\lambda(z_0+z_1-z_0)]}{\sinh(\lambda z_0)\sinh[\lambda(z_1-z_0)]}\nonumber\\
  =&\frac{1}{2}\frac{\sinh(\lambda z_0)\cosh[\lambda(z_1-z_0)]+\cosh(\lambda z_0)\sinh[\lambda(z_1-z_0)]}{\sinh(\lambda z_0)\sinh[\lambda(z_1-z_0)]}\nonumber\\
  =&\frac{1}{2}\left[\coth[\lambda(z_1-z_0)]+\coth(\lambda z_0)\right]
\end{align}
Also using Eq. \eqref{eq:KL_minus_ir}, we can rewrite Eq. \eqref{eq:odd-TM-cavity} as
\begin{equation}
  \frac{\lambda}{2}\left[\coth[\lambda(z_1-z_0)]+\coth(\lambda z_0)\right]-\frac{4\pi}{\epsilon}\frac{\lambda^2}{q^2}\left(\frac{\omega^2}{\kappa v^2 q^2}-\kappa^{-1}+\kappa_g^{-1}\right)^{-1}=0
\end{equation}
where 
\begin{equation}
  \lambda=\sqrt{q^2-\frac{\omega^2}{c^2}},\;z_0=\frac{d}{2},\;z_1=\frac{l}{2}
\end{equation}
For simplicity, we approximate $\kappa\approx\kappa_g$ and define
\begin{align}
  F(\lambda,\omega)=&\frac{\lambda}{2}\left[\coth[\lambda(z_1-z_0)]+\coth(\lambda z_0)\right]-\frac{4\pi}{\epsilon}\frac{\lambda^2}{q^2}\frac{\kappa_g v^2q^2}{\omega^2}\nonumber\\
  =&\frac{\lambda}{2}\left[\coth[\lambda(z_1-z_0)]+\coth(\lambda z_0)\right]-\frac{ v^2}{2z_0}\frac{\lambda^2 }{\omega^2}
\end{align}
Then the dispersion relation of the odd TM mode is determined by $F(\lambda,\omega)=0$.

We consider the frequency far below the cutoff frequency of the metal cavity, i.e., $\omega\ll c\pi /l$.
Thus, there will be two kinds of solutions:
\begin{itemize}
  \item Goldstone-like mode with $\lambda\in\mathbb{R}$. This mode is far below the light cone $q\gg \omega /c$, propagating along the in-plane direction and evanescent in the $z$ direction (bounded around the bilayer). In the long-wavelength limit $q\to 0,\omega\to 0,\lambda\to 0$, we have $\coth[\lambda(z_1-z_0)]\approx 1/[\lambda(z_1-z_0)]$ and $\coth(\lambda z_0)\approx 1/(\lambda z_0)$, and the dispersion relation of the Goldstone-like mode is given by
  \begin{equation}
    F(\lambda,\omega)\approx \frac{1}{d}+\frac{1}{l-d}-\frac{v^2}{2z_0}\frac{q^2-\omega^2/c^2 }{\omega^2}=0\implies \frac{\omega^2}{q^2 v^2}=\frac{1}{1+\kappa_g'/\kappa_g+v^2/c^2}
  \end{equation}
  where $\kappa_g=\epsilon/(4\pi d)$ and $\kappa_g'=\epsilon/(4\pi(l-d))$ are the geometric capacitances of the bilayer and the cavity, respectively.
  We can see that the Goldstone-like mode velocity is renormalized by the cavity effect to
  \begin{equation}
    v'=v(1+\kappa_g'/\kappa_g+v^2/c^2)^{-1/2}
  \end{equation}
  In the non-retarded limit $c\to \infty$, and assume $l \gg d$ so that $\kappa_g\gg \kappa_g'$, we have $\omega(q)\approx vq$, which recovers the Goldstone mode dispersion in the main text.
  \item Cavity-photon-like mode with $\lambda=i k_z\in i\mathbb{R}$. This mode forms standing wave in the $z$ direction characterized by $k_z=-i\lambda$. 
  Denote $F(ik_z,q)=F_0(ik_z)+\delta F(ik_z,q)$ where
  \begin{equation}
    F_0(ik_z)=\frac{k_z}{2}\left[\cot[k_z(z_1-z_0)]+\cot(k_z z_0)\right],\;\delta F(ik_z,\omega)=\frac{k_z^2 v^2}{2z_0 \omega^2}
  \end{equation}
  The dispersion relation of the cavity-photon-like mode is determined by $F(ik_z,\omega)=0$.
  In the absence of the bilayer, the cavity-photon-like mode has branches with discrete $k_z$ values determined by $F_0(ik_z)=0$:
  \begin{equation}
    \cot[k_z(z_1-z_0)]+\cot(k_z z_0)=\frac{\sin(k_z z_1)}{\sin(k_z z_0)\sin[k_z(z_1-z_0)]}=0\implies \sin(k_z l/2)=0
  \end{equation}
  This yields the bare cavity photon modes with 
  \begin{equation}
    k_{z,n}^{(0)}=\frac{2n\pi}{l},\;n=1,2,3,\cdots,\;\omega_n^{(0)}=c\sqrt{ q^2+k_{z,n}^{(0)2} }
  \end{equation}
  Denote the cavity-photon-like mode solution as $k_{z,n}=k_{z,n}^{(0)}+\delta k_{z,n}$, where $\delta k_{z,n}$ is the correction from the bilayer. 
  To first order of $\delta F$ and $\delta k_{z,n}$, we have
  \begin{equation}
    F_0(ik_{z,n}^{(0)})+\delta k_{z,n}\partial_{k_z}F_0(ik_{z,n}^{(0)})+\delta F(ik_{z,n}^{(0)},\omega_n^{(0)})=0\implies \delta k_{z,n}=-\frac{\delta F(ik_{z,n}^{(0)},\omega_n^{(0)})}{\partial_{k_z}F_0(ik_{z,n}^{(0)})}
  \end{equation}
  where 
  \begin{equation}
    \delta F(ik_{z,n}^{(0)},\omega_n^{(0)})=\frac{k_{z,n}^{(0)2}v^2}{2z_0 \omega_n^{(0)2}}
  \end{equation}
  Assume the cutoff frequency of the narrow cavity is much larger than the Goldstone mode energy scale, i.e., $c\pi/l\gg vq$, we have $\omega_n^{(0)}\approx ck_{z,n}^{(0)}\gg vq$ and we are in the regime away from the matter resonance, thus
  \begin{equation}
    \delta F(ik_{z,n}^{(0)},\omega_n^{(0)})\approx \frac{1}{2z_0}\frac{v^2}{c^2}=\mathcal{O}(v^2/c^2)
  \end{equation}
  is a very small correction.
  Thus we can safely neglect the corrections from the bilayer to the cavity-photon-like modes and approximate them by the bare cavity photon modes.
  

\end{itemize}

Natheless, for a given frequency $\omega$, we can solve $F(\lambda,\omega)=0$ to obtain $\lambda_n(\omega)$, which gives the dispersion relation of the odd TM mode as
\begin{equation}
  F(\lambda_n,\omega)=0\implies \frac{\omega^2}{v^2}=\frac{\lambda_n}{z_0}\frac{\sinh(\lambda_n z_0)\sinh[\lambda_n(z_1-z_0)]}{\sinh(\lambda_n z_1)}\label{eq:odd_TM_cavity_disp}
\end{equation}
and the corresponding in-plane momentum is given by $q_n(\omega)=\sqrt{\lambda_n^2(\omega)+\omega^2/c^2}$.
We denote $\lambda_{-1}$ as the solution with $\lambda_{-1}\in\mathbb{R}$, which corresponds to the Goldstone-like mode, and $\lambda_n$ with $n=0, 1,2,\cdots$ as the solutions with $\lambda_n=i k_{z,n}\in i\mathbb{R}$, which correspond to the cavity-photon-like modes.
Once $\lambda_n(\omega)$ is obtained, the gauge field $\bm{A}(\omega,\bm{q},z)$ can be obtained by substituting $\lambda_n(\omega)$ into Eq. \eqref{eq:bounded-vector-potential-cavity}.
For the odd TM mode, we have
\begin{align}
  \bm{A}(\omega,\bm{q}(\lambda_n),z)\propto& \hat{q} f(\lambda_n,z)\nonumber\\
  f(\lambda_n,z)=&\re^{-\lambda_n|z-z_0|}- \re^{-\lambda_n|z+z_0|}-2\frac{\re^{-\lambda_n z_1}\sinh(\lambda_n z_0)\sinh(\lambda_n z)}{\sinh(\lambda_n z_1)}\label{eq:A_profile}
\end{align}
Substituting into Maxwell's equations, we find that $f(\lambda_n,z)$ satisfies the following equations:
\begin{equation}
  \partial_z^2 f(\lambda_n,z)=\lambda_n^2 f(\lambda_n,z)-2F_0(\lambda_n)f(\lambda_n,z_0)[\delta(z-z_0)-\delta(z+z_0)]
\end{equation}
Thus
\begin{align}
  &-\int_{-l/2}^{l/2}\partial_z f(\lambda_n,z)\partial_z f(\lambda_m,z) dz\nonumber\\
=&\int_{-l/2}^{l/2} f(\lambda_n,z)\partial_z^2 f(\lambda_m,z) dz\nonumber\\
=&\int_{-l/2}^{l/2} f(\lambda_n,z)[\lambda_m^2 f(\lambda_m,z)-2F_0(\lambda_m)f(\lambda_m,z_0)[\delta(z-z_0)-\delta(z+z_0)]] dz\nonumber\\
=&\lambda_m^2 \int_{-l/2}^{l/2} f(\lambda_n,z) f(\lambda_m,z) dz-2F_0(\lambda_m)f(\lambda_m,z_0)[f(\lambda_n,z_0)-f(\lambda_n,-z_0)]\nonumber\\
=&\lambda_m^2 \int_{-l/2}^{l/2} f(\lambda_n,z) f(\lambda_m,z) dz-4F_0(\lambda_m)f(\lambda_m,z_0)f(\lambda_n,z_0)\nonumber\\
=&\lambda_m^2 \int_{-l/2}^{l/2} f(\lambda_n,z) f(\lambda_m,z) dz-\lambda_m^2\frac{2v^2}{z_0 \omega^2}f(\lambda_m,z_0)f(\lambda_n,z_0)
\end{align}
For the last step, we have used the dispersion relation of the odd TM mode Eq. \eqref{eq:odd_TM_cavity_disp} to replace $F_0(\lambda_m)$ by $\lambda_m^2 v^2/(2z_0 \omega^2)$.
Similarly, we have
\begin{equation}
  -\int_{-l/2}^{l/2}\partial_z f(\lambda_m,z)\partial_z f(\lambda_n,z) dz=\lambda_n^2 \int_{-l/2}^{l/2} f(\lambda_n,z) f(\lambda_m,z) dz-\lambda_n^2\frac{2v^2}{z_0 \omega^2}f(\lambda_m,z_0)f(\lambda_n,z_0)
\end{equation}
Thus, at the same frequency and for $\lambda_m\ne \lambda_n$, we must have
\begin{subequations}
  \begin{align}
    \int_{-l/2}^{l/2} f(\lambda_n,z) f(\lambda_m,z) dz-\frac{2v^2}{z_0 \omega^2}f(\lambda_m,z_0)f(\lambda_n,z_0)=&0\\
    \int_{-l/2}^{l/2}\partial_z f(\lambda_n,z)\partial_z f(\lambda_m,z) dz=&0\label{eq:right-cavity-B-orthogonality}
  \end{align}
\end{subequations}
In other words, the functions $\partial_zf(\lambda_n,z)$ are orthogonal to each other, and the functions $f(\lambda_n,z)$ are orthogonal to each other under the modified inner product defined by $\langle f(\lambda_n),f(\lambda_m)\rangle=\int_{-l/2}^{l/2} f(\lambda_n,z) f(\lambda_m,z) dz-(2v^2/z_0 \omega^2)f(\lambda_m,z_0)f(\lambda_n,z_0)$.

For propagation with in-plane wave vector $\bm{q}=q\hat{x}$, the nonzero electromagnetic fields associated with the odd TM mode are
\begin{subequations}
  \begin{align}
    E_x=&-i\frac{c^2\lambda_n^2}{\omega}f(\lambda_n,z)\re^{i(q_n x-\omega t)}\\
    E_z=&-\frac{c^2}{\omega}q_n\partial_z f(\lambda_n,z)\re^{i(q_n x-\omega t)}\\
    B_y=&\partial_z f(\lambda_n,z)\re^{i(q_n x-\omega t)}
  \end{align}\label{eq:EM_field_cavity_exact}
\end{subequations}

\subsection{\texorpdfstring{Transmission problem from wide to narrow cavity}{Transmission problem from wide to narrow cavity}}

\begin{figure}
  \centering
  \includegraphics[width=0.5\textwidth]{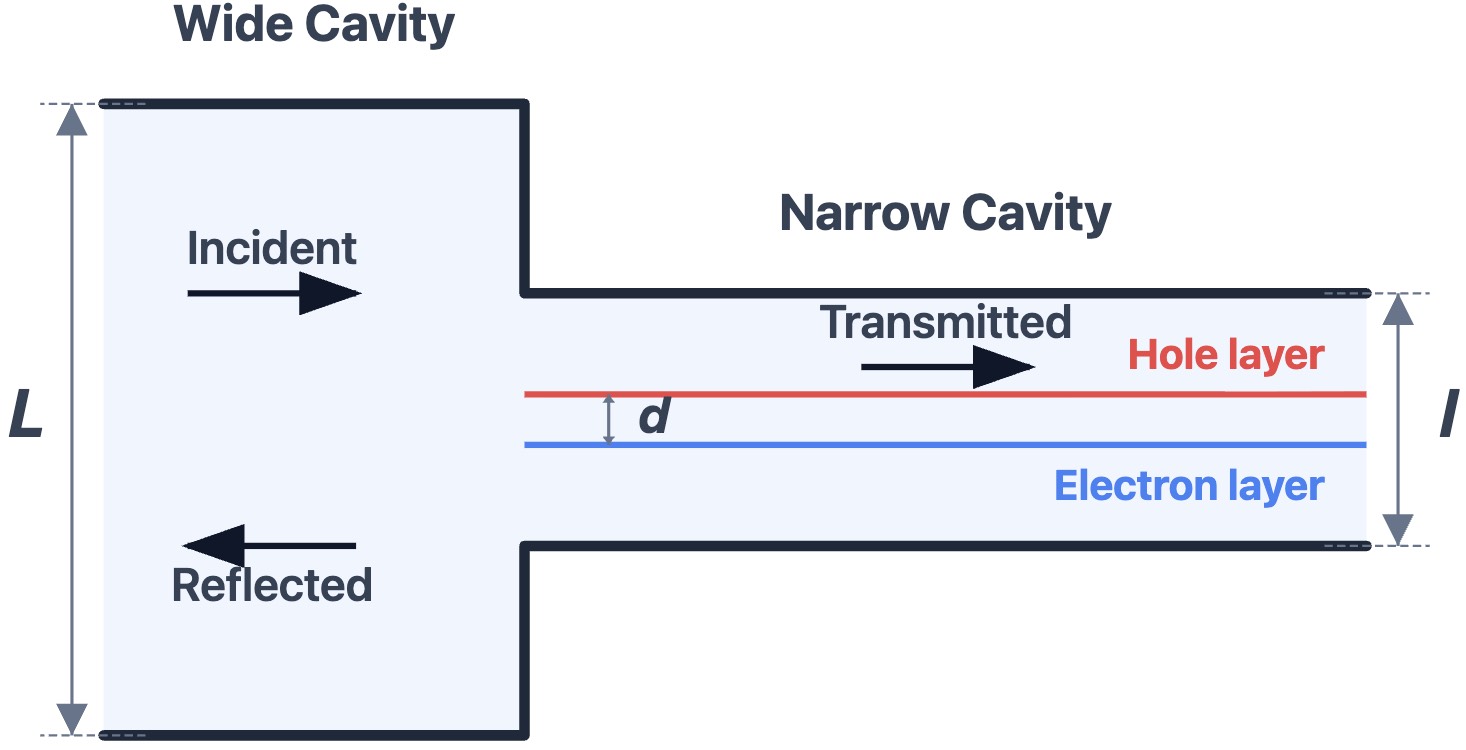}
  \caption{The transmission problem of the TM mode from a wide cavity with thickness $L$ to a narrow cavity with thickness $l<L$.}
  \label{fig:waveguide}
\end{figure}

We consider the transmission of a TM mode from a wide cavity of thickness $L$ into a narrow cavity of thickness $l<L$, as shown in Fig.~\ref{fig:waveguide}. We take the in-plane wave vector to be $\bm{q}=q\hat{x}$. For TM polarization, the nonvanishing field components are $B_y$, $E_x\propto \partial_z B_y$, and $E_z\propto \partial_x B_y$.

In the left half-space $x<0$, the system is a bare metallic cavity of thickness $L$. The transverse TM basis functions for $B_y$ are taken as
\begin{subequations}
\begin{align}
\psi_n^{(L)}(z)=\sqrt{\frac{2}{L}}\cos\!\left[n k_c^{(L)}\left(z+\frac{L}{2}\right)\right],\qquad n=1,2,\ldots,
\end{align}
\end{subequations}
with $k_c^{(L)}=\pi/L$. The corresponding longitudinal wave vector is
\begin{equation}
\kappa_n^{(L)}=\sqrt{\omega^2/c_{L}^2-n^2[k_c^{(L)}]^2}.
\end{equation}
where $c_{L}$ is the speed of light in the left half-space, which can be different from the speed of light $c$ in the right half-space if the two half-spaces have different dielectric constants.
For an incident mode $m$, the field in the left half-space is
\begin{equation}
B_y(\omega,x,z)=\psi_m^{(L)}(z)\mathrm e^{i\kappa_m^{(L)}x}
+\sum_{n=1}^\infty R_{m,n}(\omega)\psi_n^{(L)}(z)\mathrm e^{-i\kappa_n^{(L)}x},
\qquad (|z|<L/2,\; x<0).
\label{eq:By_left}
\end{equation}
Using Maxwell's equations, the corresponding electric field is
\begin{align}
E_z(\omega,x,z)=&\frac{ic_L^2}{\omega}\partial_x B_y(\omega,x,z)\nonumber\\
  =&\frac{ic_L^2}{\omega}\left[i\kappa_m^{(L)}\psi^{(L)}_{m}(z)\re^{i\kappa_m^{(L)}x}-\sum_{n=1}^{\infty}i\kappa_n^{(L)}R_{m,n}(\omega)\psi^{(L)}_{n}(z)e^{-i\kappa_n^{(L)}x}\right],\;(|z|<L/2,x<0)
  \label{eq:Ez_left}
\end{align}

In the right half-space $x>0$, the system is a narrow cavity of thickness $l$ with a bilayer EI at $z=0$. The eigenmodes of the Maxwell equation in the right half-space are modified by the bilayer EI, and the dispersion relations of the modes are given by Eq.~\eqref{eq:odd_TM_cavity_disp}.
Given a frequency $\omega$, we can solve Eq.~\eqref{eq:odd_TM_cavity_disp} to obtain the corresponding $\lambda_n(\omega)$ and $q_n(\omega)$ for the odd TM modes in the right half-space.
And the corresponding $B_y$ profiles are given by
\begin{equation}
  \psi_n^{(R)}(z)=N_n \partial_zf(\lambda_n,z),\qquad n=-1,0,1,2,\cdots
\end{equation}
where $f(\lambda_n,z)$ is given by Eq.~\eqref{eq:A_profile}, and the normalization constant $N_n$ is fixed by $\int_{-l/2}^{l/2}dz\,|\psi_n^{(R)}(z)|^2=1$.
The $n=-1$ solution is the matter-dominated Goldstone-polariton branch, with $\lambda_{-1}\in\mathbb{R}$, while $n=0,1,2,\ldots$ label the cavity-photon-like branches, with $\lambda_n=i k_{z,n}\in i\mathbb{R}$.
For frequency $\omega\ll c\pi/l$, the in-plane momentum of the Goldstone-polariton branch is given by
\begin{equation}
  \kappa_{-1}^{(R)}(\omega)=\sqrt{\lambda_{-1}^2(\omega)+\omega^2/c^2}\in \mathbb{R}
\end{equation}
which means the Goldstone-polariton branch is a propagating mode in the right half-space, while the cavity-photon-like modes are evanescent modes with in-plane momentum
\begin{equation}
  \kappa_n^{(R)}(\omega)=\sqrt{\lambda_n^2(\omega)+\omega^2/c^2}=\sqrt{-k_{z,n}^2(\omega)+\omega^2/c^2}=i\sqrt{k_{z,n}^2(\omega)-\omega^2/c^2}\in i\mathbb{R},\quad n=0,1,2,\cdots
\end{equation}
Using the eigenmodes in the right half-space as the basis, the solution of the Maxwell equation in the right half-space can be written as
\begin{equation}
  B_y(\omega,x,z)=\sum_{n=-1}^\infty T_{m,n}(\omega)\psi_n^{(R)}(z)\mathrm e^{i\kappa_n^{(R)}(\omega)x},\qquad (|z|<l/2,\; x>0)\label{eq:By_right}
\end{equation}
and 
\begin{equation}
  E_z(\omega,x,z)=\frac{ic^2}{\omega}\partial_x B_y(\omega,x,z)=\frac{ic^2}{\omega}\sum_{n=-1}^\infty i\kappa_n^{(R)}(\omega)T_{m,n}(\omega)\psi_n^{(R)}(z)\mathrm e^{i\kappa_n^{(R)}(\omega)x},\qquad (|z|<l/2,\; x>0)\label{eq:Ez_right}
\end{equation}

At the interface $x=0$, the electromagnetic boundary conditions require the tangential fields $E_z$ and $B_y$ to be continuous for $|z|<l/2$. For $l/2<|z|<L/2$, the interface is covered by a perfect conductor, so the tangential electric field vanishes there. Therefore, we have
\begin{align}
  &\kappa_m^{(L)}\psi^{(L)}_{m}(z)-\sum_{n=1}^{\infty}\kappa_n^{(L)}R_{m,n}(\omega)\psi^{(L)}_{n}(z)=\frac{c^2}{c_{L}^2}\sum_{n=-1}^{\infty}\kappa_n^{(R)}T_{m,n}(\omega)\psi^{(R)}_{n}(z),\; |z|<\frac{l}{2}\nonumber\\
  &\kappa_m^{(L)}\psi^{(L)}_{m}(z)-\sum_{n=1}^{\infty}\kappa_n^{(L)}R_{m,n}(\omega)\psi^{(L)}_{n}(z)=0,\; \frac{l}{2}<|z|<\frac{L}{2}\label{eq:continuous_Ez}
\end{align}
and
\begin{equation}
  \psi^{(L)}_{m}(z)+\sum_{n=1}^{\infty}R_{m,n}(\omega)\psi^{(L)}_{n}(z)=\sum_{n=-1}^{\infty}T_{m,n}(\omega)\psi^{(R)}_{n}(z),\; |z|<\frac{l}{2}\label{eq:continuous_By}
\end{equation}

Define the coupling matrix between the eigenmodes in the left and right half space as
\begin{equation}
  M_{m,n}=\int_{-l/2}^{l/2}dz \psi^{(L)}_{m}(z)\psi^{(R)}_{n}(z)
\end{equation}
For Eq.\eqref{eq:continuous_Ez}, multiply both sides by $\psi^{(L)}_{s}(z)$ and integrate over $z$ from $-L/2$ to $L/2$, we have
\begin{align}
  \kappa_m^{(L)}\delta_{m,s}-\kappa_s^{(L)}R_{m,s}(\omega)=&\frac{c^2}{c_{L}^2}\sum_{n=-1}^{\infty}\kappa_n^{(R)}T_{m,n}(\omega)M_{s,n}
\end{align}
{
For Eq.\eqref{eq:continuous_By}, multiply both sides by $\psi^{(R)}_{s}(z)$ and integrate over $z$ from $-l/2$ to $l/2$. Using the orthonormality of the right-cavity eigenmodes established in Eq. \eqref{eq:right-cavity-B-orthogonality}, we obtain
\begin{align}
  M_{m,s}+\sum_{n=1}^{\infty}R_{m,n}(\omega)M_{n,s}=&T_{m,s}(\omega)
\end{align}
}

Define 
\begin{align}
  Y^{(L)}=\mathrm{diag}(\kappa_1^{(L)},\kappa_2^{(L)},\cdots),\quad Y^{(R)}=\mathrm{diag}(\kappa_{-1}^{(R)},\kappa_0^{(R)},\cdots)
\end{align}
Then the above equations can be written in matrix form as
\begin{align}
  [I-R(\omega)]Y^{(L)}=&\frac{c^2}{c_{L}^2}T(\omega)Y^{(R)}M^{T}\\
  [I+R(\omega)]M =&T(\omega)
\end{align}
where $I$ is the identity matrix, and $R(\omega)$ and $T(\omega)$ are the reflection and transmission matrices, respectively.
The first equation gives
\begin{equation}
  R(\omega)=I-\frac{c^2}{c_{L}^2}T(\omega)Y^{(R)}M^{T}[Y^{(L)}]^{-1}
\end{equation}
Thus the second equation gives
\begin{align}
  &2M-\frac{c^2}{c_{L}^2}T(\omega)Y^{(R)}M^{T}[Y^{(L)}]^{-1}M=T(\omega)\nonumber\\
  \implies& T(\omega)[I+\frac{c^2}{c_{L}^2}Y^{(R)}M^{T}[Y^{(L)}]^{-1}M]=2M\nonumber\\
  \implies& T(\omega)=2M[I+\frac{c^2}{c_{L}^2}Y^{(R)}M^{T}[Y^{(L)}]^{-1}M]^{-1}
\end{align}

The physical transmission rate should be defined as the ratio between the transmitted power and the incident power. For a TM mode, the time-averaged Poynting vector is given by
\begin{equation}
\langle \bm{S}\rangle=\frac{1}{2}\mathrm{Re}(\bm{E}\times\bm{B}^*)=-\frac{1}{2}\mathrm{Re}(E_z B_y^*)\hat{x}+\frac{1}{2}\mathrm{Re}(E_x B_y^*)\hat{z}
\end{equation}
For the incident mode $m$, using Eqs. \eqref{eq:By_left} and \eqref{eq:Ez_left}, the incident power is given by
\begin{equation}
  P_{\mathrm{in}}=-\int_{-L/2}^{L/2}dz \langle S_x\rangle =\frac{c_{L}^2}{2\omega}\mathrm{Re}[\kappa_m^{(L)}]\int_{-L/2}^{L/2}dz |\psi^{(L)}_m(z)|^2=\frac{c_{L}^2}{2\omega}\mathrm{Re}[\kappa_m^{(L)}]
\end{equation}
For the transmitted electric and magnetic fields, using Eqs. \eqref{eq:By_right} and \eqref{eq:Ez_right}, the transmitted power is given by
\begin{align}
  P_{\mathrm{out}}(x)=&-\int_{-l/2}^{l/2}dz \langle S_x\rangle =\frac{c^2}{2\omega}\int_{-l/2}^{l/2}dz \mathrm{Re}\bigg[\sum_{n=-1}^{\infty}T_{m,n}\psi^{(R)}_{n}(z)e^{i\kappa_n^{(R)}x}\sum_{s=-1}^{\infty}\kappa_{s}^{(R)*}T_{m,s}^*(\omega)\psi_{s}^{(R)}(z)e^{-i\kappa_s^{(R)*}x}\bigg]\nonumber\\
  =&\frac{c^2}{2\omega}\sum_{n=-1}^{\infty}\sum_{s=-1}^{\infty}\mathrm{Re}[\kappa_{s}^{(R)*}T_{m,n}(\omega)T_{m,s}^*(\omega)\int_{-l/2}^{l/2}dz \psi^{(R)}_{n}(z)\psi_{s}^{(R)}(z)]e^{i(\kappa_n^{(R)}-\kappa_s^{(R)*})x}\nonumber\\
  =&\frac{c^2}{2\omega}\sum_{n=-1}^{\infty}\sum_{s=-1}^{\infty}\mathrm{Re}[\kappa_{s}^{(R)*}T_{m,n}(\omega)T_{m,s}^*(\omega)]\delta_{n,s}e^{i(\kappa_n^{(R)}-\kappa_s^{(R)*})x}\nonumber\\
  =&\frac{c^2}{2\omega}\sum_{n=-1}^{\infty}\mathrm{Re}[\kappa_{n}^{(R)*}]|T_{m,n}(\omega)|^2
\end{align}
which is independent of the position $x$.
Thus the total transmission rate is given by
\begin{equation}
  \mathcal{T}_m(\omega)=\frac{P_{\mathrm{out}}}{P_{\mathrm{in}}}=\frac{c^2}{c_{L}^2}\sum_{n=-1}^{\infty}\frac{\mathrm{Re}[\kappa_n^{(R)}]}{\mathrm{Re}[\kappa_m^{(L)}]}|T_{m,n}(\omega)|^2
\end{equation}

In the case without the bilayer EI, the Goldstone-polariton branch $\psi^{(R)}_{-1}(z)$ does not exist and the eigenmodes in the right half space are 
\begin{equation}
  \psi^{(R)'}_{n}(z)=\sqrt{\frac{2}{l}}\cos\!\left[n k_c^{(R)}\left(z+\frac{l}{2}\right)\right],\qquad n=1,2,\cdots
\end{equation}
with in-plane momentum
\begin{equation}
  \kappa_n^{(R)'}=\sqrt{\omega^2/c^2-n^2 [k_c^{(R)}]^2}=i\sqrt{n^2 [k_c^{(R)}]^2-\omega^2/c^2},\quad n=1,2,\cdots
\end{equation}
Denote 
\begin{equation}
  M'_{m,n}=M_{m,n},\quad n=1,2,\cdots
\end{equation}
and 
\begin{equation}
  Y'^{(R)}=\mathrm{diag}(\kappa_1^{(R)'},\kappa_2^{(R)'},\cdots)
\end{equation}
Then the transmission matrix in the absence of the bilayer EI is given by
\begin{equation}
  T'(\omega)=2M'[I+\frac{c^2}{c_{L}^2}Y'^{(R)}M'^{T}[Y^{(L)}]^{-1}M']^{-1}
\end{equation}
and the total transmission rate is given by
\begin{equation}
  \mathcal{T}'_m(\omega)=\frac{c^2}{c_{L}^2}\sum_{n=1}^{\infty}\frac{\mathrm{Re}[\kappa_n^{(R)'}]}{\mathrm{Re}[\kappa_m^{(L)}]}|T'_{m,n}(\omega)|^2
\end{equation}


Assume that the incident frequency $\omega$ satisfies $\omega < k_c^{(R)} c$. Then, for all $n=1,2,\ldots$, we have
\begin{equation}
  \mathrm{Re}[\kappa_n^{(R)'}]=0,
\end{equation}
and therefore $\mathcal{T}'_m(\omega)=0$. This is because the incident TM mode lies below the cutoff frequency of the narrow cavity and hence cannot couple to any propagating TM mode inside it.

By contrast, when the bilayer EI is present, we have
\begin{equation}
  \kappa_{-1}^{(R)}=\sqrt{\lambda_{-1}^2+\omega^2/c^2}\in \mathbb{R},\quad\kappa_{n\ge0}^{(R)}=i\sqrt{k_{z,n}^2-\omega^2/c^2}\in i\mathbb{R}
\end{equation}
Thus the Goldstone mode supports a propagating channel in the right half space, yielding a finite transmission rate 
\begin{equation}
  \mathcal{T}_m(\omega)=\frac{c^2}{c_{L}^2}\frac{\mathrm{Re}[\kappa_{-1}^{(R)}]}{\mathrm{Re}[\kappa_m^{(L)}]}|T_{m,-1}(\omega)|^2
\end{equation}

\subsubsection{\texorpdfstring{The overlap matrix}{The overlap matrix}}
In this part, we calculate the overlap matrix $M$ analytically.
In the left half space, the eigenmodes are given by
\begin{equation}
  \psi^{(L)}_{n}(z)=\sqrt{\frac{2}{L}}\cos\!\left[n k_c^{(L)}\left(z+\frac{L}{2}\right)\right],\quad n=1,2,\cdots
\end{equation}
In the right half space, the eigenmodes are given by
\begin{equation}
  \psi^{(R)}_{n}(z)=N_{n}\partial_zf(\lambda_n,z),\quad n=-1,0,1,2,\cdots
\end{equation}
where $N_n$ is the normalization constant, $\lambda_n$ is obtained by solving Eq. \eqref{eq:odd_TM_cavity_disp}, and $f(\lambda,z)$ is proportional to Eq.~\eqref{eq:A_profile} and could also be written as
\begin{equation}
f(\lambda,z)=\begin{cases}
    \dfrac{\sinh[\lambda(z_1-z)]}{\sinh[\lambda(z_1-z_0)]}, & z_0<z<z_1,\\
    -\dfrac{\sinh[\lambda(z+z_1)]}{\sinh[\lambda(z_1-z_0)]}, & -z_1<z<-z_0,\\
    \dfrac{\sinh(\lambda z)}{\sinh(\lambda z_0)}, & -z_0<z<z_0.
\end{cases}
\end{equation}
Thus we have
\begin{equation}
  \partial_z f(\lambda,z)=\begin{cases}
    -\dfrac{\lambda\cosh[\lambda(z_1-z)]}{\sinh[\lambda(z_1-z_0)]}, & z_0<z<z_1,\\
    -\dfrac{\lambda\cosh[\lambda(z+z_1)]}{\sinh[\lambda(z_1-z_0)]}, & -z_1<z<-z_0,\\
    \dfrac{\lambda\cosh(\lambda z)}{\sinh(\lambda z_0)}, & -z_0<z<z_0.
  \end{cases}
\end{equation}
and the normalization constant is given by
\begin{align}
  N_n^{-2}=&\int_{-z_1}^{-z_0}dz \bigg|\frac{\lambda_n\cosh[\lambda_n(z+z_1)]}{\sinh[\lambda_n(z_1-z_0)]}\bigg|^2+\int_{-z_0}^{z_0}dz \bigg|\frac{\lambda_n\cosh(\lambda_n z)}{\sinh(\lambda_n z_0)}\bigg|^2+\int_{z_0}^{z_1}dz \bigg|\frac{\lambda_n\cosh[\lambda_n(z_1-z)]}{\sinh[\lambda_n(z_1-z_0)]}\bigg|^2\nonumber\\
  =&2\int_0^{z_0}dz \bigg|\frac{\lambda_n\cosh(\lambda_n z)}{\sinh(\lambda_n z_0)}\bigg|^2+2\int_{z_0}^{z_1}dz \bigg|\frac{\lambda_n\cosh[\lambda_n(z_1-z)]}{\sinh[\lambda_n(z_1-z_0)]}\bigg|^2\nonumber\\
  =&\frac{1}{2}\frac{2z_0\lambda_n^2+\lambda_n\sinh(2z_0\lambda_n)}{\sinh(\lambda_n z_0)^2}+\frac{1}{2}\frac{2(z_1-z_0)\lambda_n^2+\lambda_n\sinh[2(z_1-z_0)\lambda_n]}{\sinh[\lambda_n(z_1-z_0)]^2}
\end{align}
This applies to both the Goldstone mode with $n=-1$ and the photon-like modes with $n\ge 0$ in the right half space.
For the photon-like modes with $n\ge 0$, we only need to replace $\lambda_n$ by $ik_{z,n}$.

For the overlap matrix, we further consider a non-symmetric configuration where the narrow cavity is shifted in the $z$ direction by a distance $z_s$, i.e., the cavity extends from $z=z_s-l/2$ to $z=z_s+l/2$. Then the coupling matrix elements are given by
\begin{align}
  M_{m,n}(z_s)=&\int_{z_s-l/2}^{z_s+l/2}dz \psi^{(L)}_{m}(z)\psi^{(R)}_{n}(z-z_s)\nonumber\\
  =&\int_{-l/2}^{l/2}dz \psi^{(L)}_{m}(z+z_s)\psi^{(R)}_{n}(z)\nonumber\\
  =&\sqrt{\frac{2}{L}}\int_{-z_1}^{z_1} dz\cos\!\left[m k_c^{(L)}\left(z+z_s+\frac{L}{2}\right)\right]\psi_n^{(R)}(z)
\end{align}
As $\psi_n^{(R)}(z)$ is even in $z$, we have
\begin{align}
  &M_{m,n}(z_s)\nonumber\\
  =&\sqrt{\frac{2}{L}}\cos[m k_c^{(L)}(z_s+L/2)]\int_{-z_1}^{z_1} dz\cos(m k_c^{(L)} z)\psi_n^{(R)}(z)\nonumber\\
  =&2\sqrt{\frac{2}{L}}N_n\cos[m k_c^{(L)}(z_s+L/2)]\int_{0}^{z_1} dz\cos(m k_c^{(L)} z)\partial_z f(\lambda_n,z)\nonumber\\
  =&2\sqrt{\frac{2}{L}}N_n\cos[m k_c^{(L)}(z_s+L/2)]\bigg\{\int_{0}^{z_0} dz\cos(m k_c^{(L)} z)\frac{\lambda_n\cosh(\lambda_n z)}{\sinh(\lambda_n z_0)}+\int_{z_0}^{z_1} dz\cos(m k_c^{(L)} z)\frac{-\lambda_n\cosh[\lambda_n(z_1-z)]}{\sinh[\lambda_n(z_1-z_0)]}\bigg\}\nonumber\\
  =&2\sqrt{\frac{2}{L}}N_n\frac{\cos[m k_c^{(L)}(z_s+L/2)]}{\lambda_n^2+[m k_c^{(L)}]^2}\bigg\{
    mk_c^{(L)}\sin(m k_c^{(L)} z_0)\frac{\cosh(\lambda_n z_0)\lambda_n}{\sinh(\lambda_n z_0)}+\lambda_n^2\cos(m k_c^{(L)} z_0)\nonumber\\
    &+mk_c^{(L)}\sin(m k_c^{(L)} z_0)\frac{\cosh[\lambda_n(z_1-z_0)]\lambda_n}{\sinh[\lambda_n(z_1-z_0)]}-mk_c^{(L)}\sin(m k_c^{(L)} z_1)\frac{\lambda_n}{\sinh[\lambda_n(z_1-z_0)]}-\lambda_n^2\cos(m k_c^{(L)} z_0)\bigg\}\nonumber\\
    =&\sqrt{\frac{2}{L}}N_n\frac{2\lambda_n mk_c^{(L)}\cos[m k_c^{(L)}(z_s+L/2)]}{\lambda_n^2+[m k_c^{(L)}]^2}\bigg\{\sin(m k_c^{(L)} z_0)\bigg[\coth(\lambda_n z_0)+\coth[\lambda_n(z_1-z_0)]\bigg]-\frac{\sin(m k_c^{(L)} z_1)}{\sinh[\lambda_n(z_1-z_0)]}\bigg\}
\end{align}

\subsubsection{\texorpdfstring{Transmission rate near the cutoff frequency}{Transmission rate near the cutoff frequency}}
Assume the frequency is just above the cutoff frequency of the incident mode, i.e., $\omega\approx m\omega_c^{(L)}+0^+$, where $\omega_c^{(L)}=k_c^{(L)}c_L=\pi c_L/L$.
In this limit, we have $\kappa_m^{(L)}=\kappa_m^{(L)}(\omega)=1/c_L\sqrt{\omega^2-m^2[\omega_c^{(L)}]^2}\approx 0^+$.
Thus
\begin{align}
  (M^{T}[Y^{(L)}]^{-1}M)_{n,s}=&\sum_{n'=1}^{\infty}M_{n',n}\frac{1}{\kappa_{n'}^{(L)}}M_{n',s}\nonumber\\
  =&\frac{1}{\kappa_m^{(L)}}M_{m,n}M_{m,s}+\sum_{\substack{n'=1\\n'\ne m}}^{\infty}M_{n',n}\frac{1}{\kappa_{n'}^{(L)}}M_{n',s}
\end{align}
and 
\begin{equation}
  I+\frac{c^2}{c_{L}^2}Y^{(R)}M^{T}[Y^{(L)}]^{-1}M=\frac{A+\kappa_m^{(L)}B}{\kappa_m^{(L)}}
\end{equation}
where
\begin{equation}
  A_{n,s}=\frac{c^2}{c_L^2}\kappa_n^{(R)}M_{m,n}M_{m,s},\quad B_{n,s}=\delta_{n,s}+\frac{c^2}{c_{L}^2}\kappa_n^{(R)}\sum_{\substack{n'=1\\n'\ne m}}^{\infty}M_{n',n}\frac{1}{\kappa_{n'}^{(L)}}M_{n',s}
\end{equation}
Define two column vectors $\bm{u}$ and $\bm{v}$ with elements
\begin{equation}
  u_n=\frac{c^2}{c_L^2}\kappa_n^{(R)}M_{m,n},\quad v_n=M_{m,n}
\end{equation}
Then $A$ can be written as $A=\bm{u}\bm{v}^T$.
Using the Sherman-Morrison formula, we have
\begin{equation}
  \bm{v}^T(A+\kappa_m^{(L)}B)^{-1}=\frac{\bm{v}^{T}B^{-1}}{\kappa_m^{(L)}+\bm{v}^TB^{-1}\bm{u}}
\end{equation}

Thus the transmission amplitude is given by
\begin{equation}
  T_{m,-1}(\omega)=2\kappa_m^{(L)}\sum_{s=-1}^{\infty}M_{m,s}(A+\kappa_m^{(L)}B)^{-1}_{s,-1}=2\kappa_m^{(L)}\bm{v}^T(A+\kappa_m^{(L)}B)^{-1}\bm{e}_{-1}=\frac{2\kappa_m^{(L)}\bm{v}^TB^{-1}\bm{e}_{-1}}{\kappa_m^{(L)}+\bm{v}^TB^{-1}\bm{u}}
\end{equation}
where $\bm{e}_{-1}$ is the unit vector with the element for Goldstone mode being $1$ and all other elements being $0$.
For convenience, we denote
\begin{equation}
  \alpha=\bm{v}^TB^{-1}\bm{u},\quad \beta=\bm{v}^TB^{-1}\bm{e}_{-1}
\end{equation}
which are two complex numbers dependent on $\omega$.
Then the transmission amplitude can be written as
\begin{equation}
  T_{m,-1}(\omega)=\frac{2\kappa_m^{(L)}\beta}{\kappa_m^{(L)}+\alpha}
\end{equation}
and the transmission rate is given by
\begin{equation}
  \mathcal{T}_m(\omega)=\frac{c^2}{c_{L}^2}\frac{\mathrm{Re}[\kappa_{-1}^{(R)}]}{\mathrm{Re}[\kappa_m^{(L)}]}|T_{m,-1}(\omega)|^2=4\frac{c^2}{c_{L}^2}\kappa_{-1}^{(R)}\kappa_m^{(L)}\frac{|\beta|^2}{|\kappa_m^{(L)}+\alpha|^2}
\end{equation}
where we have used the fact that $\kappa_{-1}^{(R)}$ and $\kappa_m^{(L)}$ are both real and positive when $\omega>m\omega_c^{(L)}$.

Denote $\omega=m\omega_c^{(L)}+\delta\omega$, where $\delta\omega$ is a small positive number. 
Then we have 
\begin{equation}
  \kappa_m^{(L)}\approx \frac{1}{c_L}\sqrt{2m\omega_c^{(L)}\delta\omega}
\end{equation}
In addition, as $\delta\omega\to 0^+$, the dependence of $\alpha$ and $\beta$ on $\omega$ is regular, and we can approximate them by their values at $\omega=m\omega_c^{(L)}$, which are denoted by $\alpha_0$ and $\beta_0$.
Thus the transmission rate near the cutoff frequency can be approximated by
\begin{equation}
  \mathcal{T}_m(\omega=m\omega_c^{(L)}+\delta\omega)=\mathcal{T}_m(\kappa_m^{(L)})\approx4\frac{c^2}{c_{L}^2}\kappa_{-1}^{(R)}\kappa_m^{(L)}\frac{|\beta_0|^2}{|\kappa_m^{(L)}+\alpha_0|^2}
\end{equation}
Consider the function
\begin{equation}
  f(k)=\frac{k}{|k+a+b i|^2}=\frac{k}{(k+a)^2+b^2}
\end{equation}
where $a$ and $b$ are two real numbers.
It's easy to see that $f(k)$ takes its maximum value at $k^*=\sqrt{a^2+b^2}$ and the maximum value is given by
\begin{equation}
  f(k^*)=\frac{\sqrt{a^2+b^2}}{(a+\sqrt{a^2+b^2})^2+b^2}
\end{equation}
Thus the transmission rate $\mathcal{T}_m(\omega)$ takes its maximum value at 
\begin{equation}
  \kappa_m^{(L)*}=|\alpha_0|\implies \delta\omega^*=\frac{c_L^2}{2m\omega_c^{(L)}}|\alpha_0|^2
\end{equation}
and the maximum transmission rate is given by
\begin{equation}
  \mathcal{T}_m^*=4\frac{c^2}{c_{L}^2}\kappa_{-1}^{(R)}\frac{|\beta_0|^2|\alpha_0|}{||\alpha_0|+\alpha_0|^2}=2\frac{c^2}{c_{L}^2}\kappa_{-1}^{(R)}\frac{|\beta_0|^2}{|\alpha_0|+\mathrm{Re}(\alpha_0)}
\end{equation}

}

\section{TDHF: Application to the bilayer system with finite perpendicular magnetic field}\label{app:tdhf-bilayer-magnetic}

\subsection{The Landau level basis}

When a finite static magnetic field $\bm{B}=(0,0,B)$ is applied to the bilayer system, we should not treat $B$ as a perturbation.
Instead, we should treat it strictly and work in the Landau level (LL) basis.
In the Landau gauge, the vector potential is $\bm{A}^0=(-By,0,0)$ and the non-interacting Hamiltonian becomes 
\begin{equation}
  \hat{H}_0=\int\rd\bm{r}\;\Psi^{\dagger}(\bm{r})\begin{bmatrix}
    \frac{1}{2m_e}(\bm{p}+e\bm{A}^0)^2-\frac{\mu_X}{2} & 0 \\
    0 & -\frac{1}{2m_h}(\bm{p}+e\bm{A}^0)^2+\frac{\mu_X}{2}
  \end{bmatrix}\Psi(\bm{r})\label{eq:hamiltonian-gauge-field-magnetic}
\end{equation}
Define the ladder operators
\begin{gather}
  a=i\frac{l_B}{\sqrt{2}\hbar}[p_y+i(p_x-eBy)]\\
  a^{\dagger}=-i\frac{l_B}{\sqrt{2}\hbar}[p_y-i(p_x-eBy)]
\end{gather}
where $l_B=\sqrt{\hbar/(eB)}$ is the magnetic length.
One can verify that 
\begin{equation}
  [a,a^{\dagger}]=\frac{l_B^2}{\hbar^2}[p_y,ieBy]=\frac{l_B^2eB}{\hbar}=1
\end{equation}
The LL basis $\ket{ik_x}$ ($i$ is the LL index and $k_x$ is the momentum in $x$ direction) is defined as the eigenstates of $a^{\dagger}a$ and $p_x$, such that 
\begin{equation}
  a^{\dagger}a\ket{ik_x}=i\ket{ik_x},\;p_x\ket{ik_x}=\hbar k_x\ket{ik_x}
\end{equation}
with wavefunctions
\begin{equation}
    \phi_{ik_x}(\bm{r})\equiv \inp{\bm{r}}{ik_x}=\frac{1}{\sqrt{L_xl_B}}\re^{ik_x x}\psi_i(y/l_B-l_B k_x)\label{eq:LL_wave_realspace}
\end{equation}
where 
\begin{equation}
    \psi_{i}(x)=(2^ii!\sqrt{\pi})^{-1/2}\re^{-x^2/2}H_i(x)
\end{equation}
is the $i$-th level of 1-d quantum oscillator and $H_n(x)$ is Hermite polynomial.
Define the creation and annihilation operators in the LL basis as
\begin{equation}
    \Psi_{s}(\bm{r})=\sum_{ik_x}\phi_{ik_x}(\bm{r})l_{sik_x}\Leftrightarrow l^{\dagger}_{sik_x}=\int\rd\bm{r}\;\phi_{nk_x}(\bm{r})\Psi_s^{\dagger}(\bm{r})\label{eq:LL_operator}
\end{equation}
Then the non-interacting Hamiltonian in the LL basis is written as
\begin{equation}
  \hat{H}_0=\sum_{ik_x}L_{ik_x}^{\dagger}\begin{bmatrix}
    \hbar\omega_e(i+\frac{1}{2})-\frac{\mu_X}{2} & 0 \\
    0 & -\hbar\omega_h(i+\frac{1}{2})+\frac{\mu_X}{2}
  \end{bmatrix}L_{ik_x}
\end{equation}
where $L_{ik_x}=[l_{eik_x},l_{hik_x}]$ and $\hbar\omega_s=eB/m_s$ is the cyclotron frequency in each layer.

In LL basis, the interaction Hamiltonian is written as 
\begin{align}
  \hat{H}_I=&\frac{1}{2}\sum_{ss'}\int\rd\bm{r}\rd\bm{r}'\;\Psi^{\dagger}_{s}(\bm{r})\Psi^{\dagger}_{s'}(\bm{r}')V_{ss'}(\bm{r}-\bm{r}')\Psi_{s'}(\bm{r}')\Psi_{s}(\bm{r})\nonumber\\
  =&\frac{1}{2}\sum_{ss'}\sum_{i_{1-4},k_{1-4}}\int\rd\bm{r}\rd\bm{r}'\;\inp{i_1k_1}{\bm{r}}\inp{i_2k_2}{\bm{r}'}V_{ss'}(\bm{r}-\bm{r}')\inp{\bm{r}'}{i_3k_3}\inp{\bm{r}}{i_4k_4}l^{\dagger}_{si_1k_1}l^{\dagger}_{s'i_2k_2}l_{s'i_3k_3}l_{si_4k_4}\nonumber\\
  =&\frac{1}{2\mathcal{V}}\sum_{ss'}\sum_{i_{1-4},k_{1-4}}\sum_{\bm{q}}V_{ss'}(\bm{q})\int\rd\bm{r}\;\inp{i_1k_1}{\bm{r}}\re^{i\bm{q}\cdot\bm{r}}\inp{\bm{r}}{i_4k_4}\int\rd\bm{r}'\;\inp{i_2k_2}{\bm{r}'}\re^{-i\bm{q}\cdot\bm{r}'}\inp{\bm{r}'}{i_3k_3}l^{\dagger}_{si_1k_1}l^{\dagger}_{s'i_2k_2}l_{s'i_3k_3}l_{si_4k_4}\nonumber\\
  =&\frac{1}{2\mathcal{V}}\sum_{ss'}\sum_{i_{1-4},k_{1-4}}\sum_{\bm{q}}V_{ss'}(\bm{q})\bra{i_1k_1}\re^{i\bm{q}\cdot\bm{r}}\ket{i_4k_4}\bra{i_2k_2}\re^{-i\bm{q}\cdot\bm{r}'}\ket{i_3k_3}l^{\dagger}_{si_1k_1}l^{\dagger}_{s'i_2k_2}l_{s'i_3k_3}l_{si_4k_4}\nonumber\\
  =&\frac{1}{2\mathcal{V}}\sum_{ss'}\sum_{i_{1-4},k_{1-4}}\sum_{\bm{q}}V_{ss'}(\bm{q})\delta_{k_1,k_4+q_x}\delta_{k_2,k_3-q_x}\re^{iq_y(k_1-k_2-q_x)l_B^2}F_{i_1i_4}(-\bm{q}l_B/\sqrt{2})F_{i_2i_3}(\bm{q}l_B/\sqrt{2})l^{\dagger}_{si_1k_1}l^{\dagger}_{s'i_2k_2}l_{s'i_3k_3}l_{si_4k_4}\nonumber\\
  =&\frac{1}{2\mathcal{V}}\sum_{ss'}\sum_{i_{1-4}}\sum_{k_1k_2,\bm{q}}V_{ss'}(\bm{q})F_{i_1i_4}(-\bm{q}l_B/\sqrt{2})F_{i_2i_3}(\bm{q}l_B/\sqrt{2})\re^{iq_y(k_1-k_2)l_B^2}l^{\dagger}_{si_1k_1+q_x/2}l^{\dagger}_{s'i_2k_2-q_x/2}l_{s'i_3k_2+q_x/2}l_{si_4k_1-q_x/2}
\end{align}
where the matrix element $\bra{ik_x}\re^{-i\bm{q}\cdot\bm{t}}\ket{i'k_x'}$ is evaluated as 
\begin{equation}
  \bra{ik_x}\re^{-i\bm{q}\cdot\bm{r}}\ket{i'k_x'}=\delta_{k_xk_x'-q_x}\re^{-iq_y(k_x+q_x/2)l_B^2}F_{ii'}(\bm{q}l_B/\sqrt{2})
\end{equation}
and
\begin{equation}
  F_{ii'}(\bm{q})=\begin{cases}
    \sqrt{\dfrac{i'!}{i!}}\re^{-q^2/2}(q_x-iq_y)^{i-i'}L_{i'}^{(i-i')}(q^2), & i\ge i',\\
    \sqrt{\dfrac{i!}{i'!}}\re^{-q^2/2}(-q_x-iq_y)^{i'-i}L_{i}^{(i'-i)}(q^2), & i\le i'.
  \end{cases}
\end{equation}
is the LL form factor, which satisfies $F_{ii'}^*(-\bm{q})=F_{i'i}(\bm{q})$.
It is convenient to write the LL form factor as $F_{i_1i_2}(\bm{q})=\re^{-i\Delta i_{12}\theta_{\bm{q}}}T^{\Delta i_{12}}_{i_{<,12}}(q)$, where $\Delta i_{12}=i_1-i_2$, $i_{<,12}=\min(i_1,i_2)$, and $\theta_{\bm{q}}$ is the angle between $\bm{q}$ and the $x$-axis.
By writing in this form, the function 
\begin{equation}
  T_{n}^{m}(q)\equiv \begin{cases}
    \sqrt{\dfrac{n!}{(n+|m|)!}}\re^{-q^2/2}q^{|m|}L_n^{(|m|)}(q^2), & m\geq 0,\\
    (-1)^{|m|}\sqrt{\dfrac{n!}{(n+|m|)!}}\re^{-q^2/2}q^{|m|}L_n^{(|m|)}(q^2), & m\le0.
  \end{cases}
\end{equation}
is real and satisfies $T_n^{-m}(q)=(-1)^mT_n^m(q)$.

\subsection{The mean field Hamiltonian and self-consistent equations}\label{app:LL-HF}

To get the mean field Hamiltonian, define the single particle density matrix
\begin{equation}
  \rho_{ss',ii',k_x}(q_x)\equiv \ave{l^{\dagger}_{s'i'k_x-q_x/2}l_{sik_x+q_x/2}} 
\end{equation}
and its Fourier transform
\begin{align}
  \rho_{ss',ii'}(\bm{q})=&\frac{2\pi l_B^2}{\mathcal{V}}\sum_{k_x}\rho_{ss',ii',k_x}(q_x)\re^{-iq_y k_x l_B^2}
  =\frac{l_B^2}{L_y}\int_0^{L_y/l_B^2}\rd k_x\;\rho_{ss',ii',k_x}(q_x)\re^{-iq_y k_x l_B^2}\label{eq:dst_mat_full}\\
  \rho_{ss',ii',k_x}(q_x)=&\sum_{q_y}\rho_{ss',ii'}(\bm{q})\re^{iq_y k_x l_B^2}
\end{align}

The Hartree Hamiltonian is given by
\begin{align}
  \hat{H}^{H}=&\frac{1}{\mathcal{V}}\sum_{ss'}\sum_{i_{1-4}}\sum_{k_1k_2,\bm{q}}V_{ss'}(\bm{q})F_{i_1i_4}(-\bm{q}l_B/\sqrt{2})F_{i_2i_3}(\bm{q}l_B/\sqrt{2})\re^{iq_y(k_1-k_2)l_B^2}\rho_{s's',i_3i_2,k_2}(q_x)l^{\dagger}_{si_1k_1+q_x/2}l_{si_4k_1-q_x/2}\nonumber\\
  =&\frac{1}{2\pi l_B^2}\sum_{ss'}\sum_{i_{1-4}}\sum_{k_1,\bm{q}}V_{ss'}(\bm{q})F_{i_1i_4}(-\bm{q}l_B/\sqrt{2})F_{i_2i_3}(\bm{q}l_B/\sqrt{2})\re^{iq_y k_1 l_B^2}\rho_{s's',i_3i_2}(\bm{q})l^{\dagger}_{si_1k_1+q_x/2}l_{si_4k_1-q_x/2}\nonumber\\
  =&\sum_{s,i_2i_4,k_1\bm{q}}\left[\sum_{s',i_2i_3}\frac{1}{2\pi l_B^2}V_{ss'}(\bm{q})F_{i_1i_4}(-\bm{q}l_B/\sqrt{2})F_{i_2i_3}(\bm{q}l_B/\sqrt{2})\rho_{s's',i_3i_2}(\bm{q})\right]\re^{iq_y k_1 l_B^2}l^{\dagger}_{si_1k_1+q_x/2}l_{si_4k_1-q_x/2}\nonumber\\
  =&\sum_{s,i_2i_4,k_1\bm{q}}\Sigma^{H}_{ss,i_1i_4}(\bm{q})\re^{iq_y k_1 l_B^2}l^{\dagger}_{si_1k_1+q_x/2}l_{si_4k_1-q_x/2}
\end{align}
where 
\begin{equation}
  \Sigma^{H}_{ss,i_1i_2}(\bm{q})=\sum_{s',i_3i_4}\tilde{V}^{H}_{ss';i_1i_2,i_3i_4}(\bm{q})\rho_{s's',i_3i_4}(\bm{q})\label{eq:Sigma_Hartree}
\end{equation}
is the Hartree self-energy and 
\begin{align}
  \tilde{V}^{H}_{ss';i_1i_2,i_3i_4}(\bm{q})\equiv & \frac{1}{2\pi l_B^2}V_{ss'}(q)F_{i_1i_2}(-\bm{q}l_B/\sqrt{2})F_{i_4i_3}(\bm{q}l_B/\sqrt{2})\nonumber\\
  =&\frac{1}{2\pi l_B^2}V_{ss'}(q)\re^{-i\Delta i_{12}(\theta_{\bm{q}}+\pi)-i\Delta i_{43}\theta_{\bm{q}}}T^{\Delta i_{12}}_{i_{<,12}}(ql_B/\sqrt{2})T^{\Delta i_{43}}_{i_{<,34}}(ql_B/\sqrt{2})\nonumber\\
  =&\frac{1}{2\pi l_B^2}V_{ss'}(q)\re^{-i(\Delta i_{12}-\Delta i_{34})\theta_{\bm{q}}}T^{\Delta i_{21}}_{i_{<,12}}(ql_B/\sqrt{2})T^{\Delta i_{43}}_{i_{<,34}}(ql_B/\sqrt{2})\nonumber\\
  =&\re^{-i(\Delta i_{12}-\Delta i_{34})\theta_{\bm{q}}}\tilde{V}^{H}_{ss';i_1i_2,i_3i_4}(q\hat{x})\label{eq:hartree-rotate}
\end{align}
is the direct Coulomb interaction projected to the LL basis.

The Fock Hamiltonian is given by
\begin{align}
  \hat{H}^{F}=&-\frac{1}{\mathcal{V}}\sum_{ss'}\sum_{i_{1-4}}\sum_{k_1k_2,\bm{q}}V_{ss'}(\bm{q})F_{i_1i_4}(-\bm{q}l_B/\sqrt{2})F_{i_2i_3}(\bm{q}l_B/\sqrt{2})\re^{iq_y(k_1-k_2)l_B^2}\nonumber\\
  &\qquad\qquad\times\rho_{ss',i_4i_2,(k_1+k_2-q_x)/2}(k_1-k_2)l^{\dagger}_{si_1k_1+q_x/2}l_{s'i_3k_2+q_x/2}
\end{align}
Make the substitution $q_x=k_1'-k_2'$, $k_1=(k_1'+k_2'+q_x')/2$, $k_2=(k_1'+k_2'-q_x')/2$, then we have
\begin{align}
  \hat{H}^{F}=&-\frac{1}{\mathcal{V}}\sum_{ss'}\sum_{i_{1-4}}\sum_{k_1'k_2',q_x'q_y}V_{ss'}[(k_1'-k_2',q_y)]F_{i_1i_4}[-(k_1'-k_2',q_y)l_B/\sqrt{2}]F_{i_2i_3}[(k_1'-k_2',q_y)l_B/\sqrt{2}]\re^{iq_yq_x'l_B^2}\nonumber\\
  &\qquad\qquad\times\rho_{ss',i_4i_2,k_2'}(q_x')l^{\dagger}_{si_1k_1'+q_x'/2}l_{s'i_3k_1'-q_x'/2}\nonumber\\
  =&-\frac{1}{\mathcal{V}}\sum_{ss'}\sum_{i_{1-4}}\sum_{k_1'k_2',q_x'q_y}V_{ss'}[(k_1'-k_2',q_y)]F_{i_1i_4}[-(k_1'-k_2',q_y)l_B/\sqrt{2}]F_{i_2i_3}[(k_1'-k_2',q_y)l_B/\sqrt{2}]\re^{iq_yq_x'l_B^2}\nonumber\\
  &\qquad\qquad\times\sum_{q_y'}\rho_{ss',i_4i_2}(\bm{q}')\re^{iq_y'k_2'l_B^2}l^{\dagger}_{si_1k_1'+q_x'/2}l_{s'i_3k_1'-q_x'/2}\nonumber\\
  =&-\sum_{ss'}\sum_{i_{1-4}}\sum_{k_1'k_2',\bm{q}'}\left\{\frac{1}{\mathcal{V}}\sum_{q_y}V_{ss'}[(k_1'-k_2',q_y)]F_{i_1i_3}[-(k_1'-k_2',q_y)l_B/\sqrt{2}]F_{i_4i_2}[(k_1'-k_2',q_y)l_B/\sqrt{2}]\re^{iq_yq_x'l_B^2}\right\}\nonumber\\
  &\qquad\qquad\times\rho_{ss',i_3i_4}(\bm{q}')\re^{iq_y'k_2'l_B^2}l^{\dagger}_{si_1k_1'+q_x'/2}l_{s'i_2k_1'-q_x'/2}
\end{align}
Define the exchange interaction projected to the LL basis as
\begin{align}
  \tilde{V}^F_{ss';i_1i_2,i_3i_4}(\bm{q}')\equiv &\frac{1}{\mathcal{V}}\sum_{\bm{q}}V_{ss'}(q)F_{i_1i_3}(-\bm{q}l_B/\sqrt{2})F_{i_4i_2}(\bm{q}l_B/\sqrt{2})\re^{iq_yq_x'l_B^2}\re^{-iq_xq_y'l_B^2}\nonumber\\
  =&\int_0^{\infty}\frac{q\rd q}{2\pi}\int_0^{2\pi}\frac{\rd \theta_{\bm{q}}}{2\pi}\;V_{ss'}(q)\re^{-i(\Delta i_{13}-\Delta i_{24})\theta_{\bm{q}}}T^{\Delta i_{31}}_{i_{<,13}}(ql_B/\sqrt{2})T^{\Delta i_{42}}_{i_{<,42}}(ql_B/\sqrt{2})\re^{iqq'l_B^2\sin(\theta_{\bm{q}}-\theta_{\bm{q}'})}\nonumber\\
  =&\re^{-i(\Delta i_{12}-\Delta i_{34})\theta_{\bm{q}'}}\tilde{V}^F_{i_1i_2,i_3i_4}[(q',0)]\label{eq:fock-rotate}
\end{align}
where 
\begin{align}
  \tilde{V}^F_{ss';i_1i_2,i_3i_4}[(q',0)]\equiv &\int_0^{\infty}\frac{q\rd q}{2\pi}\int_0^{2\pi}\frac{\rd \theta_{\bm{q}}}{2\pi}\;V_{ss'}(q)\re^{-i(\Delta i_{13}-\Delta i_{24})\theta_{\bm{q}}}T^{\Delta i_{31}}_{i_{<,13}}(ql_B/\sqrt{2})T^{\Delta i_{42}}_{i_{<,42}}(ql_B/\sqrt{2})\re^{iqq'l_B^2\sin\theta_{\bm{q}}}\nonumber\\
  =&\int_0^{\infty}\frac{q\rd q}{2\pi}\;V_{ss'}(q)T^{\Delta i_{31}}_{i_{<,13}}(ql_B/\sqrt{2})T^{\Delta i_{42}}_{i_{<,42}}(ql_B/\sqrt{2})J_{\Delta i_{12}-\Delta i_{34}}(qq'l_B^2)\nonumber\\
  =&\frac{1}{\pi l_B^2}\int_0^{\infty}\rd q\; qV_{ss'}(\sqrt{2}q/l_B)T^{\Delta i_{31}}_{i_{<,13}}(q)T^{\Delta i_{42}}_{i_{<,42}}(q)J_{\Delta i_{12}-\Delta i_{34}}(\sqrt{2}qq'l_B)\label{eq:LL-Fock-radial-integral}
\end{align}
Then 
\begin{align}
  &\left\{\frac{1}{\mathcal{V}}\sum_{q_y}V_{ss'}[(k_1'-k_2',q_y)]F_{i_1i_3}[-(k_1'-k_2',q_y)l_B/\sqrt{2}]F_{i_4i_2}[(k_1'-k_2',q_y)l_B/\sqrt{2}]\re^{iq_yq_x'l_B^2}\right\}\nonumber\\
  =&\sum_{q_y'}\tilde{V}^F_{ss';i_1i_2,i_3i_3}(\bm{q}')\re^{iq_y'(k_1'-k_2') l_B^2}
\end{align}
and the Fock Hamiltonian becomes 
\begin{align}
  \hat{H}^{F}=&-\sum_{ss'}\sum_{i_{1-4}}\sum_{k_1,\bm{q}}\tilde{V}^F_{ss';i_1i_2,i_3i_4}(\bm{q})\rho_{ss',i_3i_4}(\bm{q})\re^{iq_yk_1l_B^2}l^{\dagger}_{si_1k_1+q_x/2}l_{s'i_2k_1-q_x/2}\nonumber\\
  =&\sum_{ss',i_1i_2,k_1\bm{q}}\Sigma^F_{ss',i_1i_2}(\bm{q})\re^{iq_yk_1l_B^2}l^{\dagger}_{si_1k_1+q_x/2}l_{s'i_2k_1-q_x/2}
\end{align}
where the Fock self-energy is written as
\begin{equation}
  \Sigma^F_{ss',i_1i_2}(\bm{q})=-\sum_{i_3i_4}\tilde{V}^F_{ss',i_1i_2,i_3i_4}(\bm{q})\rho_{ss',i_3i_4}(\bm{q})\label{eq:Sigma_Fock}
\end{equation}
For simplicity, we will denote $\tilde{V}^{H/F}_{i_1i_2,i_3i_4}(\bm{q})\equiv \tilde{V}^{H/F}_{s=s';i_1i_2,i_3i_4}(\bm{q})$ and $\tilde{U}^{H/F}_{i_1i_2,i_3i_4}(\bm{q})\equiv \tilde{V}^{H/F}_{s\ne s';i_1i_2,i_3i_4}(\bm{q})$.

\subsubsection{The uniform EI case}\label{app:scf-uniform}

In the uniform EI phase, the single-particle density matrix is independent of $k_x$ and has the form \citep{shaoQuantumOscillationsExcitonic2024}
\begin{equation}
  \rho^{X}_{ss',ii',k_x}(q_x)\equiv \ave{l^{\dagger}_{s'i'k_x-q_x/2}l_{sik_x+q_x/2}}=\delta_{q_x,0}\delta_{ii'}\rho^{X}_{i,ss'}
\end{equation}
or equivalently
\begin{equation}
  \rho^{X}_{ss',ii'}(\bm{q})=\delta_{\bm{q},\bm{0}}\delta_{ii'}\rho^{X}_{i,ss'}
\end{equation}
The Hartree and Fock self-energies are therefore
\begin{align}
  \Sigma^{H}_{ee,i_1i_2}(\bm{q})=&\delta_{\bm{q},\bm{0}}\lim_{q\to 0}\sum_{i_3}[\tilde{V}^{H}_{i_1i_2,i_3i_3}(\bm{q})\rho^{X}_{i_3,ee}+\tilde{U}^{H}_{i_1i_2,i_3i_3}(\bm{q})(\rho^{X}_{i_3,hh}-1)]\nonumber\\
  =&\delta_{\bm{q},\bm{0}}\delta_{i_1,i_2}\lim_{q\to 0}[V(q)-U(q)]\frac{1}{2\pi l_B^2}\sum_{i_3}\rho^{X}_{i_3,ee}\\
  =&\delta_{\bm{q},\bm{0}}\delta_{i_1,i_2}\frac{2\pi e^2 d n_X}{\epsilon}\nonumber\\
  \Sigma^{F}_{ee,i_1i_2}(\bm{q})=&-\delta_{\bm{q},\bm{0}}\delta_{i_1i_2}\sum_{i_3}\tilde{V}^{F}_{i_1i_1,i_3i_3}(\bm{0})\rho^{X}_{i_3,ee}\\
  \Sigma^{F}_{eh,i_1i_2}(\bm{q})=&-\delta_{\bm{q},\bm{0}}\delta_{i_1i_2}\sum_{i_3}\tilde{U}^{F}_{i_1i_1,i_3i_3}(\bm{0})\rho^{X}_{i_3,eh}
\end{align}
where the exciton density is given by 
\begin{equation}
  n_X=\frac{1}{\mathcal{V}}\sum_{i,k_x}\rho^{X}_{i,ee}=\frac{1}{2\pi l_B^2}\sum_{i}\rho^{X}_{i,ee}.
\end{equation}
Thus the mean-field Hamiltonian is 
\begin{equation}
  \hat{H}_{MF}=\sum_{i,k_x}[l^{\dagger}_{eik_x},l^{\dagger}_{hik_x}]\begin{bmatrix}
    \varepsilon_{i} & \Delta_i \\
    \Delta_i^* & -\varepsilon_{i}
  \end{bmatrix}\begin{bmatrix}
    l_{eik_x}\\l_{hik_x}
  \end{bmatrix}
\end{equation}
where 
\begin{align}
  \varepsilon_i=&\hbar\omega_0\left(i+\frac{1}{2}\right)-\frac{\mu_X}{2}+\frac{2\pi e^2 dn_X}{\epsilon}-\sum_{i'}\tilde{V}^{F}_{ii,i'i'}(\bm{0})\rho^X_{i',ee}\\
  \Delta_i = &-\sum_{i'}\tilde{U}^{F}_{ii,i'i'}(\bm{0})\rho^X_{i',eh}
\end{align}
As the case without perpendicular magnetic field, we can choose the gauge such that $\Delta_i$ is real and negative.
Define the quasiparticle creation operators
\begin{align}
  l^{\dagger}_{vik_x}\equiv &\alpha_{i}l^{\dagger}_{eik_x}+\beta_il^{\dagger}_{hik_x}\\
  l^{\dagger}_{cik_x}\equiv& \beta_{i}l^{\dagger}_{eik_x}-\alpha_il^{\dagger}_{hik_x}
\end{align}
and the mean-field Hamiltonian is diagonalized as 
\begin{equation}
  \hat{H}_{MF}=\sum_{i,k_x}[l^{\dagger}_{cik_x},l^{\dagger}_{vik_x}]\begin{bmatrix}
    \xi_i & 0 \\ 0& -\xi_i
  \end{bmatrix}\begin{bmatrix}
    l_{cik_x}\\l_{vik_x}
  \end{bmatrix}
\end{equation}
where $\xi_{i}=\sqrt{\varepsilon_i^2+\Delta_i^2}$, $\alpha_i=\sqrt{(1-\varepsilon_i/\xi_i)/2}$ and $\beta_i=\sqrt{(1+\varepsilon_i/\xi_i)/2}$.
Then the single particle density matrix is recalculated as 
\begin{align}
  \rho^{X}_{i,ee}=\frac{1}{2}\left(1-\frac{\varepsilon_i}{\xi_i}\right),\;\rho^{X}_{i,eh}=-\frac{\Delta_i}{2\xi_i}
\end{align}

\subsubsection{The stripe order case}\label{app:scf-stripe}
In the stripe phase, the single-particle density matrix is taken to be independent of $k_x$, while finite-momentum components along the $x$ direction are retained:
\begin{equation}
  \rho^{X}_{ss',ii',k_x}(q_x)=\sum_{n}\delta_{q_x,nQ}\rho^{X}_{ii',ss'}(nQ)
\end{equation}
Substitute this form into Eq. \eqref{eq:dst_mat_full}, we have
\begin{equation}
  \rho^{X}_{ss',ii'}(\bm{q})=\sum_{n}\delta_{q_x,nQ}\delta_{q_y,0}\rho^{X}_{ii',ss'}(nQ)
\end{equation}
Then according to Eqs. \eqref{eq:Sigma_Hartree}\eqref{eq:Sigma_Fock}, the Hartree and Fock self energies also take the form as
\begin{equation}
  \Sigma^{H/F}_{ss',i_1i_2}(\bm{q})=\sum_{n}\delta_{q_x,nQ}\delta_{q_y,0}\Sigma^{H/F}_{ss',i_1i_2}(nQ)
\end{equation}
and Hartree/Fock Hamiltonian become
\begin{equation}
  \hat{H}^{H/F}=\sum_{ss',i_1i_2,k_x}\sum_{n}\Sigma^{H/F}_{ss',i_1i_2}(nQ)\re^{inQl_B^2 k_x}l^{\dagger}_{si_1k_x+nQ/2}l_{s'i_2k_x-nQ/2}
\end{equation}

The summation over $k_x$ is limited to $[0,L_y/l_B^2]$.
For any given $Q$, we will only consider the commensurate case with $Q=L_y/(Nl_B^2)$, or equivalently $L_y=Nl_B^2 Q$ where $N$ is a large integer.
Define a new set of basis as
\begin{align}
  \tilde{l}_{sik_x}(\tilde{n})=&\frac{1}{\sqrt{N}}\sum_{n=0}^{N-1}l_{sik_x+nQ}\re^{2\pi i\tilde{n}n/N},\quad \tilde{n}=0,1,\cdots,N-1\\
  l_{sik_x+nQ}=&\frac{1}{\sqrt{N}}\sum_{\tilde{n}=0}^{N-1}\tilde{l}_{sik_x}(\tilde{n})\re^{-2\pi i\tilde{n}n/N}
\end{align}
Then the mean field Hamiltonian can be written as 
\begin{align}
  \hat{H}_{MF}=&\sum_{ss',i_1i_2,k_x}\sum_{n}\left[h^0_{ss',i_1i_2}\delta_{n0}+\Sigma^{H}_{ss',i_1i_2}(nQ)+ \Sigma^{F}_{ss',i_1i_2}(nQ)\right]\re^{inQl_B^2 k_x}l^{\dagger}_{si_1k_x+nQ/2}l_{s'i_2k_x-nQ/2}\nonumber\\
  =&\sum_{ss',i_1i_2,k_x}\sum_{\tilde{n}=0}^{N-1}\left[h^0_{ss',i_1i_2}+\Sigma^{H}_{ss',i_1i_2}(\tilde{n})+ \Sigma^{F}_{ss',i_1i_2}(\tilde{n})\right]\tilde{l}^{\dagger}_{si_1k_x}(\tilde{n})\tilde{l}_{s'i_2k_x}(\tilde{n})
\end{align}
where 
\begin{align}
  h^0_{ss',i_1i_2}=&\delta_{i_1,i_2}\begin{bmatrix}
    \hbar\omega_e(i_1+1/2)-\mu_X/2 & 0 \\
    0 & -\hbar\omega_h(i_1+1/2)+\mu_X/2
  \end{bmatrix}_{ss'}\\
  \Sigma^{H/F}_{ss',i_1i_2}(\tilde{n})=&\sum_{n}\Sigma^{H/F}_{ss',i_1i_2}(nQ)\re^{-2\pi i\tilde{n} n/N}
\end{align}
The mean field Hamiltonian is block diagonalized in the new basis $\tilde{l}_{sik_x}(\tilde{n})$ with each block labeled by $\tilde{n}$.
For each block with 
\begin{equation}
  h^{MF}_{ss',i_1i_2}(\tilde{n})=h^0_{ss',i_1i_2}+\Sigma^{H}_{ss',i_1i_2}(\tilde{n})+ \Sigma^{F}_{ss',i_1i_2}(\tilde{n}),
\end{equation}
the Hamiltonian is a matrix with dimension $2N_{LL}\times 2N_{LL}$ where $N_{LL}$ is the number of Landau levels considered in each layer.
We can diagonalize it through
\begin{equation}
  h^{MF}(\tilde{n})\ket{\tilde{n},\lambda}=\xi_{\lambda}(\tilde{n})\ket{\tilde{n},\lambda},\quad \lambda=1,2,\cdots,2N_{LL}
\end{equation}
Then the new density matrix in the new basis is calculated as
\begin{equation}
  \tilde{\rho}^{X}_{ss',ii'}(\tilde{n})\equiv\ave{\tilde{l}^{\dagger}_{s'i'k_x}(\tilde{n})\tilde{l}_{sik_x}(\tilde{n})}=\sum_{\lambda}[(\ket{\tilde{n},\lambda}\bra{\tilde{n},\lambda})_{ss',ii'}-\delta_{ss'}\delta_{sh}]\Theta[\mu_e-\xi_{\lambda}(\tilde{n})]
\end{equation}
where $\Theta(x)$ is the Heaviside step function and $\mu_e$ is the electron chemical potential determined by the CNP condition:
\begin{equation}
  \sum_{s,i}\sum_{\tilde{n}}\tilde{\rho}^{X}_{ss,ii}(\tilde{n})=0
\end{equation}
Once $\tilde{\rho}(\tilde{n})$ is obtained, we can transform it back to the original basis as
\begin{equation}
  \rho^{X}_{ss',ii'}(nQ)=\frac{1}{N}\sum_{\tilde{n}=0}^{N-1}\tilde{\rho}^{X}_{ss',ii'}(\tilde{n})\re^{2\pi i\tilde{n} n/N}
\end{equation}
This iteration is repeated until convergence.

{
Using the self-consistent density matrix $\rho^X_{ss',ii'}(nQ)$, we can calculate real space density matrix as [Eqs. \eqref{eq:LL_operator} and \eqref{eq:LL_wave_realspace}]
\begin{align}
  \rho^{X}_{ss'}(\bm{r})\equiv &\ave{\Psi^{\dagger}_{s'}(\bm{r})\Psi_{s}(\bm{r})}=\sum_{ii'}\sum_{k_xk_x'}\ave{l^{\dagger}_{s' i' k_x'}l_{s i k_x}}\phi^*_{i'k_x'}(\bm{r})\phi_{ik_x}(\bm{r})\nonumber\\
  =&\sum_{ii'}\sum_{k_x'}\sum_{n}\rho^{X}_{ss',ii'}(nQ)\phi^*_{i'k_x'}(\bm{r})\phi_{ik_x'+nQ}(\bm{r})\nonumber\\
  =&\frac{1}{L_x l_B}\sum_{ii',k_x',n}\re^{i n Q x}\psi_{i'}(y/l_B-k_x' l_B)\psi_i(y/l_B-(k_x'+nQ)l_B)\rho^{X}_{ss',ii'}(nQ)\nonumber\\
  =&\frac{1}{l_B}\sum_{ii',n}\re^{i n Q x}\rho^{X}_{ss',ii'}(nQ)\int\frac{\rd k_x'}{2\pi}\psi_{i'}(y/l_B-k_x' l_B)\psi_i(y/l_B-(k_x'+nQ)l_B)\nonumber\\
  =&\frac{1}{2\pi l_B^2}\sum_{ii',n}\re^{i n Q x}\rho^{X}_{ss',ii'}(nQ)\int\rd k_x'\psi_{i'}(k_x')\psi_i(k_x'-nQl_B)\nonumber\\
  =&\frac{1}{2\pi l_B^2}\sum_{ii',n}\re^{i n Q x}\rho^{X}_{ss',ii'}(nQ)F_{i'i}[(nQl_B/\sqrt{2},0)]\label{eq:realspace-density-matrix}
\end{align}
}

\subsection{The TDHF equation and collective modes}\label{app:LL-TDHF}
Similar to Eq. \eqref{eq:tdhf-bilayer-compact}, the TDHF equation is now written as 
\begin{equation}
  i\hbar\tau_z\partial_t\begin{bmatrix}
    \rho^{(1)}_{cv,i_1i_2}(\bm{q})\\
    \rho^{(1)}_{vc,i_1i_2}(\bm{q})
  \end{bmatrix}=\sum_{i_3i_4}\mathcal{H}_{i_1i_2,i_3i_4}(\bm{q})\begin{bmatrix}
    \rho^{(1)}_{cv,i_3i_4}(\bm{q})\\\rho^{(1)}_{vc,i_3i_4}(\bm{q})
  \end{bmatrix}+\frac{1}{\mathcal{V}}\begin{bmatrix}
    o_{cv,i_1i_2}(-\bm{q}) \\ o_{vc,i_1i_2}(-\bm{q})
  \end{bmatrix}f(t,\bm{q})\label{eq:tdhf-LL}
\end{equation}
where 
\begin{align}
  \mathcal{H}_{i_1i_2,i_3i_4}(\bm{q})=&\begin{bmatrix}
    \mathcal{E}_{i_1i_2,i_3i_4}(\bm{q}) & \Gamma_{i_1i_2,i_3i_4}(\bm{q}) \\
    \Gamma_{i_3i_4,i_1i_2}^*(\bm{q}) & \mathcal{E}^*_{i_3i_4,i_1i_2}(\bm{q})
  \end{bmatrix}\\
  \mathcal{E}_{i_1i_2,i_3i_4}(\bm{q})=&\delta_{i_1,i_3}\delta_{i_2,i_4}(\xi_{i_1}+\xi_{i_2})+(\beta_{i_1}\alpha_{i_2}\beta_{i_3}\alpha_{i_4}+\alpha_{i_1}\beta_{i_2}\alpha_{i_3}\beta_{i_4})[\tilde{V}^{H}_{i_1i_2,i_3i_4}(\bm{q})-\tilde{V}^{F}_{i_1i_2,i_3i_4}(\bm{q})]\nonumber\\
  &-(\beta_{i_1}\alpha_{i_2}\alpha_{i_3}\beta_{i_4}+\alpha_{i_1}\beta_{i_2}\beta_{i_3}\alpha_{i_4})\tilde{U}^{H}_{i_1i_2,i_3i_4}(\bm{q})-(\beta_{i_1}\beta_{i_2}\beta_{i_3}\beta_{i_4}+\alpha_{i_1}\alpha_{i_2}\alpha_{i_3}\alpha_{i_4})\tilde{U}^F_{i_1i_2,i_3i_4}(\bm{q})\\
  \Gamma_{i_1i_2,i_3i_4}(\bm{q})=&(\beta_{i_1}\alpha_{i_2}\alpha_{i_3}\beta_{i_4}+\alpha_{i_1}\beta_{i_2}\beta_{i_3}\alpha_{i_4})[\tilde{V}^{H}_{i_1i_2,i_3i_4}(\bm{q})-\tilde{V}^{F}_{i_1i_2,i_3i_4}(\bm{q})]\nonumber\\
  &-(\beta_{i_1}\alpha_{i_2}\beta_{i_3}\alpha_{i_4}+\alpha_{i_1}\beta_{i_2}\alpha_{i_3}\beta_{i_4})\tilde{U}^{H}_{i_1i_2,i_3i_4}(\bm{q})-(\beta_{i_1}\beta_{i_2}\alpha_{i_3}\alpha_{i_4}+\alpha_{i_1}\alpha_{i_2}\beta_{i_3}\beta_{i_4})\tilde{U}^F_{i_1i_2,i_3i_4}(\bm{q})
\end{align}
Because the interaction matrix elements satisfy Eqs.~\eqref{eq:hartree-rotate} and \eqref{eq:fock-rotate}, the dynamical matrix $\mathcal{H}_{i_1i_2,i_3i_4}(\bm{q})$ satisfies
\begin{equation}
  \mathcal{H}_{i_1i_2,i_3i_4}(\bm{q})=\re^{-i(\Delta i_{12}-\Delta i_{34})\theta_{\bm{q}}}\mathcal{H}_{i_1i_2,i_3i_4}(q\hat{x})
\end{equation}
Assume the generalized eigenvalue equation for $\theta_{\bm{q}}=0$ is solved as 
\begin{equation}
  \sum_{i_3i_4}\mathcal{H}_{i_1i_2,i_3i_4}(q\hat{x})\Phi_{n,i_3i_4}(q\hat{x})=\tau_z\hbar\omega_n(q)\Phi_{n,i_1i_2}(q\hat{x})
\end{equation}
Substitution shows that $\Phi_{n,i_1i_2}(\bm{q})=\re^{-i\Delta i_{12}\theta_{\bm{q}}}\Phi_{n,i_1i_2}(q\hat{x})$ is the generalized eigenfunction of $\mathcal{H}_{i_1i_2,i_3i_4}(\bm{q})$ with eigenvalue $\hbar\omega_n(q)$:
\begin{align}
  \sum_{i_3i_4}\mathcal{H}_{i_1i_2,i_3i_4}(\bm{q})\Phi_{n,i_3i_4}(\bm{q})=\re^{-i\Delta i_{12}\theta_{\bm{q}}}\sum_{i_3i_4}\mathcal{H}_{i_1i_2,i_3i_4}(q\hat{x})\Phi_{n,i_3i_4}(q\hat{x})=\tau_z\hbar\omega_n(q)\Phi_{n,i_1i_2}(\bm{q})
\end{align}

\subsection{Charge- and exciton-channel density and current operators in the LL basis}\label{app:LL-current-operators}

In the presence of the static magnetic field, the gauge-field fluctuations $A^{\sigma}_{\mu}(t,\bm{r})$ produce additional terms in the layer paramagnetic charge-current operators:
\begin{subequations}
  \begin{align}
    \hat{\bm{j}}_{e,p}(\bm{r})=&-\frac{e}{2m_e}\Psi_e^{\dagger}(\bm{r})(\bm{p}+e\bm{A}^0)\Psi_{e}(\bm{r})+\mathrm{h.c.}\\
    \hat{\bm{j}}_{h,p}(\bm{r})=&\frac{e}{2m_h}\Psi_h^{\dagger}(\bm{r})(\bm{p}+e\bm{A}^0)\Psi_{h}(\bm{r})+\mathrm{h.c.}
  \end{align}
\end{subequations}
Using the relation 
\begin{equation}
  p_x-eBy=-\frac{\hbar}{\sqrt{2}l_B}(a+a^{\dagger}),\qquad p_y=\frac{\hbar}{i\sqrt{2}l_B}(a-a^{\dagger})
\end{equation}
The layer charge-current operators can be rewritten in the LL basis.
For example,
\begin{align}
  \hat{j}_{e,p,1}(\bm{r})=&-e\sum_{ii',k_xk_x'}\inp{ik_x}{\bm{r}}\bra{\bm{r}}\frac{p_x-eBy}{2m_e}\ket{i'k_x'}l^{\dagger}_{eik_x}l_{ei'k_x'}+\mathrm{h.c.}\nonumber\\
  =&-e\sum_{ii',k_xk_x'}\inp{ik_x}{\bm{r}}\bra{\bm{r}}-\frac{\hbar}{2\sqrt{2}m_el_B}(a+a^{\dagger})\ket{i'k_x'}l^{\dagger}_{eik_x}l_{ei'k_x'}+\mathrm{h.c.}\nonumber\\
  =&\frac{e\hbar}{2\sqrt{2}m_el_B}\sum_{ii',k_xk_x'}\inp{ik_x}{\bm{r}}\bra{\bm{r}}(a+a^{\dagger})\ket{i'k_x'}l^{\dagger}_{eik_x}l_{ei'k_x'}+\mathrm{h.c.}
\end{align}
The Fourier transformation is 
\begin{align}
  \hat{j}_{e,p,1}(\bm{q})=&\int\rd\bm{r}\;\re^{-i\bm{q}\cdot\bm{r}}\hat{j}_{e,p,1}(\bm{r})\nonumber\\
  =&\frac{e\hbar}{2\sqrt{2}m_el_B}\sum_{ii',k_xk_x'}\int\rd\bm{r}\;\re^{-i\bm{q}\cdot\bm{r}}\inp{ik_x}{\bm{r}}\bra{\bm{r}}(a+a^{\dagger})\ket{i'k_x'}l^{\dagger}_{eik_x}l_{ei'k_x'}+(\mathrm{h.c.},\bm{q}\to -\bm{q})\nonumber\\
  =&\frac{e\hbar}{2\sqrt{2}m_el_B}\sum_{ii',k_xk_x'}\bra{ik_x}\re^{-i\bm{q}\cdot\bm{r}}(a+a^{\dagger})\ket{i'k_x'}l^{\dagger}_{eik_x}l_{ei'k_x'}+(\mathrm{h.c.},\bm{q}\to -\bm{q})\nonumber\\
  =&\frac{e\hbar}{2\sqrt{2}m_el_B}\sum_{ii',k_xk_x'}\bra{ik_x}\re^{-i\bm{q}\cdot\bm{r}}(\sqrt{i'+1}\ket{i'+1k_x'}+\sqrt{i'}\ket{i'-1k_x'})l^{\dagger}_{eik_x}l_{ei'k_x'}+(\mathrm{h.c.},\bm{q}\to -\bm{q})\nonumber\\
  =&\frac{e\hbar}{2\sqrt{2}m_el_B}\sum_{ii'k_x}\re^{-iq_y k_x l_B^2}\left[\sqrt{i'+1}F_{ii'+1}(\bm{q}l_B/\sqrt{2})+\sqrt{i'}F_{ii'-1}(\bm{q}l_B/\sqrt{2}) \right]l^{\dagger}_{eik_x-q_x/2}l_{ei'k_x+q_x/2}\nonumber\\
  &+\frac{e\hbar}{2\sqrt{2}m_el_B}\sum_{ii'k_x}\re^{-iq_y k_x l_B^2}\left[\sqrt{i+1}F_{i+1i'}(\bm{q}l_B/\sqrt{2})+\sqrt{i}F_{i-1i'}(\bm{q}l_B/\sqrt{2}) \right]l^{\dagger}_{eik_x-q_x/2}l_{ei'k_x+q_x/2}
\end{align}
Similarly, all layer charge-density and paramagnetic charge-current operators in the LL basis can be written as
\begin{equation}
  \hat{j}_{s,p\mu=0-2}(\bm{q})=\sum_{ii'}\tilde{\gamma}^s_{\mu,ii'}(\bm{q})\sum_{k_x}\re^{-iq_y k_x l_B^2}l^{\dagger}_{sik_x-q_x/2}l_{si'k_x+q_x/2}
\end{equation}
where the bare vertices $\tilde{\gamma}^s_{\mu,ii'}(\bm{q})$ are
\begin{subequations}
  \begin{align}
    \tilde{\gamma}^{e}_{\mu=0,ii'}(\bm{q})=&eF_{ii'}(\bm{q}l_B/\sqrt{2})\\
    \tilde{\gamma}^{e}_{\mu=1,ii'}(\bm{q})=&\frac{e\hbar }{2\sqrt{2}m_el_B}\left[\sqrt{i'+1}F_{ii'+1}(\bm{q}l_B/\sqrt{2})+\sqrt{i'}F_{ii'-1}(\bm{q}l_B/\sqrt{2})+ \sqrt{i+1}F_{i+1i'}(\bm{q}l_B/\sqrt{2})+\sqrt{i}F_{i-1i'}(\bm{q}l_B/\sqrt{2})\right]\\
    \tilde{\gamma}^e_{\mu=2,ii'}(\bm{q})=&\frac{-ie\hbar }{2\sqrt{2}m_el_B}\left[\sqrt{i'+1}F_{ii'+1}(\bm{q}l_B/\sqrt{2})-\sqrt{i'}F_{ii'-1}(\bm{q}l_B/\sqrt{2})-\sqrt{i+1}F_{i+1i'}(\bm{q}l_B/\sqrt{2})+\sqrt{i}F_{i-1i'}(\bm{q}l_B/\sqrt{2})\right]\\
    \tilde{\gamma}^{h}_{\mu=0,ii'}(\bm{q})=&eF_{ii'}(\bm{q}l_B/\sqrt{2})\\
    \tilde{\gamma}^{h}_{\mu=1,ii'}(\bm{q})=&\frac{-e\hbar}{2\sqrt{2}m_hl_B}\left[\sqrt{i'+1}F_{ii'+1}(\bm{q}l_B/\sqrt{2})+\sqrt{i'}F_{ii'-1}(\bm{q}l_B/\sqrt{2})+\sqrt{i+1}F_{i+1i'}(\bm{q}l_B/\sqrt{2})+\sqrt{i}F_{i-1i'}(\bm{q}l_B/\sqrt{2})\right]\\
    \tilde{\gamma}^h_{\mu=2,ii'}(\bm{q})=&\frac{ie\hbar }{2\sqrt{2}m_hl_B}\left[\sqrt{i'+1}F_{ii'+1}(\bm{q}l_B/\sqrt{2})-\sqrt{i'}F_{ii'-1}(\bm{q}l_B/\sqrt{2})-\sqrt{i+1}F_{i+1i'}(\bm{q}l_B/\sqrt{2})+\sqrt{i}F_{i-1i'}(\bm{q}l_B/\sqrt{2})\right]
  \end{align}\label{eq:bare-vertex-LL}
\end{subequations}

When the system has additional electron--hole symmetry such that $m_e=m_h=2m$, we have $\tilde{\gamma}^{e}_{\mu=0,ii'}=\tilde{\gamma}^{h}_{\mu=0,ii'}$ and $\tilde{\gamma}^{e}_{\mu\ne 0,ii'}=-\tilde{\gamma}^{h}_{\mu\ne 0,ii'}$.
Then the paramagnetic charge- and exciton-channel four-current operators can be written as
\begin{align}
    \hat{j}^{\sigma}_{p\mu}(\bm{q})=&\sum_{k_x}\re^{-iq_y k_x l_B^2}\sum_{ii'}L^{\dagger}_{ik_x-q_x/2}\tilde{\gamma}^{\sigma}_{\mu,ii'}(\bm{q})L_{i'k_x+q_x/2}\label{eq:current-LL}
  \end{align}
where 
\begin{equation}
  \tilde{\gamma}^{+}_{\mu=0,ii'}(\bm{q})=\tilde{\gamma}^e_{\mu=0,ii'}(\bm{q})\sigma_0,\;
    \tilde{\gamma}^{+}_{\mu\ne 0,ii'}(\bm{q})=\tilde{\gamma}^e_{\mu\ne 0,ii'}(\bm{q})\sigma_z,\;
    \tilde{\gamma}^{-}_{\mu=0,ii'}(\bm{q})=\frac{1}{2}\tilde{\gamma}^e_{\mu=0,ii'}(\bm{q})\sigma_z,\;
    \tilde{\gamma}^{-}_{\mu\ne 0,ii'}(\bm{q})=\frac{1}{2}\tilde{\gamma}^e_{\mu\ne 0,ii'}(\bm{q})\sigma_0\label{eq:bare-vertex-LL-2}
\end{equation}

\subsection{The electromagnetic response functions}\label{app:LL-response-functions}

Similar to Eq. \eqref{eq:density-matrix-first-order}, solution to Eq. \eqref{eq:tdhf-LL} in frequency domain can be expressed by the generalized eigenfunctions $\Phi_{n,i_1i_2}(\bm{q})$ as 
\begin{equation}
  \begin{bmatrix}
    \rho^{(1)}_{cv,i_1i_2}(\omega,\bm{q})\\\rho^{(1)}_{vc,i_1i_2}(\omega,\bm{q})
  \end{bmatrix}=\frac{1}{2\pi l_B^2}\sum_{n}\frac{\omega_n(q)\Phi_{n,i_1i_2}(\bm{q})\mathrm{O}_{n}(\bm{q})}{\omega^{+}-\omega_n(q)}f(\omega,\bm{q})
\end{equation}
and the correlation function is 
\begin{equation}
  C_{\hat{O}'\hat{O}}(\omega,\bm{q})=\frac{1}{2\pi l_B^2}\sum_{n}\frac{\omega_n(q)[\mathrm{O}'_n(\bm{q})]^*\mathrm{O}_n(\bm{q})}{\omega^+-\omega_n(q)}.
\end{equation}
The vertex overlap between the collective-mode wavefunction and the bare vertex of the operator $\hat{O}$ is defined as
\begin{equation}
  \mathrm{O}_n(\bm{q})=\sum_{i_1i_2}\Phi^{\dagger}_{n,i_1i_2}(\bm{q})\begin{bmatrix}
    o_{cv,i_1i_2}(-\bm{q})\\ o_{vc,i_1i_2}(-\bm{q})
  \end{bmatrix}
\end{equation}
For the charge- and exciton-channel four-current operators defined by Eq. \eqref{eq:current-LL}, we have
\begin{align}
  \mathrm{J}^{\sigma}_{\mu,n}(\bm{q})=&\sum_{i_1i_2}\Phi^{\dagger}_{n,i_1i_2}(\bm{q})\begin{bmatrix}
    \tilde{\gamma}^{\sigma}_{\mu,cv,i_1i_2}(-\bm{q})\\ \tilde{\gamma}^{\sigma}_{\mu,vc,i_1i_2}(-\bm{q})
  \end{bmatrix}\\
  \tilde{\gamma}^{\sigma}_{\mu,cv,i_1i_2}(-\bm{q})=&[\beta_{i_1},-\alpha_{i_1}]\tilde{\gamma}^{\sigma}_{\mu,i_1i_2}(-\bm{q})\begin{bmatrix}
    \alpha_{i_2}\\\beta_{i_2}
  \end{bmatrix}\\
\tilde{\gamma}^{\sigma}_{\mu,vc,i_1i_2}(-\bm{q})=&[\alpha_{i_1},\beta_{i_1}]\tilde{\gamma}^{\sigma}_{\mu,i_1i_2}(-\bm{q})\begin{bmatrix}
    \beta_{i_2}\\-\alpha_{i_2}
  \end{bmatrix}
\end{align}
According to Eq. \eqref{eq:bare-vertex-LL} and  \eqref{eq:bare-vertex-LL-2}, the specific expressions are 
\begin{align}
  \tilde{\gamma}^{+}_{\mu=0,cv,i_1i_2}(-\bm{q})&=-\tilde{\gamma}^{+}_{\mu=0,vc,i_1i_2}(-\bm{q})=(\beta_{i_1}\alpha_{i_2}-\alpha_{i_1}\beta_{i_2})\tilde{\gamma}^{e}_{\mu=0,i_1i_2}(-\bm{q})\\
  \tilde{\gamma}^{+}_{\mu\ne 0,cv,i_1i_2}(-\bm{q})&=\tilde{\gamma}^{+}_{\mu \ne 0,vc,i_1i_2}(-\bm{q})=(\beta_{i_1}\alpha_{i_2}+\alpha_{i_1}\beta_{i_2})\tilde{\gamma}^{e}_{\mu\ne0,i_1i_2}(-\bm{q})\\
  \tilde{\gamma}^{-}_{\mu=0,cv,i_1i_2}(-\bm{q})&=\tilde{\gamma}^{+}_{\mu=0,vc,i_1i_2}(-\bm{q})=\frac{1}{2}(\beta_{i_1}\alpha_{i_2}+\alpha_{i_1}\beta_{i_2})\tilde{\gamma}^{e}_{\mu=0,i_1i_2}(-\bm{q})\\
  \tilde{\gamma}^{+}_{\mu\ne 0,cv,i_1i_2}(-\bm{q})&=-\tilde{\gamma}^{+}_{\mu \ne 0,vc,i_1i_2}(-\bm{q})=\frac{1}{2}(\beta_{i_1}\alpha_{i_2}-\alpha_{i_1}\beta_{i_2})\tilde{\gamma}^{e}_{\mu\ne0,i_1i_2}(-\bm{q})
\end{align}
Using $F_{i_1i_2}(\bm{q})=\re^{-i\Delta i_{12}\theta_{\bm{q}}}T^{\Delta i_{12}}_{i_{<,12}}(q)$, we could rewrite Eq. \eqref{eq:bare-vertex-LL} as 
\begin{align}
  \tilde{\gamma}^e_{0,i_1i_2}(-\bm{q})=&-e\re^{-i\Delta i_{12}\theta_{\bm{q}}}T^{\Delta i_{21}}_{i_{<,12}}(ql_B/\sqrt{2})\\
  \tilde{\gamma}^e_{1,i_1i_2}(-\bm{q})-i\tilde{\gamma}^e_{2,i_1i_2}(-\bm{q})=&\frac{e\hbar}{2\sqrt{2}ml_B}\re^{-i(\Delta i_{12}+1)\theta_{\bm{q}}}[\sqrt{i_2}T^{\Delta i_{21}-1}_{\min(i_1,i_2-1)}(ql_B/\sqrt{2})+\sqrt{i_1+1}T^{\Delta i_{21}-1}_{\min(i_1+1,i_2)}(ql_B/\sqrt{2})]\\
  \tilde{\gamma}^e_{1,i_1i_2}(-\bm{q})+i\tilde{\gamma}^e_{2,i_1i_2}(-\bm{q})=&\frac{e\hbar}{2\sqrt{2}ml_B}\re^{-i(\Delta i_{12}-1)\theta_{\bm{q}}}[\sqrt{i_2+1}T^{\Delta i_{21}+1}_{\min(i_1,i_2+1)}(ql_B/\sqrt{2})+\sqrt{i_1}T^{\Delta i_{21}+1}_{\min(i_1-1,i_2)}(ql_B/\sqrt{2})]
\end{align}
where $m=m_e/2=m_h/2$ is the reduced mass.
Thus,
\begin{equation}
  \mathrm{J}^{\sigma}_{0,n}(\bm{q})=\mathrm{J}^{\sigma}_{0,n}(q\hat{x}),\;\mathrm{J}^{\sigma}_{1,n}(\bm{q})\pm i\mathrm{J}^{\sigma}_{2,n}(\bm{q})=\re^{\pm i\theta_{\bm{q}}}\{\mathrm{J}^{\sigma}_{1,n}(q\hat{x})\pm i\mathrm{J}^{\sigma}_{2,n}(q\hat{x})\}
\end{equation}
where 
\begin{align}
  \mathrm{J}^{+}_{0,n}(q\hat{x})=&-e\sum_{i_1i_2}\Phi^{\dagger}_{n,i_1i_2}(q\hat{x})\begin{bmatrix}
    1\\-1
  \end{bmatrix}(\beta_{i_1}\alpha_{i_2}-\alpha_{i_1}\beta_{i_2})T^{\Delta i_{21}}_{i_{<,12}}(ql_B/\sqrt{2})\\
  \mathrm{J}^{+}_{1,n}(q\hat{x})-i\mathrm{J}^{+}_{2,n}(q\hat{x})=&\frac{e\hbar}{2\sqrt{2}ml_B}\sum_{i_1i_2}\Phi^{\dagger}_{n,i_1i_2}(q\hat{x})\begin{bmatrix}
    1\\1
  \end{bmatrix}(\beta_{i_1}\alpha_{i_2}+\alpha_{i_1}\beta_{i_2})\nonumber\\
  &\times[\sqrt{i_2}T^{\Delta i_{21}-1}_{\min(i_1,i_2-1)}(ql_B/\sqrt{2})+\sqrt{i_1+1}T^{\Delta i_{21}-1}_{\min(i_1+1,i_2)}(ql_B/\sqrt{2})]\\
  \mathrm{J}^{+}_{1,n}(q\hat{x})+i\mathrm{J}^{+}_{2,n}(q\hat{x})=&\frac{e\hbar}{2\sqrt{2}ml_B}\sum_{i_1i_2}\Phi^{\dagger}_{n,i_1i_2}(q\hat{x})\begin{bmatrix}
    1\\1
  \end{bmatrix}(\beta_{i_1}\alpha_{i_2}+\alpha_{i_1}\beta_{i_2})\nonumber\\
  &\times [\sqrt{i_2+1}T^{\Delta i_{21}+1}_{\min(i_1,i_2+1)}(ql_B/\sqrt{2})+\sqrt{i_1}T^{\Delta i_{21}+1}_{\min(i_1-1,i_2)}(ql_B/\sqrt{2})]
\end{align}
and
\begin{align}
  \mathrm{J}^{-}_{0,n}(q\hat{x})=&-\frac{e}{2}\sum_{i_1i_2}\Phi^{\dagger}_{n,i_1i_2}(q\hat{x})\begin{bmatrix}
    1\\1
  \end{bmatrix}(\beta_{i_1}\alpha_{i_2}+\alpha_{i_1}\beta_{i_2})T^{\Delta i_{21}}_{i_{<,12}}(ql_B/\sqrt{2})\\
  \mathrm{J}^{-}_{1,n}(q\hat{x})-i\mathrm{J}^{-}_{2,n}(q\hat{x})=&\frac{e\hbar}{4\sqrt{2}ml_B}\sum_{i_1i_2}\Phi^{\dagger}_{n,i_1i_2}(q\hat{x})\begin{bmatrix}
    1\\-1
  \end{bmatrix}(\beta_{i_1}\alpha_{i_2}-\alpha_{i_1}\beta_{i_2})\nonumber\\
  &\times[\sqrt{i_2}T^{\Delta i_{21}-1}_{\min(i_1,i_2-1)}(ql_B/\sqrt{2})+\sqrt{i_1+1}T^{\Delta i_{21}-1}_{\min(i_1+1,i_2)}(ql_B/\sqrt{2})]\\
  \mathrm{J}^{-}_{1,n}(q\hat{x})+i\mathrm{J}^{-}_{2,n}(q\hat{x})=&\frac{e\hbar}{4\sqrt{2}ml_B}\sum_{i_1i_2}\Phi^{\dagger}_{n,i_1i_2}(q\hat{x})\begin{bmatrix}
    1\\-1
  \end{bmatrix}(\beta_{i_1}\alpha_{i_2}-\alpha_{i_1}\beta_{i_2})\nonumber\\
  &\times [\sqrt{i_2+1}T^{\Delta i_{21}+1}_{\min(i_1,i_2+1)}(ql_B/\sqrt{2})+\sqrt{i_1}T^{\Delta i_{21}+1}_{\min(i_1-1,i_2)}(ql_B/\sqrt{2})]
\end{align}
The expressions above imply
\begin{equation}
  \mathrm{Im}\{\mathrm{J}^{\sigma}_{1,n}(q\hat{x})\pm i\mathrm{J}^{\sigma}_{2,n}(q\hat{x})\}=0\implies \mathrm{Im}\mathrm{J}^{\sigma}_{1,n}(q\hat{x})=0,\;\mathrm{Re}\mathrm{J}^{\sigma}_{2,n}(q\hat{x})=0.
\end{equation}
Thus we have 
\begin{align}
  \mathrm{J}^{\sigma}_{1,n}(\bm{q})=&\frac{\re^{i\theta_{\bm{q}}}}{2}\{\mathrm{J}^{\sigma}_{1,n}(q\hat{x})+ i\mathrm{J}^{\sigma}_{2,n}(q\hat{x})\}+\frac{\re^{-i\theta_{\bm{q}}}}{2}\{\mathrm{J}^{\sigma}_{1,n}(q\hat{x})- i\mathrm{J}^{\sigma}_{2,n}(q\hat{x})\}\nonumber\\
  =&\cos\theta_{\bm{q}}\mathrm{J}^{\sigma}_{1,n}(q\hat{x})-\sin\theta_{\bm{q}}\mathrm{J}^{\sigma}_{2,n}(q\hat{x})\nonumber\\
  =&\cos\theta_{\bm{q}}\mathrm{Re}\mathrm{J}^{\sigma}_{1,n}(q\hat{x})-i\sin\theta_{\bm{q}}\mathrm{Im}\mathrm{J}^{\sigma}_{2,n}(q\hat{x})\\
  \mathrm{J}^{\sigma}_{2,n}(\bm{q})=&\frac{\re^{i\theta_{\bm{q}}}}{2i}\{\mathrm{J}^{\sigma}_{1,n}(q\hat{x})+ i\mathrm{J}^{\sigma}_{2,n}(q\hat{x})\}-\frac{\re^{-i\theta_{\bm{q}}}}{2i}\{\mathrm{J}^{\sigma}_{1,n}(q\hat{x})- i\mathrm{J}^{\sigma}_{2,n}(q\hat{x})\}\nonumber\\
  =&\sin\theta_{\bm{q}}\mathrm{J}^{\sigma}_{1,n}(q\hat{x})+\cos\theta_{\bm{q}}\mathrm{J}^{\sigma}_{2,n}(q\hat{x})\nonumber\\
  =&\sin\theta_{\bm{q}}\mathrm{Re}\mathrm{J}^{\sigma}_{1,n}(q\hat{x})+i\cos\theta_{\bm{q}}\mathrm{Im}\mathrm{J}^{\sigma}_{2,n}(q\hat{x})
\end{align}

Due to the Ward identity, we only need to consider the spatial components of the correlation function.
We first consider the correlation functions which is diagonal with respect to $\sigma$ and $\sigma'$, i.e. the charge-charge and exciton-exciton correlation functions.
{Due to electron--hole symmetry of the model system, there are no Hall responses in the pure charge or exciton channel, and the correlation functions are symmetric under interchange of their spatial indices, i.e. $C_{\hat{j}^{\sigma}_{p\mu}\hat{j}^{\sigma}_{p\nu}}(\omega,\bm{q})=C_{\hat{j}^{\sigma}_{p\nu}\hat{j}^{\sigma}_{p\mu}}(\omega,\bm{q})$.}
The specific expressions are
\begin{align}
  C_{\hat{j}^{\sigma}_{pa}\hat{j}^{\sigma}_{pa}}(\omega,\bm{q})=&\frac{1}{2\pi l_B^2}\sum_{n}\frac{\omega_n(q)|\mathrm{J}^{\sigma}_{a,n}(\bm{q})|^2}{\omega^+-\omega_n(q)}\nonumber\\
  =&\frac{\cos^2\theta_{\bm{q}}}{2\pi l_B^2}\sum_{n}\frac{\omega_n(q)|\mathrm{J}^{\sigma}_{a,n}(q\hat{x})|^2}{\omega^+-\omega_n(q)}+\frac{\sin^2\theta_{\bm{q}}}{2\pi l_B^2}\sum_{n}\frac{\omega_n(q)|\mathrm{J}^{\sigma}_{\bar{a},n}(q\hat{x})|^2}{\omega^+-\omega_n(q)}      \nonumber\\
  =&\cos^2\theta_{\bm{q}}C_{\hat{j}^{\sigma}_{pa}\hat{j}^{\sigma}_{pa}}(\omega,q\hat{x})+\sin^2\theta_{\bm{q}}C_{\hat{j}^{\sigma}_{p\bar{a}}\hat{j}^{\sigma}_{p\bar{a}}}(\omega,q\hat{x})\\
  C_{\hat{j}^{\sigma}_{p1}\hat{j}^{\sigma}_{p2}}(\omega,\bm{q})=&\frac{1}{4\pi l_B^2}\sum_{n}\frac{\omega_n(q)[\mathrm{J}^{\sigma}_{1,n}(\bm{q})]^*\mathrm{J}^{\sigma}_{2,n}(\bm{q})+c.c.}{\omega^+-\omega_n(q)}\nonumber\\
  =&\frac{\cos\theta_{\bm{q}}\sin\theta_{\bm{q}}}{2\pi l_B^2}\sum_{n}\frac{\omega_n(q)\{|\mathrm{J}^{\sigma}_{1,n}(q\hat{x})|^2-|\mathrm{J}^{\sigma}_{2,n}(q\hat{x})|^2\}}{\omega^{+}-\omega_n(q)}\nonumber\\
  =&\cos\theta_{\bm{q}}\sin\theta_{\bm{q}}\left\{C_{\hat{j}^{\sigma}_{p1}\hat{j}^{\sigma}_{p1}}(\omega,q\hat{x})-C_{\hat{j}^{\sigma}_{p2}\hat{j}^{\sigma}_{p2}}(\omega,q\hat{x})\right\}
\end{align}
where $\bar{a}$ means the spatial index other than $a$.
It is clear to see that for each $\sigma$ there are only two independent components, i.e. $C_{\hat{j}^{\sigma}_{p1}\hat{j}^{\sigma}_{p1}}(\omega,q\hat{x})$ and $C_{\hat{j}^{\sigma}_{p2}\hat{j}^{\sigma}_{p2}}(\omega,q\hat{x})$, which is just the result of the rotational symmetry.

Next, we consider the correlation functions which is off-diagonal with respect to $\sigma$ and $\sigma'$, i.e. the charge-exciton correlation functions.
Due to electron--hole symmetry, these correlation functions are antisymmetric under interchange of the response and perturbation channels, i.e. $C_{\hat{j}^{\sigma}_{p\mu}\hat{j}^{-\sigma}_{p\nu}}(\omega,\bm{q})=-C_{\hat{j}^{-\sigma}_{p\nu}\hat{j}^{\sigma}_{p\mu}}(\omega,\bm{q})$.
The specific expressions are
\begin{align}
  C_{\hat{j}^{\sigma}_{p1}\hat{j}^{-\sigma}_{p2}}(\omega,\bm{q})=&\frac{1}{4\pi l_B^2}\sum_{n}\frac{\omega_n(q)[\mathrm{J}^{\sigma}_{1,n}(\bm{q})]^*\mathrm{J}^{-\sigma}_{2,n}(\bm{q})-c.c.}{\omega^+-\omega_n(q)}\nonumber\\
  =&\frac{i\cos^2\theta_{\bm{q}}}{2\pi l_B^2}\sum_{n}\frac{\omega_n(q)\mathrm{Re}\mathrm{J}^{\sigma}_{1,n}(q\hat{x})\mathrm{Im}\mathrm{J}^{-\sigma}_{2,n}(q\hat{x})}{\omega^{+}-\omega_n(q)}+\frac{i\sin^2\theta_{\bm{q}}}{2\pi l_B^2}\sum_{n}\frac{\omega_n(q)\mathrm{Re}\mathrm{J}^{-\sigma}_{1,n}(q\hat{x})\mathrm{Im}\mathrm{J}^{\sigma}_{2,n}(q\hat{x})}{\omega^{+}-\omega_n(q)}\nonumber\\
  =&\cos^2\theta_{\bm{q}}C_{\hat{j}^{\sigma}_{p1}\hat{j}^{-\sigma}_{p2}}(\omega,q\hat{x})+\sin^2\theta_{\bm{q}}C_{\hat{j}^{-\sigma}_{p1}\hat{j}^{\sigma}_{p2}}(\omega,q\hat{x})\\
  C_{\hat{j}^{\sigma}_{p1}\hat{j}^{-\sigma}_{p1}}(\omega,\bm{q})=&\frac{1}{4\pi l_B^2}\sum_{n}\frac{\omega_n(q)[\mathrm{J}^{\sigma}_{1,n}(\bm{q})]^*\mathrm{J}^{-\sigma}_{1,n}(\bm{q})-c.c}{\omega^+-\omega_n(q)}\nonumber\\
  =&\frac{i\sin\theta_{\bm{q}}\cos\theta_{\bm{q}}}{2\pi l_B^2}\sum_{n}\frac{\omega_n(q)\{\mathrm{Re}\mathrm{J}_{1,n}^{-\sigma}(q\hat{x})\mathrm{Im}\mathrm{J}^{\sigma}_{2,n}(q\hat{x})-\mathrm{Re}\mathrm{J}_{1,n}^{\sigma}(q\hat{x})\mathrm{Im}\mathrm{J}^{-\sigma}_{2,n}(q\hat{x})\}}{\omega^{+}-\omega_n(q)}\nonumber\\
  =&\cos\theta_{\bm{q}}\sin\theta_{\bm{q}}\left\{C_{\hat{j}^{-\sigma}_{p1}\hat{j}^{\sigma}_{p2}}(\omega,q\hat{x})-C_{\hat{j}^{\sigma}_{p1}\hat{j}^{-\sigma}_{p2}}(\omega,q\hat{x})\right\}\\
  C_{\hat{j}^{\sigma}_{p2}\hat{j}^{-\sigma}_{p2}}(\omega,\bm{q})=&\frac{1}{4\pi l_B^2}\sum_{n}\frac{\omega_n(q)[\mathrm{J}^{\sigma}_{2,n}(\bm{q})]^*\mathrm{J}^{-\sigma}_{2,n}(\bm{q})-c.c}{\omega^+-\omega_n(q)}\nonumber\\
  =&\frac{i\sin\theta_{\bm{q}}\cos\theta_{\bm{q}}}{2\pi l_B^2}\sum_{n}\frac{\omega_n(q)\{\mathrm{Re}\mathrm{J}_{1,n}^{\sigma}(q\hat{x})\mathrm{Im}\mathrm{J}^{-\sigma}_{2,n}(q\hat{x})-\mathrm{Re}\mathrm{J}_{1,n}^{-\sigma}(q\hat{x})\mathrm{Im}\mathrm{J}^{\sigma}_{2,n}(q\hat{x})\}}{\omega^{+}-\omega_n(q)}\nonumber\\
  =&\cos\theta_{\bm{q}}\sin\theta_{\bm{q}}\left\{C_{\hat{j}^{\sigma}_{p1}\hat{j}^{-\sigma}_{p2}}(\omega,q\hat{x})-C_{\hat{j}^{-\sigma}_{p1}\hat{j}^{\sigma}_{p2}}(\omega,q\hat{x})\right\}
\end{align}
There are only two independent components, i.e. $C_{\hat{j}^{+}_{p1}\hat{j}^{-}_{p2}}(\omega,q\hat{x})$ and $C_{\hat{j}^{-}_{p1}\hat{j}^{+}_{p2}}(\omega,q\hat{x})$, which is also the result of the rotational symmetry.

Thus, the spatial response takes the form
\begin{align}
  K^{\sigma\sigma}_{ab}(\omega,\bm{q})=&K^{\sigma}_{L}(\omega,q)\frac{q_aq_b}{q^2}+K^{\sigma}_{T}(\omega,q)\left(\delta_{ab}-\frac{q_aq_b}{q^2}\right)\\
  K^{+-}_{ab}(\omega,\bm{q})=&-K^{-+}_{ba}(\omega,\bm{q})=K_{D}(\omega,q)\frac{q_a\epsilon_{bc}q_c}{q^2}-K_{ID}(\omega,q)\left(\epsilon_{ab}+\frac{q_a\epsilon_{bc}q_c}{q^2}\right)
\end{align}
where the specific expressions of the six independent response functions are
\begin{align}
  K^{+}_{L}(\omega,q)=&-\frac{e^2n_X}{m}-C_{\hat{j}^{+}_{p1}\hat{j}^{+}_{p1}}(\omega,q\hat{x})=-\frac{e^2n_X}{m}-\frac{1}{2\pi l_B^2}\sum_{n}\frac{\omega_n(q)|\mathrm{J}^{+}_{1,n}(q\hat{x})|^2}{\omega^+-\omega_n(q)}\\
  K^{+}_{T}(\omega,q)=&-\frac{e^2n_X}{m}-C_{\hat{j}^{+}_{p2}\hat{j}^{+}_{p2}}(\omega,q\hat{x})=-\frac{e^2n_X}{m}-\frac{1}{2\pi l_B^2}\sum_{n}\frac{\omega_n(q)|\mathrm{J}^{+}_{2,n}(q\hat{x})|^2}{\omega^+-\omega_n(q)}\\
  K^{-}_{L}(\omega,q)=&-\frac{e^2n_X}{4m}-C_{\hat{j}^{-}_{p1}\hat{j}^{-}_{p1}}(\omega,q\hat{x})=-\frac{e^2n_X}{4m}-\frac{1}{2\pi l_B^2}\sum_{n}\frac{\omega_n(q)|\mathrm{J}^{-}_{1,n}(q\hat{x})|^2}{\omega^+-\omega_n(q)}\\
  K^{-}_{T}(\omega,q)=&-\frac{e^2n_X}{4m}-C_{\hat{j}^{-}_{p2}\hat{j}^{-}_{p2}}(\omega,q\hat{x})=-\frac{e^2n_X}{4m}-\frac{1}{2\pi l_B^2}\sum_{n}\frac{\omega_n(q)|\mathrm{J}^{-}_{2,n}(q\hat{x})|^2}{\omega^+-\omega_n(q)}\\
  K_{D}(\omega,q)=&-C_{\hat{j}^{-}_{p2}\hat{j}^{+}_{p1}}(\omega,q\hat{x})=\frac{i}{2\pi l_B^2}\sum_{n}\frac{\omega_n(q)\mathrm{Re}\mathrm{J}^{+}_{1,n}(q\hat{x})\mathrm{Im}\mathrm{J}^{-}_{2,n}(q\hat{x})}{\omega^{+}-\omega_n(q)}\\
  K_{ID}(\omega,q)=&-C_{\hat{j}^{+}_{p2}\hat{j}^{-}_{p1}}(\omega,q\hat{x})=\frac{i}{2\pi l_B^2}\sum_{n}\frac{\omega_n(q)\mathrm{Re}\mathrm{J}^{-}_{1,n}(q\hat{x})\mathrm{Im}\mathrm{J}^{+}_{2,n}(q\hat{x})}{\omega^{+}-\omega_n(q)}
\end{align}
Here $K^{\sigma}_{L}$ and $K^{\sigma}_{T}$ are the longitudinal and transverse response functions in the pure charge ($\sigma=+$) or exciton ($\sigma=-$) channel.
$K_D$ is the transverse exciton current induced by a longitudinal layer-symmetric gauge field, while $K_{ID}$ is the transverse charge current induced by a longitudinal layer-antisymmetric gauge field. They are the dipole and inverse dipole Hall responses, respectively.
To see this more clearly, we write down the full response function for $\bm{q}=q\hat{x}$.
Using the Ward identity, we have
\begin{equation}
  K^{\sigma\sigma}_{\mu\nu}=\begin{bmatrix}
    \frac{q^2}{\omega^2} K_{L}^{\sigma} & -\frac{q}{\omega}K^{\sigma}_{L} & 0 \\
    -\frac{q}{\omega}K^{\sigma}_{L} & K^{\sigma}_{L} & 0 \\
    0 & 0 & K^{\sigma}_{T}
  \end{bmatrix},\;
  K^{+-}_{\mu\nu}=\begin{bmatrix}
    0 & 0 & \frac{q}{\omega}K_D \\
    0 & 0 & -K_D \\
    -\frac{q}{\omega}K_{ID} & K_{ID} & 0 
  \end{bmatrix},\; K^{-+}_{\mu\nu}=\begin{bmatrix}
    0 & 0 & \frac{q}{\omega}K_{ID} \\
    0 & 0 & -K_{ID} \\
    -\frac{q}{\omega}K_D & K_D & 0
  \end{bmatrix}
\end{equation}

\section{Classical derivation of the response function of the normal exciton fluid}\label{app:normal-exciton-fluid}
Consider a normal exciton fluid with exciton number density $n_{X}$ and effective mass $m_X=m_e+m_h$, and assume that the damping is dominated by impurity scattering with relaxation time $\tau$.
With exciton-density fluctuation $\delta n_X$ and velocity field $\bm{v}$, the exciton continuity equation reads
\begin{equation}
  \partial_t \delta n_X(\bm{r},t)+n_X\nabla\cdot \bm{v}(\bm{r},t)=0
\end{equation}
The velocity equation is
\begin{equation}
  n_{X}\partial_t \bm{v}(\bm{r},t)=-v_{s}^2\nabla \delta n_X(\bm{r},t)-\frac{e\bm{E}^{-}(\bm{r},t)}{m_X}n_X  -\frac{n_{X}\bm{v}(\bm{r},t)}{\tau}
\end{equation}
where $v$ is the sound velocity of the exciton fluid, and $\bm{E}^{-}(\bm{r},t)$ is the layer-antisymmetric electric field acting on the exciton dipole.
In frequency and momentum domain, the continuum equation reads 
\begin{equation}
  -i\omega \delta n_X(\omega,\bm{q})+i n_X \bm{q}\cdot \bm{v}(\omega,\bm{q})=0
\end{equation}
and the dynamic equation for the velocity field is
\begin{equation}
  -i\omega n_X \bm{v}(\omega,\bm{q})=-i v^2 \bm{q} \delta n_X(\omega,\bm{q}) -\frac{e n_X}{m_X}\bm{E}^{-}(\omega,\bm{q}) -\frac{n_X \bm{v}(\omega,\bm{q})}{\tau}
\end{equation}

In general, the velocity field can be decomposed into longitudinal and transverse parts with respect to $\bm{q}$, i.e. $\bm{v}(\omega,\bm{q})=\bm{v}_{L}(\omega,\bm{q})+\bm{v}_{T}(\omega,\bm{q})$, where $\bm{v}_{L}(\omega,\bm{q})=\hat{\bm{q}}[\hat{\bm{q}}\cdot \bm{v}(\omega,\bm{q})]$ and $\bm{v}_{T}(\omega,\bm{q})=\hat{\bm{q}}\times [\bm{v}(\omega,\bm{q})\times \hat{\bm{q}}]$.
Then the continuum equation can be rewritten as
\begin{equation}
  -i\omega \delta n_X(\omega,\bm{q})+i n_X q v_{L}(\omega,\bm{q})=0\label{eq:continuum-eq}
\end{equation}
The velocity equation can also be decomposed into longitudinal and transverse parts:
\begin{align}
  -i\omega n_X v_{L}(\omega,\bm{q})=&-i v^2 q \delta n_X(\omega,\bm{q}) -\frac{e n_X}{m_X}E^{-}_{L}(\omega,\bm{q}) -\frac{n_X v_{L}(\omega,\bm{q})}{\tau}\label{eq:velocity-L}\\
  -i\omega n_X v_{T}(\omega,\bm{q})=& -\frac{e n_X}{m_X}E^{-}_{T}(\omega,\bm{q}) -\frac{n_X v_{T}(\omega,\bm{q})}{\tau}\label{eq:velocity-T}
\end{align}
and the relation between the longitudinal and transverse parts of the electric field and that of the velocity field is solved as
\begin{align}
  -en_Xv_{L}(\omega,\bm{q})=& \frac{e^2 n_X \omega /m_X}{\omega^2 - v^2 q^2 + i\omega/\tau}iE^{-}_{L}(\omega,\bm{q})\\
  -en_Xv_{T}(\omega,\bm{q})=& \frac{e^2 n_X/m_X}{\omega + i/\tau}iE^{-}_{T}(\omega,\bm{q})
\end{align}
Consider the definition of the exciton current $\bm{j}^{-}(\omega,\bm{q})=-e n_X \bm{v}(\omega,\bm{q})$ and the relation between electric field and gauge field $\bm{E}^{-}(\omega,\bm{q})=-iq\phi^{-}(\omega,\bm{q})+i\omega \bm{A}^{-}(\omega,\bm{q})$, we have the response function of the normal exciton fluid as
\begin{align}
  K^{-}_{L}(\omega,q)=&-\frac{e^2n_X}{m_X}\frac{\omega^2}{\omega^2 - v^2 q^2 + i\omega/\tau}\\
  K^{-}_{T}(\omega,q)=&-\frac{e^2n_X}{m_X}\frac{\omega}{\omega + i/\tau}
\end{align}

{
\section{\texorpdfstring{Analytical treatment of electron--hole symmetry and asymmetry in the nonmagnetic system}{Analytical treatment of electron--hole symmetry and asymmetry in the nonmagnetic system}}\label{app:eh_sy}

In this section, we derive the consequences of electron--hole symmetry and asymmetry for the nonmagnetic system. The corresponding numerical checks are presented in Appendix \ref{app:supplementary-data}. For the magnetic case, a complete analytical treatment of electron--hole asymmetry is beyond the scope of this work. Appendix \ref{app:supplementary-data-magnetic} presents the main numerical results for electron--hole asymmetry in the magnetic case and provides analytical understanding where possible.

\subsection{\texorpdfstring{Representation of the charge conjugation operation}{Representation of the charge conjugation operation}}
By definition, the charge conjugation operation $\mathcal{C}$ acts on the electron and hole operators as
\begin{equation}
  \mathcal{C}:\Psi_{e}(\bm{r})\to \Psi^{\dagger}_{h}(\bm{r}),\;\Psi_{h}(\bm{r})\to -\Psi^{\dagger}_{e}(\bm{r})
\end{equation}
In the momentum space, we have
\begin{equation}
  \mathcal{C}: c_{e\bm{k}}\to c^{\dagger}_{h-\bm{k}},\;c_{h\bm{k}}\to -c^{\dagger}_{e-\bm{k}}
\end{equation}
Then using the relation between the electron, hole operators and the quasi-particle operators
\begin{equation}
  c_{v\bm{k}}=\alpha_{\bm{k}}c_{e\bm{k}}+\beta_{\bm{k}}c_{h\bm{k}},\;c_{c\bm{k}}=\beta_{\bm{k}}c_{e\bm{k}}-\alpha_{\bm{k}}c_{h\bm{k}}
\end{equation}
we can get the representation of the charge conjugation operation on the quasi-particle operators as
\begin{equation}
  \mathcal{C}: c_{v\bm{k}}\to \alpha_{\bm{k}}c^{\dagger}_{h-\bm{k}}-\beta_{\bm{k}}c^{\dagger}_{e-\bm{k}}=-c^{\dagger}_{c-\bm{k}},\;c_{c\bm{k}}\to \beta_{\bm{k}}c^{\dagger}_{h-\bm{k}}+\alpha_{\bm{k}}c^{\dagger}_{e-\bm{k}}=c^{\dagger}_{v-\bm{k}}
\end{equation}
We also use the rotation symmetry such that $\alpha_{\bm{k}}=\alpha_{k}$ and $\beta_{\bm{k}}=\beta_{k}$ to get the final results.
Thus the charge conjugation operation $\mathcal{C}$ acts on the density matrix as
\begin{equation}
  \mathcal{C}:\rho_{cv\bm{k}}\equiv \ave{c^{\dagger}_{v\bm{k}}c_{c\bm{k}}}\to -\ave{c_{c-\bm{k}}c^{\dagger}_{v-\bm{k}}}=\ave{c^{\dagger}_{v-\bm{k}}c_{c-\bm{k}}}=\rho_{cv-\bm{k}}
\end{equation}
Similarly, we also have $\mathcal{C}:\rho_{vc\bm{k}}\to \rho_{vc-\bm{k}}$.
Equation \eqref{eq:tdhf-bilayer-compact} then gives the action of charge conjugation on the dynamical matrix:
\begin{equation}
  \mathcal{C}\mathcal{H}_{\bm{k},\bm{k}'}(\bm{q})\mathcal{C}^{-1}=\mathcal{H}_{-\bm{k},-\bm{k}'}(\bm{q})
\end{equation}

We then investigate the effect of the charge conjugation operation on the paramagnetic charge- and exciton-current operators:
\begin{equation}
  \hat{j}^{\sigma}_{p\mu}(\bm{q})=\sum_{\bm{k}}C^{\dagger}_{\bm{k}-\bm{q}/2}\gamma^{\sigma}_{\mu,\bm{k}}C_{\bm{k}+\bm{q}/2}
\end{equation}
where $\gamma^{\sigma}_{\mu,\bm{k}}$ is the bare vertex defined in Eq. \eqref{eq:vertex-function}.
As the charge conjugation operation $\mathcal{C}$ acts on the electron and hole operators as 
\begin{equation*}
  \mathcal{C}: C_{\bm{k}}=\begin{bmatrix}
    c_{e\bm{k}} \\ c_{h\bm{k}}
  \end{bmatrix}\to \begin{bmatrix}
    c^{\dagger}_{h-\bm{k}} \\ -c^{\dagger}_{e-\bm{k}}
  \end{bmatrix}=i\sigma_y (C^{\dagger}_{-\bm{k}})^T
\end{equation*}
we have
\begin{equation}
  \mathcal{C}:\hat{j}^{\sigma}_{p\mu}(\bm{q})\to \sum_{\bm{k}}C^{T}_{-\bm{k}+\bm{q}/2}\sigma_y\gamma^{\sigma}_{\mu,\bm{k}}\sigma_y (C^{\dagger}_{-\bm{k}-\bm{q}/2})^T=-\sum_{\bm{k}}C^{\dagger}_{\bm{k}-\bm{q}/2}\sigma_y(\gamma^{\sigma}_{\mu,-\bm{k}})^T\sigma_y C_{\bm{k}+\bm{q}/2}
\end{equation}
Thus we have
\begin{equation}
  \mathcal{C}:\gamma^{\sigma}_{\mu,\bm{k}}\to -\sigma_y(\gamma^{\sigma}_{\mu,-\bm{k}})^T\sigma_y
\end{equation}
Explicitly,
\begin{subequations}
  \begin{align}
    &\mathcal{C}:\gamma^{+}_{0,\bm{k}}=e\sigma_0\to -e\sigma_0=-\gamma^{+}_{0,\bm{k}}\\
    &\mathcal{C}:\gamma^{+}_{\mu=12,\bm{k}}=-\frac{e\hbar\bm{k}}{2m}\sigma_z-\frac{e\hbar\bm{k}\delta_m}{2m}\sigma_0\to +\frac{e\hbar\bm{k}}{2m}\sigma_z-\frac{e\hbar\bm{k}\delta_m}{2m}\sigma_0\\
    &\mathcal{C}:\gamma^{-}_{0,\bm{k}}=\frac{e}{2}\sigma_z \to \frac{e}{2}\sigma_z=\gamma^{-}_{0,\bm{k}}\\
    &\mathcal{C}:\gamma^{-}_{\mu=12,\bm{k}}=-\frac{e\hbar\bm{k}}{4m}\sigma_0-\frac{e\hbar\bm{k}\delta_m}{4m}\sigma_z \to -\frac{e\hbar\bm{k}}{4m}\sigma_0+\frac{e\hbar\bm{k}\delta_m}{4m}\sigma_z
  \end{align}\label{eq:eh_vertex}
\end{subequations}
In the presence of the electron--hole symmetry $\delta_m=0$, we have
\begin{equation}
  \mathcal{C}:\gamma^+_{\mu,\bm{k}}\to -\gamma^{+}_{\mu,\bm{k}},\;\mathcal{C}:\gamma^{-}_{\mu,\bm{k}}\to \gamma^{-}_{\mu,\bm{k}}\label{eq:eh_vertex_phsym}
\end{equation}
which means that the charge operators $\hat{j}^{+}_{p\mu}$ are odd under the charge conjugation operation $\mathcal{C}$, while the exciton operators $\hat{j}^{-}_{p\mu}$ are even under $\mathcal{C}$.

\subsection{\texorpdfstring{Consequence of electron--hole symmetry}{Consequence of electron--hole symmetry}}
In the presence of the electron--hole symmetry $\delta_{m}=0$, we must have $\mathcal{C}\mathcal{H}(\bm{q})\mathcal{C}^{-1}=\mathcal{H}(\bm{q})$, which requires
\begin{equation}
  \mathcal{H}_{\bm{k},\bm{k}'}(\bm{q})=\mathcal{H}_{-\bm{k},-\bm{k}'}(\bm{q})
\end{equation}
In addition, the collective modes are also eigenvectors of the charge conjugation operation $\mathcal{C}$ with eigenvalues $\pm 1$, i.e.,
\begin{equation}
  \mathcal{C}\Phi_{n\bm{k}}(\bm{q})=\Phi_{n-\bm{k}}(\bm{q})=c_n(\bm{q}) \Phi_{n\bm{k}}(\bm{q})
\end{equation}
where $c_n(\bm{q})=\pm 1$ is the charge conjugation eigenvalue of the $n$-th collective mode.
At $\bm{q}=\bm{0}$, the charge conjugation eigenvalue $c_n(\bm{0})$ is just the parity of the collective mode, i.e., $c_n(\bm{0})=+1$ for the even-parity modes and $c_n(\bm{0})=-1$ for the odd-parity modes.
In the vicinity of $\bm{q}=\bm{0}$, the parity is no longer a good quantum number, but the charge conjugation eigenvalue $c_n(\bm{q})$ is still well-defined and can be used to label the collective modes.
Combining Eq. \eqref{eq:eh_vertex_phsym} with the charge-conjugation parity of the collective modes, we find that the charge-density and charge-current operators $\hat{j}^{+}_{p\mu}$ can couple only to the odd-parity modes, while the exciton-density and exciton-current operators $\hat{j}^{-}_{p\mu}$ can couple only to the even-parity modes.
This is consistent with Eq. \eqref{eq:regular-mode-vertex-overlap-expansion}: to leading order in $q$, the charge-density and charge-current operators couple only to the odd-parity dipole modes, the exciton-density operator couples to the even-parity monopole modes, and the exciton-current operators couple to the even-parity monopole and quadrupole modes.

\subsection{\texorpdfstring{Effect of electron--hole asymmetry}{Effect of electron--hole asymmetry}}
There are several effects of the electron--hole asymmetry $\delta_m\ne 0$.

\subsubsection{\texorpdfstring{Effect on the collective modes}{Effect on the collective modes}}

The dynamical matrix in the presence of electron--hole asymmetry is
\begin{equation}
  \tilde{\mathcal{H}}_{\bm{k},\bm{k}'}(\bm{q})=\mathcal{H}_{\bm{k},\bm{k}'}(\bm{q})+\delta\mathcal{H}_{\bm{k},\bm{k}'}(\bm{q})
\end{equation}
where $\mathcal{H}_{\bm{k},\bm{k}'}(\bm{q})$ is the electron--hole-symmetric dynamical matrix and
\begin{equation}
  \delta\mathcal{H}_{\bm{k},\bm{k}'}(\bm{q})=\delta_{\bm{k},\bm{k}'}\frac{\delta_{m}\hbar^2\bm{k}\cdot\bm{q}}{2m}
\end{equation}
is its electron--hole-asymmetry correction. At fixed nonzero $q$, we write the perturbed eigenvector and eigenfrequency as
\begin{align}
  \tilde{\Phi}_{n\bm{k}}(\bm{q})=&\Phi_{n\bm{k}}(\bm{q})+\delta\Phi_{n\bm{k}}(\bm{q})+\mathcal{O}(\delta_m^2)\nonumber\\
  \tilde{\omega}_n(\bm{q})=&\omega_n(\bm{q})+\delta\omega_n(\bm{q})+\mathcal{O}(\delta_m^2)
\end{align}
where $\Phi_{n\bm{k}}(\bm{q})$ and $\omega_n(\bm{q})$ denote the electron--hole-symmetric-limit mode. To first order in $\delta_m$, the generalized eigenvalue equation gives
\begin{equation}
  \delta\mathcal{H}(\bm{q})\Phi_{n}(\bm{q})+\mathcal{H}(\bm{q})\delta\Phi_{n}(\bm{q})=\hbar\omega_n(\bm{q})\tau_z\delta\Phi_{n}(\bm{q})+\hbar\delta\omega_n(\bm{q})\tau_z\Phi_{n}(\bm{q})
\end{equation}
and hence
\begin{equation}
  \delta_{n,m}\frac{\delta\omega_n(\bm{q})}{\omega_n(\bm{q})}=\Phi^{\dagger}_{m}(\bm{q})\delta\mathcal{H}(\bm{q})\Phi_{n}(\bm{q})+\hbar[\omega_m(\bm{q})-\omega_n(\bm{q})]\Phi^{\dagger}_{m}(\bm{q})\tau_z\delta\Phi_{n}(\bm{q})
\end{equation}

\paragraph*{Regular gapped target mode.}
We first consider a non-degenerate gapped mode, with the degenerate $\pm l$ modes distinguished by their mirror eigenvalues. Setting $m=n$ gives
\begin{equation}
  \delta\omega_n(\bm{q})=\omega_n(\bm{q})\Phi^{\dagger}_{n}(\bm{q})\delta\mathcal{H}(\bm{q})\Phi_{n}(\bm{q})=\omega_n(\bm{q})\frac{\delta_{m}\hbar^2}{2m}\sum_{\bm{k}}|\Phi_{n\bm{k}}(\bm{q})|^2 \bm{k}\cdot\bm{q}
\end{equation}
Because $\Phi_n(\bm{q})$ is a charge-conjugation eigenvector, $|\Phi_{n\bm{k}}(\bm{q})|^2=|\Phi_{n-\bm{k}}(\bm{q})|^2$, and the odd-$\bm{k}$ matrix element vanishes:
\begin{equation}
  \delta\omega_n(\bm{q})=\omega_n(\bm{q})\frac{\delta_{m}\hbar^2}{2m}\sum_{\bm{k}}|\Phi_{n\bm{k}}(\bm{q})|^2 \bm{k}\cdot\bm{q}=\omega_n(\bm{q})\frac{\delta_{m}\hbar^2}{2m}\sum_{\bm{k}}|\Phi_{n-\bm{k}}(\bm{q})|^2 (-\bm{k})\cdot\bm{q}=0
\end{equation}
We preserve the normalization $\tilde{\Phi}_n^{\dagger}\tilde{\mathcal{H}}\tilde{\Phi}_n=1$. To first order, this requires
\begin{equation}
  2\mathrm{Re}[\Phi_n^{\dagger}(\bm{q})\mathcal{H}(\bm{q})\delta\Phi_n(\bm{q})]+\Phi_n^{\dagger}(\bm{q})\delta\mathcal{H}(\bm{q})\Phi_n(\bm{q})=0.
\end{equation}
For the non-degenerate charge-conjugation eigenstates considered here, the second term vanishes by the same odd-$\bm{k}$ argument. We choose the real normalization gauge in which $\Phi_n^{\dagger}\mathcal{H}\delta\Phi_n=0$, so $\delta\Phi_n$ has no component parallel to $\Phi_n$.

For $m\ne n$, we have
\begin{equation}
  \Phi^{\dagger}_{m}(\bm{q})\tau_z\delta\Phi_{n}(\bm{q})=\frac{1}{\hbar\omega_m(\bm{q})}\Phi^{\dagger}_{m}(\bm{q})\mathcal{H}(\bm{q})\delta\Phi_{n}(\bm{q})=-\frac{\Phi^{\dagger}_{m}(\bm{q})\delta\mathcal{H}(\bm{q})\Phi_{n}(\bm{q})}{\hbar(\omega_m(\bm{q})-\omega_n(\bm{q}))}
\end{equation}
Thus we have
\begin{equation}
  \sqrt{\mathcal{H}(\bm{q})}\delta\Phi_{n}(\bm{q})=-\sum_{m\ne n}\frac{\omega_m(\bm{q})\Phi^{\dagger}_{m}(\bm{q})\delta\mathcal{H}(\bm{q})\Phi_{n}(\bm{q})}{\omega_m(\bm{q})-\omega_n(\bm{q})}\sqrt{\mathcal{H}(\bm{q})}\Phi_{m}(\bm{q})
\end{equation}
or equivalently
\begin{equation}
  \delta\Phi_{n}(\bm{q})=-\sum_{m\ne n}\frac{\omega_m(\bm{q})\Phi^{\dagger}_{m}(\bm{q})\delta\mathcal{H}(\bm{q})\Phi_{n}(\bm{q})}{\omega_m(\bm{q})-\omega_n(\bm{q})}\Phi_{m}(\bm{q})
  \label{eq:first-order-eigenvector-correction}
\end{equation}
Consider a regular gapped mode $n$ away from a mode crossing. Each term in Eq. \eqref{eq:first-order-eigenvector-correction} describes the mixing of $n$ with one frequency branch of another mode $m$. If $m$ is also gapped, the frequency difference $\omega_m-\omega_n$ remains finite as $q\to0$, while $\delta\mathcal H\propto\delta_mq$. This term is therefore at most $\mathcal O(\delta_mq)$. If the intermediate mode is the Goldstone mode, $m=G$, its matrix element has a different $q$ dependence, but the numerator also contains $\omega_G\propto q$. A single Goldstone branch obeys the same bound, as shown below. Thus, branch by branch,
\begin{equation}
  \left.\delta\Phi_n\right|_m=\mathcal O(\delta_mq).
  \label{eq:regular-gapped-branchwise-bound}
\end{equation}
Because the first-order frequency shift of a regular gapped mode vanishes, its first nonzero frequency correction is second order:
\begin{equation}
  \tilde{\omega}_n(\bm{q})=\omega_n(\bm{q})[1+\mathcal{O}(\delta_m^2 q^2)].
  \label{eq:omega_correction}
\end{equation}
This non-degenerate expansion does not apply at an exact or near-degenerate crossing. There, $\delta\mathcal{H}$ must first be diagonalized within the degenerate subspace, and an avoided-crossing gap of order $\mathcal{O}(\delta_m q)$ can appear, as illustrated by the $1p_x$--$1d_{x^2-y^2}$ crossing in the inset of Fig. 2(c) of the main text.

\paragraph*{Goldstone target mode.}
Equation \eqref{eq:first-order-eigenvector-correction} also applies when $n=G$ at every fixed nonzero $q$. We use $\Phi_G(0)$ for the unnormalized zero mode at $q=0$. It represents a pure phase fluctuation and therefore does not couple to the nonanalytic long-range Coulomb interaction, which acts on density fluctuations. Its leading finite-$q$ energy is consequently
\[
  \Phi_G^\dagger(0)\mathcal H(\bm q)\Phi_G(0)=\mathcal O(q^2).
\]
Write the leading part of the normalized Goldstone eigenvector as $\Phi_G(\bm q)=A_G(q)\Phi_G(0)+\cdots$. Its normalization then gives
\[
  1=\Phi_G^\dagger\mathcal H\Phi_G
  \sim |A_G(q)|^2q^2,
  \qquad\text{and therefore}\qquad
  A_G(q)=\mathcal O(q^{-1}).
\]
Thus the explicit factor $q$ in $\delta\mathcal H=\mathcal O(\delta_mq)$ is canceled by the factor $q^{-1}$ in $\Phi_G$. For a regular gapped $p_x$ mode,
\begin{equation}
  \Phi_{p_x}^\dagger(\bm q)\delta\mathcal H(\bm q)\Phi_G(\bm q)
  \sim (\delta_mq)q^{-1}
  =\mathcal O(\delta_mq^0).
\end{equation}
We can now apply Eq. \eqref{eq:first-order-eigenvector-correction} to one frequency branch at a time. For $n=G$ and $m=p_x$, the frequency factor $\omega_{p_x}/(\omega_{p_x}-\omega_G)$ approaches a constant, so that branch contributes at most $\mathcal O(\delta_m)$. In the reverse direction, $n=p_x$ and $m=G$, the frequency factor contains $\omega_G\propto q$ and is $\mathcal O(q)$, so that branch contributes at most $\mathcal O(\delta_mq)$. The branchwise bounds are
\begin{equation}
  \left.\delta\Phi_G\right|_{p_x}=\mathcal O(\delta_m)\Phi_{p_x},
  \qquad
  \left.\delta\Phi_{p_x}\right|_G=\mathcal O(\delta_mq)\Phi_G.
  \label{eq:Goldstone-asymmetric-eigenvector-mixing}
\end{equation}

The first-order diagonal correction also vanishes for the Goldstone mode at every finite nonzero $q$:
\begin{equation}
  \frac{\delta\omega_G(\bm{q})}{\omega_G(\bm{q})}=\frac{\delta_{m}\hbar^2}{2m}\sum_{\bm{k}}|\Phi_{G\bm{k}}(\bm{q})|^2 \bm{k}\cdot\bm{q}=0.
\end{equation}
This cancellation does not imply an $\mathcal{O}(\delta_m^2q^2)$ leading shift of the frequency. Equation \eqref{eq:first-order-eigenvector-correction} remains valid, but the regular-mode estimate is not uniform because the normalized Goldstone eigenvector has nonuniform $q\to0$ scaling and $\omega_G(q)=vq$ vanishes. The nominal $\mathcal{O}[(\delta_m q)^2]$ correction is naturally a correction to the squared Goldstone frequency:
\begin{equation}
  \delta[\omega_G^2(q)]=\mathcal{O}(\delta_m^2q^2).
\end{equation}
Converting it to a shift of the frequency introduces the vanishing Goldstone energy in the denominator,
\begin{equation}
  \tilde{\omega}_G(q)-\omega_G(q)
  =\frac{\tilde{\omega}_G^2(q)-\omega_G^2(q)}{\tilde{\omega}_G(q)+\omega_G(q)}
  =\mathcal{O}(\delta_m^2q).
\end{equation}
Thus the perturbative expansion in $\delta_m$ remains valid, but the regular-gapped-mode power counting for $\omega_n$ is not uniform in the limit $q\to0$.

The thermodynamic analysis in the main text fixes the result directly. At fixed reduced mass $m$, the exciton mass is $m_X=4m/(1-\delta_m^2)$, while the sound velocity obeys $v\propto m_X^{-1/2}$. Therefore,
\begin{equation}
  \frac{\tilde{\omega}_G(\bm{q})}{\omega_G(\bm{q})}=\sqrt{1-\delta_m^2}
  \label{eq:omega_correction_Goldstone}
\end{equation}
and
\begin{equation}
  \tilde{\omega}_G(\bm{q})-\omega_G(\bm{q})=-\frac{vq}{2}\delta_m^2+\mathcal{O}(\delta_m^4q).
\end{equation}
Thus the leading Goldstone frequency shift is $\mathcal{O}(\delta_m^2q)$ in absolute terms and $\mathcal{O}(\delta_m^2)$ relative to the electron--hole-symmetric dispersion.

\subsubsection{\texorpdfstring{Effect on the coupling between the collective modes and the charge/exciton operators}{Effect on the coupling between the collective modes and the charge/exciton operators}}\label{app:vertex_overlap_asymmetry}

According to Eq. \eqref{eq:vertex-function}, electron--hole asymmetry leaves the bare charge- and exciton-density vertices unchanged but mixes the bare spatial charge- and exciton-current vertices. It also changes the collective-mode eigenvector from $\Phi_n$ to $\Phi_n+\delta\Phi_n$. Retaining both effects to first order in $\delta_m$ gives
\begin{equation}
  \tilde{\mathrm J}^{\sigma}_{\mu,n}
  =\mathrm J^{\sigma}_{\mu,n}+\delta\mathrm J^{\sigma}_{\mu,n}
  +\mathcal O(\delta_m^2),
  \qquad
  \delta\mathrm J^{\sigma}_{\mu,n}
  =\delta_\Phi\mathrm J^{\sigma}_{\mu,n}
  +\delta_\gamma\mathrm J^{\sigma}_{\mu,n}.
\end{equation}
The bare-vertex corrections are
\begin{subequations}\label{eq:current_correction}
  \begin{align}
    \delta_\gamma\mathrm J^{\sigma}_{0,n}=&0,\\
    \delta_\gamma\mathrm J^{+}_{\mu=12,n}=&2\delta_m\mathrm J^{-}_{\mu=12,n},\\
    \delta_\gamma\mathrm J^{-}_{\mu=12,n}=&\frac{\delta_m}{2}\mathrm J^{+}_{\mu=12,n}
    \label{eq:exciton_current_correction}
  \end{align}
\end{subequations}
Here, $\delta_\Phi\mathrm J^{\sigma}_{\mu,n}$ is the eigenvector-induced correction obtained by evaluating the electron--hole-symmetric bare vertex with $\delta\Phi_n$, while $\delta_\gamma\mathrm J^{\sigma}_{\mu,n}$ is the bare-vertex correction obtained from the change of $\gamma^\sigma_\mu$. Their sum $\delta\mathrm J^{\sigma}_{\mu,n}$ is the total first-order correction to the vertex overlap.

For a regular gapped target mode, the full wavefunction correction remains $\delta\Phi_n=\mathcal O(\delta_mq)$. This fact alone does not fix the power of $\delta_\Phi\mathrm J$: the $+m$ and $-m$ intermediate branches must be contracted with the same operator vertex before they are added, and their leading contributions can cancel.

To combine the two branches, set $X_r=x_r$ and $Y_r=iy_r$ in Eq. \eqref{eq:generalized-eigenvector}, so that
\begin{equation}
  \Phi_{\pm r}=\frac12
  \begin{bmatrix}X_r\pm Y_r\\X_r\mp Y_r\end{bmatrix}.
  \label{eq:positive-negative-pair-components}
\end{equation}
The perturbation has the form $\delta\mathcal H=\operatorname{diag}(h,h)$, with $h\propto\delta_mqk_x$. For a positive-frequency target $n$, define
\begin{equation}
  \mathcal A_{mn}=\frac12Y_m^\dagger hY_n,
  \qquad
  \mathcal B_{mn}=\frac12X_m^\dagger hX_n.\label{eq:A_B_def}
\end{equation}
The combined contribution of the intermediate pair is then
\begin{equation}
  \left.\delta\Phi_{+n}\right|_{\pm m}
  =-\omega_m\left[
  \frac{\mathcal A_{mn}+\mathcal B_{mn}}{\omega_m-\omega_n}\Phi_{+m}
  -\frac{\mathcal A_{mn}-\mathcal B_{mn}}{\omega_m+\omega_n}\Phi_{-m}
  \right].
  \label{eq:paired-wavefunction-correction}
\end{equation}
The two entries in Eq. \eqref{eq:collective-mode-vertex-overlap} contain identical real interband matrix elements, but the second bare vertex is evaluated at $-\bm k$. Density vertices are even under $\bm k\to-\bm k$, whereas current vertices are odd, so their two-component forms are
\begin{equation}
  \text{density: }\begin{bmatrix}g\\g\end{bmatrix},
  \qquad
  \text{current: }\begin{bmatrix}g\\-g\end{bmatrix}.
\end{equation}
Thus the two branches have equal density vertex overlaps and opposite current vertex overlaps. Contracting their sum gives
\begin{subequations}\label{eq:paired-eigenvector-induced-overlap}
  \begin{align}
    \left.\delta_\Phi\mathrm J_{0,n}\right|_{\pm m}
    =&-\frac{2\omega_m\mathrm J_{0,+m}}
    {\omega_m^2-\omega_n^2}
    \left(\omega_n\mathcal A_{mn}+\omega_m\mathcal B_{mn}\right),\\
    \left.\delta_\Phi\mathrm J_{\lambda,n}\right|_{\pm m}
    =&-\frac{2\omega_m\mathrm J_{\lambda,+m}}
    {\omega_m^2-\omega_n^2}
    \left(\omega_m\mathcal A_{mn}+\omega_n\mathcal B_{mn}\right),
    \qquad \lambda\in\{L,T\}.
  \end{align}
\end{subequations}
These paired formulas, rather than the factor $\delta_mq$ in the wavefunction correction by itself, determine the leading power of each induced vertex overlap. They apply to both regular and Goldstone target modes.

\paragraph{Regular gapped modes}
The remaining input is the angular sector mixed into the target mode. Since $h\propto\delta_mqk_x$ [defined around Eq.~\eqref{eq:A_B_def}], the perturbation changes angular momentum by one and preserves reflection about the $x$ axis. Keeping only the sectors with leading vertex overlaps gives
\begin{equation}
  \begin{gathered}
    \delta\Phi_{s\setminus\{G\}}\sim\delta_m q\,\Phi_{p_x},\qquad
    \delta\Phi_{d_{x^2-y^2}}\sim\delta_m q\,\Phi_{p_x},\qquad
    \delta\Phi_{d_{xy}}\sim\delta_m q\,\Phi_{p_y},\\
    \delta\Phi_{p_x}\sim\delta_m q\,(\Phi_{s\setminus\{G\}}+\Phi_G+\Phi_{d_{x^2-y^2}}),\qquad
    \delta\Phi_{p_y}\sim\delta_m q\,\Phi_{d_{xy}}.
  \end{gathered}
  \label{eq:asymmetry-angular-sector-mixing}
\end{equation}
Here the target mode $n$ is regular. In $\delta\Phi_{p_x}$, $\Phi_G$ denotes the Goldstone intermediate sector. Each branch $m=\pm G$ has a mixing coefficient $\mathcal O(\delta_mq)$, but its contribution to a vertex overlap must be combined with its frequency partner using Eq. \eqref{eq:paired-eigenvector-induced-overlap}.

To convert this sector mixing into vertex-overlap powers, we recall from Table \ref{tab:vertex-overlap-power-counting} the nonzero electron--hole-symmetric results
\begin{equation}
  \begin{gathered}
    \mathrm{J}^{+}_{0,p_x}=\mathcal{O}(q),\qquad
    \mathrm{J}^{+}_{L,p_x}=\mathrm{J}^{+}_{T,p_y}=\mathcal{O}(q^0),\\
    \mathrm{J}^{-}_{0,(s\setminus\{G\})\cup d_{x^2-y^2}}=\mathcal{O}(q^2),\quad
    \mathrm{J}^{-}_{L,(s\setminus\{G\})\cup d_{x^2-y^2}}=\mathcal{O}(q),\qquad
    \mathrm{J}^{-}_{T,d_{xy}}=\mathcal{O}(q).
  \end{gathered}
  \label{eq:symmetric-overlap-powers-asymmetry}
\end{equation}
All other regular-mode vertex overlaps either vanish by symmetry or begin at higher order in $q$. In particular, charge conjugation together with angular-momentum or reflection symmetry removes the opposite-channel overlaps at the orders displayed. A union in a mode subscript means that the stated power applies to either sector; it does not denote a sum over modes.

We now combine Eqs. \eqref{eq:asymmetry-angular-sector-mixing} and \eqref{eq:symmetric-overlap-powers-asymmetry}, organized by operator channel.
Every target mode $n$ in the following channel list and in Table \ref{tab:eigenvector-induced-overlap-power-counting} is regular. The case $n=G$ is treated separately after the table.

\begin{itemize}
\item \emph{Charge-density channel, $\tilde{\mathrm J}^+_{0,n}$.}
In the electron--hole-symmetric limit, the leading regular-mode charge-density vertex overlap is with the $p_x$ sector,
\begin{equation}
  \mathrm J^+_{0,p_x}=\mathcal O(q).
\end{equation}
Electron--hole asymmetry does not modify the charge-density bare vertex directly. A mode in $\bigl(s\setminus\{G\}\bigr)\cup d_{x^2-y^2}$ instead acquires a $p_x$ component of order $\delta_m q$, giving
\begin{equation}
  \delta_\Phi\mathrm J^+_{0,(s\setminus\{G\})\cup d_{x^2-y^2}}
  \sim(\delta_m q)\mathrm J^+_{0,p_x}
  =\mathcal O(\delta_m q^2).
  \label{eq:induced-sd-charge-density-power}
\end{equation}
Thus the electron--hole-asymmetry-induced contributions from $\bigl(s\setminus\{G\}\bigr)\cup d_{x^2-y^2}$ are higher by one power of $q$ than the leading $p_x$ contribution to the charge-density channel. They are nevertheless the leading nonzero charge-density vertex overlaps of those individual even-parity modes.
\item \emph{Charge-current channels, $\tilde{\mathrm J}^+_{L,n}$ and $\tilde{\mathrm J}^+_{T,n}$.}
The leading electron--hole-symmetric charge-current vertex overlaps are $\mathrm J^+_{L,p_x}=\mathrm J^+_{T,p_y}=\mathcal O(q^0)$. In the longitudinal channel, the modes in $\bigl(s\setminus\{G\}\bigr)\cup d_{x^2-y^2}$ receive both an eigenvector-induced correction and a bare-vertex correction,
\begin{equation}
  \delta_\Phi\mathrm J^+_{L,(s\setminus\{G\})\cup d_{x^2-y^2}}
  =\mathcal O(\delta_m q),\qquad
  \delta_\gamma\mathrm J^+_{L,(s\setminus\{G\})\cup d_{x^2-y^2}}
  =2\delta_m\mathrm J^-_{L,(s\setminus\{G\})\cup d_{x^2-y^2}}
  =\mathcal O(\delta_m q).
\end{equation}
In the transverse channel, the corresponding $d_{xy}$ corrections are
\begin{equation}
  \delta_\Phi\mathrm J^+_{T,d_{xy}}
  =\mathcal O(\delta_m q),\qquad
  \delta_\gamma\mathrm J^+_{T,d_{xy}}
  =2\delta_m\mathrm J^-_{T,d_{xy}}
  =\mathcal O(\delta_m q).
\end{equation}
Both electron--hole-asymmetry corrections must be retained to determine the finite-$q$ coefficient, but they are higher by one power of $q$ than the leading $p_x$ longitudinal and $p_y$ transverse charge-current vertex overlaps.

\item \emph{Exciton-density channel, $\tilde{\mathrm J}^-_{0,n}$.}
For regular target modes in $\bigl(s\setminus\{G\}\bigr)\cup d_{x^2-y^2}$, the exciton-density vertex overlaps begin at $\mathcal O(q^2)$. The bare exciton-density vertex is unchanged, so $\delta_\gamma\mathrm J^-_{0,n}=0$. A regular $p_x$ target mode acquires a Goldstone component of order $\delta_mq$ through the intermediate state $m=G$. Since $\mathrm J^-_{0,G}=\mathcal O(q^0)$, this gives
\begin{equation}
  \delta_\Phi\mathrm J^-_{0,p_x}=\mathcal O(\delta_m q).
\end{equation}
For fixed small nonzero $\delta_m$, this induced $p_x$ exciton-density vertex overlap is less suppressed in $q$ than the $\mathcal O(q^2)$ exciton-density vertex overlaps of modes in $\bigl(s\setminus\{G\}\bigr)\cup d_{x^2-y^2}$.

\item \emph{Exciton-current channels, $\tilde{\mathrm J}^-_{L,n}$ and $\tilde{\mathrm J}^-_{T,n}$.}
In the longitudinal channel, the exciton-current vertex overlaps of regular modes in $\bigl(s\setminus\{G\}\bigr)\cup d_{x^2-y^2}$ are $\mathcal O(q)$. For a regular $p_x$ target mode, the bare-vertex mixing gives
\begin{equation}
  \delta_\gamma\mathrm J^-_{L,p_x}
  =\frac{\delta_m}{2}\mathrm J^+_{L,p_x}
  =\mathcal O(\delta_m q^0).
\end{equation}

For the eigenvector-induced term, the Goldstone normalization gives $X_G=\mathcal O(q^0)$ and $Y_G=\mathcal O(q^{-1})$. Applying the current formula in Eq. \eqref{eq:paired-eigenvector-induced-overlap} to both intermediate Goldstone partners gives
\begin{equation}
  \left.\delta_\Phi\mathrm J^-_{L,p_x}\right|_{\pm G}
  =\mathcal O(\delta_mq^2).
  \label{eq:px-Goldstone-pair-current-power}
\end{equation}
Here $\omega_G\mathrm J^-_{L,G}=\mathcal O(q)$, while $\omega_G\mathcal A_{G,p_x}+\omega_{p_x}\mathcal B_{G,p_x}=\mathcal O(\delta_mq)$. Their product therefore starts at $q^2$ after the $\pm G$ sum. The regular intermediate modes in $\bigl(s\setminus\{G\}\bigr)\cup d_{x^2-y^2}$ also contribute only $\mathcal O(\delta_mq^2)$, because their mixing coefficients and their longitudinal exciton-current vertex overlaps are both $\mathcal O(q)$. Therefore
\begin{equation}
  \delta_\Phi\mathrm J^-_{L,p_x}=\mathcal O(\delta_mq^2).
  \label{eq:px-induced-longitudinal-current-power}
\end{equation}
The bare-vertex term $\delta_\gamma\mathrm J^-_{L,p_x}=\mathcal O(\delta_mq^0)$ is consequently leading by two powers of $q$.

In the transverse channel, $\mathrm J^-_{T,d_{xy}}=\mathcal O(q)$ in the electron--hole-symmetric limit, whereas for a $p_y$ mode
\begin{equation}
  \delta_\gamma\mathrm J^-_{T,p_y}
  =\frac{\delta_m}{2}\mathrm J^+_{T,p_y}
  =\mathcal O(\delta_m q^0),\qquad
  \delta_\Phi\mathrm J^-_{T,p_y}
  \sim(\delta_m q)\mathrm J^-_{T,d_{xy}}
  =\mathcal O(\delta_m q^2).
\end{equation}
The bare-vertex correction $\delta_\gamma\mathrm J^-_{T,p_y}$ is therefore also the leading exciton transverse-current vertex-overlap correction of the $p_y$ mode.
\end{itemize}

The exciton-density and longitudinal exciton-current results are linked by the Ward identity. Since both electron--hole-symmetric overlaps of the $p_x$ mode vanish, its first-order identity is
\begin{equation}
  \omega_{p_x}\delta\mathrm J^-_{0,p_x}
  {}+q\delta\mathrm J^-_{L,p_x}=0,
  \qquad
  \delta\mathrm J^-_{0,p_x}=\delta_\Phi\mathrm J^-_{0,p_x},
  \qquad
  \delta\mathrm J^-_{L,p_x}
  =\delta_\gamma\mathrm J^-_{L,p_x}+\delta_\Phi\mathrm J^-_{L,p_x}.
  \label{eq:px-linearized-exciton-Ward-identity}
\end{equation}
The density term and $q\delta_\gamma\mathrm J^-_{L,p_x}$ are both $\mathcal O(\delta_mq)$, whereas $q\delta_\Phi\mathrm J^-_{L,p_x}=\mathcal O(\delta_mq^3)$. Hence
\begin{equation}
  \delta_\Phi\mathrm J^-_{0,p_x}
  =-\frac{\delta_m q}{2\omega_{p_x}}\mathrm J^+_{L,p_x}
  +\mathcal O(\delta_m q^2).
  \label{eq:px-induced-density-Ward-coefficient}
\end{equation}
This relation fixes the leading coefficient of the induced density overlap and confirms that the separately derived density and current powers are consistent with charge conservation in the exciton channel.

\begin{table}[!ht]
  \centering
  \setlength{\tabcolsep}{4pt}
  \renewcommand{\arraystretch}{1.35}
  \begin{tabular}{@{}p{0.10\textwidth} p{0.25\textwidth} p{0.30\textwidth} p{0.24\textwidth}@{}}
    \hline\hline
    \raggedright Operator channel
    & \raggedright Leading electron--hole-symmetric vertex overlap
    & \raggedright Eigenvector-induced correction $\delta_\Phi\mathrm J$
    & \raggedright Bare-vertex correction $\delta_\gamma\mathrm J$ \tabularnewline
    \hline
    \raggedright $\tilde{\mathrm J}^+_0$
    & \raggedright $\mathrm J^+_{0,p_x}=\mathcal O(q)$
    & \raggedright $\delta_\Phi\mathrm J^+_{0,(s\setminus\{G\})\cup d_{x^2-y^2}}=\mathcal O(\delta_mq^2)$; one order higher in $q$ than the electron--hole-symmetric term
    & \raggedright $\delta_\gamma\mathrm J^+_0=0$ \tabularnewline
    \raggedright $\tilde{\mathrm J}^+_L$
    & \raggedright $\mathrm J^+_{L,p_x}=\mathcal O(q^0)$
    & \raggedright $\delta_\Phi\mathrm J^+_{L,(s\setminus\{G\})\cup d_{x^2-y^2}}=\mathcal O(\delta_mq)$; one order higher in $q$ than the electron--hole-symmetric term
    & \raggedright $\delta_\gamma\mathrm J^+_{L,(s\setminus\{G\})\cup d_{x^2-y^2}}=2\delta_m\mathrm J^-_{L,(s\setminus\{G\})\cup d_{x^2-y^2}}=\mathcal O(\delta_mq)$; one order higher in $q$ than the electron--hole-symmetric term \tabularnewline
    \raggedright $\tilde{\mathrm J}^+_T$
    & \raggedright $\mathrm J^+_{T,p_y}=\mathcal O(q^0)$
    & \raggedright $\delta_\Phi\mathrm J^+_{T,d_{xy}}=\mathcal O(\delta_mq)$; one order higher in $q$ than the electron--hole-symmetric term
    & \raggedright $\delta_\gamma\mathrm J^+_{T,d_{xy}}=2\delta_m\mathrm J^-_{T,d_{xy}}=\mathcal O(\delta_mq)$; one order higher in $q$ than the electron--hole-symmetric term \tabularnewline
    \raggedright $\tilde{\mathrm J}^-_0$
    & \raggedright $\mathrm J^-_{0,(s\setminus\{G\})\cup d_{x^2-y^2}}=\mathcal O(q^2)$
    & \raggedright $\delta_\Phi\mathrm J^-_{0,p_x}=\mathcal O(\delta_mq)$; one order lower in $q$ than the electron--hole-symmetric term
    & \raggedright $\delta_\gamma\mathrm J^-_0=0$ \tabularnewline
    \raggedright $\tilde{\mathrm J}^-_L$
    & \raggedright $\mathrm J^-_{L,(s\setminus\{G\})\cup d_{x^2-y^2}}=\mathcal O(q)$
    & \raggedright $\delta_\Phi\mathrm J^-_{L,p_x}=\mathcal O(\delta_mq^2)$; one order higher in $q$ than the electron--hole-symmetric regular-mode term
    & \raggedright $\delta_\gamma\mathrm J^-_{L,p_x}=\delta_m\mathrm J^+_{L,p_x}/2=\mathcal O(\delta_mq^0)$; one order lower in $q$ than the electron--hole-symmetric term and two orders lower than the eigenvector-induced term \tabularnewline
    \raggedright $\tilde{\mathrm J}^-_T$
    & \raggedright $\mathrm J^-_{T,d_{xy}}=\mathcal O(q)$
    & \raggedright $\delta_\Phi\mathrm J^-_{T,p_y}=\mathcal O(\delta_mq^2)$; one order higher in $q$ than the electron--hole-symmetric term
    & \raggedright $\delta_\gamma\mathrm J^-_{T,p_y}=\delta_m\mathrm J^+_{T,p_y}/2=\mathcal O(\delta_mq^0)$; one order lower in $q$ than the electron--hole-symmetric term \tabularnewline
    \hline\hline
  \end{tabular}
  \caption{Long-wavelength power counting for regular target modes $n$ at $\bm q=q\hat{x}$. The columns give the electron--hole-symmetric, eigenvector-induced, and bare-vertex contributions. Unlisted regular sectors vanish by symmetry or start at higher order in $q$; the Goldstone mode is treated separately.}
  \label{tab:eigenvector-induced-overlap-power-counting}
\end{table}
\FloatBarrier

\paragraph{Goldstone mode}
We now set the target mode to $n=G$. Equation \eqref{eq:first-order-eigenvector-correction} gives the Goldstone eigenvector an $\mathcal O(\delta_m)$ $p_x$ component. Its consequences are listed separately by operator channel.

\begin{itemize}
\item \emph{Charge-density and charge-current channels, $\tilde{\mathrm J}^+_{0,G}$, $\tilde{\mathrm J}^+_{L,G}$, and $\tilde{\mathrm J}^+_{T,G}$.}
For every regular intermediate $p_x$ mode, the density formula in Eq. \eqref{eq:paired-eigenvector-induced-overlap} combines its two frequency partners. Since $\mathrm J^+_{0,p_x}=\mathcal O(q)$ and $\omega_G\mathcal A_{p_x,G}+\omega_{p_x}\mathcal B_{p_x,G}=\mathcal O(\delta_mq)$,
\begin{equation}
  \left.\delta_\Phi\mathrm J^+_{0,G}\right|_{\pm p_x}
  =\mathcal O(\delta_mq^2).
  \label{eq:Goldstone-charge-density-pair-cancellation}
\end{equation}
Thus the sum over all intermediate $p_x$ modes has the same power. Since the density bare vertex has no electron--hole-asymmetry correction,
\begin{equation}
  \delta_\gamma\mathrm J^+_{0,G}=0,
  \qquad
  \delta\mathrm J^+_{0,G}=\delta_\Phi\mathrm J^+_{0,G}
  =\mathcal O(\delta_mq^2).
  \label{eq:induced-Goldstone-charge-density-power}
\end{equation}

In the electron--hole-symmetric limit, $\mathrm J^+_{L,G}=0$ because the even Goldstone mode cannot couple to the odd charge-current operator. Electron--hole asymmetry generates two first-order contributions, with the bare-vertex correction given by Eq. \eqref{eq:current_correction}:
\begin{equation}
  \tilde{\mathrm J}^+_{L,G}
  =\delta\mathrm J^+_{L,G}
  =\delta_\gamma\mathrm J^+_{L,G}+\delta_\Phi\mathrm J^+_{L,G},
  \qquad
  \delta_\gamma\mathrm J^+_{L,G}=2\delta_m\mathrm J^-_{L,G}.
\end{equation}
Both $\delta_\gamma\mathrm J^+_{L,G}$ and $\delta_\Phi\mathrm J^+_{L,G}$ are allowed to approach nonzero constants as $q\to0$. However, neither term is a physical vertex overlap by itself. Charge conservation constrains their sum through the Ward identity
\begin{equation}
  \tilde\omega_G(q)\tilde{\mathrm J}^+_{0,G}(q)
  +q\tilde{\mathrm J}^+_{L,G}(q)=0.
  \label{eq:asymmetric-Goldstone-charge-Ward-identity}
\end{equation}
The preceding charge-density result gives $\delta\mathrm J^+_{0,G}=\mathcal O(\delta_mq^2)$. Moreover, $\omega_G/q$ approaches the finite Goldstone velocity as $q\to0$. Therefore, to first order in $\delta_m$, the Ward identity gives
\begin{equation}
  \delta\mathrm J^+_{L,G}
  =-\frac{\omega_G}{q}\delta\mathrm J^+_{0,G}
  =\mathcal O(\delta_mq^2).
  \label{eq:asymmetric-Goldstone-current-cancellation}
\end{equation}
Thus the constant parts of $\delta_\gamma\mathrm J^+_{L,G}$ and $\delta_\Phi\mathrm J^+_{L,G}$ must cancel, required by charge conservation. Since the total longitudinal overlap is fixed by the quadratic density overlap, it cannot contain a term proportional to $q$ either. Reflection symmetry independently sets the transverse charge-current vertex overlap to zero. Thus, to first order in $\delta_m$,
\begin{equation}
  \tilde{\mathrm J}^+_{0,G}=\mathcal O(\delta_mq^2),
  \qquad
  \tilde{\mathrm J}^+_{L,G}=\mathcal O(\delta_mq^2),
  \qquad
  \tilde{\mathrm J}^+_{T,G}=0.
  \label{eq:asymmetric-Goldstone-charge-overlaps}
\end{equation}
The transverse result holds to all orders in $\delta_m$ as long as reflection symmetry is preserved.

\item \emph{Exciton-density and exciton-current channels, $\tilde{\mathrm J}^-_{0,G}$, $\tilde{\mathrm J}^-_{L,G}$, and $\tilde{\mathrm J}^-_{T,G}$.}
At zero temperature, the mean-field self-consistent equation shows that the exciton compressibility $\kappa$ depends on the electron and hole masses only through the reduced mass $m$. Thus, at fixed $m$, $\tilde\kappa=\kappa$, and the static compressibility gives
\begin{subequations}\label{eq:asymmetric-Goldstone-vertex-overlaps}
\begin{equation}
  \lim_{q\to0}\frac{2}{\mathcal V}
  \left|\tilde{\mathrm J}^-_{0,G}(q)\right|^2
  =\tilde\kappa=\kappa.
\end{equation}
The exciton Ward identity then fixes the longitudinal-current vertex overlap. Using $\tilde\omega_G/q=\tilde v=v\sqrt{1-\delta_m^2}$ gives
\begin{equation}
  \tilde\omega_G(q)\tilde{\mathrm J}^-_{0,G}(q)
  +q\tilde{\mathrm J}^-_{L,G}(q)=0,
  \qquad
  \tilde{\mathrm J}^-_{L,G}(q)
  =-\frac{\tilde\omega_G(q)}{q}\tilde{\mathrm J}^-_{0,G}(q)
  =-\tilde v\tilde{\mathrm J}^-_{0,G}(q).
\end{equation}
\end{subequations}
Thus the Goldstone exciton-density vertex overlap is unchanged within the zero-temperature mean-field approximation, while the longitudinal-current vertex overlap first changes at $\mathcal O(\delta_m^2)$ through $\tilde v$. Reflection symmetry keeps $\tilde{\mathrm J}^-_{T,G}=0$.
\end{itemize}

The Goldstone mode therefore produces no leading long-wavelength charge--exciton mixed pole. A subleading mixed response can remain at finite $q$ and requires the full microscopic eigenvector.

The regular-mode power counting in Table \ref{tab:eigenvector-induced-overlap-power-counting} and the Goldstone vertex-overlap results in Eqs. \eqref{eq:asymmetric-Goldstone-charge-overlaps} and \eqref{eq:asymmetric-Goldstone-vertex-overlaps} are verified numerically in Appendix \ref{app:numerical-asymmetric-overlap-power-counting}.

\subsubsection{\texorpdfstring{Effect on the correlation functions}{Effect on the correlation functions}}\label{app:eh-asymmetry-correlation-functions}

We use the spectral decomposition of Appendix \ref{app:response-kernel}.  For $\bm q=q\hat{x}$, write
\begin{equation}
  \tilde C^{\sigma\sigma'}_{\mu\nu}
  \equiv\tilde C_{\hat j^\sigma_{p\mu}\hat j^{\sigma'}_{p\nu}}
  =\tilde C^{\prime\,\sigma\sigma'}_{\mu\nu}
  +\tilde C^{G,\sigma\sigma'}_{\mu\nu},
  \label{eq:asymmetric-correlation-decomposition}
\end{equation}
where the prime denotes the regular gapped modes and the second term is the Goldstone $1s$ contribution. Pairing the positive- and negative-frequency branches gives
\begin{equation}
  \tilde C^{\prime\,\sigma\sigma'}_{\mu\nu}
  =\frac{1}{\mathcal V}\sum_{\substack{\tilde\omega_n>0\\n\ne1s}}
  (\tilde{\mathrm J}^{\sigma}_{\mu,n})^*
  \tilde{\mathrm J}^{\sigma'}_{\nu,n}\,
  \mathcal S_{\mu\nu,n},
  \qquad
  \mathcal S_{\mu\nu,n}=
  \begin{cases}
    \dfrac{2\tilde\omega_n^2}{(\omega^+)^2-\tilde\omega_n^2},
    & \mu,\nu\text{ both density or both current},\\[6pt]
    \dfrac{2\omega^+\tilde\omega_n}{(\omega^+)^2-\tilde\omega_n^2},
    & \text{one density and one current index}.
  \end{cases}
  \label{eq:asymmetric-regular-spectral-representation}
\end{equation}

\paragraph{Regular gapped modes contribution.}
According to Table \ref{tab:eigenvector-induced-overlap-power-counting}, the spatial vertex overlaps needed at leading order in $q$ are
\begin{equation}
\begin{aligned}
  \tilde{\mathrm J}^{+}_{L,n}
  &=\mathrm J^{+}_{L,n}\delta_{n,p_x}+\mathcal O(\delta_mq),
  &
  \tilde{\mathrm J}^{+}_{T,n}
  &=\mathrm J^{+}_{T,n}\delta_{n,p_y}+\mathcal O(\delta_mq),\\
  \tilde{\mathrm J}^{-}_{L,n}
  &=\mathrm J^{-}_{L,n}\delta_{n,(s\setminus\{G\})\cup d_{x^2-y^2}}
  +\frac{\delta_m}{2}\mathrm J^{+}_{L,n}\delta_{n,p_x}
  +\mathcal O(\delta_mq^2),
  &
  \tilde{\mathrm J}^{-}_{T,n}
  &=\mathrm J^{-}_{T,n}\delta_{n,d_{xy}}
  +\frac{\delta_m}{2}\mathrm J^{+}_{T,n}\delta_{n,p_y}
  +\mathcal O(\delta_mq^2).
\end{aligned}
\label{eq:leading-asymmetric-spatial-overlaps}
\end{equation}
Here $\delta_{n,\mathcal S}=1$ when mode $n$ belongs to sector $\mathcal S$ and is zero otherwise.  The first term in each line is the leading electron--hole-symmetric vertex overlap.  The second term in the exciton-current overlaps is the leading electron--hole-asymmetry correction.  We show only the spatial components because the Ward identity fixes the density and density--current components.

In Eq. \eqref{eq:leading-asymmetric-spatial-overlaps}, the two terms in each exciton-current vertex overlap couple to disjoint mode sectors. For the longitudinal response, the electron--hole-symmetric exciton-current term couples to the $s\setminus\{G\}$ and $d_{x^2-y^2}$ sectors, whereas the mass-asymmetry-induced charge-current term couples to the $p_x$ sector. For the transverse response, the corresponding sectors are $d_{xy}$ and $p_y$. Their cross terms therefore vanish in the mode sum in Eq. \eqref{eq:asymmetric-regular-spectral-representation}. The frequency correction of a regular gapped mode is higher order in the present approximation, so we set $\tilde\omega_n\simeq\omega_n$. Substituting the vertex overlaps into Eq. \eqref{eq:asymmetric-regular-spectral-representation} then gives
\begin{subequations}\label{eq:asymmetric-regular-correlation-approximation}
  \begin{align}
    \tilde C^{\prime\,++}_{\lambda\lambda}
    &\simeq C^{\prime\,++}_{\lambda\lambda},\\
    \tilde C^{\prime\,--}_{\lambda\lambda}
    &\simeq C^{\prime\,--}_{\lambda\lambda}
    +\frac{\delta_m^2}{4}C^{\prime\,++}_{\lambda\lambda},\\
    \tilde C^{\prime\,+-}_{\lambda\lambda}
    &=\tilde C^{\prime\,-+}_{\lambda\lambda}
    \simeq\frac{\delta_m}{2}C^{\prime\,++}_{\lambda\lambda}.
  \end{align}
\end{subequations}
Here $\lambda\in\{L,T\}$ labels the longitudinal and transverse components, and $\simeq$ retains the leading power of $q$.

\paragraph{Goldstone mode contribution.}
The $1s$ mode must be treated separately because $\tilde\omega_{1s}=\tilde vq$, with $\tilde v=v\sqrt{1-\delta_m^2}$.  Equation \eqref{eq:asymmetric-Goldstone-vertex-overlaps} gives
\begin{subequations}\label{eq:asymmetric-Goldstone-correlation}
  \begin{align}
    \tilde C^{G,--}_{00}
    &=\frac{\kappa\tilde v^2q^2}{(\omega^+)^2-\tilde v^2q^2},\\
    \tilde C^{G,--}_{L0}
    =\tilde C^{G,--}_{0L}
    &=-\frac{\kappa\tilde v^2\omega^+q}{(\omega^+)^2-\tilde v^2q^2},\\
    \tilde C^{G,--}_{LL}
    &=\frac{\kappa\tilde v^4q^2}{(\omega^+)^2-\tilde v^2q^2},
    \qquad \tilde C^{G,--}_{TT}=0.
  \end{align}
\end{subequations}
By Eq. \eqref{eq:asymmetric-Goldstone-charge-overlaps}, the Goldstone charge vertex overlaps begin at $\mathcal O(\delta_mq^2)$.  They therefore produce neither a charge nor a charge--exciton Goldstone pole at the leading powers retained above.

Combining the gapped and Goldstone sectors yields
\begin{subequations}\label{eq:asymmetric-full-correlation-approximation}
  \begin{align}
    \tilde C^{++}_{\lambda\lambda}
    &\simeq C^{++}_{\lambda\lambda},\\
    \tilde C^{--}_{\lambda\lambda}
    &\simeq C^{--}_{\lambda\lambda}
    +\frac{\delta_m^2}{4}C^{++}_{\lambda\lambda}
    +\Delta C^{G,--}_{\lambda\lambda},\\
    \tilde C^{+-}_{\lambda\lambda}
    &=\tilde C^{-+}_{\lambda\lambda}
    \simeq\frac{\delta_m}{2}C^{++}_{\lambda\lambda},
    \qquad \lambda\in\{L,T\},
  \end{align}
\end{subequations}
Here $\Delta C^{G,--}_{\lambda\lambda}\equiv
\tilde C^{G,--}_{\lambda\lambda}-C^{G,--}_{\lambda\lambda}$ is explicitly written as
\begin{equation}
  \Delta C^{G,--}_{LL}
  =\kappa q^2\left[
  \frac{\tilde v^4}{(\omega^+)^2-\tilde v^2q^2}
  -\frac{v^4}{(\omega^+)^2-v^2q^2}
  \right],
  \qquad
  \Delta C^{G,--}_{TT}=0.
  \label{eq:Goldstone-correlation-shift}
\end{equation}
Equation \eqref{eq:Goldstone-correlation-shift} is the $q\to0$ form; its finite-$q$ numerical evaluation is replaced by the mode-resolved Goldstone pole and vertex overlaps in Appendix \ref{app:numerical-asymmetric-response}.

\paragraph{The total response functions.}
For diamagnetic response, equations \eqref{eq:current-diamagnetic-charge} and \eqref{eq:current-diamagnetic-exciton} give
$D_+=e^2n_X/m$, $D_-=e^2n_X/(4m)$, and
$\tilde D^{++}=D_+$, $\tilde D^{--}=D_-$, and
$\tilde D^{+-}=\tilde D^{-+}=\delta_mD_+/2$.  Defining
$P_{\mu\nu}=(1-\delta_{\mu0})(1-\delta_{\nu0})\delta_{\mu\nu}$, the response kernel is
\begin{equation}
  \tilde K^{\sigma\sigma'}_{\mu\nu}
  =-\tilde C^{\sigma\sigma'}_{\mu\nu}
  -\tilde D^{\sigma\sigma'}P_{\mu\nu}.
  \label{eq:asymmetric-response-from-correlation}
\end{equation}
Combining Eq. \eqref{eq:asymmetric-full-correlation-approximation} with the diamagnetic terms gives, for $\lambda\in\{L,T\}$,
\begin{subequations}\label{eq:K_tilde_approx}
  \begin{align}
    \tilde K^{++}_{\lambda,\mathrm{approx}}
    &=K^{++}_{\lambda},\\
    \tilde K^{--}_{\lambda,\mathrm{approx}}
    &=K^{--}_{\lambda}+\frac{\delta_m^2}{4}K^{++}_{\lambda}
    +\frac{\delta_m^2D_+}{4}-\Delta C^{G,--}_{\lambda\lambda},\label{eq:K_tilde_approx_exciton}\\
    \tilde K^{+-}_{\lambda,\mathrm{approx}}
    &=\tilde K^{-+}_{\lambda,\mathrm{approx}}
    =\frac{\delta_m}{2}K^{++}_{\lambda}.
  \end{align}
\end{subequations}
The gauge-invariant Goldstone contribution is
\begin{equation}
  \tilde K^{G,--}_{\mu\nu}(\omega,q\hat x)
  =-\frac{\kappa\tilde v^2}{(\omega^+)^2-\tilde v^2q^2}
  \begin{bmatrix}
    q^2&-\omega^+q&0\\
    -\omega^+q&(\omega^+)^2&0\\
    0&0&0
  \end{bmatrix}_{\mu\nu}.
  \label{eq:asymmetric-Goldstone-response-matrix}
\end{equation}
The Goldstone shift vanishes in the transverse channel but is essential in the longitudinal channel.  Using $D_+/4=\kappa v^2$ and $\tilde v^2=v^2(1-\delta_m^2)$,
\begin{equation}
  \frac{\delta_m^2D_+}{4}-\Delta C^{G,--}_{LL}
  =\frac{\kappa v^2\delta_m^2(\omega^+)^4}
  {[(\omega^+)^2-\tilde v^2q^2][(\omega^+)^2-v^2q^2]}.
  \label{eq:Goldstone-Ward-cancellation-correlation}
\end{equation}
The Goldstone correction $\Delta C^{G,--}_{LL}$ is required by the longitudinal Ward identity. Without it, Eq. \eqref{eq:K_tilde_approx_exciton} would approach the nonzero constant $\delta_m^2D_+/4$ as $\omega\to0$. With this correction, the last two terms in Eq. \eqref{eq:K_tilde_approx_exciton} combine as shown above and scale as $\mathcal O(\omega^4)$ at fixed nonzero $q$. Consequently, $\tilde K^{--}_L=\mathcal O(\omega^2)$. The exciton-current responses are therefore
\begin{subequations}\label{eq:asymmetric-exciton-response-LT}
  \begin{align}
    \tilde K^{--}_{L,\mathrm{approx}}
    &=K^{--}_{L}+\frac{\delta_m^2}{4}K^{++}_{L}
    +\frac{\kappa v^2\delta_m^2(\omega^+)^4}
    {[(\omega^+)^2-\tilde v^2q^2][(\omega^+)^2-v^2q^2]},\\
    \tilde K^{--}_{T,\mathrm{approx}}
    &=K^{--}_{T}+\frac{\delta_m^2}{4}K^{++}_{T}
    +\frac{e^2n_X\delta_m^2}{4m}.
  \end{align}
\end{subequations}
All kernels on the right-hand side have arguments $(\omega,q)$. For $\bm q=q\hat x$, the remaining components follow from the Ward identity:
\begin{equation}
  \tilde K^{\sigma\sigma'}_{\mu\nu}=
  \begin{bmatrix}
    \dfrac{q^2}{\omega^2}\tilde K^{\sigma\sigma'}_L&
    -\dfrac{q}{\omega}\tilde K^{\sigma\sigma'}_L&0\\
    -\dfrac{q}{\omega}\tilde K^{\sigma\sigma'}_L&
    \tilde K^{\sigma\sigma'}_L&0\\
    0&0&\tilde K^{\sigma\sigma'}_T
  \end{bmatrix}.
  \label{eq:asymmetric-response-long-wavelength-matrix}
\end{equation}

}

\clearpage
{
\section{\texorpdfstring{Supplementary data for the nonmagnetic case}{Supplementary data for the nonmagnetic case}}\label{app:supplementary-data}

\subsection{\texorpdfstring{Numerical parameters}{Numerical parameters}}
Unless stated otherwise, all nonmagnetic calculations use an $80\times80$ momentum mesh with $k_{\rm cut}a_B^*=2.5$ and $\Delta k a_B^*=0.0625$.

\subsection{\texorpdfstring{Long-wavelength vertex-overlap power counting}{Long-wavelength vertex-overlap power counting}}\label{app:numerical-overlap-power-counting}

The power-counting calculations are an exception to the default mesh: we use $N_k=128$, $k_{\rm cut}a_B^*=4$, and $\Delta k a_B^*=0.0625$ to suppress finite-cutoff errors in the Ward identity. We test Table \ref{tab:vertex-overlap-power-counting} in the electron--hole-symmetric limit. For $\bm q=q\hat{x}$, we sample 25 momenta over $0.0015625\leq q a_B^*\leq0.05$ and fit the $1s$ Goldstone mode and the regular $1p_x$, $1p_y$, $2s$, $1d_{x^2-y^2}$, and $1d_{xy}$ modes over $q a_B^*\leq0.02$ to
\begin{equation}
  \ln\!\left|\mathrm{J}^{\sigma}_{\mu,n}(q)\right|
  =m\ln q+b .
  \label{eq:numerical-overlap-fit}
\end{equation}
Table \ref{tab:numerical-overlap-power-counting} and Fig. \ref{fig:numerical-overlap-power-counting} summarize the results.

\begin{table}[!ht]
  
  \centering
  \small
  \setlength{\tabcolsep}{3.5pt}
  \renewcommand{\arraystretch}{1.08}
  \begin{minipage}[t]{0.62\textwidth}
    \vspace{0pt}\centering
    \begin{tabular}{@{}c c c@{}}
      \multicolumn{3}{c}{Regular gapped modes} \\
      \hline\hline
      Vertex overlap & Predicted & Fitted $m$ \\
      \hline
      $\left|\mathrm{J}^{+}_{0,1p_x}\right|$ & $1$ & $0.985$ \\
      $\left|\mathrm{J}^{+}_{L,1p_x}\right|$ & $0$ & $0.000$ \\
      $\left|\mathrm{J}^{+}_{T,1p_y}\right|$ & $0$ & $0.000$ \\
      $\left|\mathrm{J}^{-}_{0,2s}\right|$ & $2$ & $2.000$ \\
      $\left|\mathrm{J}^{-}_{L,2s}\right|$ & $1$ & $1.000$ \\
      $\left|\mathrm{J}^{-}_{0,1d_{x^2-y^2}}\right|$ & $2$ & $2.000$ \\
      $\left|\mathrm{J}^{-}_{L,1d_{x^2-y^2}}\right|$ & $1$ & $1.000$ \\
      $\left|\mathrm{J}^{-}_{T,1d_{xy}}\right|$ & $1$ & $1.000$ \\
      \hline\hline
    \end{tabular}
  \end{minipage}%
  \hfill
  \begin{minipage}[t]{0.34\textwidth}
    \vspace{0pt}\centering
    \begin{tabular}{@{}c c c@{}}
      \multicolumn{3}{c}{$1s$ Goldstone mode} \\
      \hline\hline
      Vertex overlap & Predicted & Fitted $m$ \\
      \hline
      $\left|\mathrm{J}^{-}_{0,1s}\right|$ & $0$ & $0.002$ \\
      $\left|\mathrm{J}^{-}_{L,1s}\right|$ & $0$ & $0.000$ \\
      \hline\hline
    \end{tabular}
  \end{minipage}
  \caption{Long-wavelength powers in the electron--hole-symmetric limit. Predictions follow from Table \ref{tab:vertex-overlap-power-counting}; fits use Eq. \eqref{eq:numerical-overlap-fit}. The $2s$ and $1d_{x^2-y^2}$ modes share the same predicted powers but are listed separately. For $2s$, orthogonality removes the symmetry-allowed constant density term, and the Ward identity gives $\mathrm J^-_{0,2s}=\mathcal O(q^2)$.}
  \label{tab:numerical-overlap-power-counting}
\end{table}

\begin{figure}[!ht]
  
  \centering
  \includegraphics[width=0.90\textwidth]{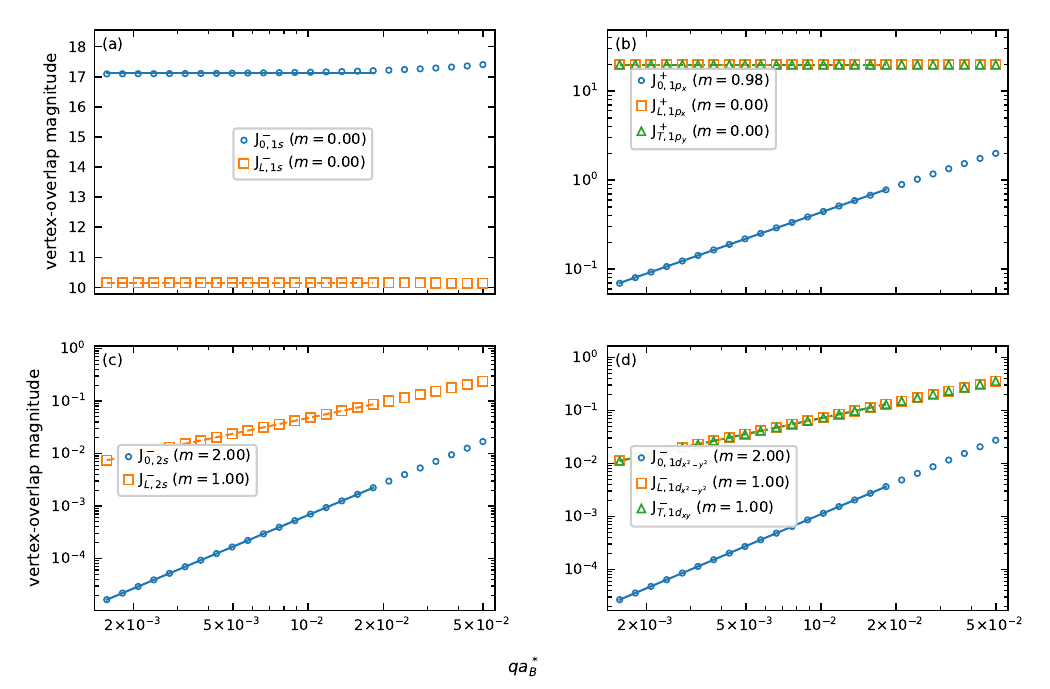}
  \caption{Electron--hole-symmetric vertex-overlap scaling at $N_k=128$ and $k_{\rm cut}a_B^*=4$. Open markers show all 25 momenta; lines are fits over $q a_B^*\leq0.02$. (a) $1s$ Goldstone exciton vertices. (b) $1p$ charge vertices. (c) $2s$ exciton vertices. (d) $1d$ exciton vertices. Parentheses give the powers $m$.}
  \label{fig:numerical-overlap-power-counting}
\end{figure}

\subsection{\texorpdfstring{Vertex overlaps with electron--hole asymmetry}{Vertex overlaps with electron--hole asymmetry}}\label{app:numerical-asymmetric-overlap-power-counting}

For $\delta_m=0.2$, we use $N_k=128$, $k_{\rm cut}a_B^*=4$, and $\Delta k a_B^*=0.0625$, with the same momentum range as above. The fits use the 18 points with $q a_B^*\leq0.02$. Table \ref{tab:numerical-asymmetric-overlap-power-counting} and Fig. \ref{fig:numerical-asymmetric-overlap-power-counting} confirm the powers derived in Appendix \ref{app:vertex_overlap_asymmetry}.


\begin{table}[!ht]
  
  \centering
  \small
  \setlength{\tabcolsep}{3pt}
  \renewcommand{\arraystretch}{1.06}
  \begin{minipage}[t]{0.62\textwidth}
    \vspace{0pt}\centering
    \begin{tabular}{@{}c c c@{}}
      \multicolumn{3}{c}{Regular gapped modes} \\
      \hline\hline
      Vertex overlap & Predicted & Fitted $m$ \\
      \hline
      \multicolumn{3}{c}{Eigenvector correction} \\
      $|\delta_\Phi\mathrm J^+_{0,2s}|$ & $2$ & $2.020$ \\
      $|\delta_\Phi\mathrm J^+_{L,2s}|$ & $1$ & $1.079$ \\
      $|\delta_\Phi\mathrm J^+_{0,1d_{x^2-y^2}}|$ & $2$ & $2.027$ \\
      $|\delta_\Phi\mathrm J^+_{L,1d_{x^2-y^2}}|$ & $1$ & $1.039$ \\
      $|\delta_\Phi\mathrm J^+_{T,1d_{xy}}|$ & $1$ & $1.000$ \\
      $|\delta_\Phi\mathrm J^-_{0,1p_x}|$ & $1$ & $0.985$ \\
      $|\delta_\Phi\mathrm J^-_{L,1p_x}|$ & $2$ & $2.000$ \\
      $|\delta_\Phi\mathrm J^-_{T,1p_y}|$ & $2$ & $2.000$ \\
      \hline
      \multicolumn{3}{c}{Bare-vertex correction} \\
      $|\delta_\gamma\mathrm J^+_{L,2s}|$ & $1$ & $1.000$ \\
      $|\delta_\gamma\mathrm J^+_{L,1d_{x^2-y^2}}|$ & $1$ & $1.000$ \\
      $|\delta_\gamma\mathrm J^+_{T,1d_{xy}}|$ & $1$ & $1.000$ \\
      $|\delta_\gamma\mathrm J^-_{L,1p_x}|$ & $0$ & $0.000$ \\
      $|\delta_\gamma\mathrm J^-_{T,1p_y}|$ & $0$ & $0.000$ \\
      \hline\hline
    \end{tabular}
  \end{minipage}%
  \hfill
  \begin{minipage}[t]{0.34\textwidth}
    \vspace{0pt}\centering
    \begin{tabular}{@{}c c c@{}}
      \multicolumn{3}{c}{$1s$ Goldstone mode} \\
      \hline\hline
      Vertex overlap & Predicted & Fitted $m$ \\
      \hline
      $|\delta_\Phi\mathrm J^+_{L,1s}|$ & $0$ & $0.000$ \\
      $|\delta_\gamma\mathrm J^+_{L,1s}|$ & $0$ & $0.000$ \\
      $|\delta\mathrm J^+_{L,1s}|$ & $2$ & $1.927$ \\
      $|\delta\mathrm J^+_{0,1s}|$ & $2$ & $1.970$ \\
      $|\tilde{\mathrm J}^-_{0,1s}|$ & $0$ & $0.002$ \\
      $|\tilde{\mathrm J}^-_{L,1s}|$ & $0$ & $0.000$ \\
      \hline\hline
    \end{tabular}
  \end{minipage}
  \caption{Long-wavelength powers of the electron--hole-asymmetry-induced vertex overlaps. In the $1s$ charge channel, the constant terms in $\delta_\Phi\mathrm J^+_{L,1s}$ and $\delta_\gamma\mathrm J^+_{L,1s}$ cancel in the total correction, as required by the Ward identity.}
  \label{tab:numerical-asymmetric-overlap-power-counting}
\end{table}

\begin{figure}[!ht]
  
  \centering
  \includegraphics[width=0.96\textwidth]{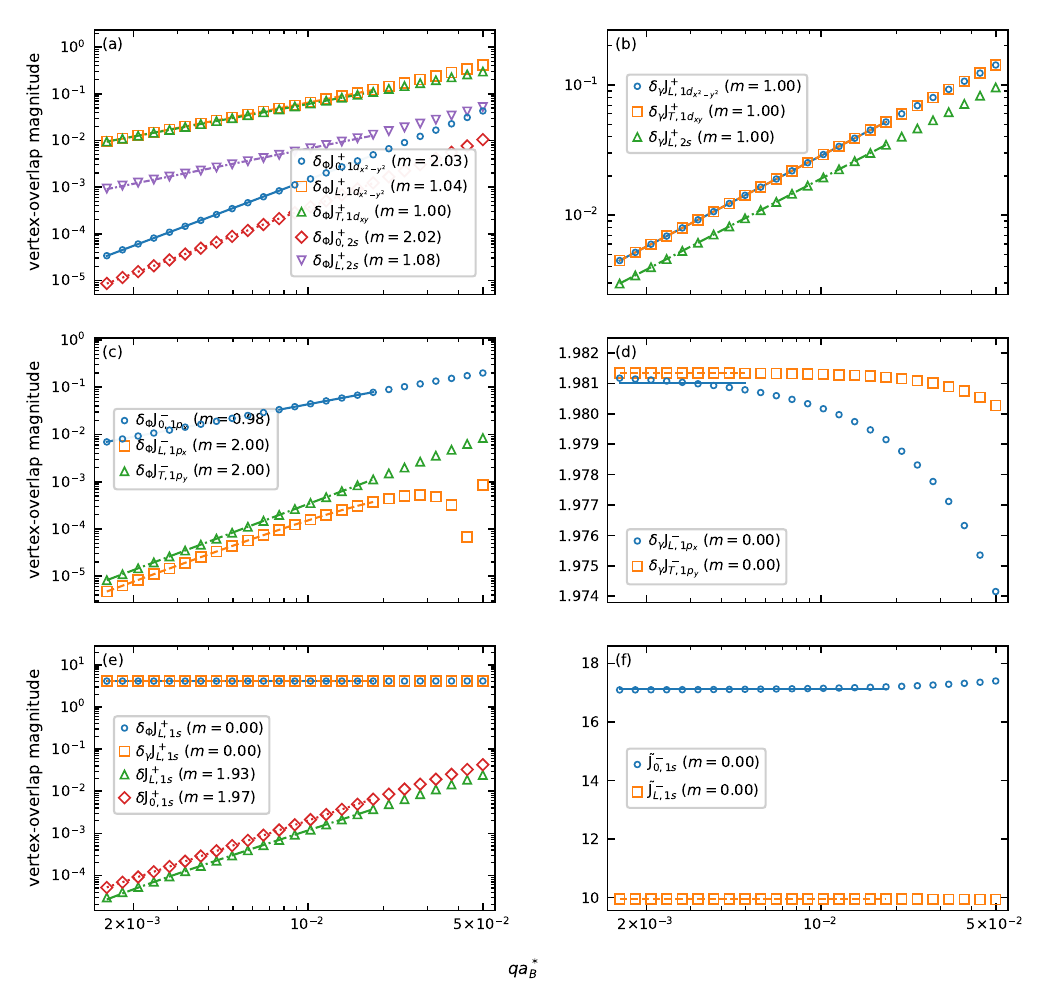}
  \caption{Electron--hole-asymmetry-induced vertex-overlap scaling at $\delta_m=0.2$. Open markers show all 25 momenta; lines are fits over $q a_B^*\leq0.02$. The rows show regular $\mathrm J^+$, regular $\mathrm J^-$, and $1s$ Goldstone contributions. For the regular modes, the left and right columns give $\delta_\Phi\mathrm J$ and $\delta_\gamma\mathrm J$, respectively; the last row gives the Goldstone charge and exciton channels. The horizontal $q^0$ guides in (d) use $q a_B^*\leq0.005$. Parentheses give the powers $m$.}
  \label{fig:numerical-asymmetric-overlap-power-counting}
\end{figure}

\FloatBarrier
\subsection{\texorpdfstring{Response functions with electron--hole asymmetry}{Response functions with electron--hole asymmetry}}\label{app:numerical-asymmetric-response}

We compare the full response at $\delta_m=0.2$ with Eqs. \eqref{eq:K_tilde_approx}, using the complete electron--hole-symmetric response at each $(\omega,q)$ rather than its $q\to0$ limit. At finite $q$, the Goldstone contribution is reconstructed from the numerical pole and vertex overlap,
\begin{equation}
  K^{G,--}_{\lambda}(\omega,q)
  =-\frac{1}{\mathcal V}
  \frac{2\omega_G^2(q)\left|\mathrm J^-_{\lambda,G}(q)\right|^2}
  {(\omega^+)^2-\omega_G^2(q)},
  \qquad \lambda\in\{L,T\}.
  \label{eq:numerical-Goldstone-response}
\end{equation}
The corresponding electron--hole-asymmetric term $\tilde K^{G,--}_{\lambda}$ is obtained by replacing $\omega_G$ and $\mathrm J^-_{\lambda,G}$ with their electron--hole-asymmetric values. With all kernels evaluated at $(\omega,q)$, the electron--hole-asymmetric response is approximated by
\begin{subequations}\label{eq:numerical-response-approximation}
  \begin{align}
    \tilde K^{++}_{\lambda,\mathrm{approx}}
    &=K^{++}_{\lambda},\\
    \tilde K^{--}_{\lambda,\mathrm{approx}}
    &=K^{--}_{\lambda}+\frac{\delta_m^2}{4}K^{++}_{\lambda}
    +\frac{\delta_m^2D_+}{4}
    +\tilde K^{G,--}_{\lambda}
    -K^{G,--}_{\lambda},\\
    \tilde K^{+-}_{\lambda,\mathrm{approx}}
    &=\tilde K^{-+}_{\lambda,\mathrm{approx}}
    =\frac{\delta_m}{2}K^{++}_{\lambda},\\
    \tilde K^{--}_{00,\mathrm{approx}}
    &=\frac{q^2}{(\omega^+)^2}\tilde K^{--}_{L,\mathrm{approx}}.
    \label{eq:numerical-density-response-approximation}
  \end{align}
\end{subequations}

Figures \ref{fig:charge_LT_approx}--\ref{fig:charge_exciton_LT_approx} show that Eq. \eqref{eq:numerical-response-approximation} reproduces the full electron--hole-asymmetric response at small $q$.

\begin{figure}[!ht]
  
  \centering
  \includegraphics[width=\textwidth]{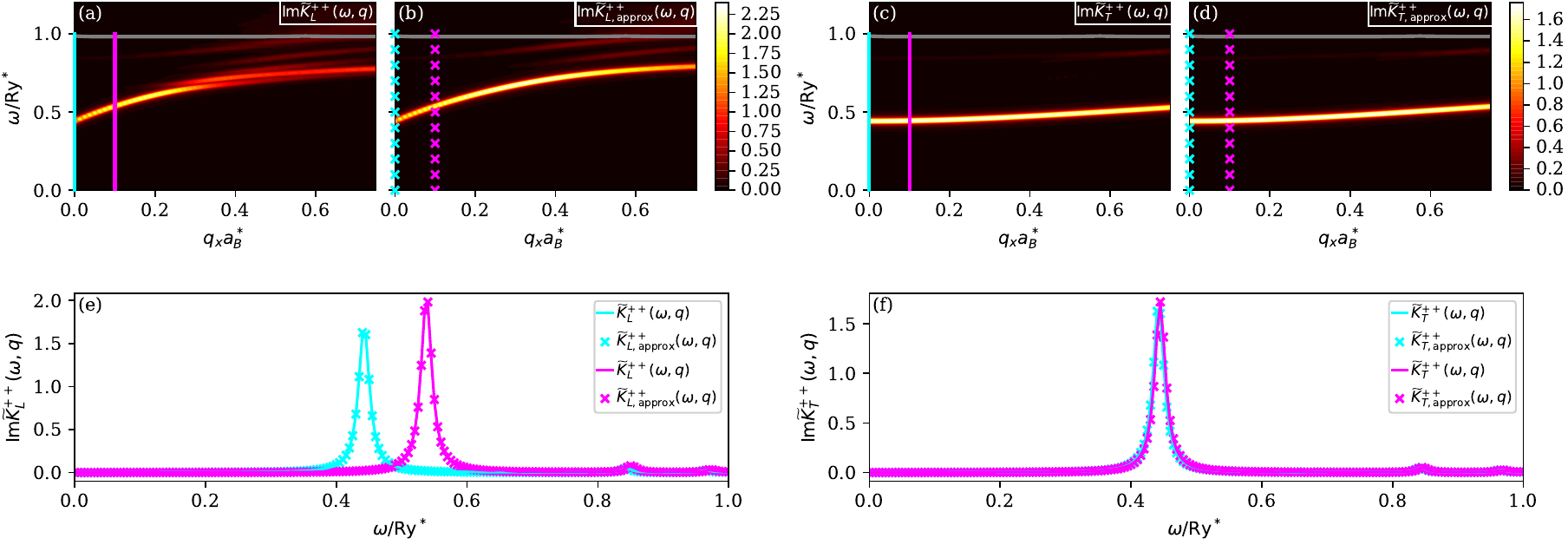}
  \caption{Charge-current response at $\delta_m=0.2$. (a,b) Full and approximate longitudinal spectra; (c,d) transverse spectra. (e,f) Corresponding cuts at $q a_B^*=0$ (cyan) and $0.1$ (magenta). Solid lines show the full response; crosses show Eq. \eqref{eq:numerical-response-approximation}.}
  \label{fig:charge_LT_approx}
\end{figure}

\begin{figure}[!ht]
  
  \centering
  \includegraphics[width=\textwidth]{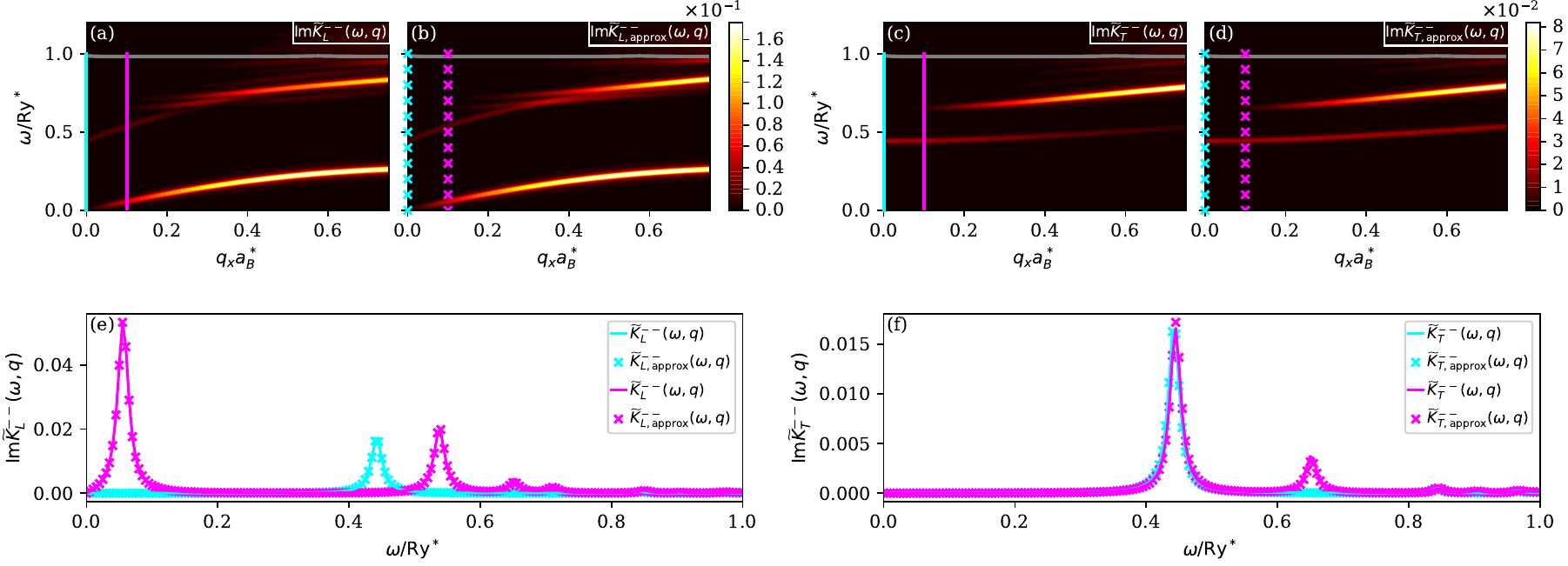}
  \caption{Exciton-current response at $\delta_m=0.2$, with the Goldstone contribution reconstructed from Eq. \eqref{eq:numerical-Goldstone-response}. Panels and symbols follow Fig. \ref{fig:charge_LT_approx}.}
  \label{fig:exciton_LT_approx}
\end{figure}

\begin{figure}[!ht]
  
  \centering
  \includegraphics[width=\textwidth]{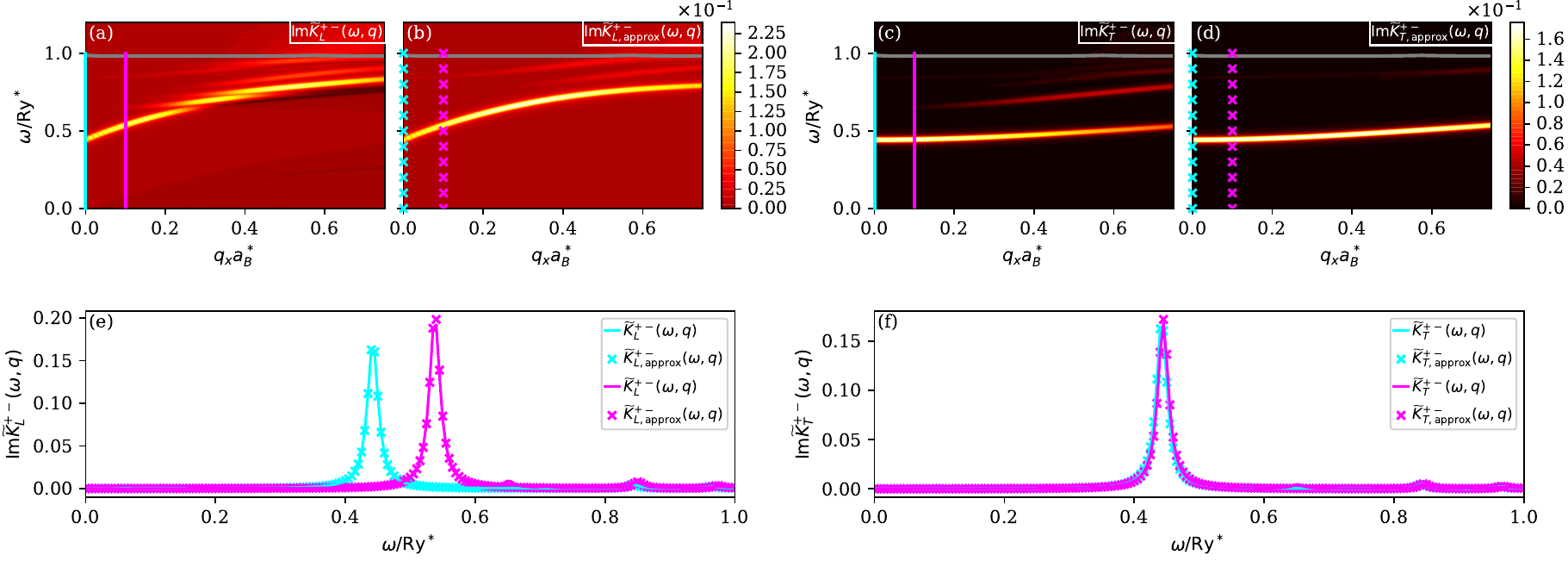}
  \caption{Charge--exciton response at $\delta_m=0.2$. Panels and symbols follow Fig. \ref{fig:charge_LT_approx}.}
  \label{fig:charge_exciton_LT_approx}
\end{figure}

Figure \ref{fig:exciton_00_approx} compares the full exciton-density response with the approximation in Eq. \eqref{eq:numerical-density-response-approximation}.

\begin{figure}[!ht]
  
  \centering
  \includegraphics[width=0.55\textwidth]{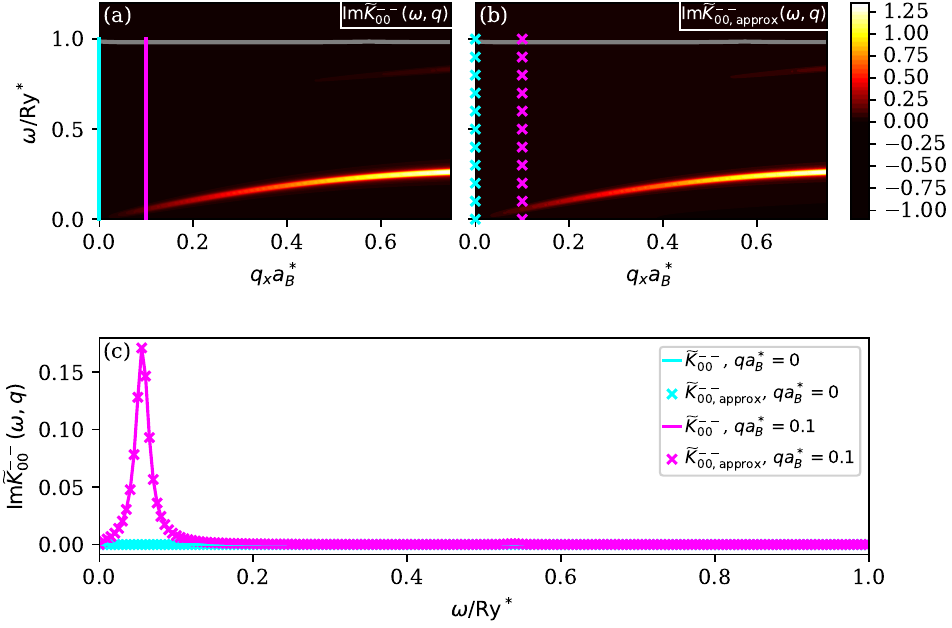}
  \caption{Exciton-density response at $\delta_m=0.2$. (a,b) Full response and the approximation \eqref{eq:numerical-density-response-approximation}; (c) cuts at $q a_B^*=0$ (cyan) and $0.1$ (magenta). Solid lines show the full response; crosses show the approximation.}
  \label{fig:exciton_00_approx}
\end{figure}

\FloatBarrier
\subsection{\texorpdfstring{Roton minimum of the nonmagnetic Goldstone mode}{Roton minimum of the nonmagnetic Goldstone mode}}\label{app:nonmagnetic-Goldstone-roton}

The main text shows that a perpendicular magnetic field produces a finite-$q$ roton in the Goldstone branch.  Its softening signals an instability toward a stripe-ordered EI, analogous to the stripe phase of a partially filled Landau level \cite{foglerGroundStateTwodimensional1996}.  Here we show that a Goldstone roton can also occur at zero magnetic field.

For $d/a_B^*=2$ and $\mu_X/\mathrm{Ry}^*=0.5$, the self-consistent density is $n_X(a_B^*)^2=0.032272$ and $k_Fa_B^*=0.63682$.  Figure \ref{fig:nonmagnetic-Goldstone-roton} shows a maxon followed by a roton at
\begin{equation}
  q_{\rm rot}a_B^*=1.21274,
  \qquad
  \omega_{\rm rot}/\mathrm{Ry}^*=0.04528 .
  \label{eq:nonmagnetic-Goldstone-roton}
\end{equation}
The roton wave vector also agrees with the shortest reciprocal-lattice vector of a triangular crystal at density $n_X$,
\begin{equation}
  G_{\triangle}=\sqrt{\frac{8\pi^2n_X}{\sqrt{3}}},
  \qquad
  G_{\triangle}a_B^*=1.21291 .
  \label{eq:triangular-crystal-wavevector}
\end{equation}
This agreement suggests that the roton is a finite-wave-vector precursor of translation-symmetry breaking.  Related bilayer calculations find a Wigner crystal at low density and large layer separation, a phase-coherent Wigner crystal of excitons, and an exciton crystal in appropriate parameter regimes \cite{depaloExcitonicCondensationSymmetric2002,joglekarWignerSupersolidExcitons2006,schleedePhaseDiagramBilayer2012}.

\begin{figure}[!ht]
  
  \centering
  \includegraphics[width=0.88\textwidth]{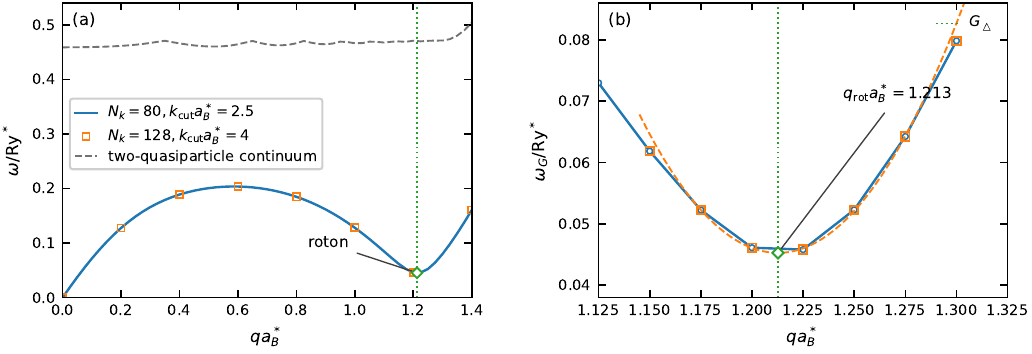}
  \caption{Nonmagnetic Goldstone dispersion at $d/a_B^*=2$ and $\mu_X/\mathrm{Ry}^*=0.5$. (a) The dense $N_k=80$ branch (blue), the two-quasiparticle continuum (gray dashed), and the $N_k=128$ cutoff check (orange squares). The green dotted line marks $G_{\triangle}$, the magnitude of the shortest reciprocal-lattice vector of a triangular crystal at density $n_X$. (b) The roton region. The orange dashed line is a local quadratic fit to the $N_k=128$ results.}
  \label{fig:nonmagnetic-Goldstone-roton}
\end{figure}

\FloatBarrier
}

\clearpage
{
\section{\texorpdfstring{Supplementary data for the magnetic case}{Supplementary data for the magnetic case}}\label{app:supplementary-data-magnetic}

We use $\delta_m=(m_h-m_e)/(m_e+m_h)$ and keep the reduced mass $m=m_em_h/(m_e+m_h)$ fixed. This is the same convention as in Appendix \ref{app:supplementary-data}.

\subsection{\texorpdfstring{Numerical parameters}{Numerical parameters}}\label{app:magnetic-asymmetric-parameters}

Unless stated otherwise, $B/B^*=1$, $\nu_X=4.5$ in each layer, $d/a_B^*=1$, and $N_{\rm LL}=16$. {The Landau-level-projected exchange matrix elements are evaluated using the radial momentum integral in Eq. \eqref{eq:LL-Fock-radial-integral}. Writing its dimensionless integration variable as $k=q_{\rm phys}\ell_B/\sqrt{2}$, we use $k_{\max}=5\sqrt{N_{\rm LL}}$ and $N_k=400N_{\rm LL}$ uniformly spaced points.} The self-consistent chemical potential is $\mu_X/\mathrm{Ry}^*=8.10228194$. Calculations were performed for $\delta_m=0,\,+0.2$, and $-0.2$. The figures show $\delta_m=0.2$; the result at $-0.2$ is used to check electron--hole interchange.

\subsection{\texorpdfstring{Mode splitting at zero momentum}{Mode splitting at zero momentum}}\label{app:magnetic-asymmetric-q0-splitting}

At fixed reduced mass, the mass-dependent shift of the single-particle Hamiltonian in Landau level $i$ is
\begin{equation}
  h_{0,i}(\delta_m)
  =h_{0,i}(0)
  +\frac{\delta_m}{2}\,\mathrm{Ry}^*\frac{B}{B^*}
  \left(i+\frac{1}{2}\right)\mathbbm 1 .
  \label{eq:magnetic-asymmetric-H0-shift}
\end{equation}
Because this term is proportional to the identity in layer space, the same-level mean-field density, eigenvectors, and band gap are independent of $\delta_m$. At $q=0$, the Landau-level difference $l=i_c-i_v$ is a good angular-momentum label, and the mode energies satisfy
\begin{equation}
  \omega_l(\delta_m)
  =\omega_l(0)+\frac{l\delta_m}{2}\,\mathrm{Ry}^*\frac{B}{B^*},
  \qquad
  \omega_{+|l|}-\omega_{-|l|}
  =|l|\delta_m\,\mathrm{Ry}^*\frac{B}{B^*} .
  \label{eq:magnetic-asymmetric-angular-shift}
\end{equation}
{Figure \ref{fig:magnetic-asymmetric-q0-splitting} verifies Eq.~\eqref{eq:magnetic-asymmetric-angular-shift}: the $l=\pm1$ and $\pm2$ branches split linearly with $\delta_m$, while the $l=0$ branches are independent of $\delta_m$. The Goldstone branch therefore remains at zero energy.}

\begin{figure}[!ht]
  
  \centering
  \includegraphics[width=0.94\textwidth]{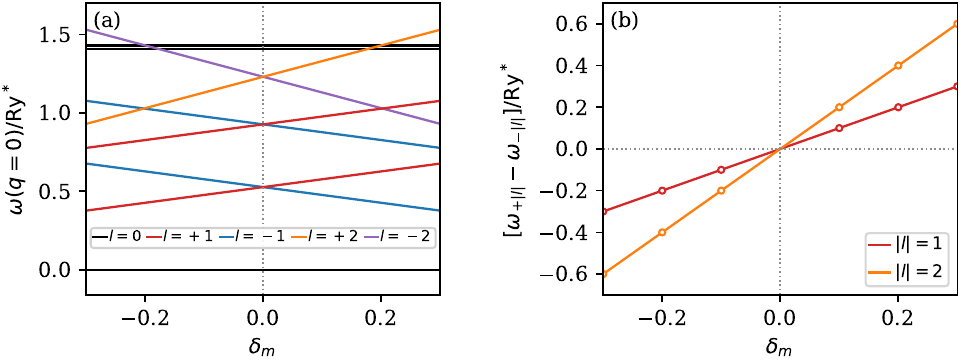}
  \caption{Angular-momentum-resolved spectrum at $q=0$, $B/B^*=1$, and $\nu_X=4.5$. (a) Low-energy $l=0,\pm1,\pm2$ modes. Opposite-angular-momentum partners split linearly with $\delta_m$. (b) Splittings $\omega_{+|l|}-\omega_{-|l|}$. In units of $\mathrm{Ry}^*$, the solid lines show $|l|\delta_m B/B^*$; open circles show direct diagonalization.}
  \label{fig:magnetic-asymmetric-q0-splitting}
\end{figure}

\FloatBarrier
}
{
\subsection{\texorpdfstring{Dispersion and spatial response functions}{Dispersion and spatial response functions}}\label{app:magnetic-asymmetric-response}

The finite-momentum spectrum at $\delta_m=0.2$ is shown in Fig. \ref{fig:magnetic-asymmetric-spectrum}. At $q=0$, the lowest branch remains gapless. The higher branches with nonzero angular momentum are split by the unequal electron and hole cyclotron energies.

\begin{figure}[!ht]
  
  \centering
  \includegraphics[width=0.56\textwidth]{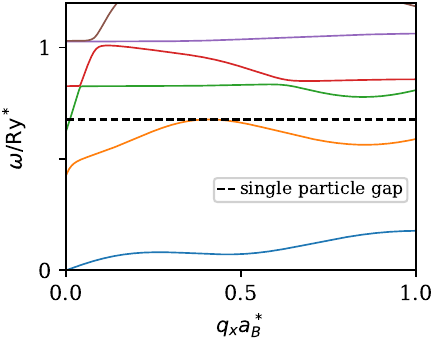}
  \caption{Collective-mode spectrum at $B/B^*=1$, $\nu_X=4.5$, and $\delta_m=0.2$. The dashed line is the minimum same-Landau-level single-particle excitation energy.}
  \label{fig:magnetic-asymmetric-spectrum}
\end{figure}

For $\bm q=q\hat{x}$, the same-channel response is written as
\begin{equation}
  \tilde K^{\sigma\sigma}_{ab}
  =
  \begin{pmatrix}
    \tilde K_L^\sigma & \tilde K_H^{\sigma\sigma}\\
    -\tilde K_H^{\sigma\sigma} & \tilde K_T^\sigma
  \end{pmatrix},
  \qquad \sigma\in\{+,-\}.
  \label{eq:magnetic-asymmetric-diagonal-tensor}
\end{equation}
The mixed response separates into
\begin{equation}
  \tilde K^{+-,S}_{ab}
  =
  \begin{pmatrix}
    \tilde K_L^{+-} & 0\\
    0 & \tilde K_T^{+-}
  \end{pmatrix},
  \qquad
  \tilde K^{+-,A}_{ab}
  =
  \begin{pmatrix}
    0 & -\tilde K_D\\
    \tilde K_{ID} & 0
  \end{pmatrix},
  \label{eq:magnetic-asymmetric-mixed-tensor}
\end{equation}
where
$\tilde K^{+-,S/A}_{ab}
=[\tilde K^{+-}_{ab}\pm\tilde K^{-+}_{ba}]/2$.
The longitudinal and transverse charge responses $\tilde K_L^+$ and $\tilde K_T^+$, the longitudinal and transverse exciton responses $\tilde K_L^-$ and $\tilde K_T^-$, and the dipole and inverse dipole Hall responses $\tilde K_D$ and $\tilde K_{ID}$ are even under $\delta_m\to-\delta_m$. The charge and exciton Hall responses $\tilde K_H^{++}$ and $\tilde K_H^{--}$ and the mixed longitudinal and transverse responses $\tilde K_L^{+-}$ and $\tilde K_T^{+-}$ are odd under $\delta_m\to-\delta_m$ and therefore vanish at equal mass.

Figures \ref{fig:magnetic-asymmetric-diagonal-response}--\ref{fig:magnetic-asymmetric-induced-response} show the spatial response at $\delta_m=0.2$. Mass asymmetry shifts the collective branches, redistributes their spectral weights, and produces finite mixed longitudinal, mixed transverse, and same-channel Hall responses.

\begin{figure}[!ht]
  
  \centering
  \includegraphics[width=0.88\textwidth]{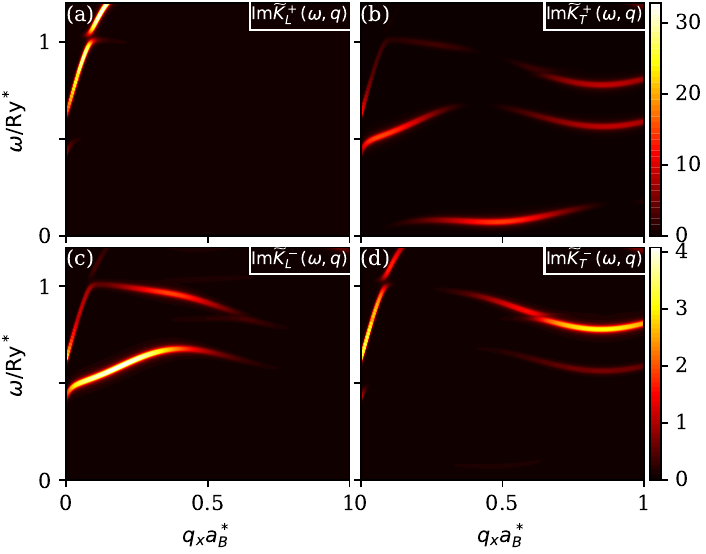}
  \caption{Diagonal current response at $B/B^*=1$, $\nu_X=4.5$, and $\delta_m=0.2$. (a,b) Longitudinal and transverse charge responses $\mathrm{Im}\,\tilde K_L^+$ and $\mathrm{Im}\,\tilde K_T^+$. (c,d) Longitudinal and transverse exciton responses $\mathrm{Im}\,\tilde K_L^-$ and $\mathrm{Im}\,\tilde K_T^-$.}
  \label{fig:magnetic-asymmetric-diagonal-response}
\end{figure}

\begin{figure}[!ht]
  
  \centering
  \includegraphics[width=0.88\textwidth]{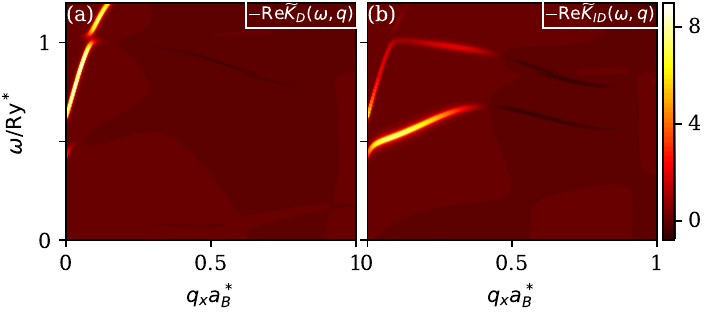}
  \caption{Dipole and inverse dipole Hall responses at $B/B^*=1$, $\nu_X=4.5$, and $\delta_m=0.2$. (a) $-\mathrm{Re}\,\tilde K_D$. (b) $-\mathrm{Re}\,\tilde K_{ID}$.}
  \label{fig:magnetic-asymmetric-hall-response}
\end{figure}

\begin{figure}[!ht]
  
  \centering
  \includegraphics[width=0.88\textwidth]{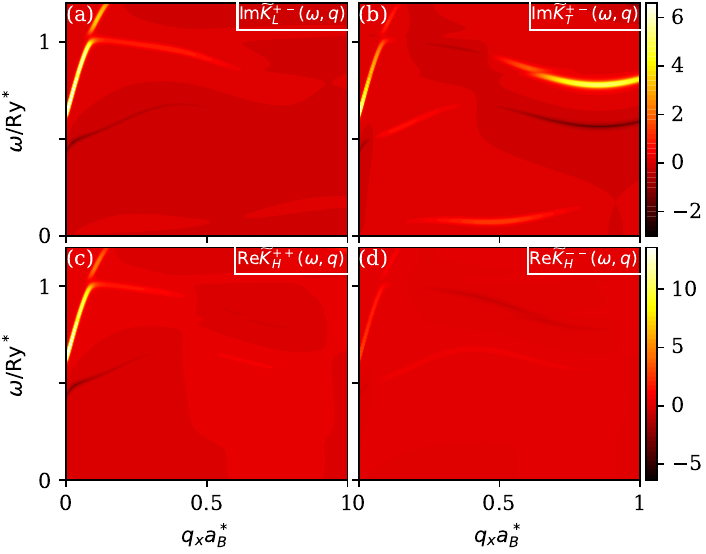}
  \caption{Responses generated by electron--hole mass asymmetry. (a,b) Mixed longitudinal and transverse responses $\mathrm{Im}\,\tilde K_L^{+-}$ and $\mathrm{Im}\,\tilde K_T^{+-}$. (c,d) Charge and exciton Hall responses $\mathrm{Re}\,\tilde K_H^{++}$ and $\mathrm{Re}\,\tilde K_H^{--}$. All four components reverse sign under $\delta_m\to-\delta_m$.}
  \label{fig:magnetic-asymmetric-induced-response}
\end{figure}

\FloatBarrier
}
{
\subsection{\texorpdfstring{Inverse dipole Hall response}{Inverse dipole Hall response}}\label{app:magnetic-asymmetric-inverse-Hall}

At unequal masses the mixed response also has an exchange-symmetric part. We therefore extract the inverse dipole Hall conductance from its exchange-antisymmetric part:
\begin{equation}
  \sigma_{ID}
  =
  \lim_{\omega\to0}
  \frac{1}{i\omega}
  \frac{K^{+-}_{yx}(\omega,0)-K^{-+}_{xy}(\omega,0)}{2}.
  \label{eq:magnetic-asymmetric-sigma-ID}
\end{equation}
This limit defines a linear-response coefficient. It does not describe a steady response to a finite static layer-antisymmetric field, which would continuously accelerate the counterflow and eventually exceed the critical current.
To isolate the effect of mass asymmetry on the response, we calculate $\sigma_{ID}$ at $\delta_m=0.2$ while keeping the reduced mass fixed. At zero temperature, the mean-field self-consistency equation for the $s$-wave ground state depends on $m_e$ and $m_h$ only through the reduced mass. The phase boundaries in Fig. \ref{fig:magnetic-asymmetric-inverse-Hall-map}(a) therefore coincide with those in Fig. 14(a) of the main text. The collective-mode spectrum and current vertices, however, retain the electron--hole mass asymmetry. Their changes give the small difference from the equal-mass conductance shown in Fig. \ref{fig:magnetic-asymmetric-inverse-Hall-map}(c).

\begin{figure}[!ht]
  
  \centering
  \includegraphics[width=0.72\textwidth]{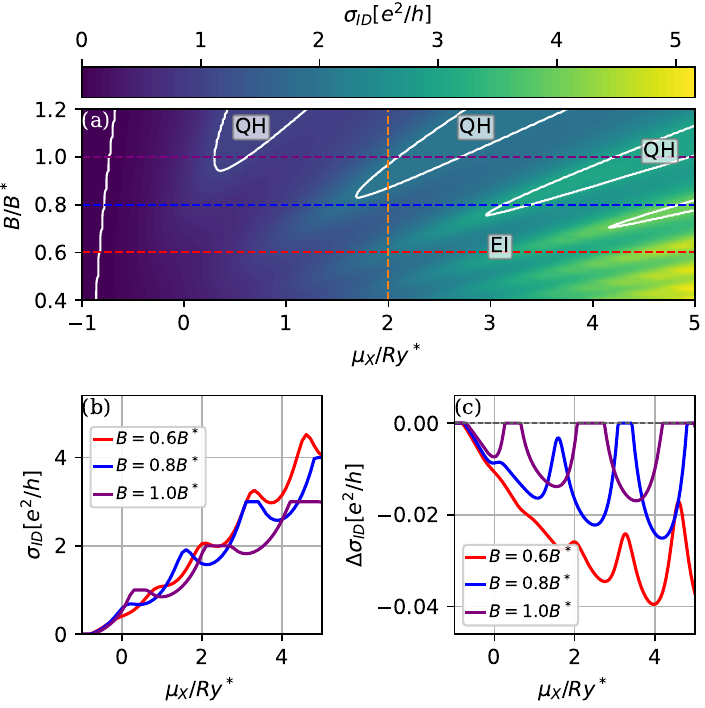}
  \caption{Inverse dipole Hall conductance at $\delta_m=0.2$ and fixed reduced mass. (a) Conductance as a function of $\mu_X$ and $B$. The white curves are the zero-temperature mean-field phase boundaries. (b) Cuts at $B/B^*=0.6$, $0.8$, and $1.0$. (c) Difference from the equal-mass conductance at the same $\mu_X$ and $B$.}
  \label{fig:magnetic-asymmetric-inverse-Hall-map}
\end{figure}

Semiclassically, the inverse dipole Hall conductance is written as
\begin{equation}
  \sigma_{ID}=\frac{eB\chi}{m_X},
  \label{eq:magnetic-asymmetric-semiclassical}
\end{equation}
where
\begin{equation}
  \chi=\lim_{\omega\to0}\frac{K_L^+(\omega,0)}{\omega^2}
  \label{eq:magnetic-asymmetric-polarization}
\end{equation}
is the two-dimensional polarization susceptibility. The remaining parameters are the exciton mass $m_X=m_e+m_h=4m/(1-\delta_m^2)$, the reduced mass $m=m_em_h/(m_e+m_h)$, and the mass-asymmetry parameter $\delta_m=(m_h-m_e)/(m_e+m_h)$.
In Fig. \ref{fig:magnetic-asymmetric-semiclassical-check}, we compare $\sigma_{ID}$ from Eq. \eqref{eq:magnetic-asymmetric-sigma-ID} with $eB\chi/m_X$. The two results agree, indicating that Eq. \eqref{eq:magnetic-asymmetric-semiclassical} also holds at $\delta_m\neq0$ when both $\chi$ and $\sigma_{ID}$ are evaluated from the full quantum response.

\begin{figure}[!ht]
  
  \centering
  \includegraphics[width=0.46\textwidth]{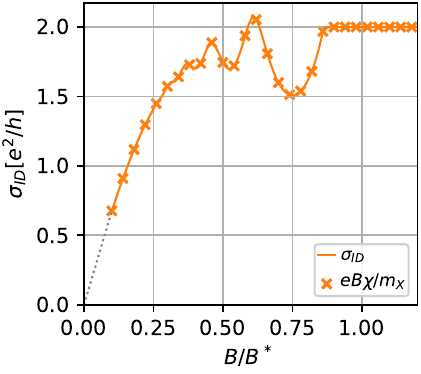}
  \caption{Inverse dipole Hall conductance at $\delta_m=0.2$ and $\mu_X/\mathrm{Ry}^*=2$. The solid line is the full quantum result from Eq. \eqref{eq:magnetic-asymmetric-sigma-ID}; crosses show $eB\chi/m_X$, with $\chi$ obtained independently from the longitudinal charge response in Eq. \eqref{eq:magnetic-asymmetric-polarization}.}
  \label{fig:magnetic-asymmetric-semiclassical-check}
\end{figure}

The dependence of $\chi$ on $\delta_m$ is constrained by electron--hole interchange. This operation sends $\delta_m\to-\delta_m$ and exchanges the $l=+1$ and $l=-1$ dipolar modes. Since both modes contribute to $\chi$, their interchange leaves the susceptibility unchanged, $\chi(B,\delta_m)=\chi(B,-\delta_m)$. We can therefore write
\begin{equation}
  \frac{\chi(B,\delta_m)}{\chi(B,0)}
  =1+c_2(B)\delta_m^2+\mathcal O(\delta_m^4).
  \label{eq:magnetic-asymmetric-chi-expansion}
\end{equation}
At zero magnetic field and zero temperature, the mean-field equation depends on the electron and hole masses only through the reduced mass. Hence, at fixed $m$, $\chi(0,\delta_m)=\chi(0,0)$ and $c_2(0)=0$ \cite{shaoElectricalBreakdownExcitonic2024}. The longitudinal susceptibility is also even under $B\to-B$, so the weak-field expansion of $c_2(B)$ starts at order $B^2$. Figure \ref{fig:magnetic-asymmetric-chi-coefficient}(a) shows $c_2(B)$ obtained from Eqs. \eqref{eq:magnetic-asymmetric-polarization} and \eqref{eq:magnetic-asymmetric-chi-expansion}, while Fig. \ref{fig:magnetic-asymmetric-chi-coefficient}(b) shows that $c_2(B)/(B/B^*)^2$ approaches a constant at weak field.

\begin{figure}[!ht]
  
  \centering
  \includegraphics[width=0.78\textwidth]{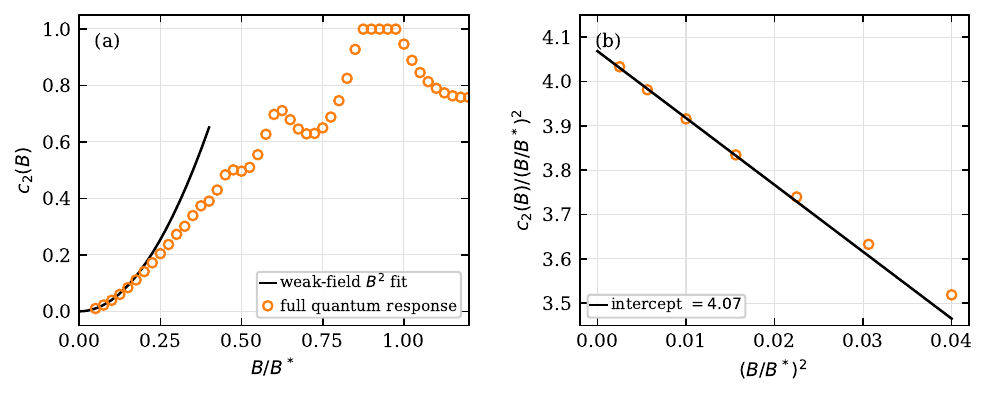}
  \caption{(a) Mass-asymmetry coefficient $c_2(B)$ at $\mu_X/\mathrm{Ry}^*=2$ and $d/a_B^*=1$. A larger Landau-level cutoff $N_{\rm LL}=240$ is used to access the low-field data. The orange circles show $c_2(B)$ from Eqs. \eqref{eq:magnetic-asymmetric-polarization} and \eqref{eq:magnetic-asymmetric-chi-expansion}, and the black line shows the $B^2$ fit. (b) $c_2(B)/(B/B^*)^2$ as a function of $(B/B^*)^2$. The constant intercept at weak field shows that $c_2(B)\sim B^2$.}
  \label{fig:magnetic-asymmetric-chi-coefficient}
\end{figure}

This result also fixes the weak-field mass dependence of the inverse dipole Hall conductance. Substituting Eq. \eqref{eq:magnetic-asymmetric-chi-expansion} and $m_X=4m/(1-\delta_m^2)$ into Eq. \eqref{eq:magnetic-asymmetric-semiclassical} gives
\begin{equation}
  \frac{\sigma_{ID}(B,\delta_m)}{\sigma_{ID}(B,0)}
  =\left(1-\delta_m^2\right)
  \frac{\chi(B,\delta_m)}{\chi(B,0)}
  \simeq 1-\delta_m^2
  +\mathcal{O}\!\left[\delta_m^2(B/B^*)^2\right]
  \qquad (B/B^*\ll1).
  \label{eq:magnetic-asymmetric-normalized-conductance}
\end{equation}
The correction from $\chi$ is higher order in $B$, leaving the exciton-mass factor $1-\delta_m^2$ as the leading weak-field dependence. The full response at $B/B^*=0.05$ follows this form closely, as shown in Fig. \ref{fig:magnetic-asymmetric-inverse-Hall}.

\begin{figure}[!ht]
  
  \centering
  \includegraphics[width=0.58\textwidth]{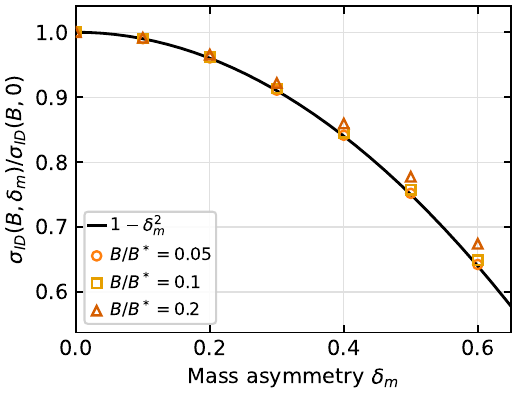}
  \caption{Normalized inverse dipole Hall conductance at $B/B^*=0.05$, $0.1$, and $0.2$, with $\mu_X/\mathrm{Ry}^*=2$, $d/a_B^*=1$, and $N_{\rm LL}=240$. Symbols show the numerical results obtained from Eq. \eqref{eq:magnetic-asymmetric-sigma-ID}, and the solid line shows the weak-field limit $1-\delta_m^2$.}
  \label{fig:magnetic-asymmetric-inverse-Hall}
\end{figure}
}


\begin{thebibliography}{51}%
\makeatletter
\providecommand \@ifxundefined [1]{%
 \@ifx{#1\undefined}
}%
\providecommand \@ifnum [1]{%
 \ifnum #1\expandafter \@firstoftwo
 \else \expandafter \@secondoftwo
 \fi
}%
\providecommand \@ifx [1]{%
 \ifx #1\expandafter \@firstoftwo
 \else \expandafter \@secondoftwo
 \fi
}%
\providecommand \natexlab [1]{#1}%
\providecommand \enquote  [1]{``#1''}%
\providecommand \bibnamefont  [1]{#1}%
\providecommand \bibfnamefont [1]{#1}%
\providecommand \citenamefont [1]{#1}%
\providecommand \href@noop [0]{\@secondoftwo}%
\providecommand \href [0]{\begingroup \@sanitize@url \@href}%
\providecommand \@href[1]{\@@startlink{#1}\@@href}%
\providecommand \@@href[1]{\endgroup#1\@@endlink}%
\providecommand \@sanitize@url [0]{\catcode `\\12\catcode `\$12\catcode `\&12\catcode `\#12\catcode `\^12\catcode `\_12\catcode `\%12\relax}%
\providecommand \@@startlink[1]{}%
\providecommand \@@endlink[0]{}%
\providecommand \url  [0]{\begingroup\@sanitize@url \@url }%
\providecommand \@url [1]{\endgroup\@href {#1}{\urlprefix }}%
\providecommand \urlprefix  [0]{URL }%
\providecommand \Eprint [0]{\href }%
\providecommand \doibase [0]{https://doi.org/}%
\providecommand \selectlanguage [0]{\@gobble}%
\providecommand \bibinfo  [0]{\@secondoftwo}%
\providecommand \bibfield  [0]{\@secondoftwo}%
\providecommand \translation [1]{[#1]}%
\providecommand \BibitemOpen [0]{}%
\providecommand \bibitemStop [0]{}%
\providecommand \bibitemNoStop [0]{.\EOS\space}%
\providecommand \EOS [0]{\spacefactor3000\relax}%
\providecommand \BibitemShut  [1]{\csname bibitem#1\endcsname}%
\let\auto@bib@innerbib\@empty
\bibitem [{\citenamefont {Mott}(1961)}]{mottTransitionMetallicState1961}%
  \BibitemOpen
  \bibfield  {author} {\bibinfo {author} {\bibfnamefont {N.~F.}\ \bibnamefont {Mott}},\ }\bibfield  {title} {\bibinfo {title} {The transition to the metallic state},\ }\href {https://doi.org/10.1080/14786436108243318} {\bibfield  {journal} {\bibinfo  {journal} {The Philosophical Magazine: A Journal of Theoretical Experimental and Applied Physics}\ }\textbf {\bibinfo {volume} {6}},\ \bibinfo {pages} {287} (\bibinfo {year} {1961})}\BibitemShut {NoStop}%
\bibitem [{\citenamefont {J{\'e}rome}\ \emph {et~al.}(1967)\citenamefont {J{\'e}rome}, \citenamefont {Rice},\ and\ \citenamefont {Kohn}}]{Jerome1967}%
  \BibitemOpen
  \bibfield  {author} {\bibinfo {author} {\bibfnamefont {D.}~\bibnamefont {J{\'e}rome}}, \bibinfo {author} {\bibfnamefont {T.~M.}\ \bibnamefont {Rice}},\ and\ \bibinfo {author} {\bibfnamefont {W.}~\bibnamefont {Kohn}},\ }\bibfield  {title} {\bibinfo {title} {Excitonic insulator},\ }\href {https://doi.org/10.1103/PhysRev.158.462} {\bibfield  {journal} {\bibinfo  {journal} {Physical Review}\ }\textbf {\bibinfo {volume} {158}},\ \bibinfo {pages} {462} (\bibinfo {year} {1967})}\BibitemShut {NoStop}%
\bibitem [{\citenamefont {Kohn}(1967)}]{kohnExcitonicPhases1967}%
  \BibitemOpen
  \bibfield  {author} {\bibinfo {author} {\bibfnamefont {W.}~\bibnamefont {Kohn}},\ }\bibfield  {title} {\bibinfo {title} {Excitonic {{Phases}}},\ }\href {https://doi.org/10.1103/PhysRevLett.19.439} {\bibfield  {journal} {\bibinfo  {journal} {Physical Review Letters}\ }\textbf {\bibinfo {volume} {19}},\ \bibinfo {pages} {439} (\bibinfo {year} {1967})}\BibitemShut {NoStop}%
\bibitem [{\citenamefont {Littlewood}\ \emph {et~al.}(2004)\citenamefont {Littlewood}, \citenamefont {Eastham}, \citenamefont {Keeling}, \citenamefont {Marchetti}, \citenamefont {Simons},\ and\ \citenamefont {Szymanska}}]{littlewoodModelsCoherentExciton2004}%
  \BibitemOpen
  \bibfield  {author} {\bibinfo {author} {\bibfnamefont {P.~B.}\ \bibnamefont {Littlewood}}, \bibinfo {author} {\bibfnamefont {P.~R.}\ \bibnamefont {Eastham}}, \bibinfo {author} {\bibfnamefont {J.~M.~J.}\ \bibnamefont {Keeling}}, \bibinfo {author} {\bibfnamefont {F.~M.}\ \bibnamefont {Marchetti}}, \bibinfo {author} {\bibfnamefont {B.~D.}\ \bibnamefont {Simons}},\ and\ \bibinfo {author} {\bibfnamefont {M.~H.}\ \bibnamefont {Szymanska}},\ }\bibfield  {title} {\bibinfo {title} {Models of coherent exciton condensation},\ }\href {https://doi.org/10.1088/0953-8984/16/35/003} {\bibfield  {journal} {\bibinfo  {journal} {Journal of Physics: Condensed Matter}\ }\textbf {\bibinfo {volume} {16}},\ \bibinfo {pages} {S3597} (\bibinfo {year} {2004})}\BibitemShut {NoStop}%
\bibitem [{\citenamefont {Balatsky}\ \emph {et~al.}(2004)\citenamefont {Balatsky}, \citenamefont {Joglekar},\ and\ \citenamefont {Littlewood}}]{balatskyDipolarSuperfluidityElectronHole2004}%
  \BibitemOpen
  \bibfield  {author} {\bibinfo {author} {\bibfnamefont {A.~V.}\ \bibnamefont {Balatsky}}, \bibinfo {author} {\bibfnamefont {Y.~N.}\ \bibnamefont {Joglekar}},\ and\ \bibinfo {author} {\bibfnamefont {P.~B.}\ \bibnamefont {Littlewood}},\ }\bibfield  {title} {\bibinfo {title} {Dipolar {{Superfluidity}} in {{Electron-Hole Bilayer Systems}}},\ }\href {https://doi.org/10.1103/PhysRevLett.93.266801} {\bibfield  {journal} {\bibinfo  {journal} {Physical Review Letters}\ }\textbf {\bibinfo {volume} {93}},\ \bibinfo {pages} {266801} (\bibinfo {year} {2004})}\BibitemShut {NoStop}%
\bibitem [{\citenamefont {Su}\ and\ \citenamefont {MacDonald}(2008)}]{suHowMakeBilayer2008}%
  \BibitemOpen
  \bibfield  {author} {\bibinfo {author} {\bibfnamefont {J.-J.}\ \bibnamefont {Su}}\ and\ \bibinfo {author} {\bibfnamefont {A.~H.}\ \bibnamefont {MacDonald}},\ }\bibfield  {title} {\bibinfo {title} {How to make a bilayer exciton condensate flow},\ }\href {https://doi.org/10.1038/nphys1055} {\bibfield  {journal} {\bibinfo  {journal} {Nature Physics}\ }\textbf {\bibinfo {volume} {4}},\ \bibinfo {pages} {799} (\bibinfo {year} {2008})}\BibitemShut {NoStop}%
\bibitem [{\citenamefont {Zhu}\ \emph {et~al.}(1995)\citenamefont {Zhu}, \citenamefont {Littlewood}, \citenamefont {Hybertsen},\ and\ \citenamefont {Rice}}]{zhuExcitonCondensateSemiconductor1995}%
  \BibitemOpen
  \bibfield  {author} {\bibinfo {author} {\bibfnamefont {X.}~\bibnamefont {Zhu}}, \bibinfo {author} {\bibfnamefont {P.~B.}\ \bibnamefont {Littlewood}}, \bibinfo {author} {\bibfnamefont {M.~S.}\ \bibnamefont {Hybertsen}},\ and\ \bibinfo {author} {\bibfnamefont {T.~M.}\ \bibnamefont {Rice}},\ }\bibfield  {title} {\bibinfo {title} {Exciton {{Condensate}} in {{Semiconductor Quantum Well Structures}}},\ }\href {https://doi.org/10.1103/PhysRevLett.74.1633} {\bibfield  {journal} {\bibinfo  {journal} {Physical Review Letters}\ }\textbf {\bibinfo {volume} {74}},\ \bibinfo {pages} {1633} (\bibinfo {year} {1995})}\BibitemShut {NoStop}%
\bibitem [{\citenamefont {Fogler}\ \emph {et~al.}(2014)\citenamefont {Fogler}, \citenamefont {Butov},\ and\ \citenamefont {Novoselov}}]{foglerHightemperatureSuperfluidityIndirect2014a}%
  \BibitemOpen
  \bibfield  {author} {\bibinfo {author} {\bibfnamefont {M.~M.}\ \bibnamefont {Fogler}}, \bibinfo {author} {\bibfnamefont {L.~V.}\ \bibnamefont {Butov}},\ and\ \bibinfo {author} {\bibfnamefont {K.~S.}\ \bibnamefont {Novoselov}},\ }\bibfield  {title} {\bibinfo {title} {High-temperature superfluidity with indirect excitons in van der {{Waals}} heterostructures},\ }\href {https://doi.org/10.1038/ncomms5555} {\bibfield  {journal} {\bibinfo  {journal} {Nature Communications}\ }\textbf {\bibinfo {volume} {5}},\ \bibinfo {pages} {4555} (\bibinfo {year} {2014})}\BibitemShut {NoStop}%
\bibitem [{\citenamefont {Wu}\ \emph {et~al.}(2015)\citenamefont {Wu}, \citenamefont {Xue},\ and\ \citenamefont {MacDonald}}]{wuTheoryTwodimensionalSpatially2015}%
  \BibitemOpen
  \bibfield  {author} {\bibinfo {author} {\bibfnamefont {F.-C.}\ \bibnamefont {Wu}}, \bibinfo {author} {\bibfnamefont {F.}~\bibnamefont {Xue}},\ and\ \bibinfo {author} {\bibfnamefont {A.~H.}\ \bibnamefont {MacDonald}},\ }\bibfield  {title} {\bibinfo {title} {Theory of two-dimensional spatially indirect equilibrium exciton condensates},\ }\href {https://doi.org/10.1103/PhysRevB.92.165121} {\bibfield  {journal} {\bibinfo  {journal} {Physical Review B}\ }\textbf {\bibinfo {volume} {92}},\ \bibinfo {pages} {165121} (\bibinfo {year} {2015})}\BibitemShut {NoStop}%
\bibitem [{\citenamefont {Zeng}\ and\ \citenamefont {MacDonald}(2020)}]{zengElectricallyControlledTwodimensional2020a}%
  \BibitemOpen
  \bibfield  {author} {\bibinfo {author} {\bibfnamefont {Y.}~\bibnamefont {Zeng}}\ and\ \bibinfo {author} {\bibfnamefont {A.~H.}\ \bibnamefont {MacDonald}},\ }\bibfield  {title} {\bibinfo {title} {Electrically controlled two-dimensional electron-hole fluids},\ }\href {https://doi.org/10.1103/PhysRevB.102.085154} {\bibfield  {journal} {\bibinfo  {journal} {Physical Review B}\ }\textbf {\bibinfo {volume} {102}},\ \bibinfo {pages} {085154} (\bibinfo {year} {2020})}\BibitemShut {NoStop}%
\bibitem [{\citenamefont {Du}\ \emph {et~al.}(2017)\citenamefont {Du}, \citenamefont {Li}, \citenamefont {Lou}, \citenamefont {Sullivan}, \citenamefont {Chang}, \citenamefont {Kono},\ and\ \citenamefont {Du}}]{duEvidenceTopologicalExcitonic2017}%
  \BibitemOpen
  \bibfield  {author} {\bibinfo {author} {\bibfnamefont {L.}~\bibnamefont {Du}}, \bibinfo {author} {\bibfnamefont {X.}~\bibnamefont {Li}}, \bibinfo {author} {\bibfnamefont {W.}~\bibnamefont {Lou}}, \bibinfo {author} {\bibfnamefont {G.}~\bibnamefont {Sullivan}}, \bibinfo {author} {\bibfnamefont {K.}~\bibnamefont {Chang}}, \bibinfo {author} {\bibfnamefont {J.}~\bibnamefont {Kono}},\ and\ \bibinfo {author} {\bibfnamefont {R.-R.}\ \bibnamefont {Du}},\ }\bibfield  {title} {\bibinfo {title} {Evidence for a topological excitonic insulator in {{InAs}}/{{GaSb}} bilayers},\ }\href {https://doi.org/10.1038/s41467-017-01988-1} {\bibfield  {journal} {\bibinfo  {journal} {Nature Communications}\ }\textbf {\bibinfo {volume} {8}},\ \bibinfo {pages} {1971} (\bibinfo {year} {2017})}\BibitemShut {NoStop}%
\bibitem [{\citenamefont {Wu}\ \emph {et~al.}(2019{\natexlab{a}})\citenamefont {Wu}, \citenamefont {Lou}, \citenamefont {Chang}, \citenamefont {Sullivan}, \citenamefont {Ikhlassi},\ and\ \citenamefont {Du}}]{wuElectricallyTuningManybody2019}%
  \BibitemOpen
  \bibfield  {author} {\bibinfo {author} {\bibfnamefont {X.-J.}\ \bibnamefont {Wu}}, \bibinfo {author} {\bibfnamefont {W.}~\bibnamefont {Lou}}, \bibinfo {author} {\bibfnamefont {K.}~\bibnamefont {Chang}}, \bibinfo {author} {\bibfnamefont {G.}~\bibnamefont {Sullivan}}, \bibinfo {author} {\bibfnamefont {A.}~\bibnamefont {Ikhlassi}},\ and\ \bibinfo {author} {\bibfnamefont {R.-R.}\ \bibnamefont {Du}},\ }\bibfield  {title} {\bibinfo {title} {Electrically tuning many-body states in a {{Coulomb-coupled InAs}}/{{InGaSb}} double layer},\ }\href {https://doi.org/10.1103/PhysRevB.100.165309} {\bibfield  {journal} {\bibinfo  {journal} {Physical Review B}\ }\textbf {\bibinfo {volume} {100}},\ \bibinfo {pages} {165309} (\bibinfo {year} {2019}{\natexlab{a}})}\BibitemShut {NoStop}%
\bibitem [{\citenamefont {Wu}\ \emph {et~al.}(2019{\natexlab{b}})\citenamefont {Wu}, \citenamefont {Lou}, \citenamefont {Chang}, \citenamefont {Sullivan},\ and\ \citenamefont {Du}}]{wuResistiveSignatureExcitonic2019}%
  \BibitemOpen
  \bibfield  {author} {\bibinfo {author} {\bibfnamefont {X.}~\bibnamefont {Wu}}, \bibinfo {author} {\bibfnamefont {W.}~\bibnamefont {Lou}}, \bibinfo {author} {\bibfnamefont {K.}~\bibnamefont {Chang}}, \bibinfo {author} {\bibfnamefont {G.}~\bibnamefont {Sullivan}},\ and\ \bibinfo {author} {\bibfnamefont {R.-R.}\ \bibnamefont {Du}},\ }\bibfield  {title} {\bibinfo {title} {Resistive signature of excitonic coupling in an electron-hole double layer with a middle barrier},\ }\href {https://doi.org/10.1103/PhysRevB.99.085307} {\bibfield  {journal} {\bibinfo  {journal} {Physical Review B}\ }\textbf {\bibinfo {volume} {99}},\ \bibinfo {pages} {085307} (\bibinfo {year} {2019}{\natexlab{b}})}\BibitemShut {NoStop}%
\bibitem [{\citenamefont {Wang}\ \emph {et~al.}(2019)\citenamefont {Wang}, \citenamefont {Rhodes}, \citenamefont {Watanabe}, \citenamefont {Taniguchi}, \citenamefont {Hone}, \citenamefont {Shan},\ and\ \citenamefont {Mak}}]{wangEvidenceHightemperatureExciton2019a}%
  \BibitemOpen
  \bibfield  {author} {\bibinfo {author} {\bibfnamefont {Z.}~\bibnamefont {Wang}}, \bibinfo {author} {\bibfnamefont {D.~A.}\ \bibnamefont {Rhodes}}, \bibinfo {author} {\bibfnamefont {K.}~\bibnamefont {Watanabe}}, \bibinfo {author} {\bibfnamefont {T.}~\bibnamefont {Taniguchi}}, \bibinfo {author} {\bibfnamefont {J.~C.}\ \bibnamefont {Hone}}, \bibinfo {author} {\bibfnamefont {J.}~\bibnamefont {Shan}},\ and\ \bibinfo {author} {\bibfnamefont {K.~F.}\ \bibnamefont {Mak}},\ }\bibfield  {title} {\bibinfo {title} {Evidence of high-temperature exciton condensation in two-dimensional atomic double layers},\ }\href {https://doi.org/10.1038/s41586-019-1591-7} {\bibfield  {journal} {\bibinfo  {journal} {Nature}\ }\textbf {\bibinfo {volume} {574}},\ \bibinfo {pages} {76} (\bibinfo {year} {2019})}\BibitemShut {NoStop}%
\bibitem [{\citenamefont {Ma}\ \emph {et~al.}(2021)\citenamefont {Ma}, \citenamefont {Nguyen}, \citenamefont {Wang}, \citenamefont {Zeng}, \citenamefont {Watanabe}, \citenamefont {Taniguchi}, \citenamefont {MacDonald}, \citenamefont {Mak},\ and\ \citenamefont {Shan}}]{maStronglyCorrelatedExcitonic2021}%
  \BibitemOpen
  \bibfield  {author} {\bibinfo {author} {\bibfnamefont {L.}~\bibnamefont {Ma}}, \bibinfo {author} {\bibfnamefont {P.~X.}\ \bibnamefont {Nguyen}}, \bibinfo {author} {\bibfnamefont {Z.}~\bibnamefont {Wang}}, \bibinfo {author} {\bibfnamefont {Y.}~\bibnamefont {Zeng}}, \bibinfo {author} {\bibfnamefont {K.}~\bibnamefont {Watanabe}}, \bibinfo {author} {\bibfnamefont {T.}~\bibnamefont {Taniguchi}}, \bibinfo {author} {\bibfnamefont {A.~H.}\ \bibnamefont {MacDonald}}, \bibinfo {author} {\bibfnamefont {K.~F.}\ \bibnamefont {Mak}},\ and\ \bibinfo {author} {\bibfnamefont {J.}~\bibnamefont {Shan}},\ }\bibfield  {title} {\bibinfo {title} {Strongly correlated excitonic insulator in atomic double layers},\ }\href {https://doi.org/10.1038/s41586-021-03947-9} {\bibfield  {journal} {\bibinfo  {journal} {Nature}\ }\textbf {\bibinfo {volume} {598}},\ \bibinfo {pages} {585} (\bibinfo {year} {2021})}\BibitemShut {NoStop}%
\bibitem [{\citenamefont {Qi}\ \emph {et~al.}(2023)\citenamefont {Qi}, \citenamefont {Joe}, \citenamefont {Zhang}, \citenamefont {Zeng}, \citenamefont {Zheng}, \citenamefont {Feng}, \citenamefont {Xie}, \citenamefont {Regan}, \citenamefont {Lu}, \citenamefont {Taniguchi}, \citenamefont {Watanabe}, \citenamefont {Tongay}, \citenamefont {Crommie}, \citenamefont {MacDonald},\ and\ \citenamefont {Wang}}]{qiThermodynamicBehaviorCorrelated2023a}%
  \BibitemOpen
  \bibfield  {author} {\bibinfo {author} {\bibfnamefont {R.}~\bibnamefont {Qi}}, \bibinfo {author} {\bibfnamefont {A.~Y.}\ \bibnamefont {Joe}}, \bibinfo {author} {\bibfnamefont {Z.}~\bibnamefont {Zhang}}, \bibinfo {author} {\bibfnamefont {Y.}~\bibnamefont {Zeng}}, \bibinfo {author} {\bibfnamefont {T.}~\bibnamefont {Zheng}}, \bibinfo {author} {\bibfnamefont {Q.}~\bibnamefont {Feng}}, \bibinfo {author} {\bibfnamefont {J.}~\bibnamefont {Xie}}, \bibinfo {author} {\bibfnamefont {E.}~\bibnamefont {Regan}}, \bibinfo {author} {\bibfnamefont {Z.}~\bibnamefont {Lu}}, \bibinfo {author} {\bibfnamefont {T.}~\bibnamefont {Taniguchi}}, \bibinfo {author} {\bibfnamefont {K.}~\bibnamefont {Watanabe}}, \bibinfo {author} {\bibfnamefont {S.}~\bibnamefont {Tongay}}, \bibinfo {author} {\bibfnamefont {M.~F.}\ \bibnamefont {Crommie}}, \bibinfo {author} {\bibfnamefont {A.~H.}\ \bibnamefont {MacDonald}},\ and\ \bibinfo {author} {\bibfnamefont {F.}~\bibnamefont {Wang}},\ }\bibfield  {title} {\bibinfo {title} {Thermodynamic behavior of correlated electron-hole fluids in van der {{Waals}} heterostructures},\ }\href {https://doi.org/10.1038/s41467-023-43799-7} {\bibfield  {journal} {\bibinfo  {journal} {Nature Communications}\ }\textbf {\bibinfo {volume} {14}},\ \bibinfo {pages} {8264} (\bibinfo {year} {2023})}\BibitemShut {NoStop}%
\bibitem [{\citenamefont {Qi}\ \emph {et~al.}(2025)\citenamefont {Qi}, \citenamefont {Joe}, \citenamefont {Zhang}, \citenamefont {Xie}, \citenamefont {Feng}, \citenamefont {Lu}, \citenamefont {Wang}, \citenamefont {Taniguchi}, \citenamefont {Watanabe}, \citenamefont {Tongay},\ and\ \citenamefont {Wang}}]{qiPerfectCoulombDrag2025}%
  \BibitemOpen
  \bibfield  {author} {\bibinfo {author} {\bibfnamefont {R.}~\bibnamefont {Qi}}, \bibinfo {author} {\bibfnamefont {A.~Y.}\ \bibnamefont {Joe}}, \bibinfo {author} {\bibfnamefont {Z.}~\bibnamefont {Zhang}}, \bibinfo {author} {\bibfnamefont {J.}~\bibnamefont {Xie}}, \bibinfo {author} {\bibfnamefont {Q.}~\bibnamefont {Feng}}, \bibinfo {author} {\bibfnamefont {Z.}~\bibnamefont {Lu}}, \bibinfo {author} {\bibfnamefont {Z.}~\bibnamefont {Wang}}, \bibinfo {author} {\bibfnamefont {T.}~\bibnamefont {Taniguchi}}, \bibinfo {author} {\bibfnamefont {K.}~\bibnamefont {Watanabe}}, \bibinfo {author} {\bibfnamefont {S.}~\bibnamefont {Tongay}},\ and\ \bibinfo {author} {\bibfnamefont {F.}~\bibnamefont {Wang}},\ }\bibfield  {title} {\bibinfo {title} {Perfect {{Coulomb}} drag and exciton transport in an excitonic insulator},\ }\href {https://doi.org/10.1126/science.adl1839} {\bibfield  {journal} {\bibinfo  {journal} {Science}\ }\textbf {\bibinfo {volume} {388}},\ \bibinfo {pages} {278} (\bibinfo {year} {2025})}\BibitemShut {NoStop}%
\bibitem [{\citenamefont {Nguyen}\ \emph {et~al.}(2025)\citenamefont {Nguyen}, \citenamefont {Ma}, \citenamefont {Chaturvedi}, \citenamefont {Watanabe}, \citenamefont {Taniguchi}, \citenamefont {Shan},\ and\ \citenamefont {Mak}}]{nguyenPerfectCoulombDrag2025}%
  \BibitemOpen
  \bibfield  {author} {\bibinfo {author} {\bibfnamefont {P.~X.}\ \bibnamefont {Nguyen}}, \bibinfo {author} {\bibfnamefont {L.}~\bibnamefont {Ma}}, \bibinfo {author} {\bibfnamefont {R.}~\bibnamefont {Chaturvedi}}, \bibinfo {author} {\bibfnamefont {K.}~\bibnamefont {Watanabe}}, \bibinfo {author} {\bibfnamefont {T.}~\bibnamefont {Taniguchi}}, \bibinfo {author} {\bibfnamefont {J.}~\bibnamefont {Shan}},\ and\ \bibinfo {author} {\bibfnamefont {K.~F.}\ \bibnamefont {Mak}},\ }\bibfield  {title} {\bibinfo {title} {Perfect {{Coulomb}} drag in a dipolar excitonic insulator},\ }\href {https://doi.org/10.1126/science.adl1829} {\bibfield  {journal} {\bibinfo  {journal} {Science}\ }\textbf {\bibinfo {volume} {388}},\ \bibinfo {pages} {274} (\bibinfo {year} {2025})}\BibitemShut {NoStop}%
\bibitem [{\citenamefont {Cutshall}\ \emph {et~al.}(2025)\citenamefont {Cutshall}, \citenamefont {Mahdikhany}, \citenamefont {Roche}, \citenamefont {Shanks}, \citenamefont {Koehler}, \citenamefont {Mandrus}, \citenamefont {Taniguchi}, \citenamefont {Watanabe}, \citenamefont {Zhu}, \citenamefont {LeRoy},\ and\ \citenamefont {Schaibley}}]{cutshallImagingInterlayerExciton2025}%
  \BibitemOpen
  \bibfield  {author} {\bibinfo {author} {\bibfnamefont {J.}~\bibnamefont {Cutshall}}, \bibinfo {author} {\bibfnamefont {F.}~\bibnamefont {Mahdikhany}}, \bibinfo {author} {\bibfnamefont {A.}~\bibnamefont {Roche}}, \bibinfo {author} {\bibfnamefont {D.~N.}\ \bibnamefont {Shanks}}, \bibinfo {author} {\bibfnamefont {M.~R.}\ \bibnamefont {Koehler}}, \bibinfo {author} {\bibfnamefont {D.~G.}\ \bibnamefont {Mandrus}}, \bibinfo {author} {\bibfnamefont {T.}~\bibnamefont {Taniguchi}}, \bibinfo {author} {\bibfnamefont {K.}~\bibnamefont {Watanabe}}, \bibinfo {author} {\bibfnamefont {Q.}~\bibnamefont {Zhu}}, \bibinfo {author} {\bibfnamefont {B.~J.}\ \bibnamefont {LeRoy}},\ and\ \bibinfo {author} {\bibfnamefont {J.~R.}\ \bibnamefont {Schaibley}},\ }\bibfield  {title} {\bibinfo {title} {Imaging interlayer exciton superfluidity in a {{2D}} semiconductor heterostructure},\ }\href@noop {} {\bibfield  {journal} {\bibinfo  {journal} {Science Advances}\ } (\bibinfo {year} {2025})}\BibitemShut {NoStop}%
\bibitem [{\citenamefont {Kumar}\ \emph {et~al.}(2024)\citenamefont {Kumar}, \citenamefont {Patri},\ and\ \citenamefont {Senthil}}]{kumarUnconventionalSuperconductivityMediated2024}%
  \BibitemOpen
  \bibfield  {author} {\bibinfo {author} {\bibfnamefont {A.}~\bibnamefont {Kumar}}, \bibinfo {author} {\bibfnamefont {A.~S.}\ \bibnamefont {Patri}},\ and\ \bibinfo {author} {\bibfnamefont {T.}~\bibnamefont {Senthil}},\ }\href {https://doi.org/10.48550/arXiv.2410.09148} {\bibinfo {title} {Unconventional superconductivity mediated by exciton density wave fluctuations}} (\bibinfo {year} {2024}),\ \Eprint {https://arxiv.org/abs/2410.09148} {arXiv:2410.09148 [cond-mat]} \BibitemShut {NoStop}%
\bibitem [{\citenamefont {Panigrahi}\ and\ \citenamefont {Kumar}(2025)}]{panigrahiNonFermiLiquidsSubsystem2025}%
  \BibitemOpen
  \bibfield  {author} {\bibinfo {author} {\bibfnamefont {A.}~\bibnamefont {Panigrahi}}\ and\ \bibinfo {author} {\bibfnamefont {A.}~\bibnamefont {Kumar}},\ }\bibfield  {title} {\bibinfo {title} {Non-{{Fermi Liquids}} from {{Subsystem Symmetry Breaking}} in van der {{Waals Multilayers}}},\ }\href {https://doi.org/10.1103/v6r7-4ph9} {\bibfield  {journal} {\bibinfo  {journal} {Physical Review Letters}\ }\textbf {\bibinfo {volume} {134}},\ \bibinfo {pages} {236502} (\bibinfo {year} {2025})}\BibitemShut {NoStop}%
\bibitem [{\citenamefont {Gole{\v z}}\ \emph {et~al.}(2020)\citenamefont {Gole{\v z}}, \citenamefont {Sun}, \citenamefont {Murakami}, \citenamefont {Georges},\ and\ \citenamefont {Millis}}]{golezNonlinearSpectroscopyCollective2020}%
  \BibitemOpen
  \bibfield  {author} {\bibinfo {author} {\bibfnamefont {D.}~\bibnamefont {Gole{\v z}}}, \bibinfo {author} {\bibfnamefont {Z.}~\bibnamefont {Sun}}, \bibinfo {author} {\bibfnamefont {Y.}~\bibnamefont {Murakami}}, \bibinfo {author} {\bibfnamefont {A.}~\bibnamefont {Georges}},\ and\ \bibinfo {author} {\bibfnamefont {A.~J.}\ \bibnamefont {Millis}},\ }\bibfield  {title} {\bibinfo {title} {Nonlinear {{Spectroscopy}} of {{Collective Modes}} in an {{Excitonic Insulator}}},\ }\href {https://doi.org/10.1103/PhysRevLett.125.257601} {\bibfield  {journal} {\bibinfo  {journal} {Physical Review Letters}\ }\textbf {\bibinfo {volume} {125}},\ \bibinfo {pages} {257601} (\bibinfo {year} {2020})}\BibitemShut {NoStop}%
\bibitem [{\citenamefont {Murakami}\ \emph {et~al.}(2020)\citenamefont {Murakami}, \citenamefont {Gole{\v z}}, \citenamefont {Kaneko}, \citenamefont {Koga}, \citenamefont {Millis},\ and\ \citenamefont {Werner}}]{murakamiCollectiveModesExcitonic2020}%
  \BibitemOpen
  \bibfield  {author} {\bibinfo {author} {\bibfnamefont {Y.}~\bibnamefont {Murakami}}, \bibinfo {author} {\bibfnamefont {D.}~\bibnamefont {Gole{\v z}}}, \bibinfo {author} {\bibfnamefont {T.}~\bibnamefont {Kaneko}}, \bibinfo {author} {\bibfnamefont {A.}~\bibnamefont {Koga}}, \bibinfo {author} {\bibfnamefont {A.~J.}\ \bibnamefont {Millis}},\ and\ \bibinfo {author} {\bibfnamefont {P.}~\bibnamefont {Werner}},\ }\bibfield  {title} {\bibinfo {title} {Collective modes in excitonic insulators: {{Effects}} of electron-phonon coupling and signatures in the optical response},\ }\href {https://doi.org/10.1103/PhysRevB.101.195118} {\bibfield  {journal} {\bibinfo  {journal} {Physical Review B}\ }\textbf {\bibinfo {volume} {101}},\ \bibinfo {pages} {195118} (\bibinfo {year} {2020})}\BibitemShut {NoStop}%
\bibitem [{\citenamefont {Kaneko}\ and\ \citenamefont {Ohta}(2025)}]{kanekoNewEraExcitonic2025}%
  \BibitemOpen
  \bibfield  {author} {\bibinfo {author} {\bibfnamefont {T.}~\bibnamefont {Kaneko}}\ and\ \bibinfo {author} {\bibfnamefont {Y.}~\bibnamefont {Ohta}},\ }\bibfield  {title} {\bibinfo {title} {A {{New Era}} of {{Excitonic Insulators}}},\ }\href {https://doi.org/10.7566/JPSJ.94.012001} {\bibfield  {journal} {\bibinfo  {journal} {Journal of the Physical Society of Japan}\ }\textbf {\bibinfo {volume} {94}},\ \bibinfo {pages} {012001} (\bibinfo {year} {2025})}\BibitemShut {NoStop}%
\bibitem [{\citenamefont {Xue}\ \emph {et~al.}(2020)\citenamefont {Xue}, \citenamefont {Wu},\ and\ \citenamefont {MacDonald}}]{xueHiggslikeModesTwodimensional2020}%
  \BibitemOpen
  \bibfield  {author} {\bibinfo {author} {\bibfnamefont {F.}~\bibnamefont {Xue}}, \bibinfo {author} {\bibfnamefont {F.}~\bibnamefont {Wu}},\ and\ \bibinfo {author} {\bibfnamefont {A.~H.}\ \bibnamefont {MacDonald}},\ }\bibfield  {title} {\bibinfo {title} {Higgs-like modes in two-dimensional spatially indirect exciton condensates},\ }\href {https://doi.org/10.1103/PhysRevB.102.075136} {\bibfield  {journal} {\bibinfo  {journal} {Physical Review B}\ }\textbf {\bibinfo {volume} {102}},\ \bibinfo {pages} {075136} (\bibinfo {year} {2020})}\BibitemShut {NoStop}%
\bibitem [{\citenamefont {Sun}\ and\ \citenamefont {Millis}(2020)}]{sunBardasisSchriefferPolaritonsExcitonic2020}%
  \BibitemOpen
  \bibfield  {author} {\bibinfo {author} {\bibfnamefont {Z.}~\bibnamefont {Sun}}\ and\ \bibinfo {author} {\bibfnamefont {A.~J.}\ \bibnamefont {Millis}},\ }\bibfield  {title} {\bibinfo {title} {Bardasis-{{Schrieffer}} polaritons in excitonic insulators},\ }\href {https://doi.org/10.1103/PhysRevB.102.041110} {\bibfield  {journal} {\bibinfo  {journal} {Physical Review B}\ }\textbf {\bibinfo {volume} {102}},\ \bibinfo {pages} {041110} (\bibinfo {year} {2020})}\BibitemShut {NoStop}%
\bibitem [{\citenamefont {Shao}\ and\ \citenamefont {Dai}(2024{\natexlab{a}})}]{shaoElectricalBreakdownExcitonic2024}%
  \BibitemOpen
  \bibfield  {author} {\bibinfo {author} {\bibfnamefont {Y.}~\bibnamefont {Shao}}\ and\ \bibinfo {author} {\bibfnamefont {X.}~\bibnamefont {Dai}},\ }\bibfield  {title} {\bibinfo {title} {Electrical {{Breakdown}} of {{Excitonic Insulators}}},\ }\href {https://doi.org/10.1103/PhysRevX.14.021047} {\bibfield  {journal} {\bibinfo  {journal} {Physical Review X}\ }\textbf {\bibinfo {volume} {14}},\ \bibinfo {pages} {021047} (\bibinfo {year} {2024}{\natexlab{a}})}\BibitemShut {NoStop}%
\bibitem [{\citenamefont {Das~Sarma}\ and\ \citenamefont {Hanke}(1983)}]{dassarmaCommentsTimedependentHartreeFock1983}%
  \BibitemOpen
  \bibfield  {author} {\bibinfo {author} {\bibfnamefont {S.}~\bibnamefont {Das~Sarma}}\ and\ \bibinfo {author} {\bibfnamefont {W.}~\bibnamefont {Hanke}},\ }\bibfield  {title} {\bibinfo {title} {Comments on "{{Time-dependent Hartree-Fock}} formalism for the dielectric function"},\ }\href {https://doi.org/10.1103/PhysRevB.28.1134} {\bibfield  {journal} {\bibinfo  {journal} {Physical Review B}\ }\textbf {\bibinfo {volume} {28}},\ \bibinfo {pages} {1134} (\bibinfo {year} {1983})}\BibitemShut {NoStop}%
\bibitem [{\citenamefont {Bardasis}\ and\ \citenamefont {Schrieffer}(1961)}]{bardasisExcitonsPlasmonsSuperconductors1961}%
  \BibitemOpen
  \bibfield  {author} {\bibinfo {author} {\bibfnamefont {A.}~\bibnamefont {Bardasis}}\ and\ \bibinfo {author} {\bibfnamefont {J.~R.}\ \bibnamefont {Schrieffer}},\ }\bibfield  {title} {\bibinfo {title} {Excitons and {{Plasmons}} in {{Superconductors}}},\ }\href {https://doi.org/10.1103/PhysRev.121.1050} {\bibfield  {journal} {\bibinfo  {journal} {Physical Review}\ }\textbf {\bibinfo {volume} {121}},\ \bibinfo {pages} {1050} (\bibinfo {year} {1961})}\BibitemShut {NoStop}%
\bibitem [{\citenamefont {Fan}\ \emph {et~al.}(2024)\citenamefont {Fan}, \citenamefont {Xiao},\ and\ \citenamefont {Yao}}]{fanIntrinsicDipoleHall2024}%
  \BibitemOpen
  \bibfield  {author} {\bibinfo {author} {\bibfnamefont {F.-R.}\ \bibnamefont {Fan}}, \bibinfo {author} {\bibfnamefont {C.}~\bibnamefont {Xiao}},\ and\ \bibinfo {author} {\bibfnamefont {W.}~\bibnamefont {Yao}},\ }\bibfield  {title} {\bibinfo {title} {Intrinsic dipole {{Hall}} effect in twisted {{MoTe2}}: Magnetoelectricity and contact-free signatures of topological transitions},\ }\href {https://doi.org/10.1038/s41467-024-52314-5} {\bibfield  {journal} {\bibinfo  {journal} {Nature Communications}\ }\textbf {\bibinfo {volume} {15}},\ \bibinfo {pages} {7997} (\bibinfo {year} {2024})}\BibitemShut {NoStop}%
\bibitem [{SM()}]{SM}%
  \BibitemOpen
  \href@noop {} {}\bibinfo {note} {See Supplemental Material at URL-will-be-inserted-by-publisher for technical details of the TDHF formalism, electrostatics, response-function calculations, and additional numerical results, which includes Refs.~\cite{depaloExcitonicCondensationSymmetric2002,joglekarWignerSupersolidExcitons2006,schleedePhaseDiagramBilayer2012}.}\BibitemShut {Stop}%
\bibitem [{\citenamefont {De~Palo}\ \emph {et~al.}(2002)\citenamefont
  {De~Palo}, \citenamefont {Rapisarda},\ and\ \citenamefont
  {Senatore}}]{depaloExcitonicCondensationSymmetric2002}%
  \BibitemOpen
  \bibfield  {author} {\bibinfo {author} {\bibfnamefont {S.}~\bibnamefont
  {De~Palo}}, \bibinfo {author} {\bibfnamefont {F.}~\bibnamefont {Rapisarda}},\
  and\ \bibinfo {author} {\bibfnamefont {G.}~\bibnamefont {Senatore}},\
  }\bibfield  {title} {\bibinfo {title} {Excitonic condensation in a symmetric
  electron--hole bilayer},\ }\href
  {https://doi.org/10.1103/PhysRevLett.88.206401} {\bibfield  {journal}
  {\bibinfo  {journal} {Physical Review Letters}\ }\textbf {\bibinfo {volume}
  {88}},\ \bibinfo {pages} {206401} (\bibinfo {year} {2002})}\BibitemShut
  {NoStop}%
\bibitem [{\citenamefont {Joglekar}\ \emph {et~al.}(2006)\citenamefont
  {Joglekar}, \citenamefont {Balatsky},\ and\ \citenamefont
  {Das~Sarma}}]{joglekarWignerSupersolidExcitons2006}%
  \BibitemOpen
  \bibfield  {author} {\bibinfo {author} {\bibfnamefont {Y.~N.}\ \bibnamefont
  {Joglekar}}, \bibinfo {author} {\bibfnamefont {A.~V.}\ \bibnamefont
  {Balatsky}},\ and\ \bibinfo {author} {\bibfnamefont {S.}~\bibnamefont
  {Das~Sarma}},\ }\bibfield  {title} {\bibinfo {title} {Wigner supersolid of
  excitons in electron--hole bilayers},\ }\href
  {https://doi.org/10.1103/PhysRevB.74.233302} {\bibfield  {journal} {\bibinfo
  {journal} {Physical Review B}\ }\textbf {\bibinfo {volume} {74}},\ \bibinfo
  {pages} {233302} (\bibinfo {year} {2006})}\BibitemShut {NoStop}%
\bibitem [{\citenamefont {Schleede}\ \emph {et~al.}(2012)\citenamefont
  {Schleede}, \citenamefont {Filinov}, \citenamefont {Bonitz},\ and\
  \citenamefont {Fehske}}]{schleedePhaseDiagramBilayer2012}%
  \BibitemOpen
  \bibfield  {author} {\bibinfo {author} {\bibfnamefont {J.}~\bibnamefont
  {Schleede}}, \bibinfo {author} {\bibfnamefont {A.}~\bibnamefont {Filinov}},
  \bibinfo {author} {\bibfnamefont {M.}~\bibnamefont {Bonitz}},\ and\ \bibinfo
  {author} {\bibfnamefont {H.}~\bibnamefont {Fehske}},\ }\bibfield  {title}
  {\bibinfo {title} {Phase diagram of bilayer electron--hole plasmas},\ }\href
  {https://doi.org/10.1002/ctpp.201200045} {\bibfield  {journal} {\bibinfo
  {journal} {Contributions to Plasma Physics}\ }\textbf {\bibinfo {volume}
  {52}},\ \bibinfo {pages} {819} (\bibinfo {year} {2012})}\BibitemShut
  {NoStop}%
\bibitem [{\citenamefont {Sun}\ \emph {et~al.}(2021)\citenamefont {Sun}, \citenamefont {Kaneko}, \citenamefont {Gole{\v z}},\ and\ \citenamefont {Millis}}]{sunSecondOrderJosephsonEffect2021}%
  \BibitemOpen
  \bibfield  {author} {\bibinfo {author} {\bibfnamefont {Z.}~\bibnamefont {Sun}}, \bibinfo {author} {\bibfnamefont {T.}~\bibnamefont {Kaneko}}, \bibinfo {author} {\bibfnamefont {D.}~\bibnamefont {Gole{\v z}}},\ and\ \bibinfo {author} {\bibfnamefont {A.~J.}\ \bibnamefont {Millis}},\ }\bibfield  {title} {\bibinfo {title} {Second-{{Order Josephson Effect}} in {{Excitonic Insulators}}},\ }\href {https://doi.org/10.1103/PhysRevLett.127.127702} {\bibfield  {journal} {\bibinfo  {journal} {Physical Review Letters}\ }\textbf {\bibinfo {volume} {127}},\ \bibinfo {pages} {127702} (\bibinfo {year} {2021})}\BibitemShut {NoStop}%
\bibitem [{\citenamefont {Sun}\ \emph {et~al.}(2024)\citenamefont {Sun}, \citenamefont {Murakami}, \citenamefont {Xuan}, \citenamefont {Kaneko}, \citenamefont {Gole{\v z}},\ and\ \citenamefont {Millis}}]{sunDynamicalExcitonCondensates2024}%
  \BibitemOpen
  \bibfield  {author} {\bibinfo {author} {\bibfnamefont {Z.}~\bibnamefont {Sun}}, \bibinfo {author} {\bibfnamefont {Y.}~\bibnamefont {Murakami}}, \bibinfo {author} {\bibfnamefont {F.}~\bibnamefont {Xuan}}, \bibinfo {author} {\bibfnamefont {T.}~\bibnamefont {Kaneko}}, \bibinfo {author} {\bibfnamefont {D.}~\bibnamefont {Gole{\v z}}},\ and\ \bibinfo {author} {\bibfnamefont {A.~J.}\ \bibnamefont {Millis}},\ }\bibfield  {title} {\bibinfo {title} {Dynamical {{Exciton Condensates}} in {{Biased Electron-Hole Bilayers}}},\ }\href {https://doi.org/10.1103/PhysRevLett.133.217002} {\bibfield  {journal} {\bibinfo  {journal} {Physical Review Letters}\ }\textbf {\bibinfo {volume} {133}},\ \bibinfo {pages} {217002} (\bibinfo {year} {2024})}\BibitemShut {NoStop}%
\bibitem [{\citenamefont {Zeng}\ \emph {et~al.}(2024)\citenamefont {Zeng}, \citenamefont {Cr{\'e}pel},\ and\ \citenamefont {Millis}}]{zengKeldyshFieldTheory2024}%
  \BibitemOpen
  \bibfield  {author} {\bibinfo {author} {\bibfnamefont {Y.}~\bibnamefont {Zeng}}, \bibinfo {author} {\bibfnamefont {V.}~\bibnamefont {Cr{\'e}pel}},\ and\ \bibinfo {author} {\bibfnamefont {A.~J.}\ \bibnamefont {Millis}},\ }\bibfield  {title} {\bibinfo {title} {Keldysh {{Field Theory}} of {{Dynamical Exciton Condensation Transitions}} in {{Nonequilibrium Electron-Hole Bilayers}}},\ }\href {https://doi.org/10.1103/PhysRevLett.132.266001} {\bibfield  {journal} {\bibinfo  {journal} {Physical Review Letters}\ }\textbf {\bibinfo {volume} {132}},\ \bibinfo {pages} {266001} (\bibinfo {year} {2024})}\BibitemShut {NoStop}%
\bibitem [{\citenamefont {Korm{\'a}nyos}\ \emph {et~al.}(2015)\citenamefont {Korm{\'a}nyos}, \citenamefont {Burkard}, \citenamefont {Gmitra}, \citenamefont {Fabian}, \citenamefont {Z{\'o}lyomi}, \citenamefont {Drummond},\ and\ \citenamefont {Fal'ko}}]{kormanyosKpTheoryTwodimensional2015}%
  \BibitemOpen
  \bibfield  {author} {\bibinfo {author} {\bibfnamefont {A.}~\bibnamefont {Korm{\'a}nyos}}, \bibinfo {author} {\bibfnamefont {G.}~\bibnamefont {Burkard}}, \bibinfo {author} {\bibfnamefont {M.}~\bibnamefont {Gmitra}}, \bibinfo {author} {\bibfnamefont {J.}~\bibnamefont {Fabian}}, \bibinfo {author} {\bibfnamefont {V.}~\bibnamefont {Z{\'o}lyomi}}, \bibinfo {author} {\bibfnamefont {N.~D.}\ \bibnamefont {Drummond}},\ and\ \bibinfo {author} {\bibfnamefont {V.}~\bibnamefont {Fal'ko}},\ }\bibfield  {title} {\bibinfo {title} {K{$\cdot$}p theory for two-dimensional transition metal dichalcogenide semiconductors},\ }\href {https://doi.org/10.1088/2053-1583/2/2/022001} {\bibfield  {journal} {\bibinfo  {journal} {2D Materials}\ }\textbf {\bibinfo {volume} {2}},\ \bibinfo {pages} {022001} (\bibinfo {year} {2015})}\BibitemShut {NoStop}%
\bibitem [{\citenamefont {Cai}\ \emph {et~al.}(2007)\citenamefont {Cai}, \citenamefont {Zhang}, \citenamefont {Zeng}, \citenamefont {Cheng},\ and\ \citenamefont {Xu}}]{caiInfraredReflectanceSpectrum2007}%
  \BibitemOpen
  \bibfield  {author} {\bibinfo {author} {\bibfnamefont {Y.}~\bibnamefont {Cai}}, \bibinfo {author} {\bibfnamefont {L.}~\bibnamefont {Zhang}}, \bibinfo {author} {\bibfnamefont {Q.}~\bibnamefont {Zeng}}, \bibinfo {author} {\bibfnamefont {L.}~\bibnamefont {Cheng}},\ and\ \bibinfo {author} {\bibfnamefont {Y.}~\bibnamefont {Xu}},\ }\bibfield  {title} {\bibinfo {title} {Infrared reflectance spectrum of {{BN}} calculated from first principles},\ }\href {https://doi.org/10.1016/j.ssc.2006.10.040} {\bibfield  {journal} {\bibinfo  {journal} {Solid State Communications}\ }\textbf {\bibinfo {volume} {141}},\ \bibinfo {pages} {262} (\bibinfo {year} {2007})}\BibitemShut {NoStop}%
\bibitem [{\citenamefont {Sohier}\ \emph {et~al.}(2017)\citenamefont {Sohier}, \citenamefont {Gibertini}, \citenamefont {Calandra}, \citenamefont {Mauri},\ and\ \citenamefont {Marzari}}]{sohierBreakdownOpticalPhonons2017}%
  \BibitemOpen
  \bibfield  {author} {\bibinfo {author} {\bibfnamefont {T.}~\bibnamefont {Sohier}}, \bibinfo {author} {\bibfnamefont {M.}~\bibnamefont {Gibertini}}, \bibinfo {author} {\bibfnamefont {M.}~\bibnamefont {Calandra}}, \bibinfo {author} {\bibfnamefont {F.}~\bibnamefont {Mauri}},\ and\ \bibinfo {author} {\bibfnamefont {N.}~\bibnamefont {Marzari}},\ }\href {https://doi.org/10.1021/acs.nanolett.7b01090} {\bibinfo {title} {Breakdown of {{Optical Phonons}}' {{Splitting}} in {{Two-Dimensional Materials}}}},\ \bibinfo {howpublished} {https://pubs.acs.org/doi/abs/10.1021/acs.nanolett.7b01090} (\bibinfo {year} {2017})\BibitemShut {NoStop}%
\bibitem [{\citenamefont {Rivera}\ \emph {et~al.}(2019)\citenamefont {Rivera}, \citenamefont {Christensen},\ and\ \citenamefont {Narang}}]{riveraPhononPolaritonicsTwoDimensional2019}%
  \BibitemOpen
  \bibfield  {author} {\bibinfo {author} {\bibfnamefont {N.}~\bibnamefont {Rivera}}, \bibinfo {author} {\bibfnamefont {T.}~\bibnamefont {Christensen}},\ and\ \bibinfo {author} {\bibfnamefont {P.}~\bibnamefont {Narang}},\ }\bibfield  {title} {\bibinfo {title} {Phonon {{Polaritonics}} in {{Two-Dimensional Materials}}},\ }\bibfield  {journal} {\bibinfo  {journal} {Nano Letters}\ }\href {https://doi.org/10.1021/acs.nanolett.9b00518} {10.1021/acs.nanolett.9b00518} (\bibinfo {year} {2019})\BibitemShut {NoStop}%
\bibitem [{\citenamefont {Shi}\ \emph {et~al.}(2025)\citenamefont {Shi}, \citenamefont {Li}, \citenamefont {Pan},\ and\ \citenamefont {Dai}}]{shiTwodimensionalMoirePhonon2025}%
  \BibitemOpen
  \bibfield  {author} {\bibinfo {author} {\bibfnamefont {H.}~\bibnamefont {Shi}}, \bibinfo {author} {\bibfnamefont {C.}~\bibnamefont {Li}}, \bibinfo {author} {\bibfnamefont {D.}~\bibnamefont {Pan}},\ and\ \bibinfo {author} {\bibfnamefont {X.}~\bibnamefont {Dai}},\ }\bibfield  {title} {\bibinfo {title} {Two-{{Dimensional Moir\'e Phonon Polaritons}}},\ }\href {https://doi.org/10.1021/acs.nanolett.5c03046} {\bibfield  {journal} {\bibinfo  {journal} {Nano Letters}\ }\textbf {\bibinfo {volume} {25}},\ \bibinfo {pages} {15460} (\bibinfo {year} {2025})}\BibitemShut {NoStop}%
\bibitem [{\citenamefont {Li}\ \emph {et~al.}(2024)\citenamefont {Li}, \citenamefont {Wang}, \citenamefont {Wang}, \citenamefont {Tao}, \citenamefont {Zhong}, \citenamefont {Su}, \citenamefont {Xue}, \citenamefont {Miao}, \citenamefont {Wang}, \citenamefont {Peng}, \citenamefont {Guo},\ and\ \citenamefont {Zhu}}]{liObservationNonanalyticBehavior2024}%
  \BibitemOpen
  \bibfield  {author} {\bibinfo {author} {\bibfnamefont {J.}~\bibnamefont {Li}}, \bibinfo {author} {\bibfnamefont {L.}~\bibnamefont {Wang}}, \bibinfo {author} {\bibfnamefont {Y.}~\bibnamefont {Wang}}, \bibinfo {author} {\bibfnamefont {Z.}~\bibnamefont {Tao}}, \bibinfo {author} {\bibfnamefont {W.}~\bibnamefont {Zhong}}, \bibinfo {author} {\bibfnamefont {Z.}~\bibnamefont {Su}}, \bibinfo {author} {\bibfnamefont {S.}~\bibnamefont {Xue}}, \bibinfo {author} {\bibfnamefont {G.}~\bibnamefont {Miao}}, \bibinfo {author} {\bibfnamefont {W.}~\bibnamefont {Wang}}, \bibinfo {author} {\bibfnamefont {H.}~\bibnamefont {Peng}}, \bibinfo {author} {\bibfnamefont {J.}~\bibnamefont {Guo}},\ and\ \bibinfo {author} {\bibfnamefont {X.}~\bibnamefont {Zhu}},\ }\bibfield  {title} {\bibinfo {title} {Observation of the nonanalytic behavior of optical phonons in monolayer hexagonal boron nitride},\ }\href {https://doi.org/10.1038/s41467-024-46229-4} {\bibfield  {journal} {\bibinfo  {journal} {Nature Communications}\ }\textbf {\bibinfo {volume} {15}},\ \bibinfo {pages} {1938} (\bibinfo {year} {2024})}\BibitemShut {NoStop}%
\bibitem [{\citenamefont {Landau}\ and\ \citenamefont {Lifshitz}(1980)}]{landauStatisticalPhysicsThird1980}%
  \BibitemOpen
  \bibfield  {author} {\bibinfo {author} {\bibfnamefont {L.~D.}\ \bibnamefont {Landau}}\ and\ \bibinfo {author} {\bibfnamefont {E.~M.}\ \bibnamefont {Lifshitz}},\ }\href@noop {} {\emph {\bibinfo {title} {Statistical {{Physics}}, {{Third Edition}}, {{Part}} 1: {{Volume}} 5}}},\ \bibinfo {edition} {3rd}\ ed.\ (\bibinfo  {publisher} {Butterworth-Heinemann},\ \bibinfo {address} {Amsterdam Heidelberg},\ \bibinfo {year} {1980})\BibitemShut {NoStop}%
\bibitem [{\citenamefont {Pearl}(1964)}]{pearlCURRENTDISTRIBUTIONSUPERCONDUCTING1964}%
  \BibitemOpen
  \bibfield  {author} {\bibinfo {author} {\bibfnamefont {J.}~\bibnamefont {Pearl}},\ }\bibfield  {title} {\bibinfo {title} {{{CURRENT DISTRIBUTION IN SUPERCONDUCTING FILMS CARRYING QUANTIZED FLUXOIDS}}},\ }\href {https://doi.org/10.1063/1.1754056} {\bibfield  {journal} {\bibinfo  {journal} {Applied Physics Letters}\ }\textbf {\bibinfo {volume} {5}},\ \bibinfo {pages} {65} (\bibinfo {year} {1964})}\BibitemShut {NoStop}%
\bibitem [{\citenamefont {Shao}\ and\ \citenamefont {Dai}(2024{\natexlab{b}})}]{shaoQuantumOscillationsExcitonic2024}%
  \BibitemOpen
  \bibfield  {author} {\bibinfo {author} {\bibfnamefont {Y.}~\bibnamefont {Shao}}\ and\ \bibinfo {author} {\bibfnamefont {X.}~\bibnamefont {Dai}},\ }\bibfield  {title} {\bibinfo {title} {Quantum oscillations in an excitonic insulating electron-hole bilayer},\ }\href {https://doi.org/10.1103/PhysRevB.109.155107} {\bibfield  {journal} {\bibinfo  {journal} {Physical Review B}\ }\textbf {\bibinfo {volume} {109}},\ \bibinfo {pages} {155107} (\bibinfo {year} {2024}{\natexlab{b}})}\BibitemShut {NoStop}%
\bibitem [{\citenamefont {Zou}\ \emph {et~al.}(2024)\citenamefont {Zou}, \citenamefont {Zeng}, \citenamefont {MacDonald},\ and\ \citenamefont {Strashko}}]{zouElectricalControlTwodimensional2024}%
  \BibitemOpen
  \bibfield  {author} {\bibinfo {author} {\bibfnamefont {B.}~\bibnamefont {Zou}}, \bibinfo {author} {\bibfnamefont {Y.}~\bibnamefont {Zeng}}, \bibinfo {author} {\bibfnamefont {A.~H.}\ \bibnamefont {MacDonald}},\ and\ \bibinfo {author} {\bibfnamefont {A.}~\bibnamefont {Strashko}},\ }\bibfield  {title} {\bibinfo {title} {Electrical control of two-dimensional electron-hole fluids in the quantum {{Hall}} regime},\ }\href {https://doi.org/10.1103/PhysRevB.109.085416} {\bibfield  {journal} {\bibinfo  {journal} {Physical Review B}\ }\textbf {\bibinfo {volume} {109}},\ \bibinfo {pages} {085416} (\bibinfo {year} {2024})}\BibitemShut {NoStop}%
\bibitem [{\citenamefont {Nguyen}\ \emph {et~al.}(2026)\citenamefont {Nguyen}, \citenamefont {Chaturvedi}, \citenamefont {Zou}, \citenamefont {Watanabe}, \citenamefont {Taniguchi}, \citenamefont {MacDonald}, \citenamefont {Mak},\ and\ \citenamefont {Shan}}]{nguyenQuantumOscillationsDipolar2026}%
  \BibitemOpen
  \bibfield  {author} {\bibinfo {author} {\bibfnamefont {P.~X.}\ \bibnamefont {Nguyen}}, \bibinfo {author} {\bibfnamefont {R.}~\bibnamefont {Chaturvedi}}, \bibinfo {author} {\bibfnamefont {B.}~\bibnamefont {Zou}}, \bibinfo {author} {\bibfnamefont {K.}~\bibnamefont {Watanabe}}, \bibinfo {author} {\bibfnamefont {T.}~\bibnamefont {Taniguchi}}, \bibinfo {author} {\bibfnamefont {A.~H.}\ \bibnamefont {MacDonald}}, \bibinfo {author} {\bibfnamefont {K.~F.}\ \bibnamefont {Mak}},\ and\ \bibinfo {author} {\bibfnamefont {J.}~\bibnamefont {Shan}},\ }\bibfield  {title} {\bibinfo {title} {Quantum oscillations in a dipolar excitonic insulator},\ }\href {https://doi.org/10.1038/s41563-025-02334-3} {\bibfield  {journal} {\bibinfo  {journal} {Nature Materials}\ }\textbf {\bibinfo {volume} {25}},\ \bibinfo {pages} {42} (\bibinfo {year} {2026})}\BibitemShut {NoStop}%
\bibitem [{\citenamefont {Qi}\ \emph {et~al.}(2026)\citenamefont {Qi}, \citenamefont {Li}, \citenamefont {Zhang}, \citenamefont {Nie}, \citenamefont {Zou}, \citenamefont {Cui}, \citenamefont {Kim}, \citenamefont {Sanborn}, \citenamefont {Chen}, \citenamefont {Xie}, \citenamefont {Taniguchi}, \citenamefont {Watanabe}, \citenamefont {Crommie}, \citenamefont {MacDonald},\ and\ \citenamefont {Wang}}]{qiCompetitionExcitonicInsulators2026}%
  \BibitemOpen
  \bibfield  {author} {\bibinfo {author} {\bibfnamefont {R.}~\bibnamefont {Qi}}, \bibinfo {author} {\bibfnamefont {Q.}~\bibnamefont {Li}}, \bibinfo {author} {\bibfnamefont {Z.}~\bibnamefont {Zhang}}, \bibinfo {author} {\bibfnamefont {J.}~\bibnamefont {Nie}}, \bibinfo {author} {\bibfnamefont {B.}~\bibnamefont {Zou}}, \bibinfo {author} {\bibfnamefont {Z.}~\bibnamefont {Cui}}, \bibinfo {author} {\bibfnamefont {H.}~\bibnamefont {Kim}}, \bibinfo {author} {\bibfnamefont {C.}~\bibnamefont {Sanborn}}, \bibinfo {author} {\bibfnamefont {S.}~\bibnamefont {Chen}}, \bibinfo {author} {\bibfnamefont {J.}~\bibnamefont {Xie}}, \bibinfo {author} {\bibfnamefont {T.}~\bibnamefont {Taniguchi}}, \bibinfo {author} {\bibfnamefont {K.}~\bibnamefont {Watanabe}}, \bibinfo {author} {\bibfnamefont {M.~F.}\ \bibnamefont {Crommie}}, \bibinfo {author} {\bibfnamefont {A.~H.}\ \bibnamefont {MacDonald}},\ and\ \bibinfo {author} {\bibfnamefont {F.}~\bibnamefont {Wang}},\ }\bibfield  {title} {\bibinfo {title} {Competition between excitonic insulators and quantum {{Hall}} states in correlated electron--hole bilayers},\ }\href {https://doi.org/10.1038/s41563-025-02316-5} {\bibfield  {journal} {\bibinfo  {journal} {Nature Materials}\ }\textbf {\bibinfo {volume} {25}},\ \bibinfo {pages} {35} (\bibinfo {year} {2026})}\BibitemShut {NoStop}%
\bibitem [{\citenamefont {Fogler}\ \emph {et~al.}(1996)\citenamefont {Fogler}, \citenamefont {Koulakov},\ and\ \citenamefont {Shklovskii}}]{foglerGroundStateTwodimensional1996}%
  \BibitemOpen
  \bibfield  {author} {\bibinfo {author} {\bibfnamefont {M.~M.}\ \bibnamefont {Fogler}}, \bibinfo {author} {\bibfnamefont {A.~A.}\ \bibnamefont {Koulakov}},\ and\ \bibinfo {author} {\bibfnamefont {B.~I.}\ \bibnamefont {Shklovskii}},\ }\bibfield  {title} {\bibinfo {title} {Ground state of a two-dimensional electron liquid in a weak magnetic field},\ }\href {https://doi.org/10.1103/PhysRevB.54.1853} {\bibfield  {journal} {\bibinfo  {journal} {Physical Review B}\ }\textbf {\bibinfo {volume} {54}},\ \bibinfo {pages} {1853} (\bibinfo {year} {1996})}\BibitemShut {NoStop}%
\bibitem [{\citenamefont {Dupas}\ \emph {et~al.}(2018)\citenamefont {Dupas}, \citenamefont {{Guillemet-Fritsch}}, \citenamefont {Geffroy}, \citenamefont {Chartier}, \citenamefont {Baillergeau}, \citenamefont {Mangeney}, \citenamefont {Roux}, \citenamefont {Ganne}, \citenamefont {Marcellin}, \citenamefont {Degiron},\ and\ \citenamefont {Akmansoy}}]{dupasHighPermittivityProcessed2018}%
  \BibitemOpen
  \bibfield {author} {\bibinfo {author} {\bibfnamefont {C.}~\bibnamefont {Dupas}}, \bibinfo {author} {\bibfnamefont {S.}~\bibnamefont {{Guillemet-Fritsch}}}, \bibinfo {author} {\bibfnamefont {P.-M.}\ \bibnamefont {Geffroy}}, \bibinfo {author} {\bibfnamefont {T.}~\bibnamefont {Chartier}}, \bibinfo {author} {\bibfnamefont {M.}~\bibnamefont {Baillergeau}}, \bibinfo {author} {\bibfnamefont {J.}~\bibnamefont {Mangeney}}, \bibinfo {author} {\bibfnamefont {J.-F.}\ \bibnamefont {Roux}}, \bibinfo {author} {\bibfnamefont {J.-P.}\ \bibnamefont {Ganne}}, \bibinfo {author} {\bibfnamefont {S.}~\bibnamefont {Marcellin}}, \bibinfo {author} {\bibfnamefont {A.}~\bibnamefont {Degiron}},\ and\ \bibinfo {author} {\bibfnamefont {{\'E}.}~\bibnamefont {Akmansoy}},\ }\bibfield {title} {\bibinfo {title} {High permittivity processed {{SrTiO3}} for metamaterials applications at terahertz frequencies},\ }\href {https://doi.org/10.1038/s41598-018-33251-y} {\bibfield {journal} {\bibinfo {journal} {Scientific Reports}\ }\textbf {\bibinfo {volume} {8}},\ \bibinfo {pages} {15275} (\bibinfo {year} {2018})}\BibitemShut {NoStop}%
\end{thebibliography}
\end{document}